\documentclass[A4paper,fleqn,usenatbib]{mnras}

\usepackage[T1]{fontenc}

\DeclareRobustCommand{\VAN}[3]{#2}
\let\VANthebibliography\thebibliography
\def\thebibliography{\DeclareRobustCommand{\VAN}[3]{##3}\VANthebibliography}

\usepackage{graphicx}	% Including figure files
\usepackage{amsmath}	% Advanced maths commands
\usepackage{amssymb}	% Extra maths symbols
\usepackage{comment}    %comment out a section of the writing
\usepackage{rotating}   %rotate a table
\usepackage{hyperref}   %hyperlink
\usepackage{caption}
\usepackage{subcaption}
\usepackage{soul}
\usepackage[normalem]{ulem}
\usepackage{dirtytalk}
\usepackage[dvipsnames]{xcolor}
\usepackage{float}
\title[The GMRT archive atomic gas survey – I]{The GMRT archive atomic gas survey – I. Survey definition, methodology, and initial results from the pilot sample}

\author[Biswas et al.]{
Prerana Biswas$^{1,3}$ \thanks{E-mail: preranab@iisc.ac.in},
Narendra Nath Patra$^{2}$,
Nirupam Roy$^{3}$,
Md. Rashid $^{1,3}$
\\
$^{1}$Joint Astronomy Programme, Indian Institute of Science, Bangalore 560012, India \\
$^{2}$Raman Research Institute,  C. V. Raman Avenue, Bangalore 560 080, India\\
$^{3}$Department of Physics, Indian Institute of Science, Bangalore 560012, India\\
}

\date{Accepted 2020 March 18. Received 2022 March 18; in original form 2021 September 9}

\pubyear{2015}

\begin{document}
\label{firstpage}
\pagerange{\pageref{firstpage}--\pageref{lastpage}}
\maketitle

\begin{abstract}
Interferometric observations of HI in galaxies played a pivotal role in studies of nearby galaxies. Compared to single-dish observation, it provides resolved distribution of gas in galaxies with unprecedented resolution. Several extensive HI surveys of nearby galaxies have been performed in the past; however, most of them consist of less than 100 galaxies due to individual efforts. On the other hand, present-day archives of the radio telescopes include data for at least several hundred galaxies. To utilize these data sets to their full potential, we construct a sample including all galaxies observed by the Giant Meter wave Radio Telescope (GMRT) in HI. This results in a total of 515 galaxies, the largest sample to date. We intend to analyze all the data uniformly and carry out different exciting science. As a pilot project, we analyze data from 11 galaxies and present the data products in this paper. We further investigate the neutral ISM in these galaxies and extract cold and warm phases using a multi-Gaussian decomposition method. This pilot project assures the quality of the data, which will enable us to perform critical science investigations using the full sample. 
\end{abstract}


%%%%%%%%%%%%%%%%%%%%%%%%%%%%%%%%%%%%%%%%%%%%%%%%%%

%%%%%%%%%%%%%%%%% BODY OF PAPER %%%%%%%%%%%%%%%%%%

\section{Introduction}
During the last four decades, several HI surveys have been done aiming to study the atomic content of the galaxies through different radio telescopes of the world. In the earlier times, many large surveys were done with the single dishes  \citep[e.g.][]{Tully1975, Fisher1981, hipass, alpha3, alpha2, alpha1, cosmicflow4}. Single dish observations for its large beam sizes yield the unresolved maps and only the global properties of the galaxies, i.e., HI line flux, peak intensities,  velocity widths ($W_{p20}$ \citep{Tully_1977}, $W_{m50}$ \citep{wm50_def} ), HI mass etc. Only after the interferometers started functioning \citep{interfero1971} in the 1970s was it possible through the spectral line observation to study the resolve HI images and the kinematics of the galaxies.
      
      The HI interferometric observations are of immense importance and useful as they estimate those galaxy properties that can not be explored through single dish observation or observation through other wavelengths. Unlike the optical and UV emission, this line does not suffer extinction due to interstellar dust. The largest extent of the atomic hydrogen in the galaxies makes it usable to probe various galaxies' properties up to very large radii. Interferometric observations are important to explore a broad range of scientific problems. The global HI spectra obtained from the interferometric observations may be more accurate than that obtained from single-dish observations if the larger beam size of the single-dish compared to the smaller synthesized beam size include other sources' fluxes in the background or the foreground and if they lie in the same velocity range.    \citet{BFTR1} pointed out that velocity at the flat part of the rotation curve ($V_{flat}$) gives tighter Baryonic Tully-Fisher Relation (BTFR) \cite[see][]{Tully_1977,  mcgaugh2000, BFTR1} in comparison to global spectral widths. As the rotation curves of the galaxies can be obtained only through the interferometric observation, in the studies \citep[e.g.,][]{cosmicflow4} where the widths of these spectra are used to calculate the distances of the galaxies through BTFR, if the interferometric observations are used instead of the single-dish observations, it is possible to obtain more accurate measurements of the distances. 
      
      Besides that, the synthesized data cubes obtained from the interferometric observations directly provide the spatial distribution of the atomic gas, its velocity field and velocity dispersion \citep{whisp_main, Begum2006, things, figgs1,  figgs2,  wallaby, little_things, mighty_hi}. The spatially resolved maps are convenient to investigate the fine-scale structure of the ISM, which in turn, when correlated with other parameters, gives valuable information about the galaxy structure. \citet{feedback1} detected more than 1000 HI holes in 20 galaxies from the THINGS survey \citep{things}, measured their sizes, expansion velocities and estimated their ages. Upon extrapolating the observed number of holes to include the holes created by a single supernova, \citet{feedback1} showed that they are in good agreement with the hypothesis that the star formation is the cause for the HI holes inside galaxies.
      
      Furthermore, the HI cubes with its 3D kinematic modeling done with the recently developed tools to fit the tilted ring model \citep{tiltd_ring1974} by \citet{fat1} and \citet{BBarolo} gives us the detailed kinematics and dynamics of the galaxies. The rotation curve, when combined with optical or infrared observation, provides the mass model of the galaxies, i.e., the radial distribution of different components (star, gas and the dark matter) of galaxies and the scope to test different dark matter halo profiles \citep[see][]{dark_matter2, mass_model_sushma, mass_model}. 
      The interferometric cubes are also helpful to find the 3D distribution of the atomic gas and its volumetric properties. \citet{naren_hi_height} found the 3D distribution and hence the vertical scale height of the atomic gas by solving the Poisson-Boltzmann equation numerically of seven nearby spiral galaxies. By calculating the radial growth of the thickness of galaxy discs, \citet{volmetric_star_form} found a tight correlation between the volume density of the gas ($HI+H_2$) and Star Formation Rate (SFR). This relation has a significantly smaller scatter than Kennicutt-Schmidt (KS) law \citep{kslaw} which relates SFR with surface densities. \citet{volmetric_star_form} also showed that their correlation is unbroken over five orders of magnitude of SFR and suggests that the break seen in the KS law is due to the disc's flaring rather than the drop of star formation efficiency at low surface densities. 
      
      The specific angular momentum, a crucial property to understand galaxy formation and evolution, can also be derived from interferometric observations. This area has been explored in detail with more and more sources during the last few years but still requires further investigations. \citet{anguler_mont3} and later \citet{bryn_spect_anguler_montm} showed that angular momentum - mass relation for different components show unbroken power-law unlike what was found by \citet{anguler_momnt_sushma} for the low mass galaxies. Their study suggests that galaxies with higher specific angular momentum tend to retain more atomic gas.  Again, \citet{anguler_mont2} and  \citet{anglur_momnt_envrn_effct} showed a tight correlation between atomic gas fraction and global atomic stability parameter $(q=\frac{j\sigma}{GM}$, where j is the specific byronic angular momentum, $\sigma$ is  Warm Neutral Medium (WNM) velocity dispersion of ISM and $M=M_{\star}+ M_{gas}$ ) in different galaxy environments and finds a trend of increasing disc stability and HI fraction with decreasing bulge to total mass ratio. 
      
      Another aspect of this type of observation is that it yields the scope to study the interface of the galaxies with the surrounding medium. From the HI interferometric observation of fifteen galaxies, \citet{halogas} found that the extra-planner HI gas is ubiquitous in nearly every sample; the mass of the extra-planner HI gas matches well with the predictions from the simple model of galactic fountain powered by stellar feedback.  Using the HI observations of 148 nearby spiral galaxies from the WHISP \citep{whisp_main} sample,  \citet{gas_accretion_minor_merger} measured the gas accretion rate of minor mergers onto star-forming galaxies; their results strongly imply that minor mergers do not have a significant role in local galaxies' gas accretion budget. \citet{ram_pressure2} studied the effect of ram pressure from inter-cluster in stripping out the atomic gas from the galaxies.  
      
      All these topics are scientifically vast, and here in this paper, we explore some of them. The HI spectral line observations also allow us to differentiate the two phases of the ISM \citep{Warren_2012, naren2016}. In this paper, we explore this part in detail to pick up the different components of the ISM in the case of a representative sample of galaxies.
      
    The various interferometric HI surveys \citep[e.g.][]{bosoma1981, bosoma1981b, brinks_bajaja1986,  whisp_main, Begum2006, Braun2007, things, figgs1,   figgs2, shild, wsrt_end, lvhis,   wallaby, little_things, mighty_hi} done till now, through different interferometers functioning globally,  comprise a large number of data set of the nearby galaxies.  The surveys have been performed in different set-ups, and thus the sources have been observed in different sensitivity, spectral and spatial resolutions. It has become essential to appraise the existing archival data so that this large sample volume can be used to examine the completion and coverage of sources that have already been observed through different groups and different interferometers at different settings and sensitivity. This study is further helpful for sample selection to explore various science cases for any future observations.

      This paper presents a preliminary study of the extra-galactic systems observed with GMRT in the L band and channel resolution less than $35$ kHz. The Giant Metrewave Radio Telescope (GMRT) \citep{gmrt_1st}, now the uGMRT, is one of the best interferometers among the interferometers functioning globally. Since it started operating, it has been used extensively for observing continuum,  spectral lines, and pulsars. The GMRT has a hybrid antennae configuration with 14 out of 30 antennae located in the central square within the distance $\sim 1.1$ km. The other dishes are roughly distributed in a Y shape with a maximum baseline distance of $25$ km. This makes GMRT very useful in probing highly resolved and also diffuse emissions with adequate sensitivity at the same time. We find out the coverage and the distribution of the GMRT archival HI extra-galactic systems in terms of their position, optical diameter, distance, B band magnitude, inclination angles, position angles, projected rotation velocity, and mass of HI. While detailed analysis and results based on these data are published for some of the targets, there are no reported results for a significant fraction of the sources. Also, for the sources for which the results are published, the analysis is very heterogeneous, the data products are not readily available, and often, for some of the key analysis, better, more accurate and well-established methods compared to the ones used (e.g., rotation curve estimation from the image cube instead of using the 2-D velocity field) are now available. Considering the above points, it is clear that (re-) analyzing these archival data uniformly will potentially result in a very useful large sample of nearby galaxies observed in HI; hence we have started carrying out this exercise of a GMRT ARChIve Atomic gas survey (GARCIA). We plan to make the science-ready data products (integrated HI column density image, rotation curve data and model etc.) available publicly in a phased manner for the entire sample. Although our study is limited to the sample of galaxies observed through GMRT only, the same can be done in the case of the other big interferometers.
      
      As a prefatory work, here we have selected a sub-sample of these galaxies and presented a detailed study of these galaxies (Paper - I ). For this sample of galaxies, we present the comparison of the global HI spectra derived from the resolved HI cubes with the single-dish observation; the comparison of two different velocity widths of the HI spectra: $W_{p20}$ \citep{Tully_1977} and $W_{m50}$ \citep{wm50_def}, which is a representation of the projected rotation velocity of the gas; HI masses of the galaxies; their moment maps and the column density distribution. Further, equipped with the resolved maps, we extended our work to differentiate the two atomic phases of the ISM, which is professed to exist theoretically as explored by \citet{Field2_1965, Field1969, wolfire1995, Wolfire2003} with the method developed by \citet{naren2016}. Additionally, the 3D modeling of these sources with the derived rotation curves along with their mass models and the different aspects of Baryonic Tully-Fisher relation \citep{Tully_1977, BFTR1, mcgaugh2000} will be presented in a companion paper ( Paper-II; Biswas et al. in prep.). 
      
      In section \ref{sample_selection}, we have described the process of sample selection and the sub-sample of galaxies that are selected for our prefatory work. The following section \ref{data_reduction} describes the process of data reduction followed by us, section \ref{data_product} presents the data products deducted in the analysis and finally, in the section \ref{phase_ism_intro}, we have discussed our process to differentiate the phases of the ISM, the efficiency of the process and presented the result of decomposition. Furthermore, in conclusion, section \ref{conclusion}, we summarise the results from our study and explain the importance of our work.

\section{Sample Selection}
\label{sample_selection}
\begin{figure}
    \centering
    \includegraphics[height=6cm]{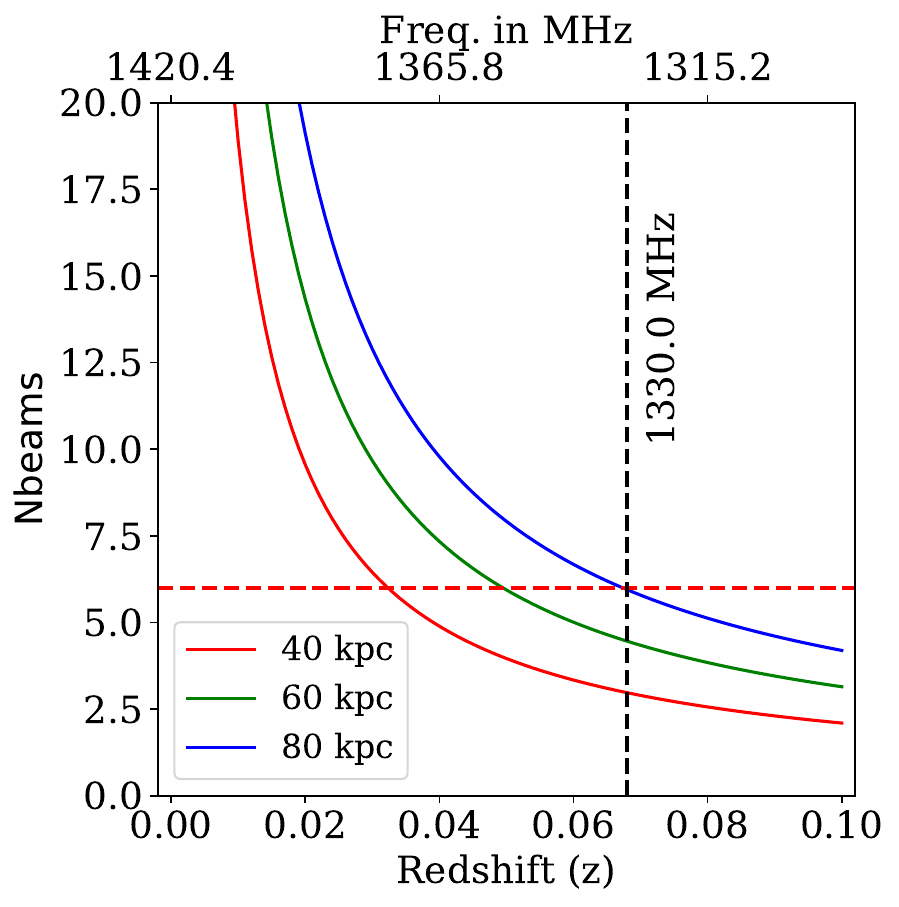}
    \caption{The number of synthesized beams of $10^{\prime\prime}$ required to cover galaxies with diameter $40$ kpc (red line), $60$ kpc (green line) and $80$ kpc (blue line) versus their redshift or observing frequency. The red horizontal dotted line represents six beams and the black vertical dotted line represents the frequency $1330.0$ MHz. }
    \label{fig:nbeams_vs_freq}
\end{figure}

\begin{figure*}
    \centering
    \begin{subfigure}[b]{0.3\textwidth}
        \centering
        \includegraphics[width=\textwidth]{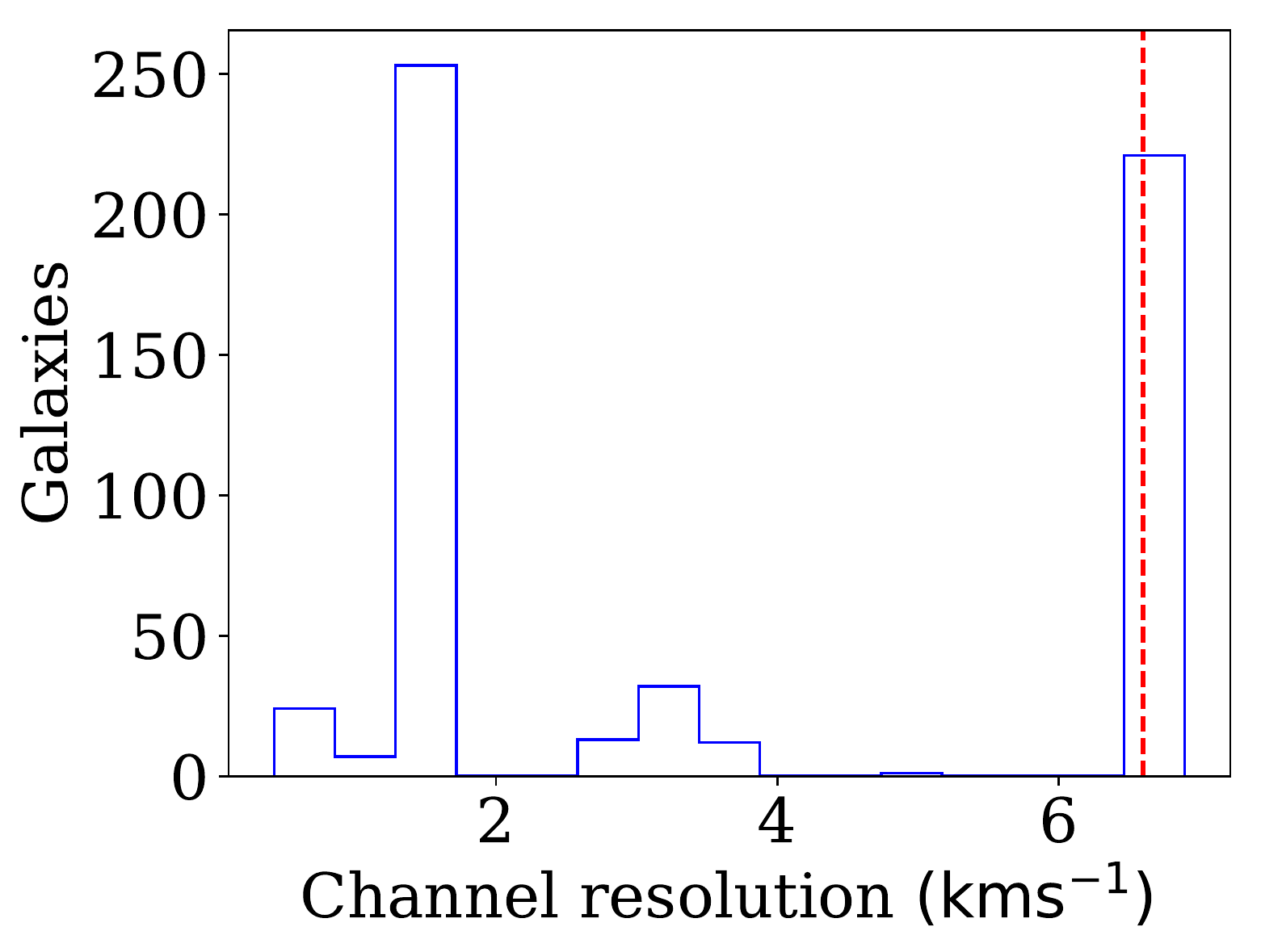}
        \caption{}
        \label{fig:hist_velres}
    \end{subfigure}
    \begin{subfigure}[b]{0.3\textwidth}
       \centering
       \includegraphics[width=\textwidth]{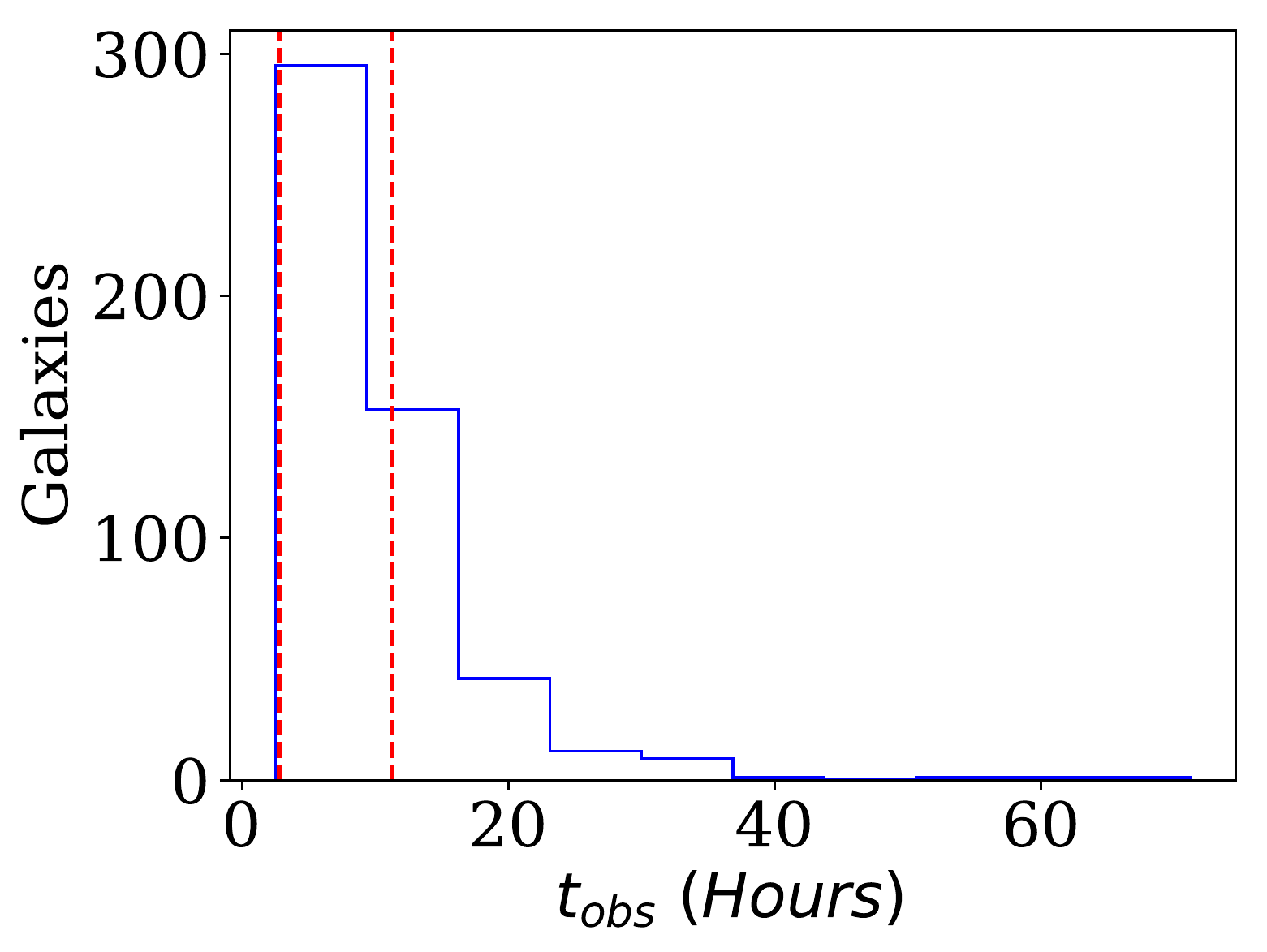}
       \caption{}
       \label{fig:hist_time}
    \end{subfigure}
            \begin{subfigure}[b]{0.3\textwidth}
       \centering
       \includegraphics[width=\textwidth]{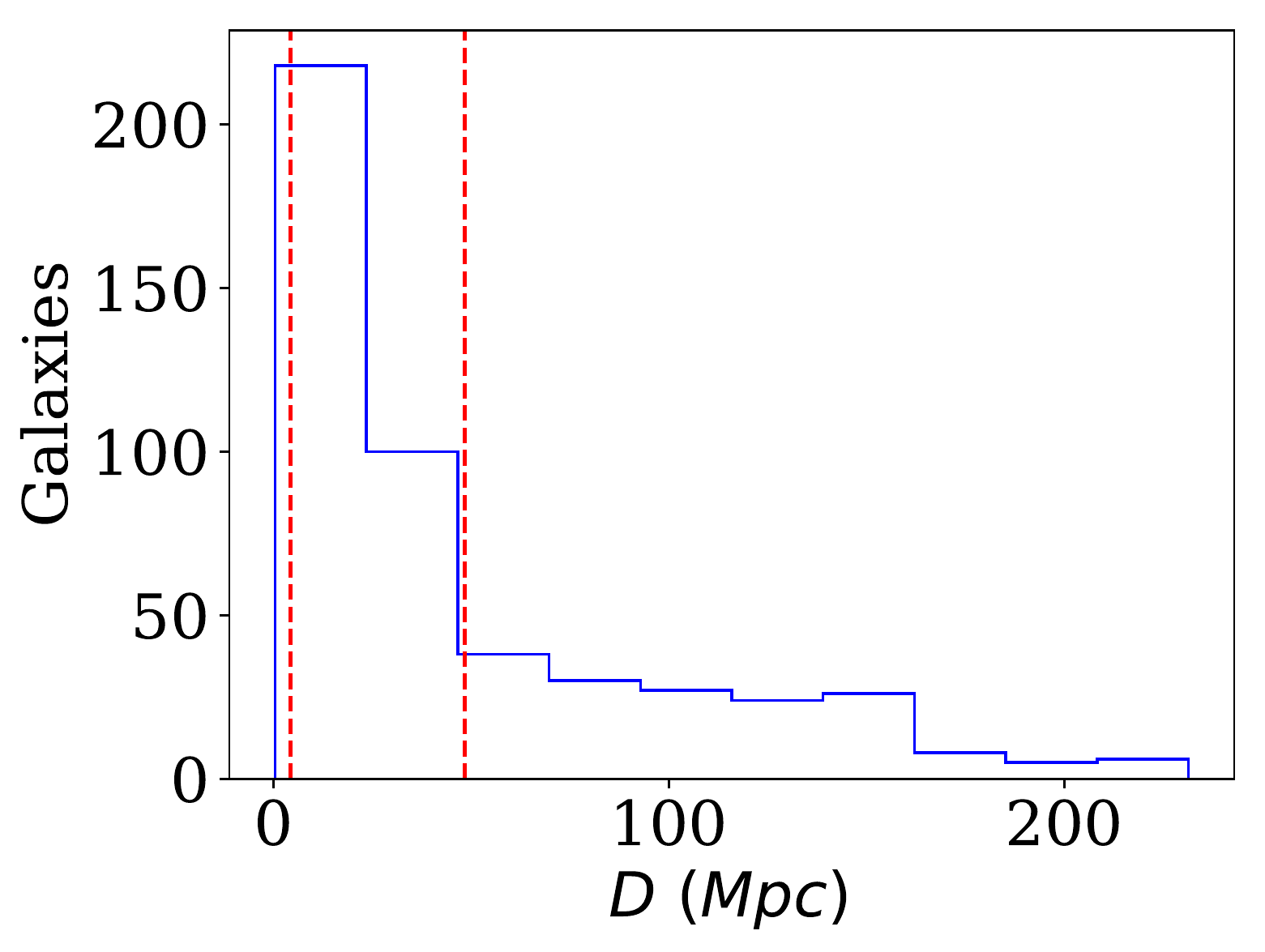}
       \caption{}
       \label{fig:hist_dist}
    \end{subfigure}
    \caption{The distribution of the different observational properties of the sources.  From the left to right, figures respectively show the distributions of the observation channel resolution, on source observation time and the distance of the sources. The red dashed lines represent the range of the respective parameters for the selected sub-sample.}
    \label{fig:scatter_plots_obs}
\end{figure*}

\begin{figure*}
    \centering
    \begin{subfigure}[b]{0.3\textwidth}
        \centering
        \includegraphics[width=\textwidth]{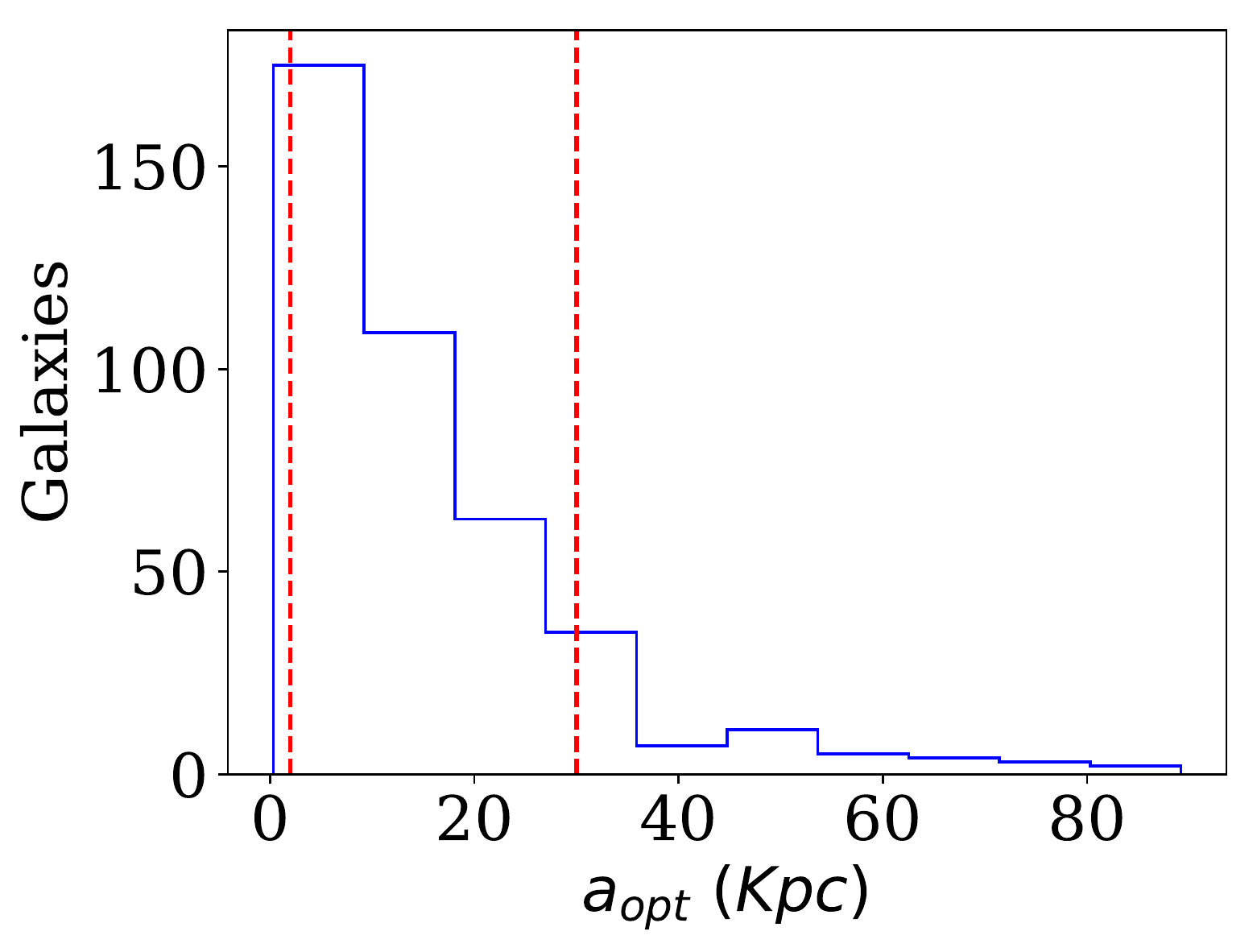}
        \caption{}
        \label{fig:hist_bkpc}
    \end{subfigure}
    \begin{subfigure}[b]{0.3\textwidth}
       \centering
       \includegraphics[width=\textwidth]{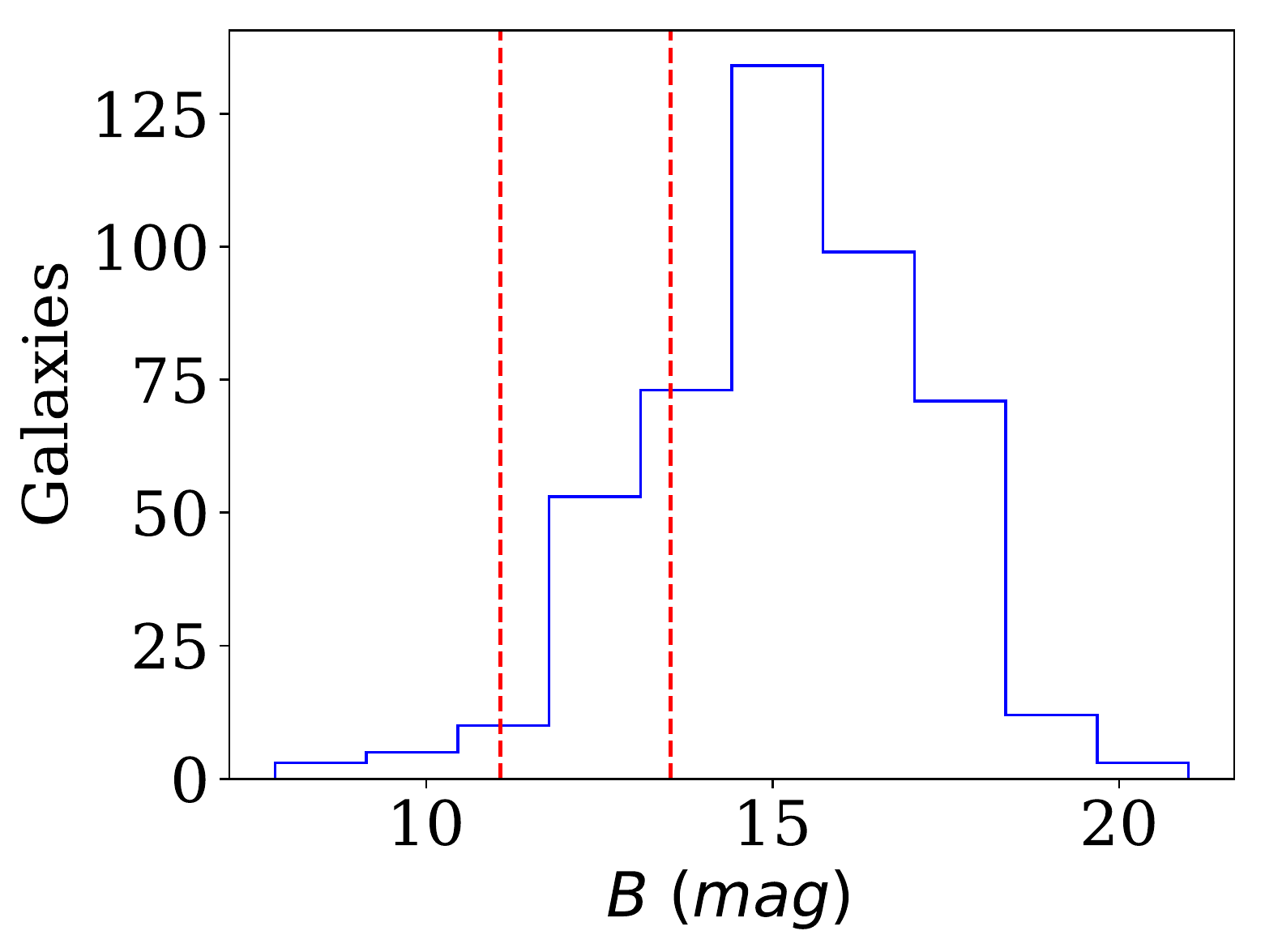}
       \caption{}
       \label{fig:hist_bmag}
    \end{subfigure}
    \begin{subfigure}[b]{0.3\textwidth}
       \centering
       \includegraphics[width=\textwidth]{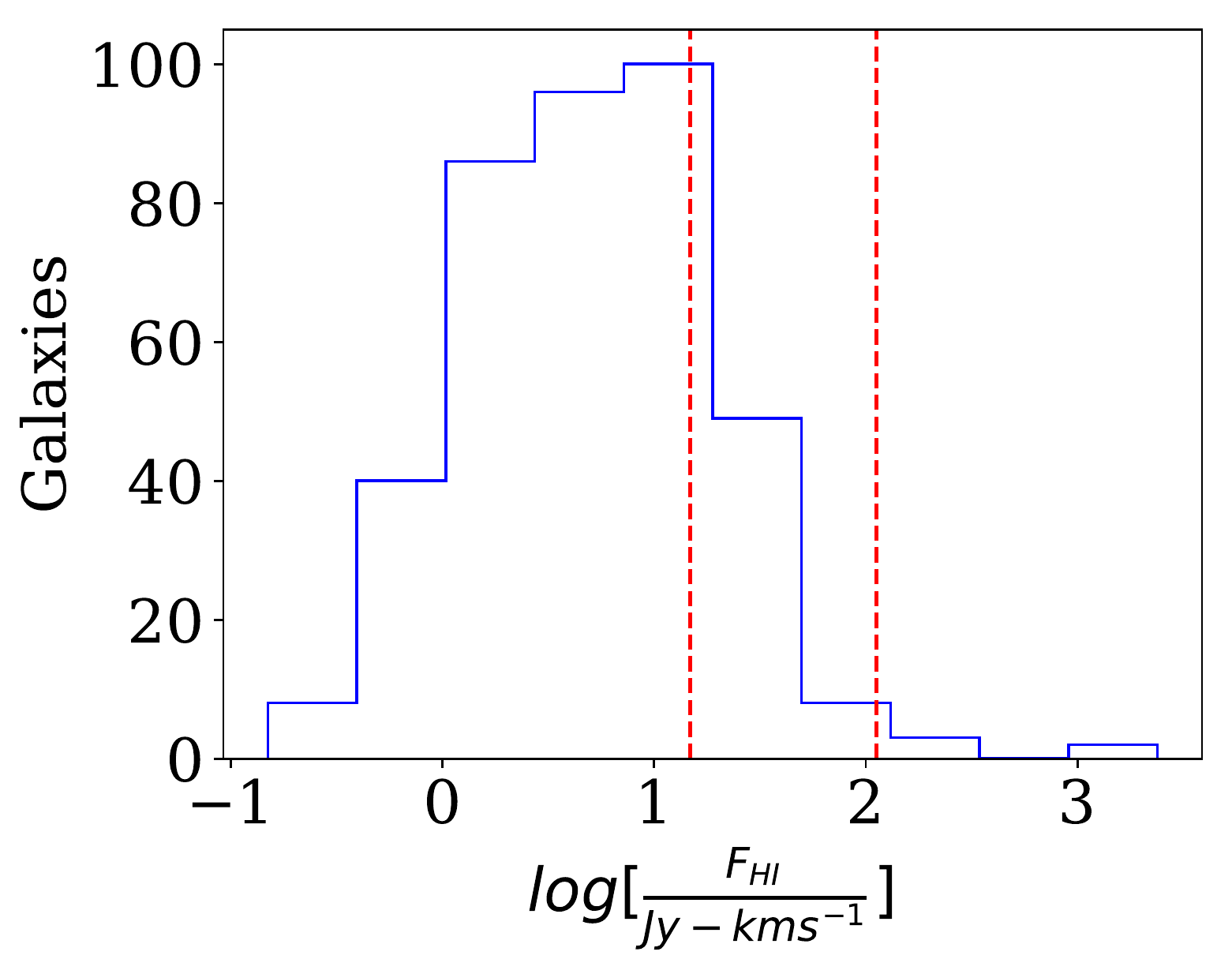}
       \caption{}
       \label{fig:hist_lineflux}
    \end{subfigure}
    \begin{subfigure}[b]{0.3\textwidth}
       \centering
       \includegraphics[width=\textwidth]{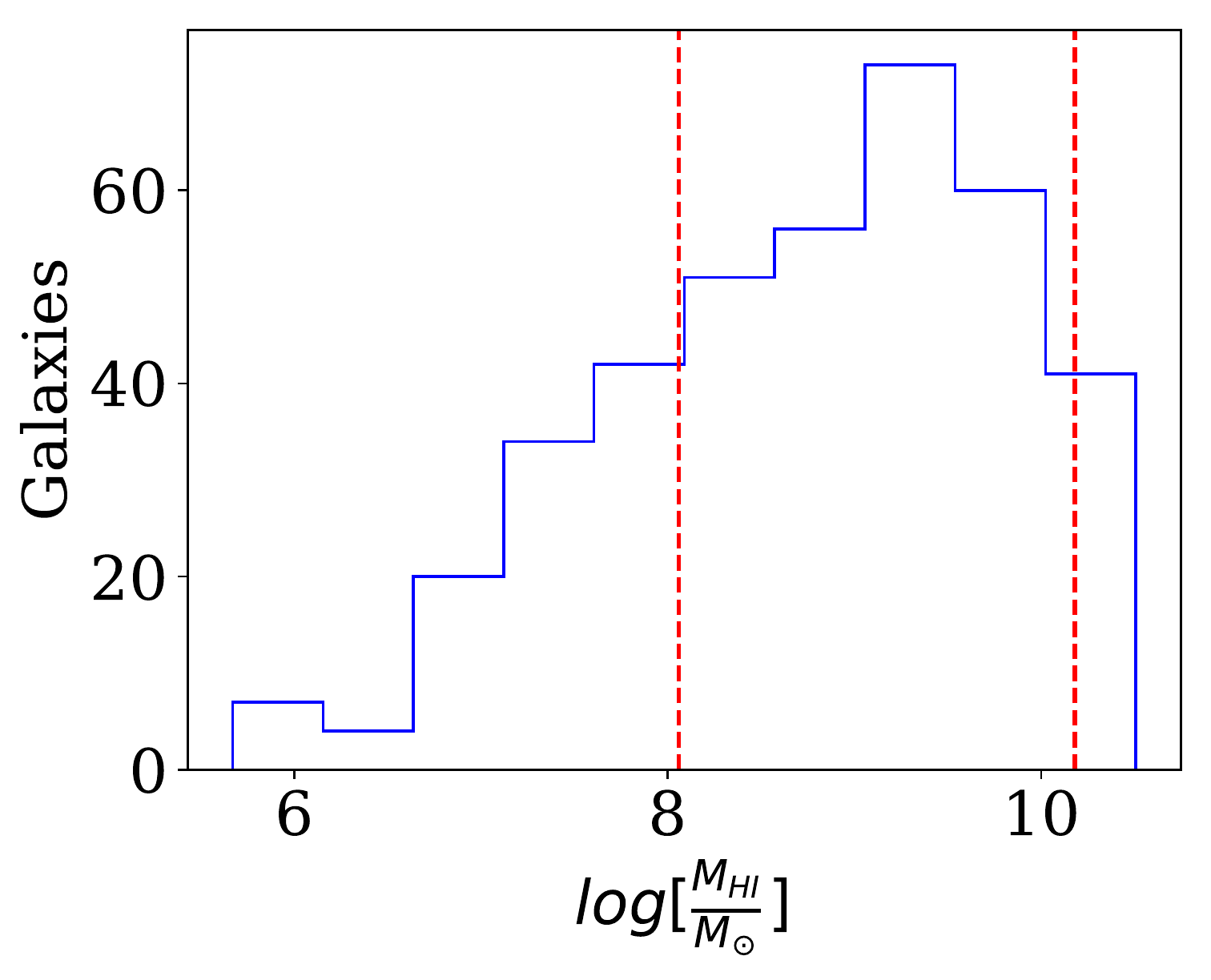}
       \caption{}
       \label{fig:hist_himass}
    \end{subfigure}
    \begin{subfigure}[b]{0.3\textwidth}
        \centering
        \includegraphics[width=\textwidth]{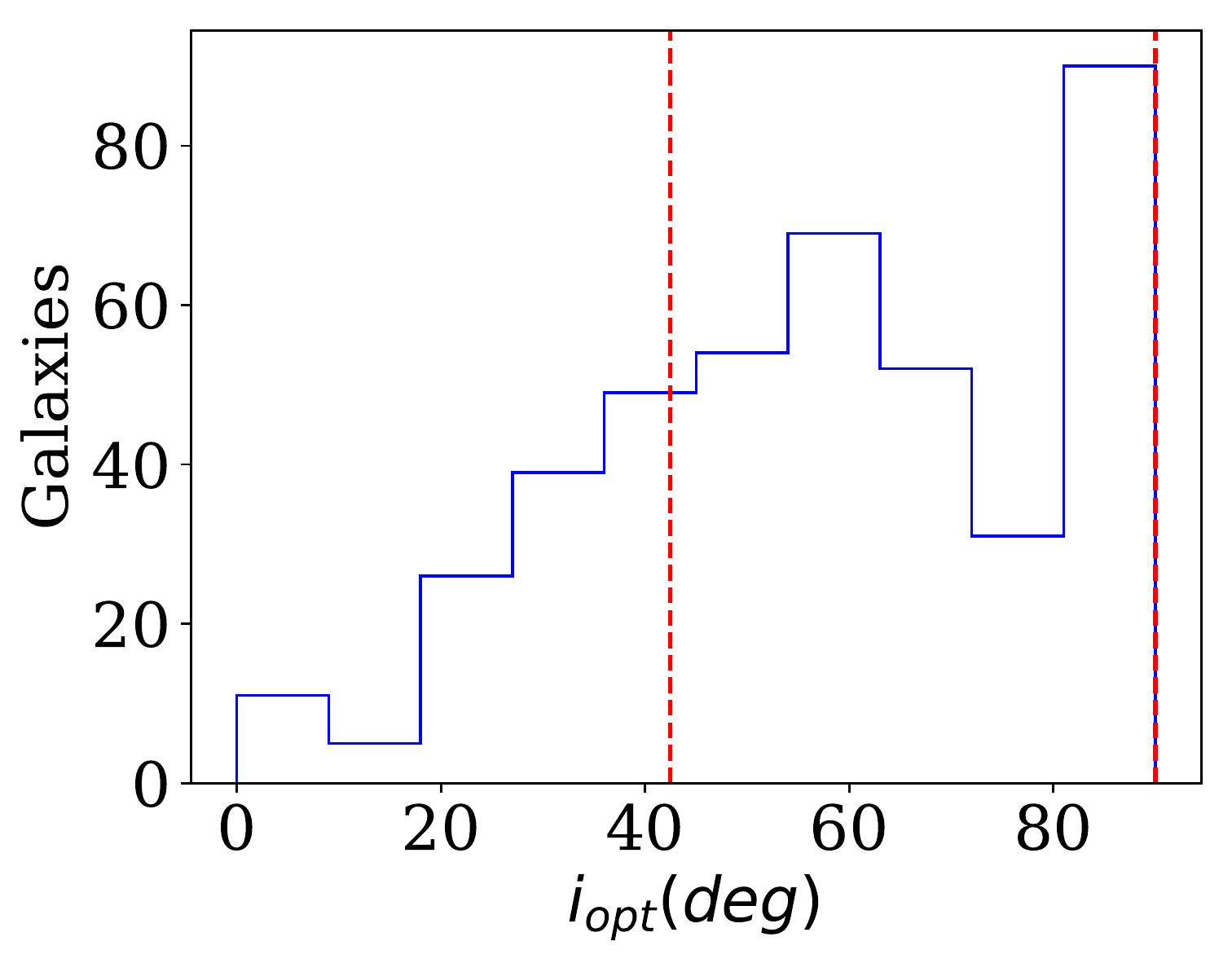}
        \caption{}
        \label{fig:hist_incl}
    \end{subfigure}
    \begin{subfigure}[b]{0.3\textwidth}
       \centering
       \includegraphics[width=\textwidth]{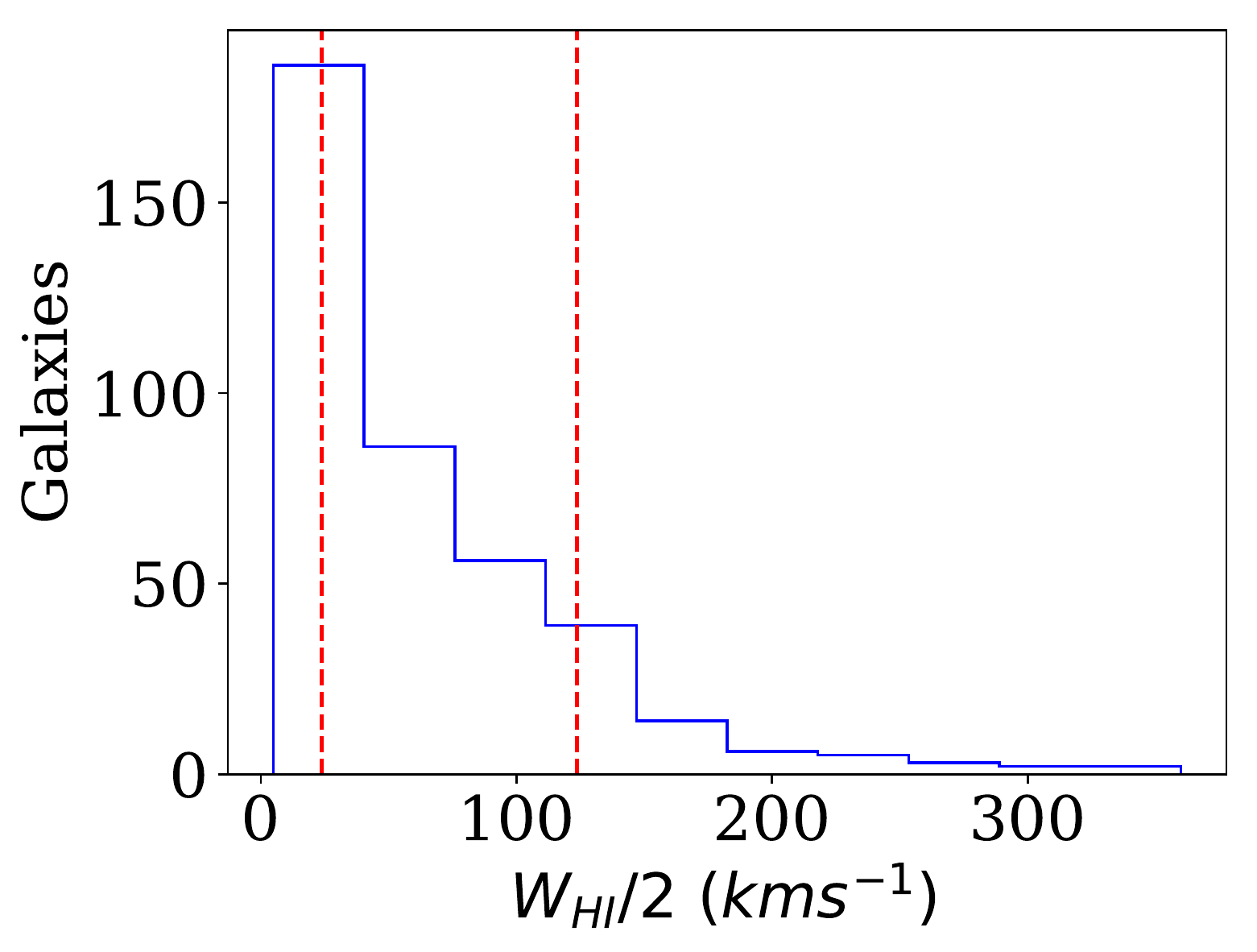}
       \caption{}
       \label{fig:hist_vort}
    \end{subfigure}
     \caption{The distributions of different parameters of the sources. From left-to-right and then top-to-bottom, the x-axis represents respectively the optical diameter ($a_{opt}$) in Kpc,  B-band magnitude ($B$), HI line flux ($log[\frac{F_{HI}}{Jy-Kms^{-1}}]$), HI mass ($log[\frac{M_{HI}}{M_{\sun}}]$),  optical inclination angle ($i_{opt}$) and the width of the HI spectra ($W_{HI}/2$).  The red dashed lines represent the range of the respective parameters for the selected sub-sample.}
    \label{fig:scatter_plots}
\end{figure*}

   This study features the extra-galactic systems observed through GMRT in the L band using spectral line mode. This include external galaxies(G), pair of galaxies(M2), triple galaxies(M3), multiple galaxies(M), group of galaxies(MG) and galaxy clusters(MC). This study reveals the parameter spaces of different properties for these sources and includes the different scaling relations. We intended to list those objects for which the 3D modeling can be done satisfactorily using the recently developed pipelines \citep[e.g.][]{BBarolo, fat2}. The procedure described below has been followed to build up the source catalog from the archival data of GMRT. First, all the sources observed through GMRT in the frequency range between $1000$ MHz and $1500$ MHz are listed. Then, the observations of the same object with the same observation settings with the same central frequency and channel width but observed by different observers or on different days under different observation IDs are merged. From this, objects observed with channel width less than $35$ kHz and on source observation time greater than two and half hour are selected for the main list. The constraint of $35$ kHz in channel width ($\sim 7$ km\thinspace s$^{-1}$ velocity resolution)  is set to ensure that the observation has been done in spectral line mode. The constraint on the channel width along with the constraint on the observation time also ensures that the velocity resolution of the interferometric data and signal-to-noise ratio in each channel will be adequate so that upon analyzing the data, the 3D modeling of the sources can be done adequately. In the next step, the primary calibrators and the secondary calibrators are removed from this list by matching the source name from the VLA-calibrator list. As a next step, we determine the distance up to which the HI emission of galaxies can be observed to take at least three synthesized beams to cover its angular radius. This study is important because for doing the 3D modeling with the best pipelines available \citep{BBarolo, fat2}, it needs to construct at least three independent rings in the data to fit the tilted ring model adequately. Now, taking three types of galaxies with diameter $40$ kpc, $60$ kpc and $80$ kpc, and considering a synthesized beam size of $10^{\prime\prime}$, we check how the number of the beams decrease to cover these three radii of galaxies with the increase of their redshift or distance, see Figure \ref{fig:nbeams_vs_freq}. Based on this study, we impose a constraint on the observing frequency and exclude all the sources from the list with their observing frequency less than $1325$ MHz ($ z \sim 0.07 $). After that, the different parameters of these sources are extracted from the HyperLeda database \footnote{\href{http://leda.univ-lyon1.fr/}{HyperLeda:http://leda.univ-lyon1.fr/}} \citep[see][]{leda}. These parameters include the object types, morphologies, optical diameters, optical inclination angles, optical position angles, B-band magnitudes, the width of the HI spectra, which correspond to the projected rotation velocities of the gas, HI line fluxes, heliocentric optical velocities and their distances. After knowing the object types of these sources, only the galaxies, pair of galaxies, triple galaxies, multiple galaxies, group of galaxies and the cluster of galaxies are retained for the final list. Then, the sources with heliocentric optical velocities greater than $15645$ km\thinspace $s^{-1}$, which corresponds to the frequency $1325$ MHz, have been excluded. This list included some Galactic sources, so we exclude sources with galactic latitude within $\pm10^o$. This gave us a list containing a total of  561 sources . As the elliptical galaxies are deprived of atomic hydrogen or sometimes contain a very little amount of atomic hydrogen, we excluded 46 elliptical galaxies from the list to achieve the final list containing 515 sources. This is the target sample for the GMRT ARChIve Atomic gas survey (GARCIA).   Figure \ref{fig:hist_velres}, \ref{fig:hist_time} and {\ref{fig:hist_dist}} respectively show the distribution of the channel resolution, on-source observation time and the distance for the various finalised sources. We note that out of the 515 finalized sources, 68 sources  ($\sim 13.2\%$) were observed between observation cycle No. 0 to cycle No. 06. As all the thirty dishes of GMRT were not functional in these observing periods, data of these sources may not have the adequate sensitivity for doing the kinematic modeling. Figure \ref{fig:scatter_plots} shows the distribution of different observed properties of these sources i.e., optical diameter ($a_{opt}$),  B-band magnitude ($B$), HI line flux (${F_{HI}}$), HI mass (${M_{HI}}$),  optical inclination angle ($i_{opt}$) and the width of the HI spectra ($W_{HI}/2$). The atomic gas masses ($M_{HI}$) of these galaxies have been derived using the procedure described below. If we consider the HI emission from a galaxy is optically thin, then the total integrated flux of the global HI spectra or the line flux is directly proportional to the HI mass of the galaxies. Now, if a galaxy is located at a luminosity distance $D_L$ with redshift $z$, then its HI mass can be calculated using the following equation \citep{highz_hi2017}:
   
  \begin{equation}
    \left( \frac{M_{HI}}{h_c^{-2} M_{\sun}}\right) \equiv \frac{2.35 \times 10^5}{1+z} \left( \frac{D_L}{h_c^{-1}Mpc} \right)^2 \left( \frac{S^{V_{rest}}}{Jy-kms^{-1}} \right)
    \label{hi_mass_ori}
 \end{equation}
 
   where $h_c$  is the Hubble parameter and is given by $h=\frac{H_o}{100 km s^{-1}{Mpc}^{-1}}$ \citep[see][]{highz_hi2017} and the $S^{V_{rest}}$ is the integrated line flux calculated in the rest frame of the galaxy and hence, $S^{V_{rest}} = \int \left( \frac{S(v)}{Jy} \right) \left( \frac{dv}{km s^{-1}} \right)$; therefore, $S^{V_{rest}}$ is the HI line flux ${F_{HI}}$. 
   The red dashed lines in figure \ref{fig:scatter_plots} represent the range of the respective parameters for the galaxies of the selected sub-sample.  The full sample and its properties are given in a table that can be found in the link: \href{http://www.physics.iisc.ac.in/~nroy/gmrt_paper1_table1.pdf}{GARCIA full sample} \footnote{\href{http://www.physics.iisc.ac.in/~nroy/gmrt_paper1v3_main_sources.pdf}{http://www.physics.iisc.ac.in/~nroy/gmrttable1.pdf}} . A separate table containing the properties of the 46 elliptical galaxies, which were excluded from the final list of the GARCIA sample, can also be found in the given link: \href{http://www.physics.iisc.ac.in/~nroy/gmrt_paper1_table2.pdf}{Elliptical galaxies} \footnote{\href{http://www.physics.iisc.ac.in/~nroy/gmrt_paper1v3_table2_elpgal.pdf}{http://www.physics.iisc.ac.in/~nroy/gmrttable2.pdf}} .
    
\begin{figure*}
    \centering
    \begin{subfigure}[b]{0.32\textwidth}
         \centering
         \includegraphics[width=\textwidth]{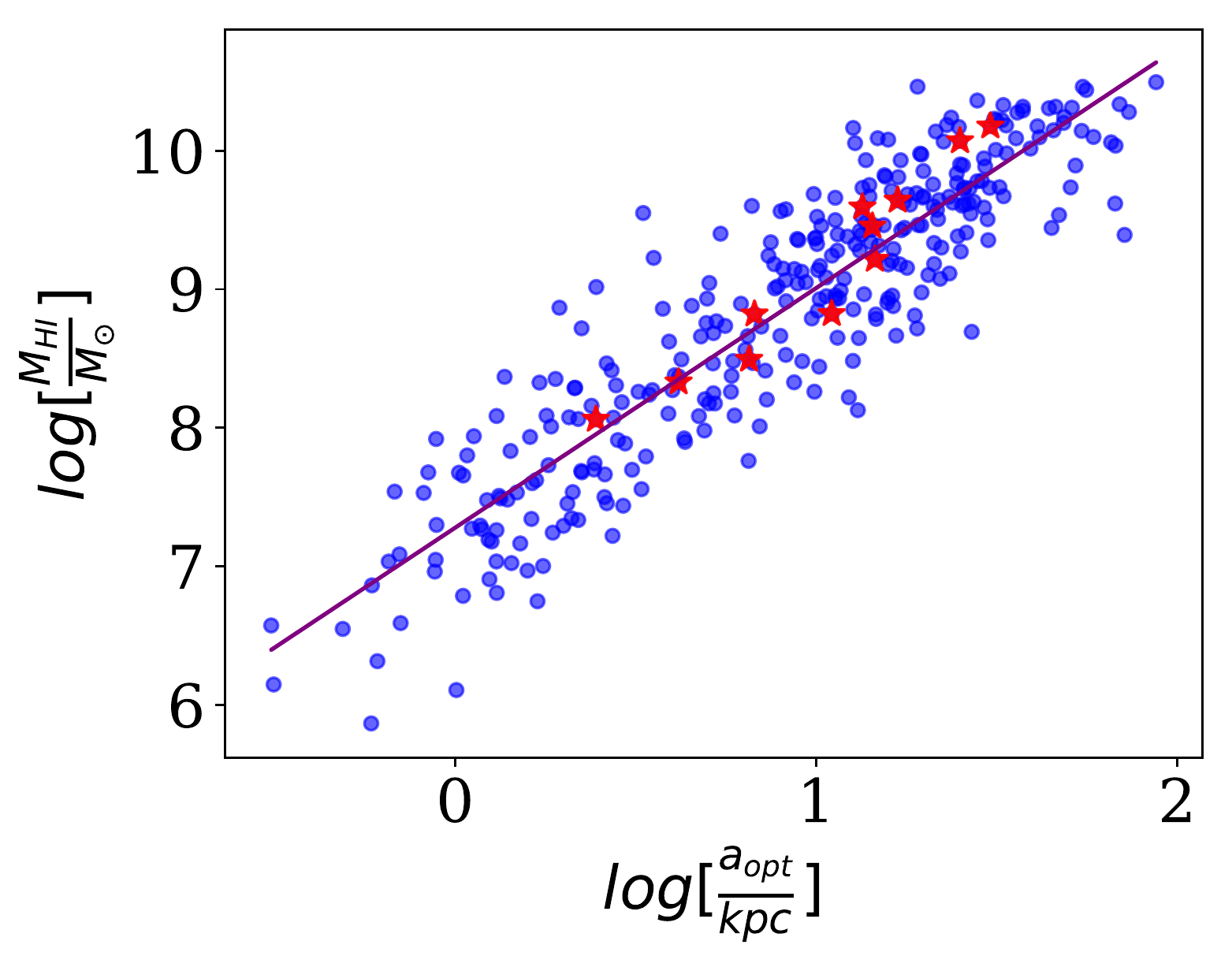}
         \caption{}
         \label{fig:opt_dia_mass}
     \end{subfigure}
    \begin{subfigure}[b]{0.32\textwidth}
         \centering
         \includegraphics[width=\textwidth]{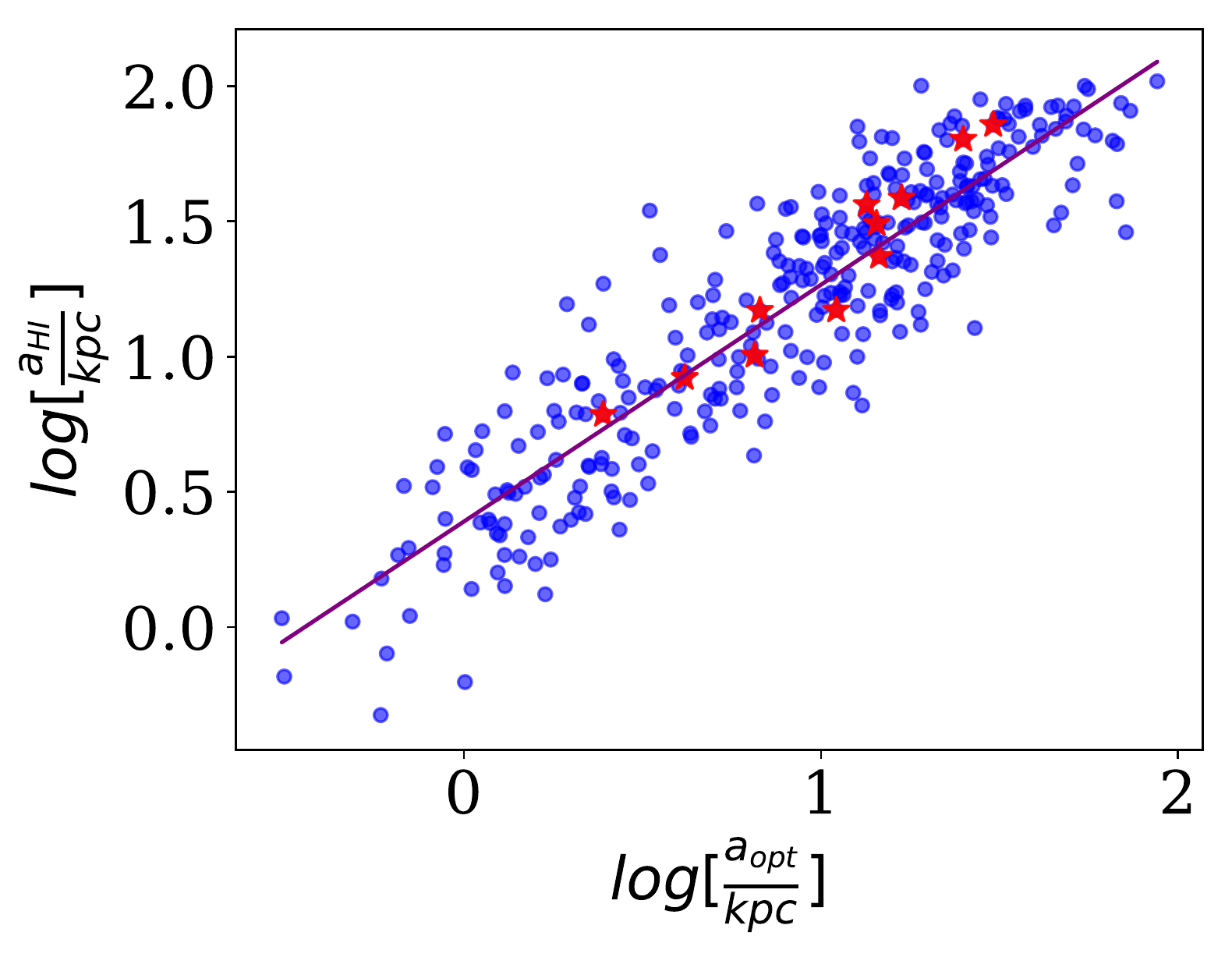}
         \caption{}
         \label{fig:opt_hi_dia}
     \end{subfigure}
          \begin{subfigure}[b]{0.32\textwidth}
         \centering
         \includegraphics[width=\textwidth]{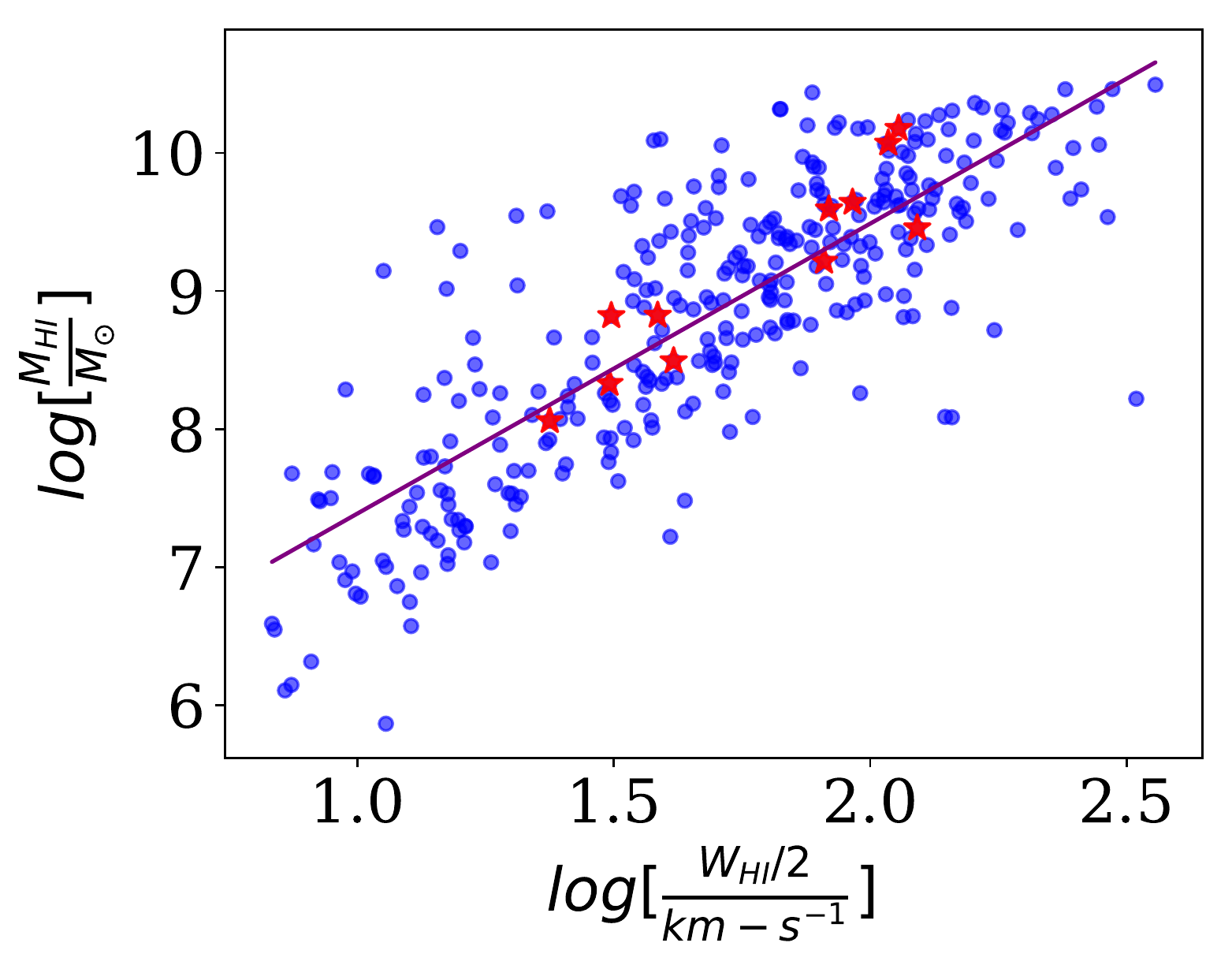}
         \caption{}
         \label{fig:gass_TFR}
     \end{subfigure}
          \begin{subfigure}[b]{0.33\textwidth}
         \centering
         \includegraphics[width=\textwidth]{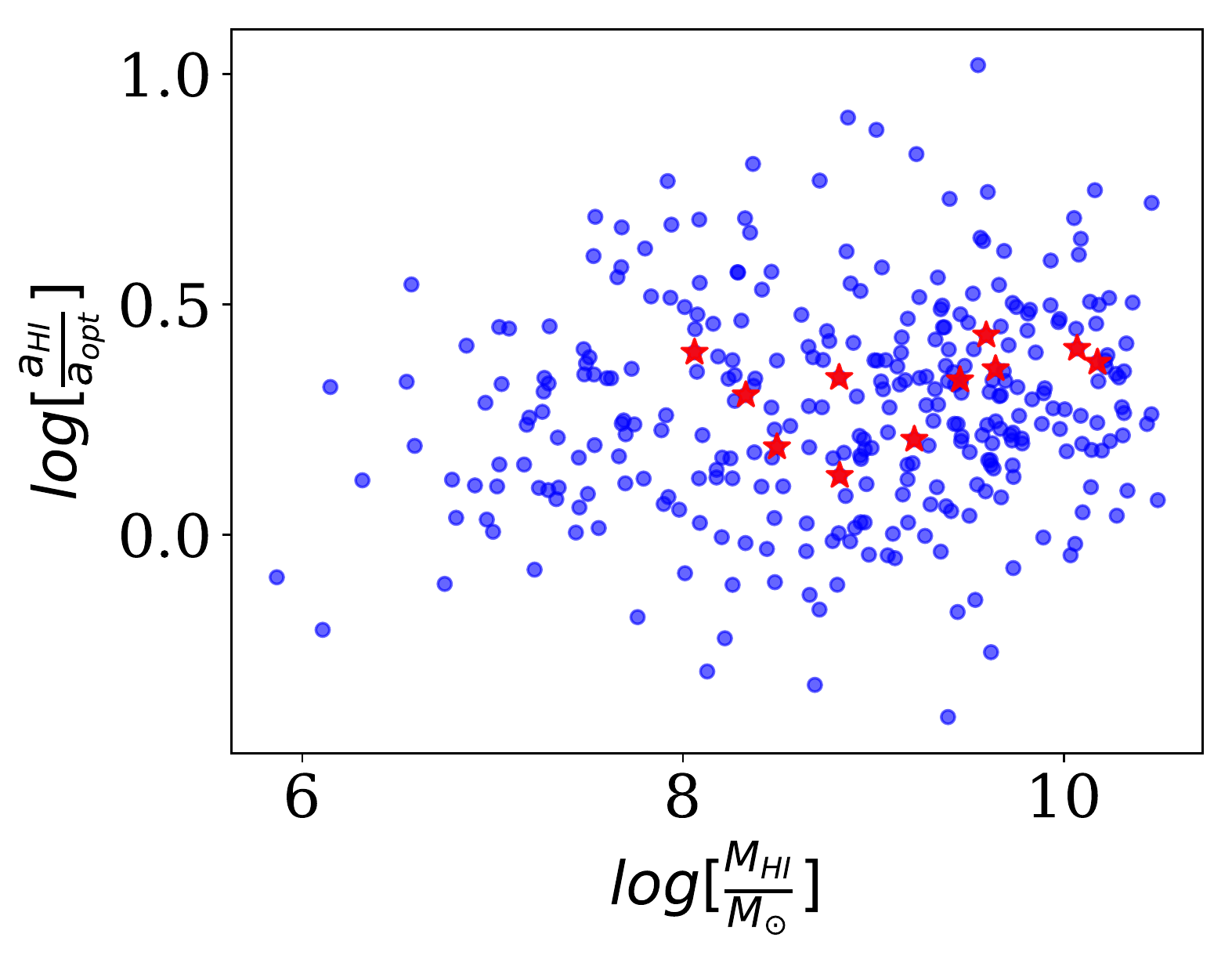}
         \caption{}
         \label{fig:hiopt_hmi}
     \end{subfigure}
     \begin{subfigure}[b]{0.33\textwidth}
         \centering
         \includegraphics[width=\textwidth]{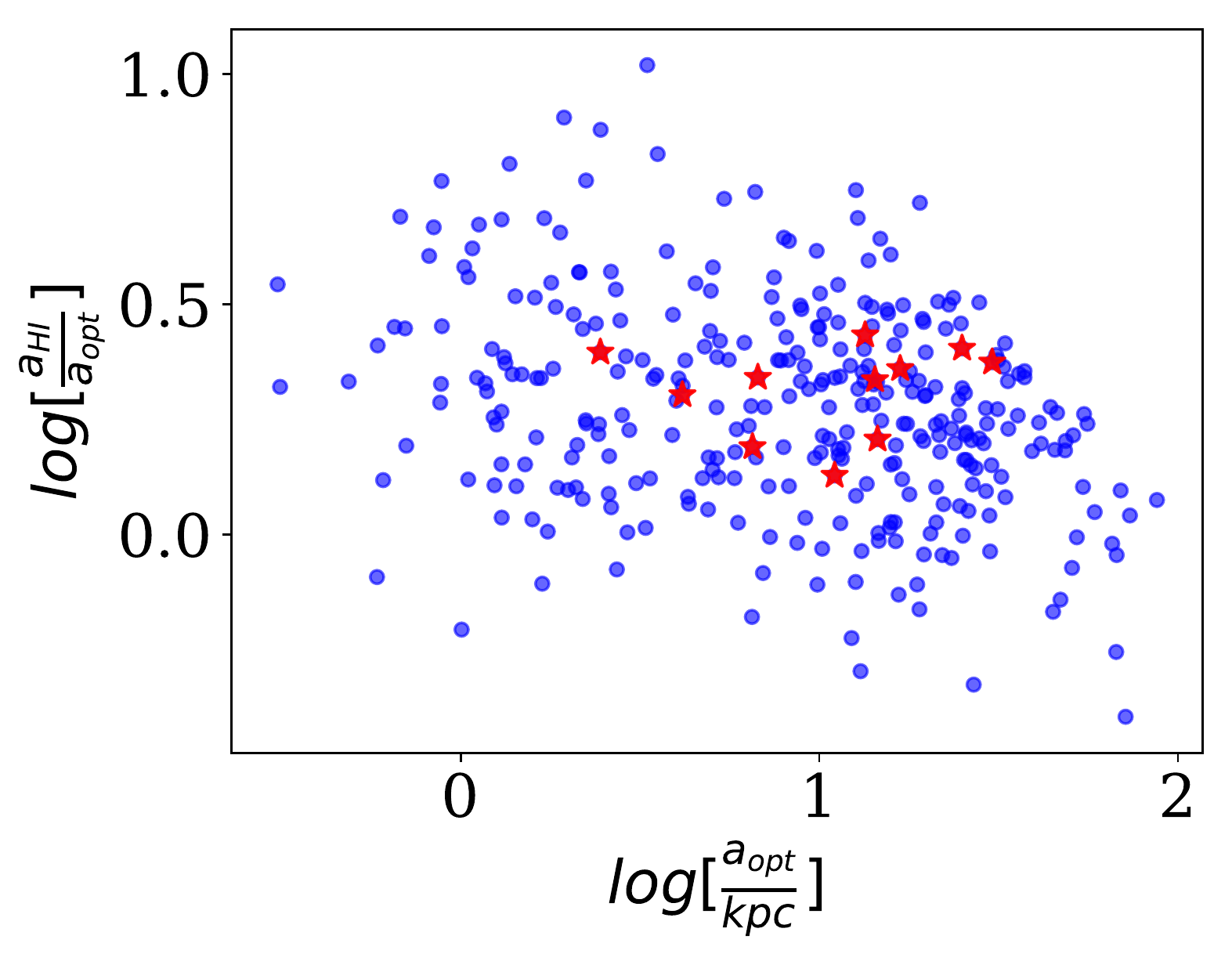}
         \caption{}
         \label{fig:hiopt_bkpc}
     \end{subfigure}
     \begin{subfigure}[b]{0.31\textwidth}  
     \centering
       \includegraphics[width=\textwidth]{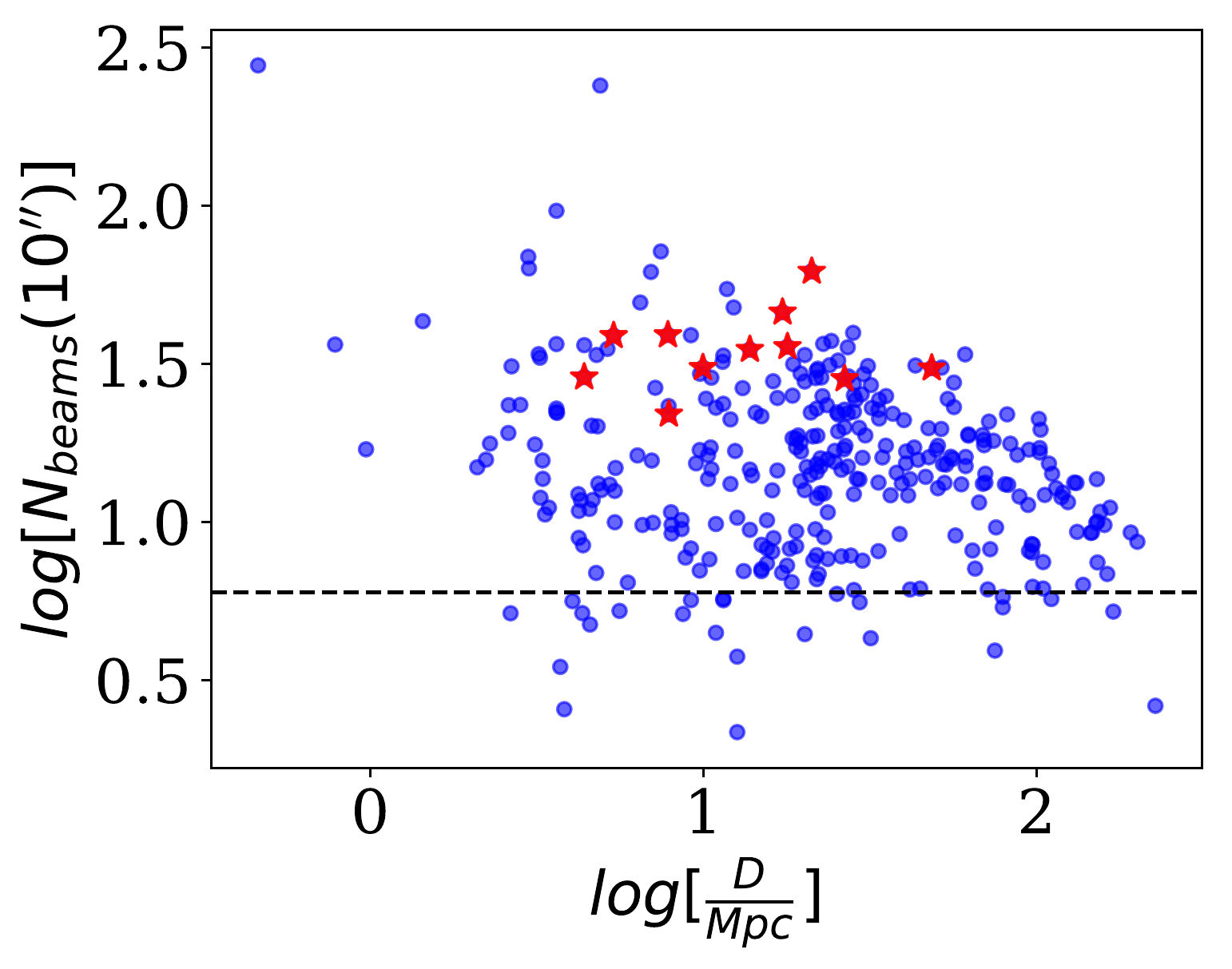}
        \caption{}
        \label{fig:nbeam_dist6}
     \end{subfigure}
    \caption{The scaling relations between the different parameters for the sources. The blue circles denote finalised sources selected for the GARCIA survey. The red stars in each plot represent the galaxies from the selected sub-sample. The solid violet lines in figure \ref{fig:opt_dia_mass}, \ref{fig:opt_hi_dia}, \ref{fig:gass_TFR} show the best fitted line. The black dashed line in figure \ref{fig:nbeam_dist6} represents the line where the number of beams of beam size $10^{\prime\prime}$ is equal to six. }
    \label{fig:scal_rel}
\end{figure*}

     Figure \ref{fig:scal_rel} shows the different scaling relations that exist between the different properties of the sources. Figure \ref{fig:opt_dia_mass} shows the HI mass - optical diameter relation of the galaxies. This correlation was primarily investigated by \citet{mass_size_first}. Upon doing the least square fitting \citet{mass_size_first} found a slope of $(1.95\pm0.06)$ and intercept $(7.00\pm0.08)$ to this relation. In Our results of the least square fitting, we found a slope of $(1.73\pm 0.04)$ and intercept of $(7.28\pm0.05 )$, which is in considerable agreement with \citet{mass_size_first}. \citet{mass_size_first} also showed that, although both the optical and HI diameter follows a tight correlation with the HI mass, the correlation between the HI mass and HI size of the galaxies have a significantly smaller scatter than the scatter on the HI mass-optical radius relation. 
     Figure \ref{fig:opt_hi_dia} shows the correlation between the HI diameter and optical diameter of the galaxies. The HI diameter $a_{HI}$ of these galaxies were found using the mass-size relation by \citet{mass_size}. \citet{mass_size_first} found a slope of $(1.00\pm0.03)$ and intercept of $(0.23\pm0.04)$, while we found the slope and the intercept of this fit are to be $(0.87\pm0.02)$ and $(0.39\pm0.02)$ respectively. Figure \ref{fig:gass_TFR} reveals the correlation between the rotation velocity of the galaxies with its atomic mass and can be called as Gas Tully-Fisher relation \citep[see][]{gas_TFR}. The slope and intercept of the best fitting line for this relation are found to be $(2.099\pm0.081)$ and $(5.29\pm0.14)$. The figure \ref{fig:hiopt_hmi} and  figure \ref{fig:hiopt_bkpc} show the ratio of the atomic diameter to HI diameter with HI mass and optical diameter respectively. From these plots, we see that the HI diameter to optical diameter ratio of the galaxies does not vary significantly within six orders of magnitude in HI mass ( $\sim 10^6 M_{\sun}$ to $\sim 10^{11} M_{\sun}$) the galaxies. However, we can find a dependence of this ratio with the morphology of the galaxies \citep{sparc}. We see from figure \ref{fig:hiopt_bkpc} that there is a weak dependence of this ratio with the optical diameter, as the optical diameter of the galaxies increases, the atomic diameter does not increase with the same rate; resulting in a decreasing slope in the ratio with the optical diameter.  Figure \ref{fig:nbeam_dist6} presents the number of beams of size $10^{\prime\prime}$ that will be fitted across the HI diameters ($a_{HI}$) of the respective sources versus the distance to the sources. This last figure \ref{fig:nbeam_dist6} in comparison with figure \ref{fig:nbeams_vs_freq} suggest that most of the selected samples are adequate for doing the 3D kinematic modeling. The range of the parameters and expected correlations assure that these galaxies, for which GMRT archival HI 21 cm data exist, can serve as a useful large sample of galaxies for detailed HI studies through a uniform analysis.  
    
    From these finalized sources, we have selected a representative sample of eleven galaxies (see table \ref{tab:cross_id}) for our preliminary study to establish the analysis method that will be applied uniformly to the full archival sample. The selected sample galaxies are mainly bright ($m_B < 14.10$) nearby ( distance < $50$ Mpc) spiral/irregular galaxies, each of which has a  different combination of inclination and position angles and is distributed over a broad range of masses ($\sim 10^8M_{\sun}$ to $\sim 10^{10}M_{\sun}$). This representative sample is well distributed in the parameter space of optical diameter versus HI mass and HI mass versus rotation velocity (see Figure \ref{fig:scal_rel}). The former relation, which is called the gas Tully-Fisher relation \citep{gas_TFR} along with the Baryonic Tully-Fisher relation, will be further explored in our upcoming paper.  The observation of these sources is done with a channel width $31.25$ kHz or equivalently $6.6$ kms$^{-1}$. Although some of these galaxies have their earlier HI interferometric observations, their observing setup and sensitivity are different. These studies helped us to compare the data products from our analysis and hence validate our analysis procedure applied to the GMRT sample of eleven galaxies. In a \say{short} 21cm line observation with WSRT (Westerbork Synthesis Radio Telescope), \citet{paper3_short_wsrt1}, \citet{paper4_short_wsrt2} respectively observed about 50 and 108 galaxies among which NGC0784, NGC1156, NGC3027, NGC3359, NGC4068, NGC4861, NGC7741 \citep{n7497_wsrt} and N7497 \citep{n7497_wsrt} are common from our studies. But their observation is carried out only along the major axis of the galaxies for around 2 to 3 hours and with a low velocity resolution of $17$ kms$^{-1}$ or $34$ kms$^{-1}$ and synthesized beam size in between $13^{\prime\prime}$ to $40^{\prime\prime}$. This type of observation yields only the position-velocity diagram (PV diagram) along the major axis of the galaxies. From the PV diagram, they further derived the global HI profile and the surface density profile by integrating the PV diagram along the resolution axis and velocity axis, respectively. On other hand, \citet{paper5_whisp1}, \citet{paper6_whisp2} and \citet{paper7_whisp3}  in WHISP \footnote{\href{https://www.astro.rug.nl/~whisp/}{https://www.astro.rug.nl/~whisp/s}} (Westerbork observations of neutral Hydrogen in Irregular and SPiral galaxies) survey observed around 382 objects and their companions; among them NGC0784, NGC1156, NGC4068 and NGC7741 are common from our sample. They observed with 14 telescopes (40 interferometers, 2 abundant) with around 12 hours observation on each source, reaching to a sensitivity of $1.5$ mJy/beam, $3.0$ mJy/beam and $4.0$ mJy/beam for channel resolution $16.4$ kms$^{-1}$, $4.1$ kms$^{-1}$ and $2.1$ kms$^{-1}$ respectively. They produced data cubes with three different resolutions: full resolution, $30^{\prime\prime}$ and $60^{\prime\prime}$ and provided the global HI spectra and the moment maps.  Besides that, \citet{n0784_paper1} provided the synthesis maps of NGC0784 from an early VLA observation when it was in the provisional state of operation. The observation was carried out in VLA-C configuration resulting in a resolution of  $15^{\prime\prime}$ and with a velocity resolution of  $11$ kms$^{-1}$. This galaxy, along with NGC7292, was also being observed by \citet{n0784_n7292_paper1} in a WSRT wide-field HI survey with low sensitivity of   $\sim 18$ mJy/beam and channel resolution $17$ kms$^{-1}$; they also provided the global HI spectra of these galaxies. Further, NGC1156 along with NGC7610 was observed by \citet{n1156_paper1} with VLA-D configuration where the synthesized beam size is $\sim 44^{\prime\prime}$, velocity resolution $20.5$ kms$^{-1}$ and sensitivity reaching to $\sim 1.35$ mJy/beam. \citet{n3027_paper1} made again a \say{short} observation (described earlier) of NGC3027 using WSRT with a different velocity resolution of $33$ kms$^{-1}$; these also provided only the PV diagram, spectra and the surface density profiles of the galaxy. \citet{n3359_paper2,n3359_paper3} observed NGC3359 using VLA C and D configuration at resolution  $\sim 18^{\prime\prime}$ and velocity resolution $25$ kms$^{-1}$ and studied its dynamics using hydrodynamical simulation.  \citet{n3359_paper6, n3359_paper7} observed this source with WSRT and studied its structure and dynamics in great detail. Further \citet{n3359_paper9} used the findings of \citet{n3359_paper7} and studied the stellar and gas dynamics for this galaxy. \citet{n4068_paper3} in their study of the baryonic distributions in galaxy dark matter halo, observed NGC4068 through VLA-C configuration with velocity resolution $0.825$ kms$^{-1}$. \citet{n4861_paper1} and \citet{n4861_paper8} observed NGC4861 using VLA-D and VLA-C configuration with channel resolution  $5.2$ kms$^{-1}$ and $2.6$ kms$^{-1}$ respectively and sensitivity $\sim 1$ mJy/beam for both cases. for NGC7800, no earlier HI interferometric observation was found as per the best of our knowledge. Most of these previous observations have velocity resolution coarser than $6.6$ kms$^{-1}$ and is not suitable for doing the 3D kinematic modeling by fitting the tilted ring model to this galaxy. On the other hand, due to the hybrid configuration of GMRT, the angular resolution in $1420$ MHz can be varied from $\sim 2^{\prime\prime}$ to $\sim 50^{\prime\prime}$. Thus, both the diffuse emission and fine structures can be studied with adequate sensitivity at this velocity resolution of $6.6$ kms$^{-1}$. 
    The selection of this set of galaxies was also made to compare the inclination angles measured from optical or single dish observation with the kinematic inclination angles found by fitting the 3D Tilted Ring model \citep{tiltd_ring1974} on the interferometric spectral line data. The inclination angles are the source of the dominant error \citep{dominant_err_in_BTFR2, dominant_err_in_BTFR} in correcting the Global HI line width. On the other hand, this HI line width is a beneficial quantity for measuring HI mass and estimating its rotation velocity. Both of these quantities imprint their effect in the Baryonic Tully-Fisher relation. All these aspects will be discussed in greater detail in the companion paper (Paper-II). This initial sample is also selected to verify how good the galaxies' dynamical modeling can be done with a velocity resolution as low as $6.6$ kms$^{-1}$, the lower end of the spectral resolution of the full sample. Table \ref{tab:obs_sample} presents the basic properties of these sources . The distances used in this study and mentioned in table \ref{tab:obs_sample} are not derived from the Tully-Fisher relation, except for NGC7741 (where the method of Sosies is used \citep[see][]{sosies}, this uses TF relation indirectly to calculate the distance. This distance is consistent with the distance of the group associated with this galaxy.),  as this sample will be further used to study the validity of the mentioned relation itself for galaxies with different inclination angles. The morphology type mentioned in this table is taken from LEDA database \citep{leda}; $m_B$ represents the B band apparent magnitude; the optical inclination ($i_{opt}^o$ ) angles mentioned here are found by the axial ratio of the isophote at $25$ mag/arcsec$^2$ in the B-band for the galaxies \citep{leda}; $a_{HI}$ and $b_{HI}$ respectively denote the major axis and the minor axis of the ellipse fitted to the derived moment zero maps'  $1.25\times10^{20}$  cm$^{-2}$ column density contour. The column density of $1.25\times10^{20}$  cm$^{-2}$ is equivalent to the surface density ($\Sigma_{HI}$) of $1$ M$_{\sun}$\thinspace pc$^{-2}$.  The HI inclination angle mentioned in this table are derived from the ratio of major axis to minor axis of the fitted ellipse ( $ \sin(i_{HI})= \sqrt{\frac{1-(b_{HI}/a_{HI})^2} {1-q_o^2}}$ , where $q_o$ is the intrinsic axial ratio: for spirals, $q_o \sim 0.2$ and for dwarfs, $q_o \sim 0.6$) ; the position angle ($PA_{HI}$) is measured as the angle obtained by $a_{HI}$ anticlockwise from north at the center of the ellipse fitted to the $1.25 \times 10^{20}$  cm$^{-2}$ column density contours . The kinetic inclination angles and position angles found by 3D kinematic modeling of the observed interferometric data are mentioned and compared in Paper-II.

\begin{table} 
\caption{Pilot sample name and their cross-identification }  \label{sample}
\begin{center}
\begin{tabular}{||c|c||}
\hline
    Galaxy name & Cross Identification  \\
    \hline
     NGC0784 & UGC01501; PGC007671 \\
    NGC1156 &  UGC02455; PGC011329	  \\
    NGC3027 & UGC05316; PGC028636\\
    NGC3359 & UGC05873; PGC032183\\
    NGC4068 & IC0757; UGC07047; PGC038148; PGC2418706 \\
    NGC4861 & IC3961; UGC08098; PGC044532; PGC044536 \\
   NGC7292 &  UGC12048; PGC068941 \\
   NGC7497 & UGC12392; PGC070569  \\
   NGC7610 &  NGC7616; UGC12511; PGC071087; PGC1375849 \\
   NGC7741 & UGC12754; PGC072237; PGC1761654 \\
   NGC7800 &  UGC12885; PGC073177 \\
   \hline
\end{tabular}
\end{center}
 \label{tab:cross_id}
\end{table}

\begin{table*}
    %\caption{Observation Sample}
    \centering
    \begin{tabular}{||c|c|c|c|c|c|c|c|c|c|c|c||}
\hline
Galaxy  & $RA$           & $DEC$          & Distance        & Method used           & Morphology     & $m_B$ & $ i_{opt} $ & $a_{HI}$ & $b_{HI}$   & $i_{HI}$  & $PA_{HI}$ \\
name    & hr-m-s       & d-m-s            & (Mpc)           & for Distance          & Type           & Mag   &  (deg)      & kpc         & kpc      & (deg)   & (deg)  \\
\hline
NGC0784 & 02h01m16.93s & +28d50m14.1s & 5.45 & 1 & SBd & 11.91 & 90.0 & 13.47 & 4.39 & 75 & 4 \\
NGC1156 & 02h59m42.300s & +25d14m16.20s & 6.79 & 2 & IB & 12.08 & 42.4 & 12.67 & 7.76 & 54 & 103 \\
NGC3027 & 09h55m40.60s & +72d12m12.8s & 16.54 & 3 & Sc & 12.17 & 69.7 & 47.76 & 24.18 & 62 & 131 \\
NGC3359 & 10h46m36.863s & +63d13m27.25s & 16.79 & 3 & Sc & 11.03 & 47.2 & 65.28 & 45.49 & 47 & 177 \\
NGC4068 & 12h04m00.776s & +52d35m17.78s & 4.39 & 1 & I & 13.0 & 61.3 & 6.52 & 4.63 & 46 & 9 \\
NGC4861 & 12h59m02.34s & +34d51m34.0s & 9.95 & 1 & Sm & 12.54 & 90.0 & 17.44 & 8.21 & 64 & 11 \\
NGC7292 & 22h28m25.91s & +30d17m32.3s & 9.60 & 3 & I & 12.94 & 54.4 & 11.03 & 8.06 & 44 & 9 \\
NGC7497 & 23h09m03.41s & +18d10m37.9s & 19.82 & 3 & Sc & 13.30 & 71.9 & 42.26 & 13.69 & 75 & 43 \\
NGC7610 & 23h19m41.37s & +10d11m06.0s & 46.96 & 3 & SABc & 14.10 & 67.6 & 60.94 & 48.81 & 38 & 74 \\
NGC7741 & 23h43m54.37s & +26d04m32.2s & 12.60 & 4 & SBc & 11.66 & 50.1 & 20.73 & 13.25 & 52 & 176 \\
NGC7800 & 23h59m36.32s & +14d48m20.1s & 20.56 & 3 & IB & 13.06 & 90.0 & 33.57 & 21.51 & 52 & 48 \\
\hline
    \end{tabular}
    \caption{Observation Sample Properties. References used for methods of distance measurements are given by the following. 1: Tip of the red giant branch (TRGB), 2:Brightest Stars, 3:Cosmological Distance, 4:Sosies.  }
    \label{tab:obs_sample}
\end{table*}

\section{Observation and Data Reduction}
\label{data_reduction}
The details of the observation of the representative sample are given in Table \ref{tab:obs_detail2}. All the sources were observed with bandwidth $4$ MHz and channel resolution of $31.25$ kHz with adequate observation time so that we get $\sim$ 1 mJy/beam noise at the line-free channels. The radio interferometric data from GMRT were calibrated and imaged using the standard routines in AIPS. We used our own AIPS script to carry out the calibration and bandpass. The bad data points and RFI's were found and flagged with several tasks in AIPS (e.g., QUACK, UVFLG, TVFLG, SPFLG).  From the UV data, some line-free channels were used to make a high-resolution image of the continuum with a large field of view using the IMAGR task and the continuum image is then subtracted from the UV data using UVSUB. To correct the frequency of the observed data due to the effect of the Earth's rotation around the sun and its motion, the entire spectra are shifted with respect to the channels so that we get the red-shifted frequency of the source to the central channel with the velocity equals to zero, i.e., transferring to a frame where the systemic velocity of the galaxy is equal to zero. This part has been performed using the CVEL task in AIPS. Then this UV data is used for making the 3D image cube of the data using the task IMAGR. The imaging is done manually with the task IMAGR.  Further, for some cases, if there were some residual continuum emission in the image cubes, the task IMLIN was run to fit it with a first-order polynomial to some line-free channels and remove it from the image cube.  To construct the image cubes of each galaxy, as mentioned above,  different UV cut-off (starting from $5$ $k\lambda$ to $20$  $k\lambda$ ), UV tapering and different ROBUST values (starting from $0$ to $4$, where ROBUST = $0$ is no weighting  and ROBUST = $+4$ is nearly pure natural weighting to the cubes \citep[see][]{briggs_robust}) have been used. In the different cases for all the galaxies, we found image cubes with beam size varying from $ \sim 50^{\prime\prime} \times 45^{\prime\prime}$ to $ \sim 15^{\prime\prime} \times 10^{\prime\prime}$ and RMS noise at line free channel to vary in between $1.4$ and $0.7$ mJy/beam per $31.25$ kHz channel width. These different image cubes are used for the deduction of different results.

\begin{table*}
    \caption{Observational details}
    \centering
    \begin{tabular}{|c|c|c|c|c|c||}
      \hline
Galaxy & Observation date & Time on source & Phase          & Observing frequency & Velocity (Heliocentric, \\
name &               & (minutes)        & calibrator   & (GHz) &  Optical) (km\thinspace s$^{-1}$) \\
\hline
NGC0784 & 31stDec,2004 & 200 & 0237+288 & 1.41746765 & 197.86 \\
NGC1156 & 30thDec,2004 & 220 & 0237+288 & 1.41663002 & 375.04 \\
NGC3027 & 30thDec,2004 & 248 & 1035+564 & 1.41339290 & 1057.97 \\
NGC3359 & 01stJan,2005 & 168 & 1035+564 & 1.41360141 & 1013.9 \\
NGC4068 & 02ndJan,2005 & 169 & 1035+564 & 1.41741082 & 209.85 \\
NGC4861 & 02ndJan,2005 & 251 & 1331+305 & 1.41445914 & 834.92 \\
NGC7292 & 28thJul,2005 & 504 & 2236+284 & 1.41337011 & 337.11 \\
NGC7497 & 30thJul,2005 & 623 & 2251+188 & 1.40994995 & 1707.91 \\
NGC7610 & 31stJul,2005 & 577 & 2330+110 & 1.40156698 & 3554.04 \\
NGC7741 & 30thJul,2005 & 675 & 2254+247 & 1.41448000 & 750.08 \\
NGC7800 & 01stAug,2005 & 672 & 2330+110 & 1.41009549 & 1754.09 \\

     \hline    
    \end{tabular}
    \label{tab:obs_detail2}
\end{table*}

\section{Data Products} 
\label{data_product}
  
\begin{figure*}
    \centering
    \begin{subfigure}[b]{0.3\textwidth}
        \centering
        \includegraphics[width=\textwidth]{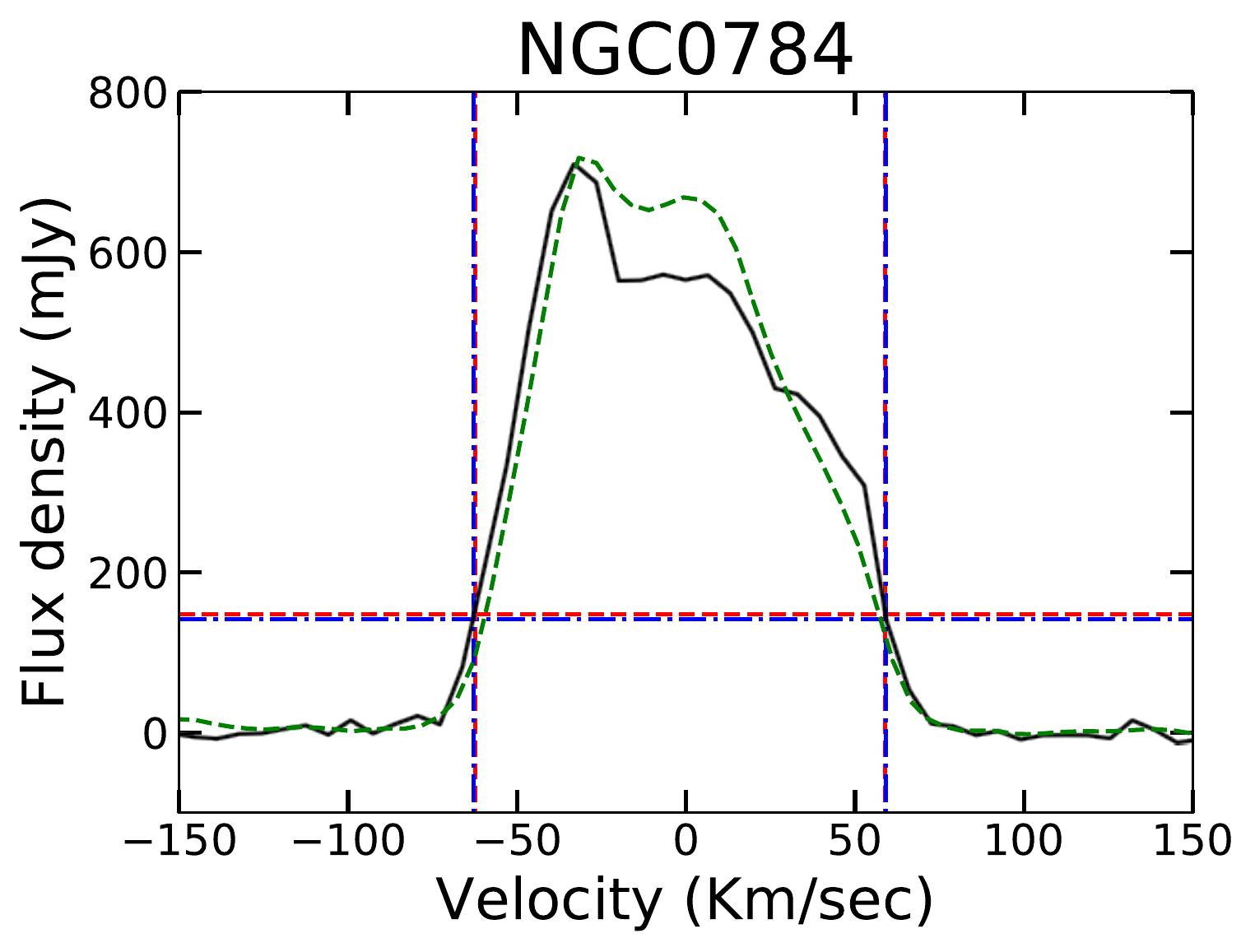}
        \caption{}
        \label{}
    \end{subfigure}
    \begin{subfigure}[b]{0.3\textwidth}
        \centering
        \includegraphics[width=\textwidth]{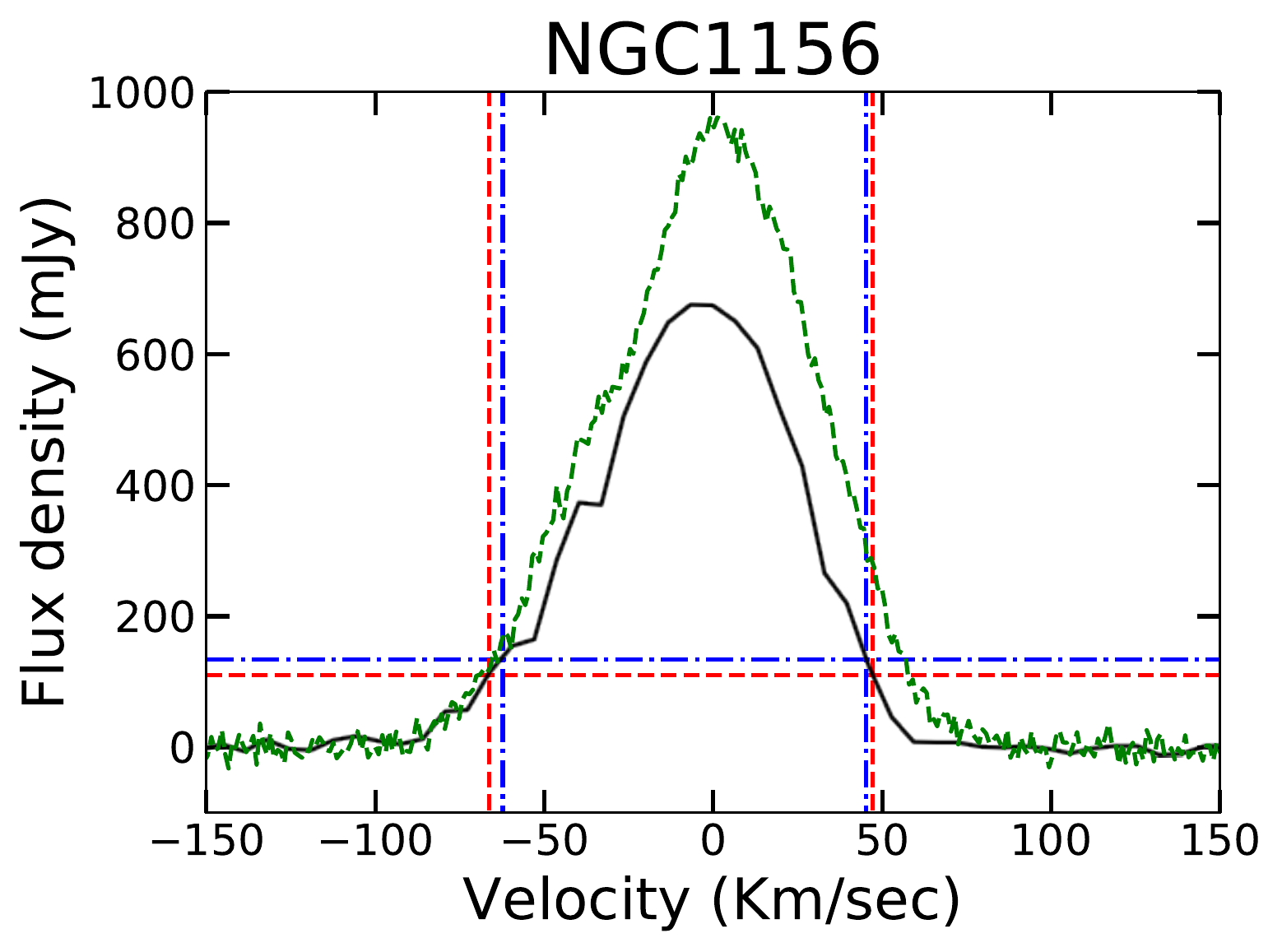}
        \caption{}
        \label{}
    \end{subfigure}
    \begin{subfigure}[b]{0.3\textwidth}
        \centering
        \includegraphics[width=\textwidth]{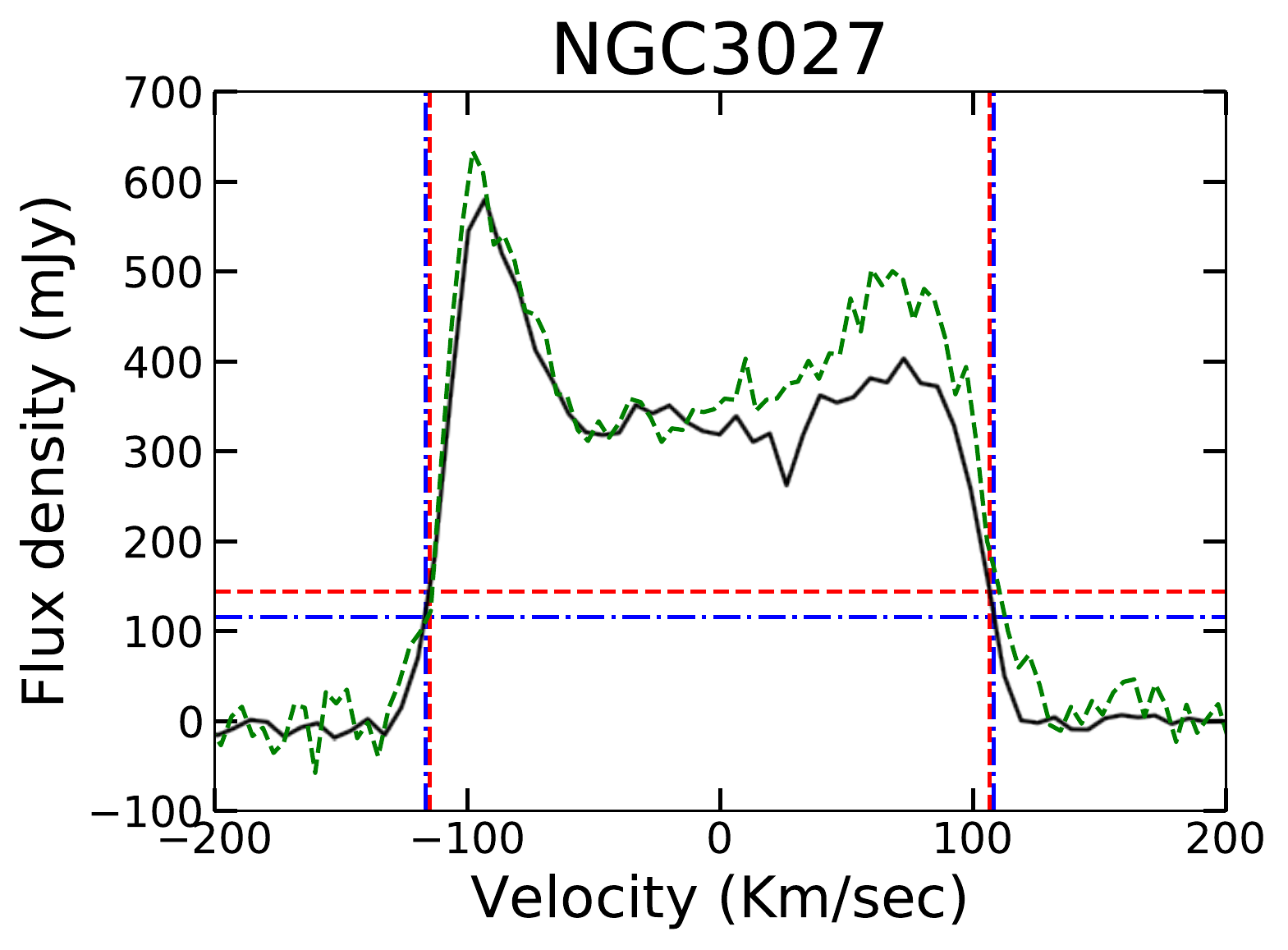}
        \caption{}
        \label{}
    \end{subfigure}
    
    \begin{subfigure}[b]{0.3\textwidth}
        \centering
        \includegraphics[width=\textwidth]{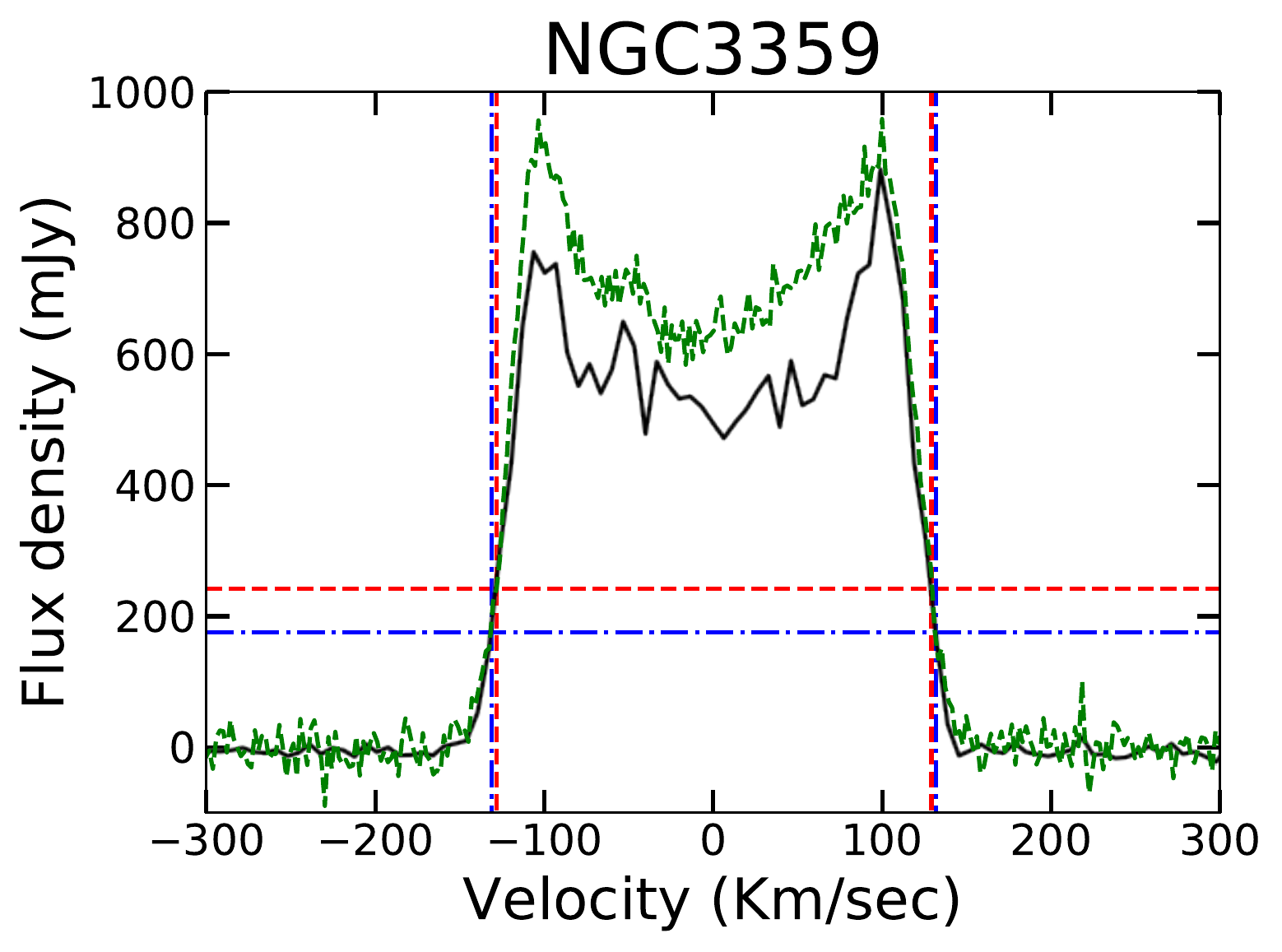}
        \caption{}
        \label{}
    \end{subfigure}
    \begin{subfigure}[b]{0.3\textwidth}
        \centering
        \includegraphics[width=\textwidth]{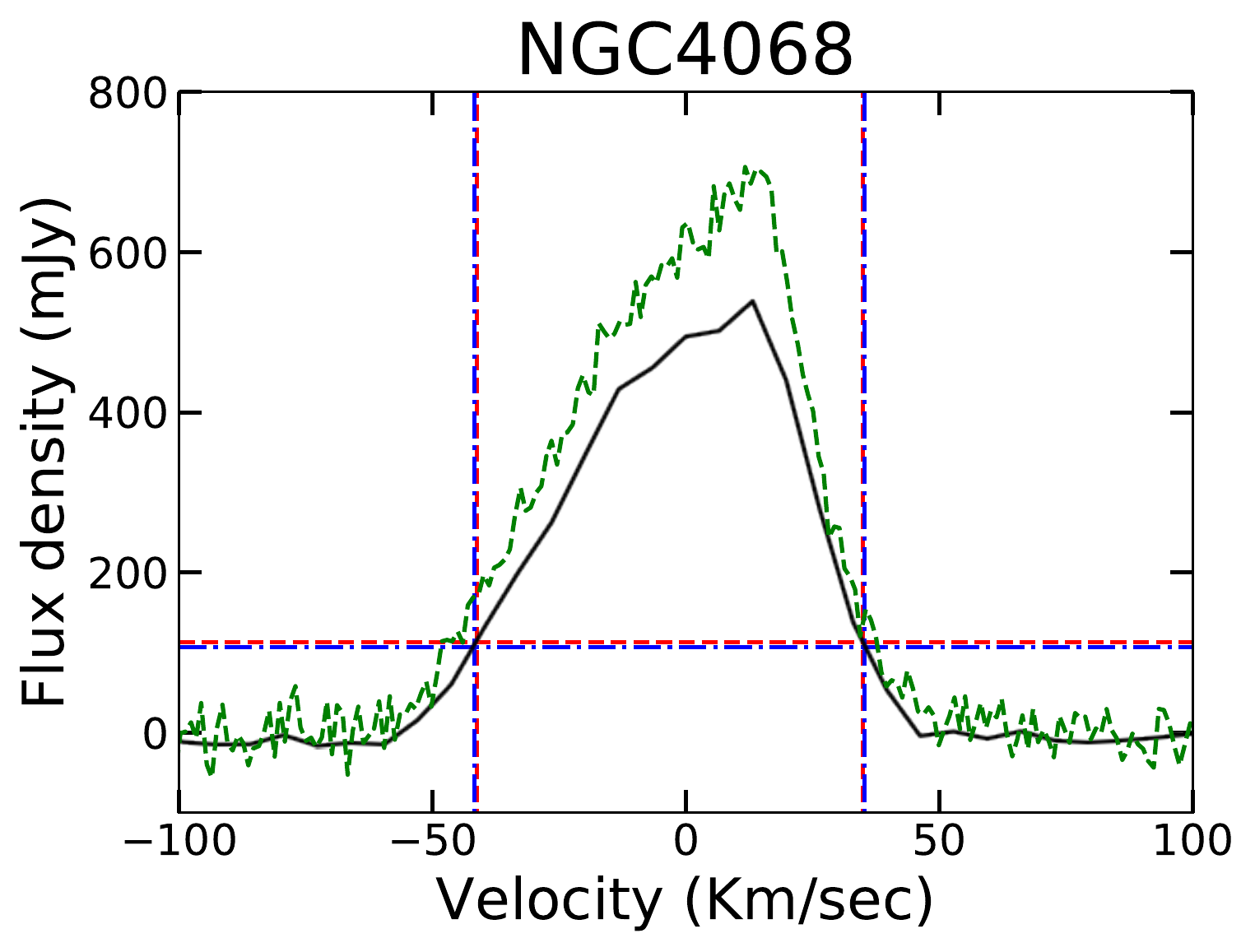}
        \caption{}
        \label{}
    \end{subfigure}
    \begin{subfigure}[b]{0.3\textwidth}
        \centering
        \includegraphics[width=\textwidth]{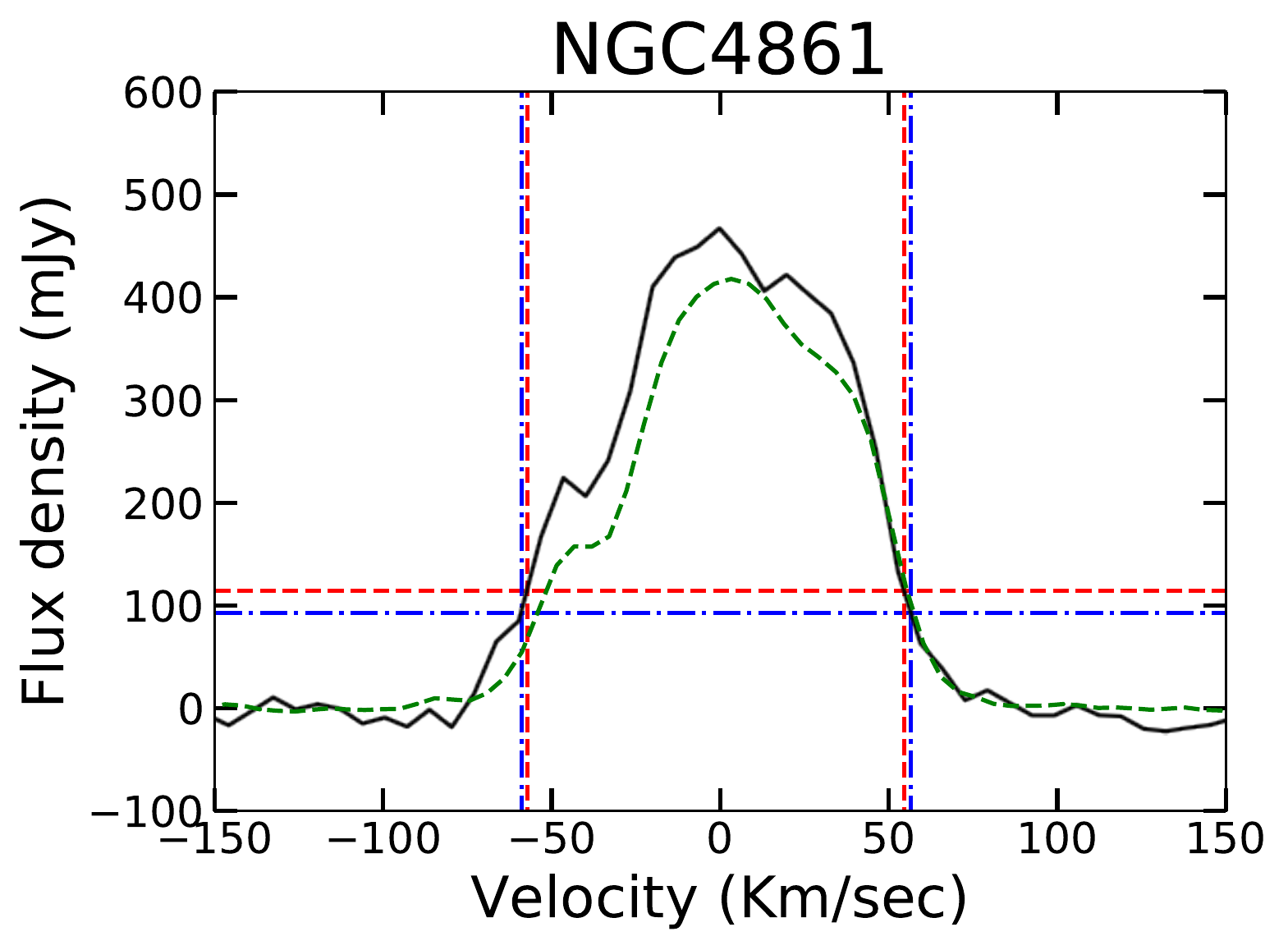}
        \caption{}
        \label{}
    \end{subfigure}
    
    \begin{subfigure}[b]{0.3\textwidth}
        \centering
        \includegraphics[width=\textwidth]{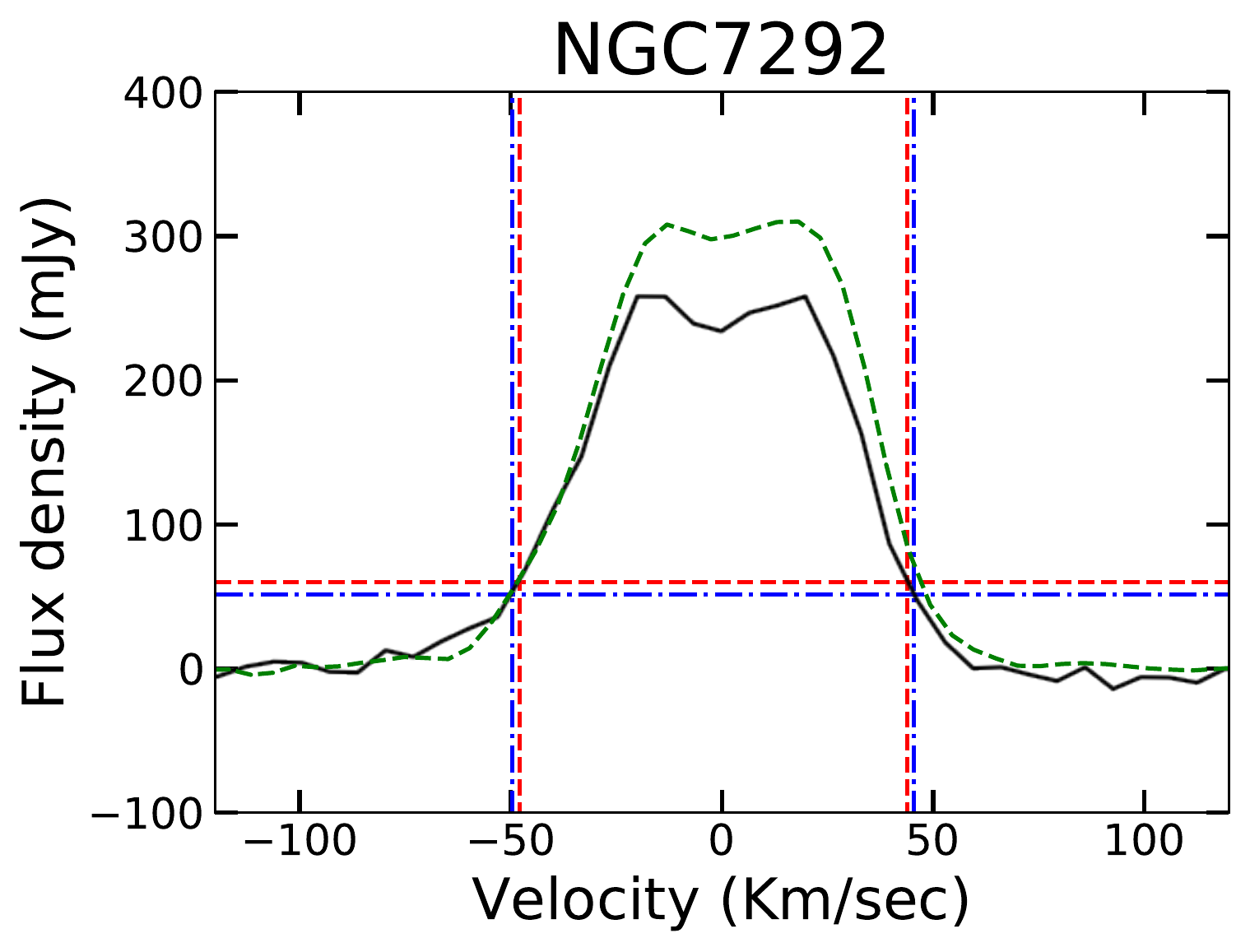}
        \caption{}
        \label{}
    \end{subfigure}
    \begin{subfigure}[b]{0.3\textwidth}
        \centering
        \includegraphics[width=\textwidth]{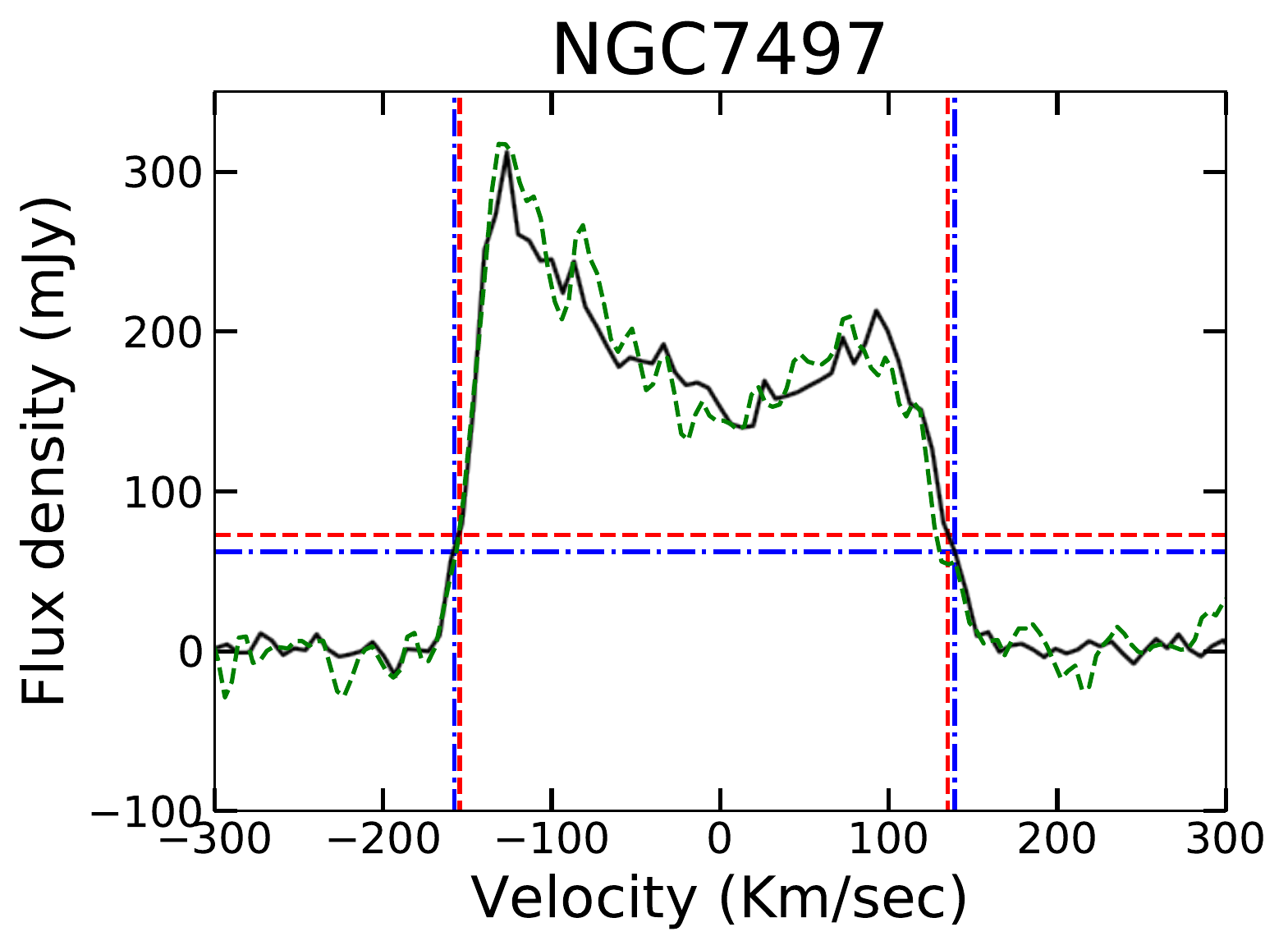}
        \caption{}
        \label{}
    \end{subfigure}
    \begin{subfigure}[b]{0.3\textwidth}
        \centering
        \includegraphics[width=\textwidth]{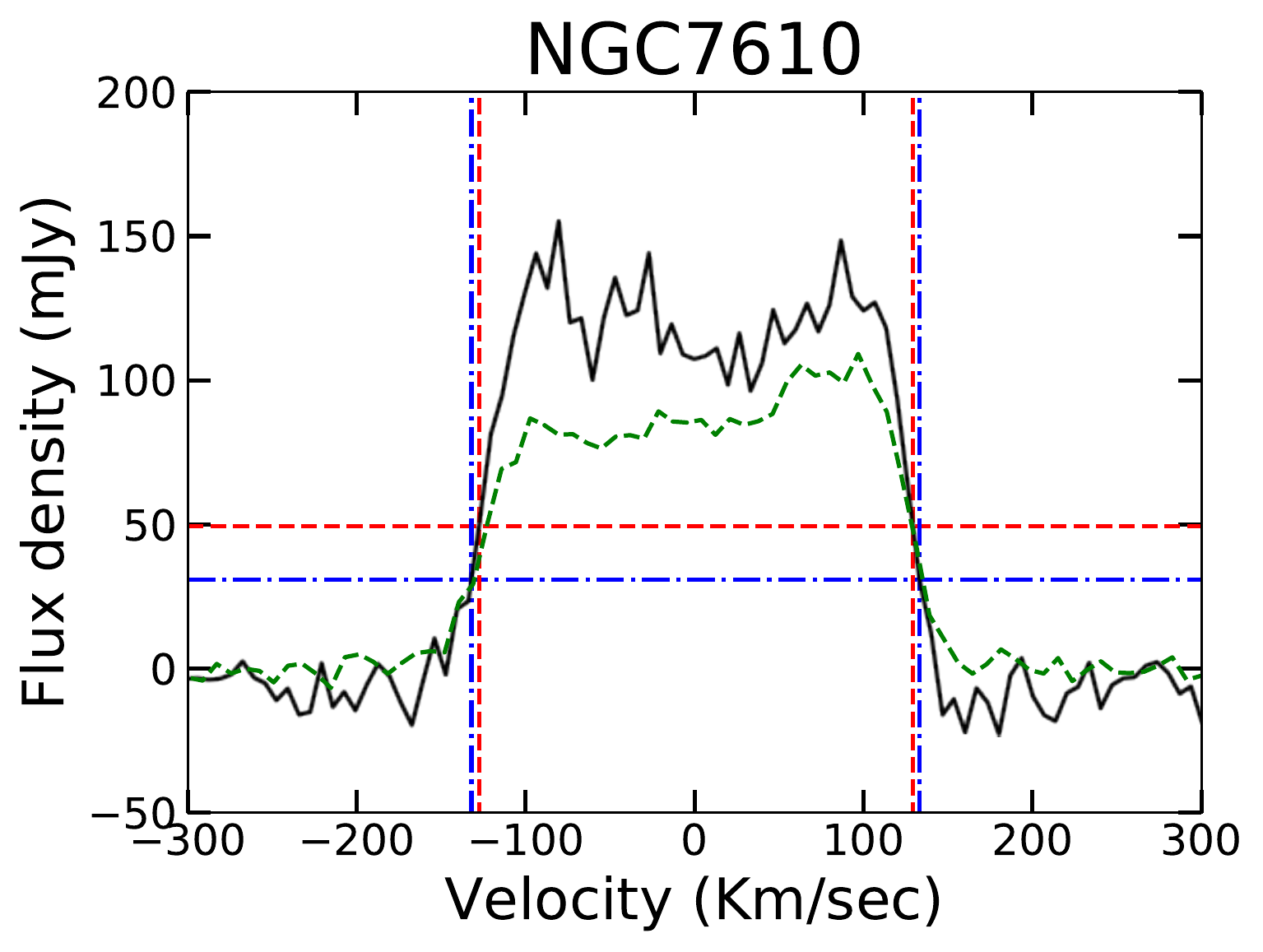}
        \caption{}
        \label{}
    \end{subfigure}
    
    \begin{subfigure}[b]{0.3\textwidth}
        \centering
        \includegraphics[width=\textwidth]{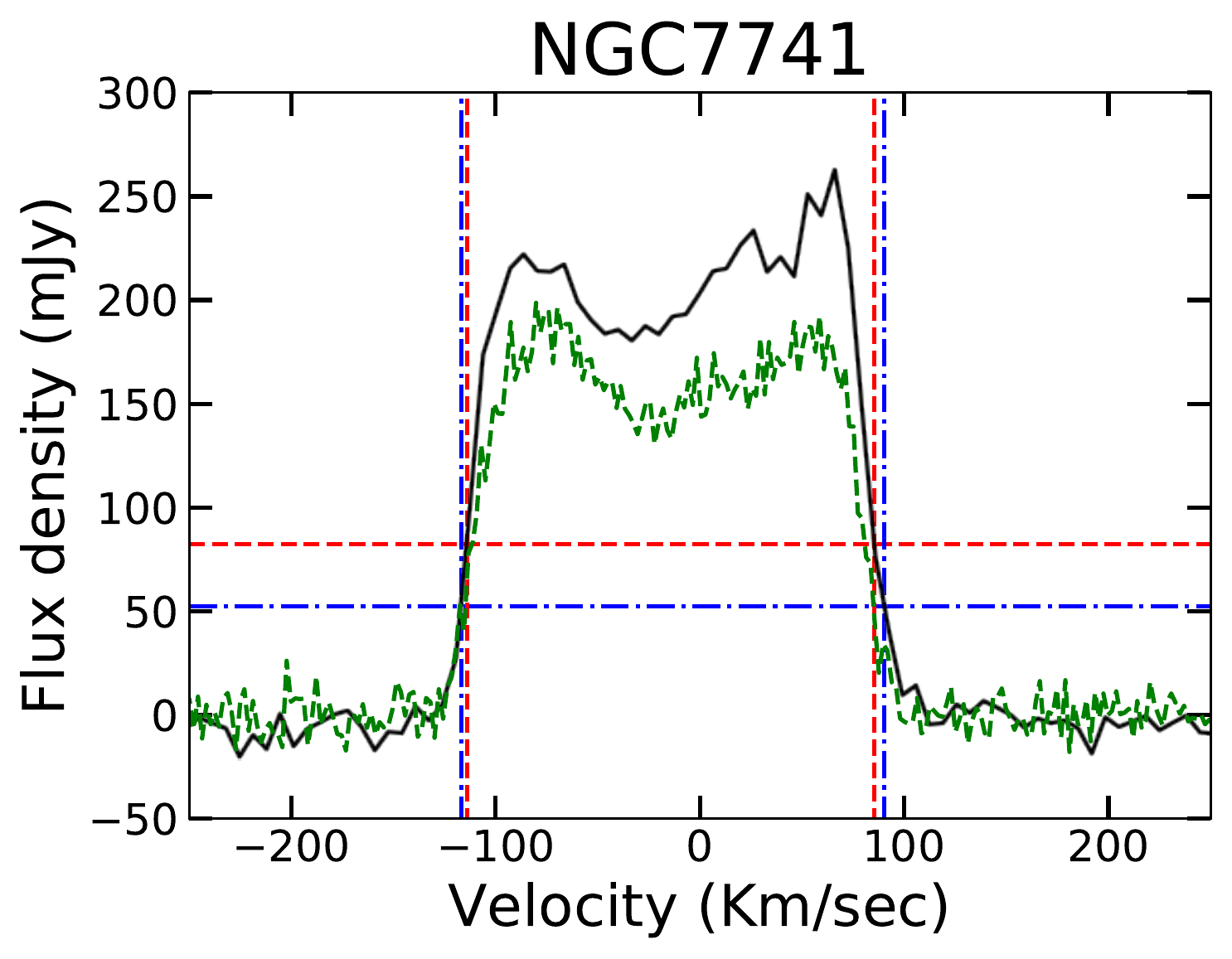}
        \caption{}
        \label{}
    \end{subfigure}
    \begin{subfigure}[b]{0.3\textwidth}
        \centering
        \includegraphics[width=\textwidth]{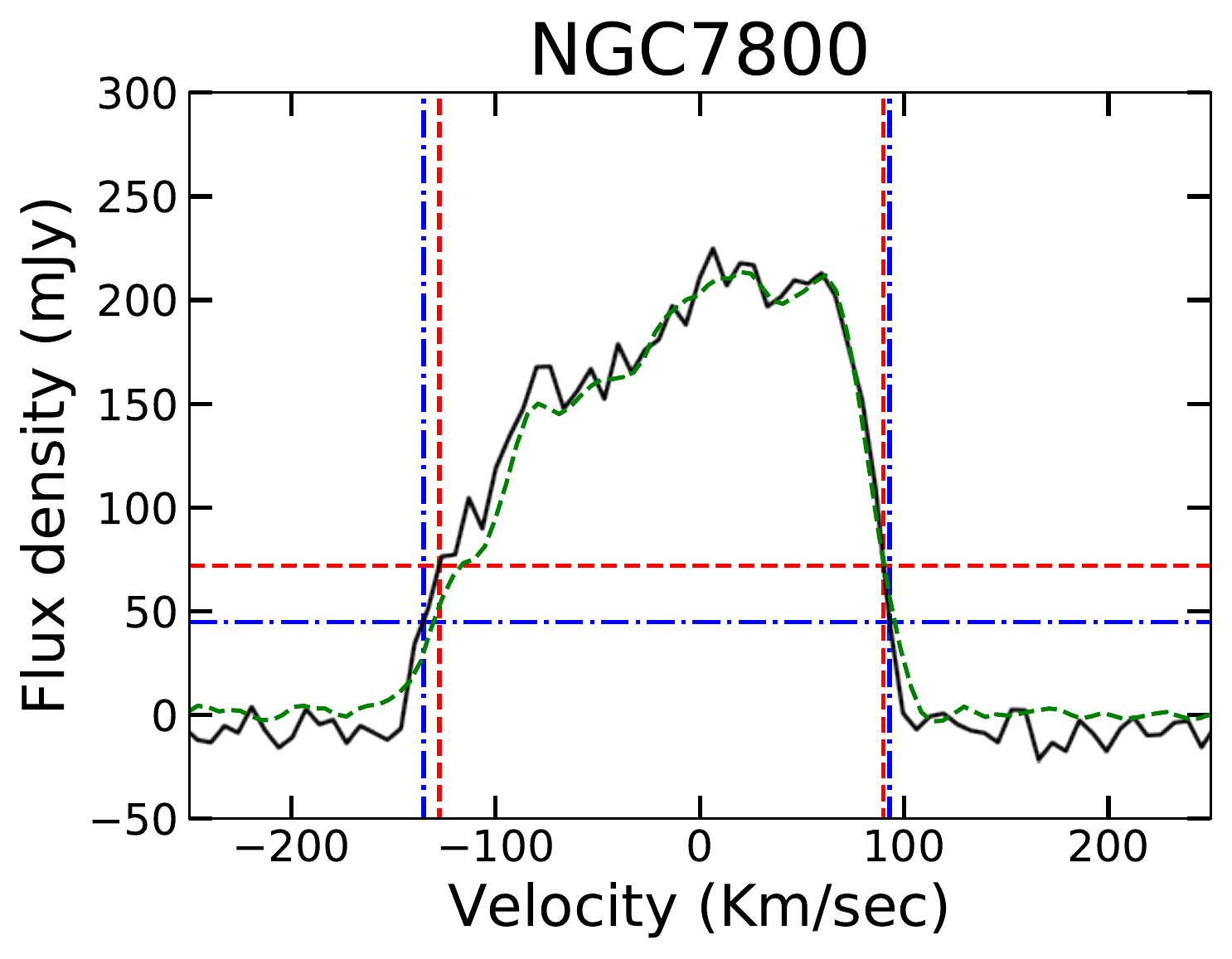}
        \caption{}
        \label{}
    \end{subfigure}
    \caption{The global HI Spectra of the galaxies. The black solid line denotes the spectra extracted from the interferometric cube. The Green dashed line denotes single-dish spectra. The red dashed line and the blue dash-dotted line respectively denote $50\%$ of the mean flux and $20\%$ of the peak flux of the interferometric spectra.}
    \label{fig:hi_spectra}
\end{figure*}

 Due to the galaxy's rotation and dynamic motion, the HI signal coming from different parts has a different line of sight velocities. Galaxy's Global HI spectra is the HI flux measured as a function of the frequency or its corresponding Doppler velocity. The spectra's central part roughly corresponds to the galaxy's systemic velocity, while half the width of the spectra roughly corresponds to the component of the maximum rotation velocity towards our line of sight. Figure \ref{fig:hi_spectra} show the Global HI spectra obtained from our analysis in comparison to the spectra from the single-dish observation for the selected sample of eleven galaxies. The single dish spectra for NGC1156, NGC3027, NGC3359, NGC4068, NGC7497, NGC7610, NGC7741 are taken from \citet{wm50_def}; and for NGC0784, NGC4861, NGC7292, NGC7800 are taken from \citet{sngl_dish2}. To extract the GMRT spectra of these galaxies, we have used the lowest resolution image cubes made using UV-cut off of $5$ $k\lambda$ ( with beam size varying from roughly $35^{\prime\prime} \times 40^{\prime\prime}$ to $45^{\prime\prime} \times 52^{\prime\prime}$ for different galaxies)   so that it picks up maximum diffuse emission from the galaxies. The image cubes are masked with the moment zero maps so that the spectra do not include any leftover noises or sources coming from the other locations in the field. In our analysis, we have shifted the spectra in the frequency domain such that the systemic velocity of the galaxies becomes equal to zero as our primary interest is on the galaxies' rotation velocities. The line width, which represents the galaxy's circular velocity at the outer part, has been defined in various ways in the literature. $W_{p20}$, first used by \citet{Tully_1977}, is the width measured at the  $20\%$ of the peak of the HI spectra. Another definition of the velocity that is broadly used in the literature is $W_{m50}$. It is the width measured at the $50\%$ of the mean HI line flux \citep{wm50_def}. A comparison of these line widths and the total line flux obtained from our analysis for each of the galaxies are noted in table \ref{tab:hi_spectra}.
 
 We note that for all the sources, the flux recovered from the global HI spectra obtained in the interferometric data analysis is similar to the flux obtained from the spectra in the single-dish observation. For some galaxies, the interferometric spectra have less flux in comparison to the single-dish spectra. One apparent reason for getting less flux in the interferometric observation is due to the zero spacing problem. Due to the finite distance between the antenna, we can never measure visibility ($V$) at ($u=0,v=0$). So, for interferometric observation, this term $V(0,0)$ will not contribute to the calculation of sky brightness, which is the Fourier transform of the visibility. For some sources, the interferometric flux is larger than the single dish spectra; this mismatch in the flux can be an effect of the source position in the single-dish beam in the case of the single-dish observation. For example, as found from \citet{wm50_def} the observed line flux in the single-dish spectra for NGC7610 and NGC7741 are $22.88$ Jy\thinspace kms$^{-1}$ and $31.80$ Jy\thinspace kms$^{-1}$ respectively. But after correcting the pointing offset, they found the line flux to be respectively $34.64$ Jy\thinspace kms$^{-1}$ and $68.94$  Jy\thinspace kms$^{-1}$. While these corrected spectra were not available, we compared the spectra with the observed ones.
  
 From these interferometric global HI profiles, we calculate the HI line Fluxes and hence the atomic masses $M_{HI}$ of these galaxies using equation \ref{hi_mass_ori}. The derived HI masses for each of the galaxies are mentioned in Table \ref{tab:hi_spectra}. The errors for HI mass here do not include the contribution due to distance uncertainty. Please note that the uncertainties mentioned for both line flux and HI mass are $3\sigma$ errors.

   \begin{table*}
       \centering
       \begin{tabular}{||c|c|c|c|c||}
       \hline
            Galaxy name & $W_{m50}/2$ (km\thinspace s$^{-1}$) & $W_{p20}/2$ (km\thinspace s$^{-1}$) & Line flux (Jy\thinspace kms$^{-1}$) & $M_{HI}$ (M$_{\sun}$) \\
            \hline
NGC 0784 & 60.7 & 61.025 & $60.5\pm0.8$ & $(4.22\pm0.06)\times10^{8}$ \\
NGC 1156 & 56.74 & 53.755 & $49.7\pm0.6$ & $(5.38\pm0.07)\times10^{8}$ \\
NGC 3027 & 110.71 & 112.365 & $80.2\pm0.9$ & $(5.14\pm0.06)\times10^{9}$ \\
NGC 3359 & 128.71 & 131.405 & $150.8\pm1.1$ & $(9.95\pm0.07)\times10^{9}$ \\
NGC 4068 & 38.025 & 38.495 & $28.4\pm0.8$ & $(1.284\pm0.035)\times10^{8}$ \\
NGC 4861 & 55.815 & 57.685 & $39.3\pm1.0$ & $(9.12\pm0.22)\times10^{8}$ \\
NGC 7292 & 45.875 & 47.545 & $19.1\pm0.5$ & $(4.13\pm0.11)\times10^{8}$ \\
NGC 7497 & 144.835 & 148.345 & $55.0\pm0.7$ & $(5.05\pm0.06)\times10^{9}$ \\
NGC 7610 & 128.23 & 132.545 & $30.4\pm0.8$ & $(1.56\pm0.04)\times10^{10}$ \\
NGC 7741 & 99.635 & 103.465 & $41.3\pm0.8$ & $(1.536\pm0.031)\times10^{9}$ \\
NGC 7800 & 108.72 & 114.045 & $37.4\pm0.9$ & $(3.69\pm0.08)\times10^{9}$ \\
\hline
       \end{tabular}
        \caption{Parameters found from the global HI spectra of the galaxies.}
       \label{tab:hi_spectra}
   \end{table*}

\subsection{Moment maps}
\label{subsec:mom_maps}
    The 3D image cubes of the galaxies produced in the analysis are the intensities as a function of position (R.A. and Dec.) and velocity.  Adding these intensities over the velocity axis gives us the integrated HI emission, i.e., the "Moment zero" maps. From the moment zero maps, we directly get the spatial distribution of the atomic gas and its column density. For optically thin media, the column density $N_{HI}$ can be directly determined  from the integrated line brightness $T(\nu)$ through the following equation:
 
     \begin{equation}
    \left( \frac{N_{HI}}{cm^{-2}} \right) = 1.82 \times 10^{18} \int \left( \frac{T(v)}{K} \right) \left( \frac{dv}{km s^{-1}} \right)
    \label{equ:column_den}
    \end{equation}
 
  In the radio frequency limit, the  brightness temperature of the system is proportional to the flux density for a fixed wavelength ( Rayleigh-Jeans law). So, from the moment zero map and using equation  \ref{equ:column_den}, the column density distribution of the galaxies can be derived through the equation given below:
 
     \begin{equation}
        \left(\frac{N_{HI}}{cm^{-2}} \right) = \frac{2.224 \times 10^{21} }{(\frac{\nu_0}{GHz})^2 \theta_{maj} \theta_{minor}} \left(\frac{I_{HI}}{mJy/beam-km/s}\right)
        \label{equ:clmnden}
    \end{equation}
    
   Where $\nu_o$ is the rest frequency of HI, $\theta_{maj}$ and $\theta_{minor}$ are beam sizes expressed in arc-second and $I_{HI}$ is the integrated flux from moment zero map.  Using equation \ref{equ:clmnden}, we derived the column density distribution of the sample galaxies and saw that for all the sources, the HI column density roughly varies from $\sim 10^{19}$ cm$^{-2}$ to $\sim 10^{21.5}$ cm$^{-2}$.

    The Moment maps of the galaxies are computed using the MOMNT task of AIPS.  To minimize the contribution from noise in the moment zero map, we apply a threshold cutoff to the flux to select pixels with high significance. This threshold was chosen on a trial and error basis as we obtain a clean MOM0 map without stray fluxes in the outer region where emission of the galaxy is not expected. The Moment one maps i.e., the intensity-weighted average of the velocity and the Moment two maps  i.e., the intensity-weighted dispersion map  of the galaxies are created in the same way, i.e., using the MOMNT task of the AIPS.

    Moment one map gives an estimate of the line of sight rotation velocity of the galaxies at each pixel. Thus, the galaxies' velocity map gives us a first estimate of the overall kinematics. However, the galaxies' detailed kinematics can only be derived by fitting the cube with models of the galaxies, which incorporates the inclination and position angle of the galaxies and channel by channel distribution of intensity. These are discussed in Paper-II. 
    
     Moment two maps represents the dispersion of the HI gas at each pixel of the galaxies. The velocity dispersion represented by Moment two maps include the effect of rotational gradient (within a beam), local non-circular motion, and intrinsic gas velocity dispersion. However, a more sophisticated method to determine the velocity dispersion is discussed in detail later in this paper. The moment maps of the galaxies are shown in the figure \ref{fig:mom_maps}.  In the left panel, the moment zero contours are plotted over the optical images of the galaxies. We used the highest resolution data cubes to produce these moment maps.  From these images, we can clearly see that HI in galaxies extends much further than the optical disks. Hence it serves as a powerful tracer to probe galaxy dynamics up to large radii.
    
    \begin{figure*}
    \centering
    \begin{tabular}{cccc}
        \includegraphics[height=4cm]{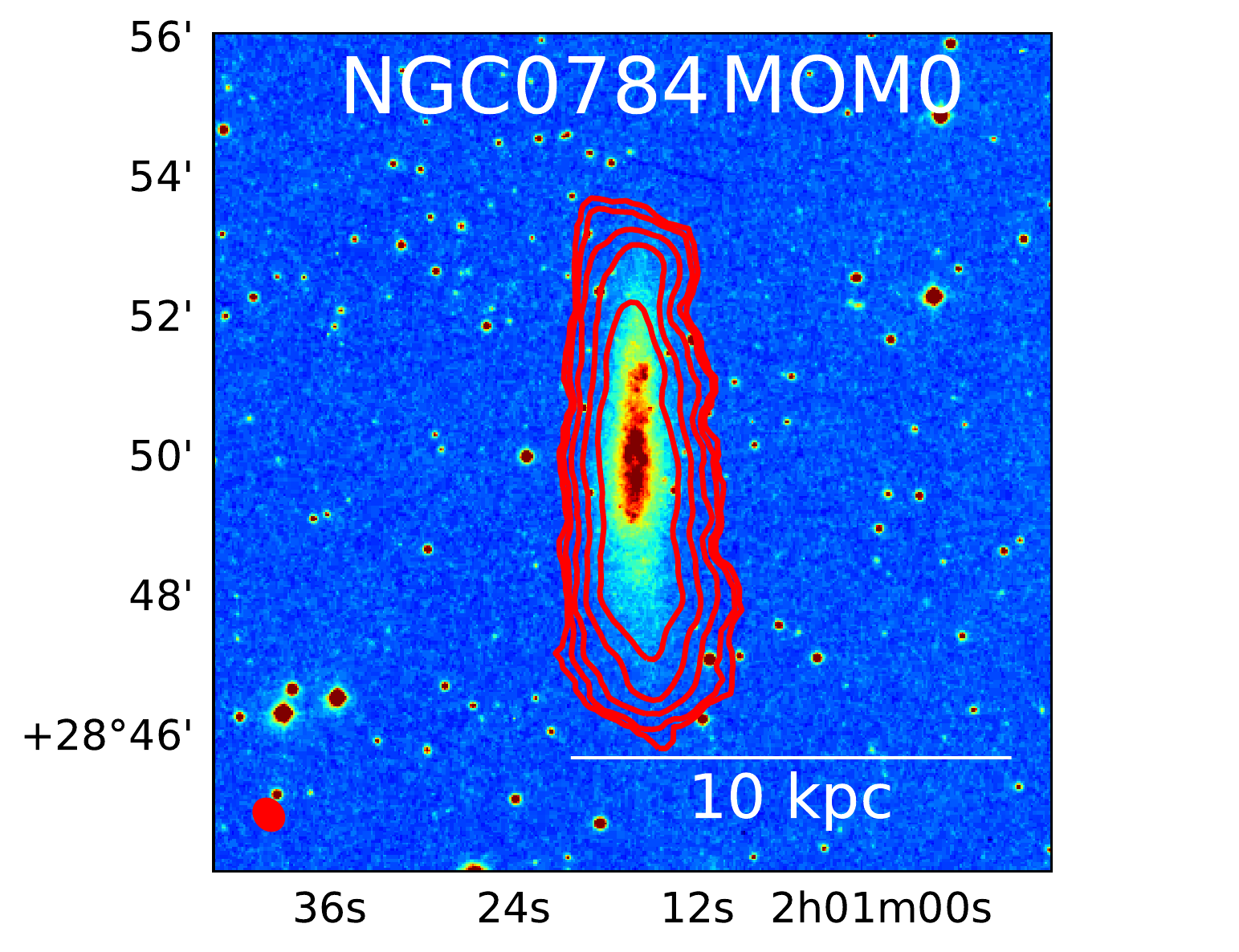} & 
       \includegraphics[height=3.8cm]{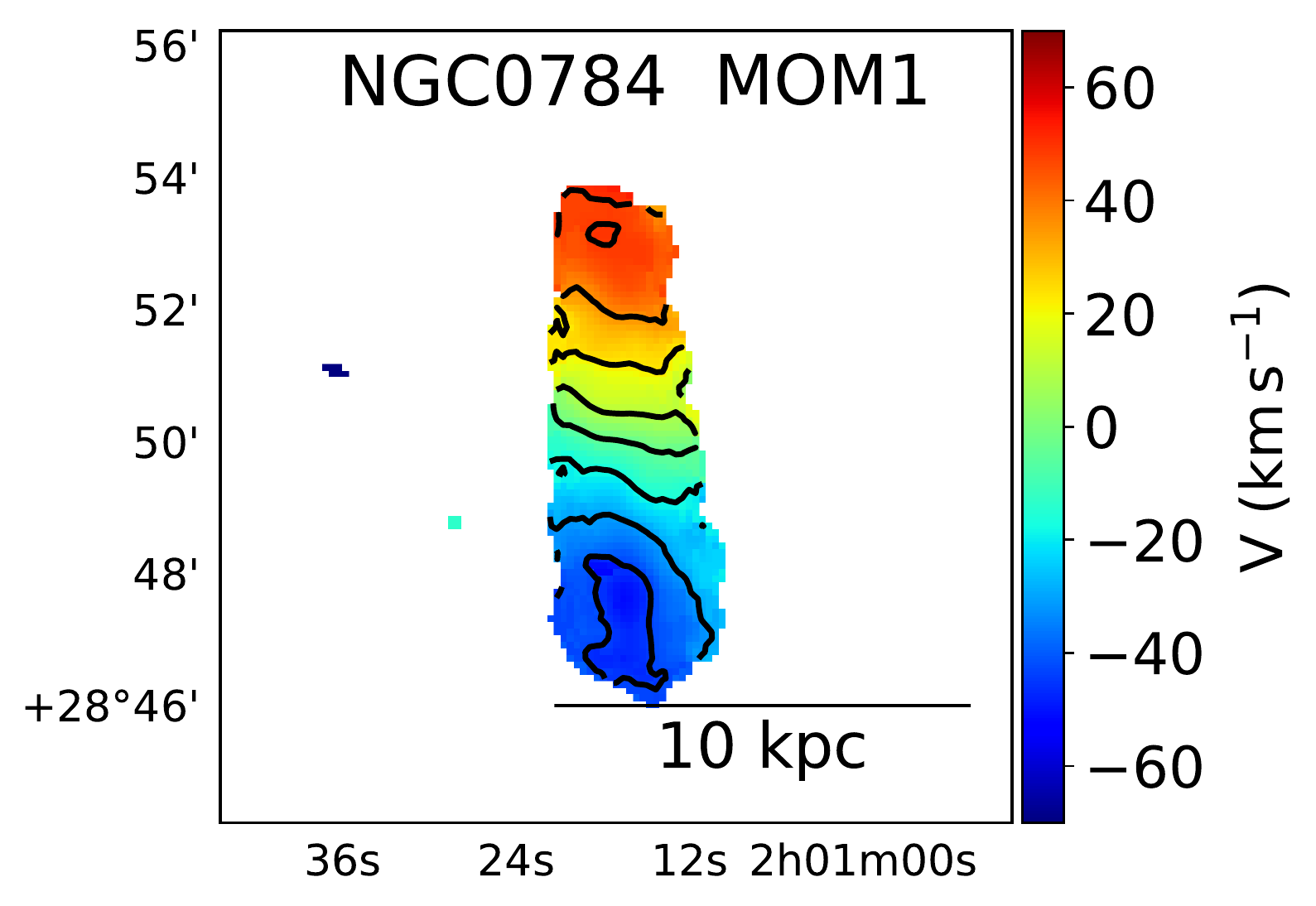} &
       \includegraphics[height=3.8cm]{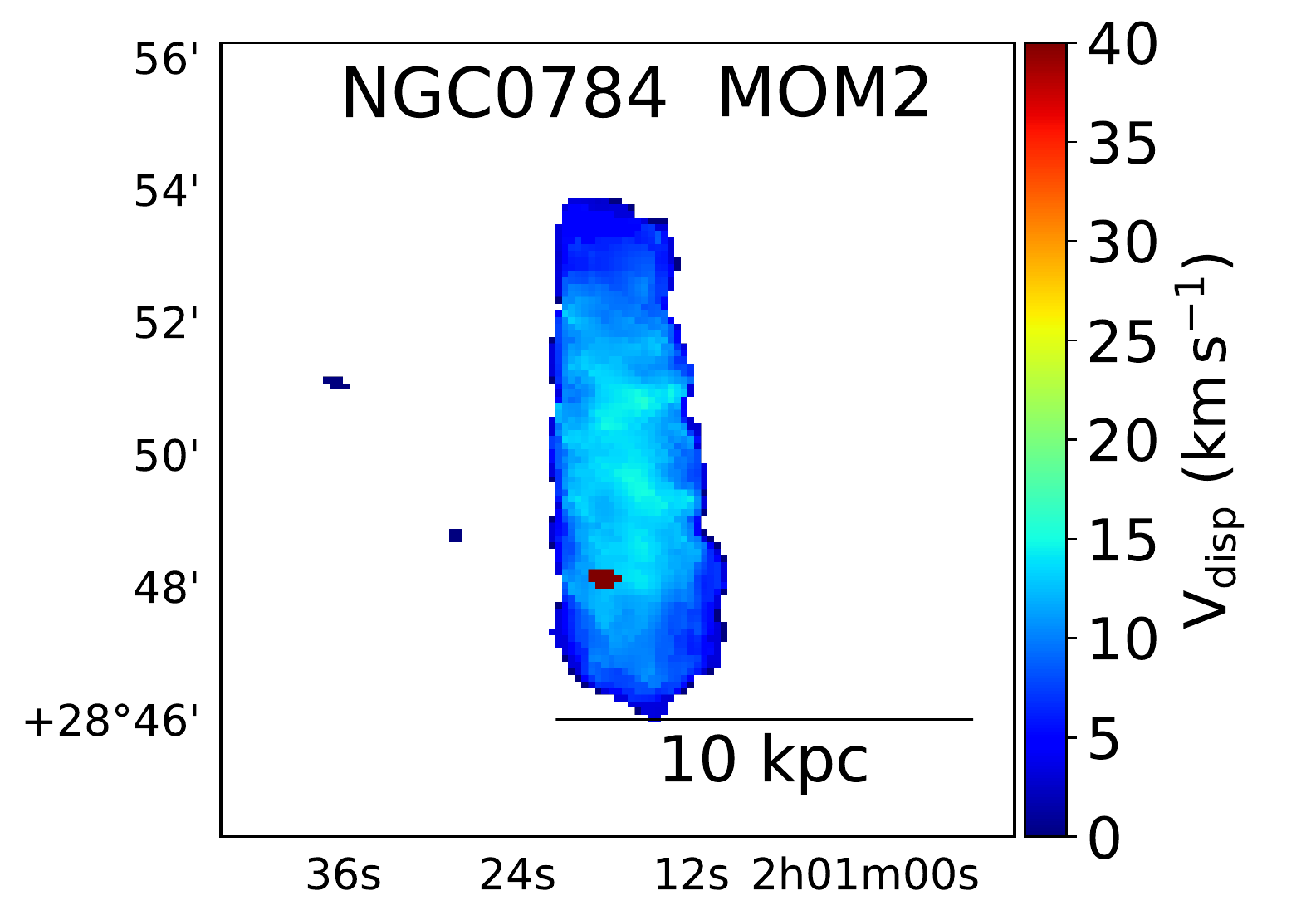} \\
    
       \includegraphics[height=4cm]{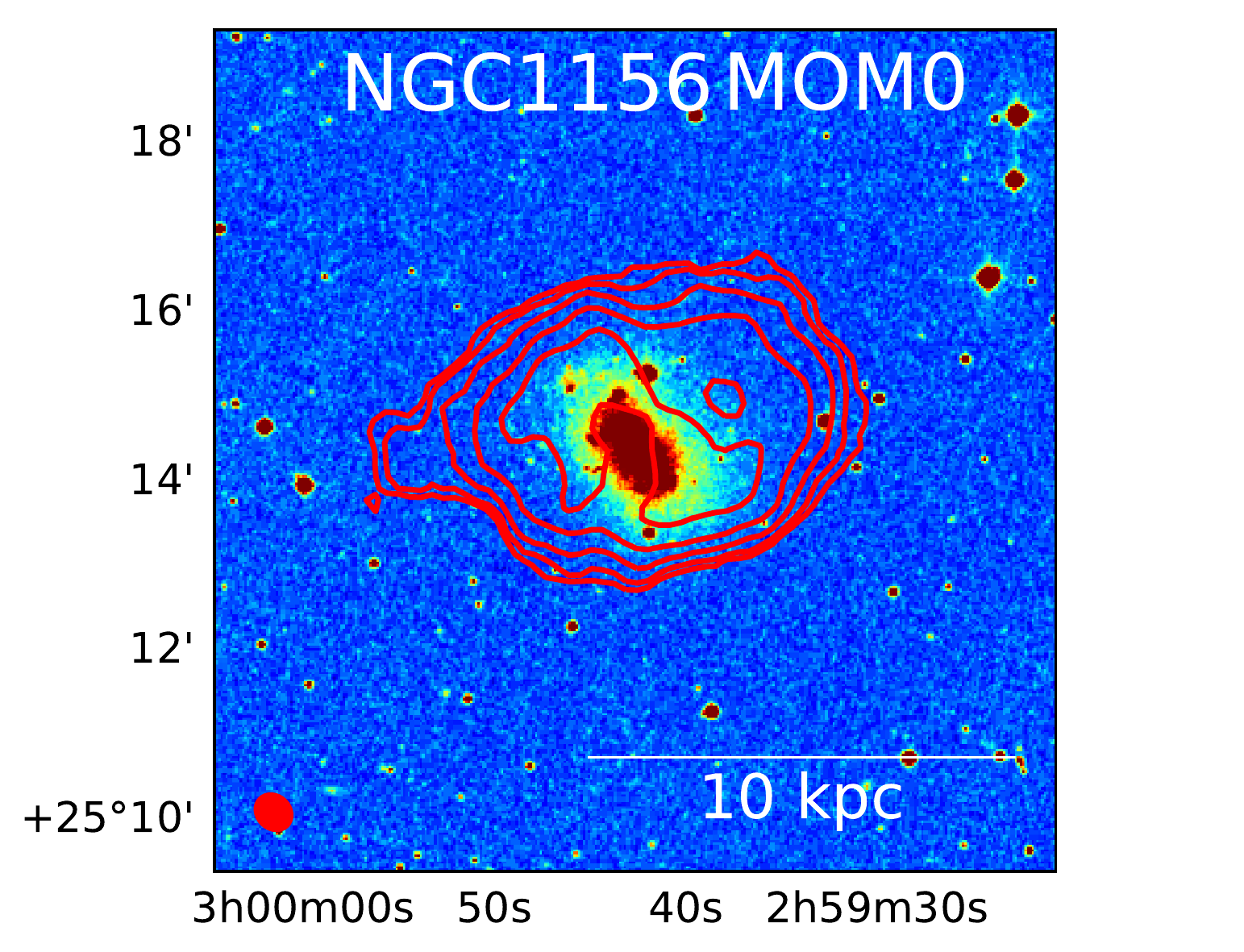} & 
       \includegraphics[height=3.8cm]{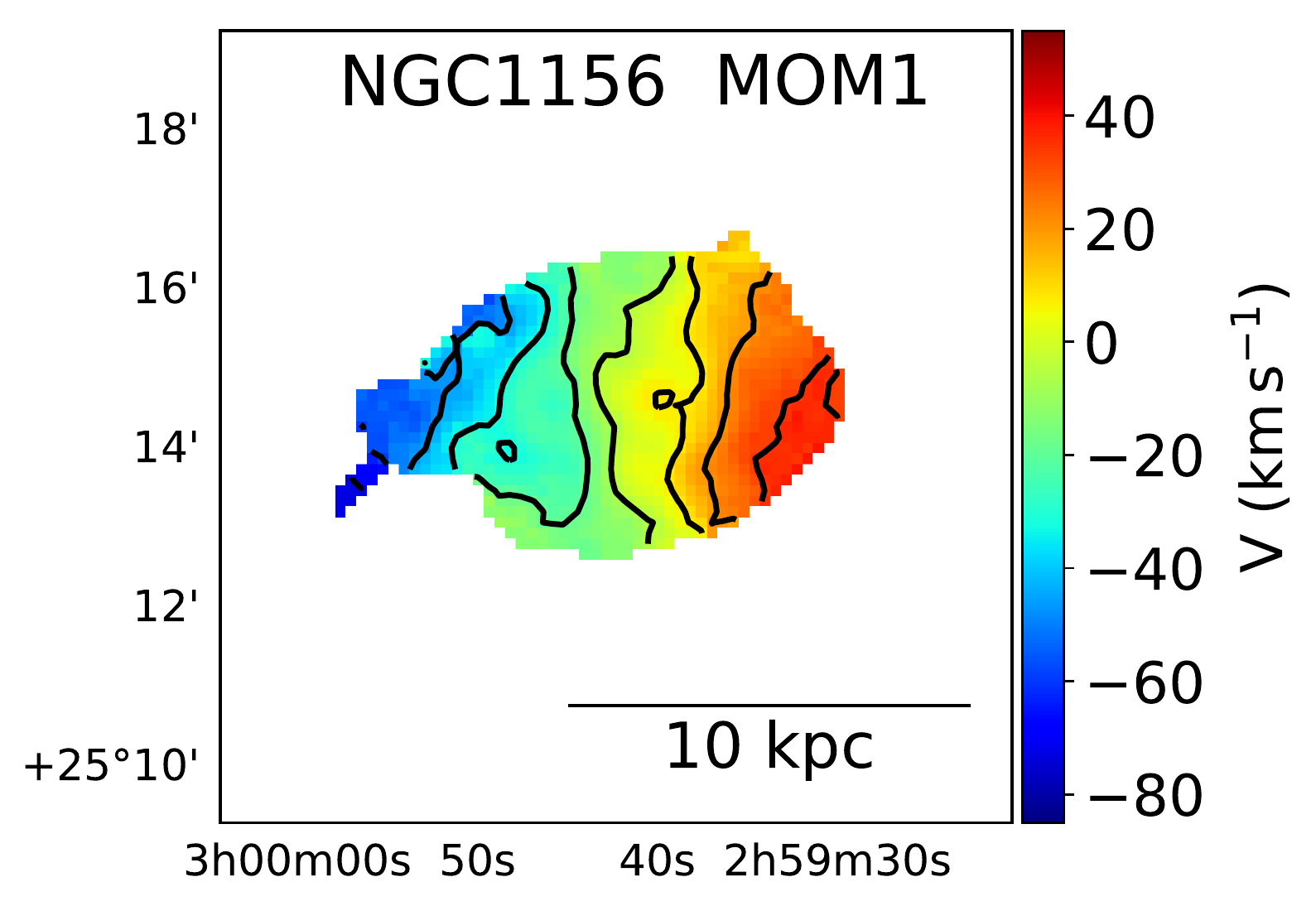} &
       \includegraphics[height=3.8cm]{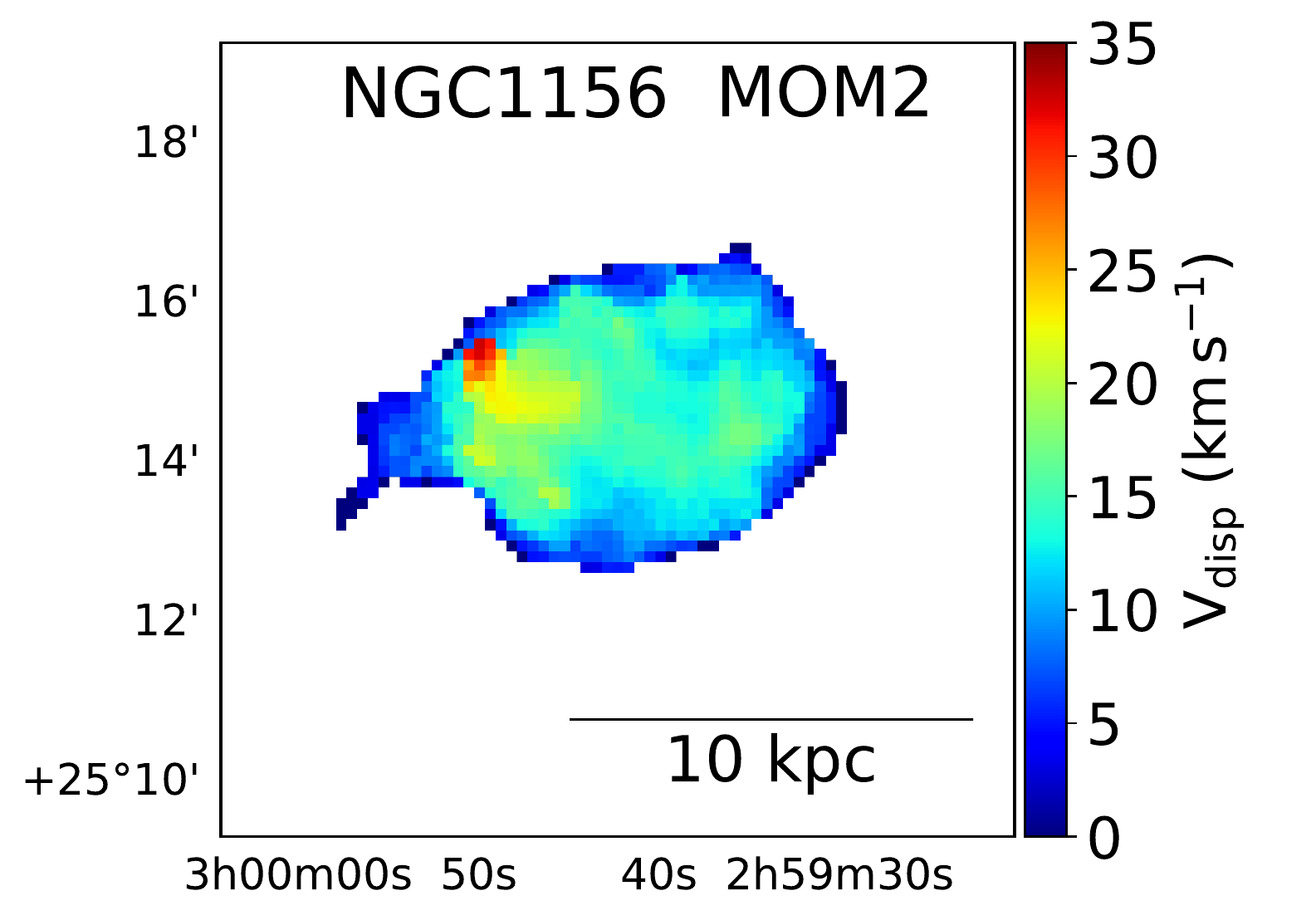} \\
       
       \includegraphics[height=4cm]{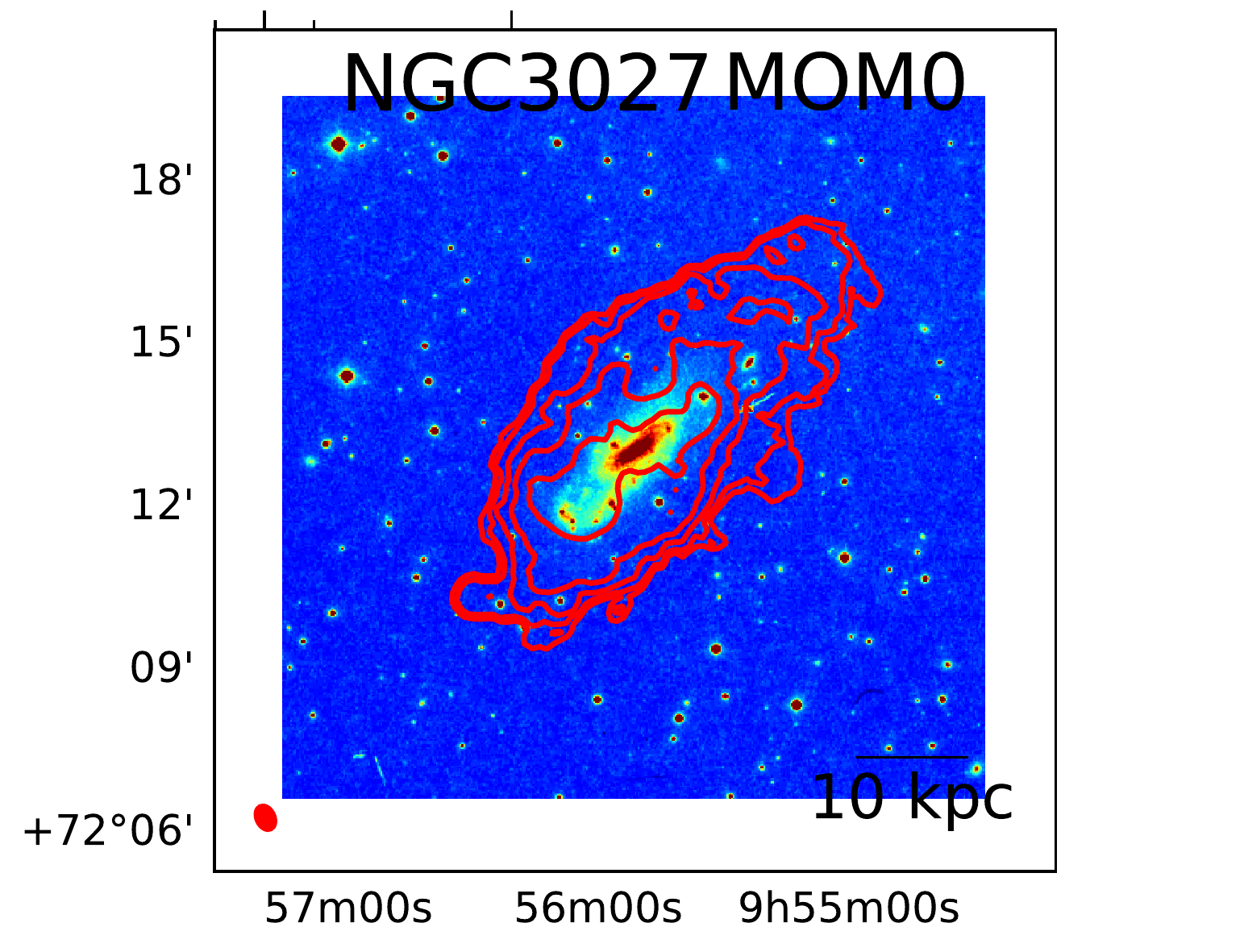} & 
       \includegraphics[height=3.8cm]{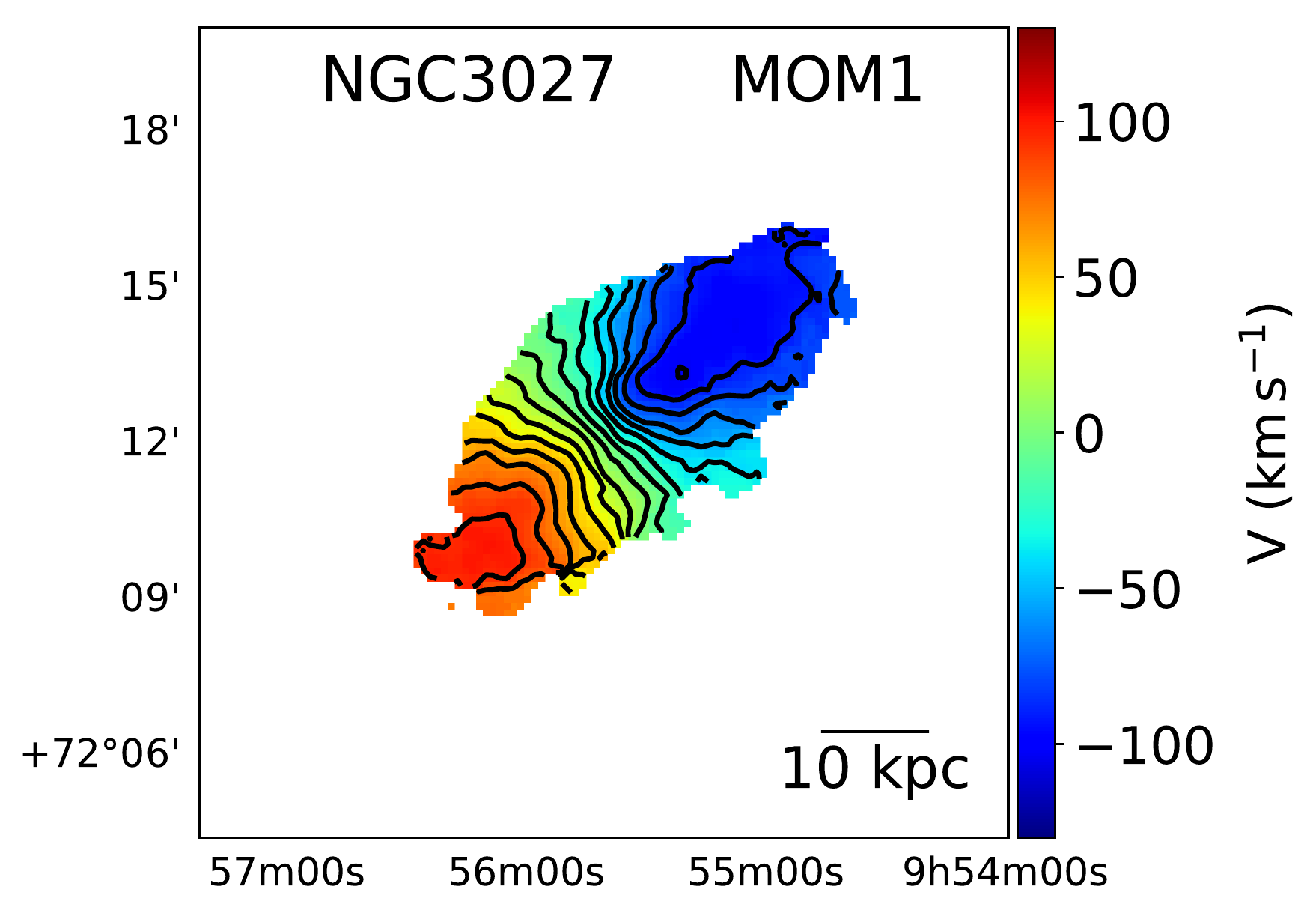} &
       \includegraphics[height=3.8cm]{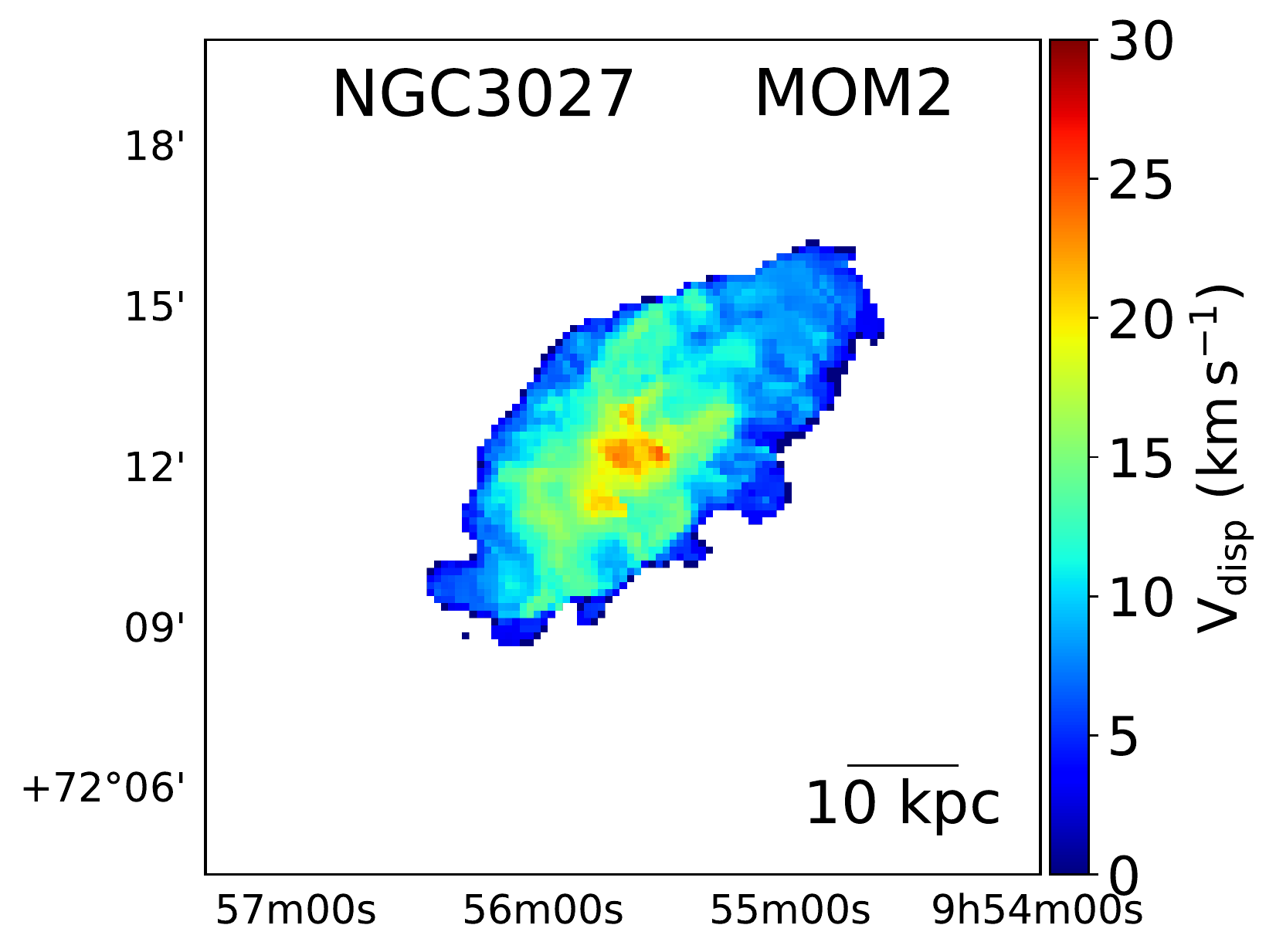} \\
       
       \includegraphics[height=4cm]{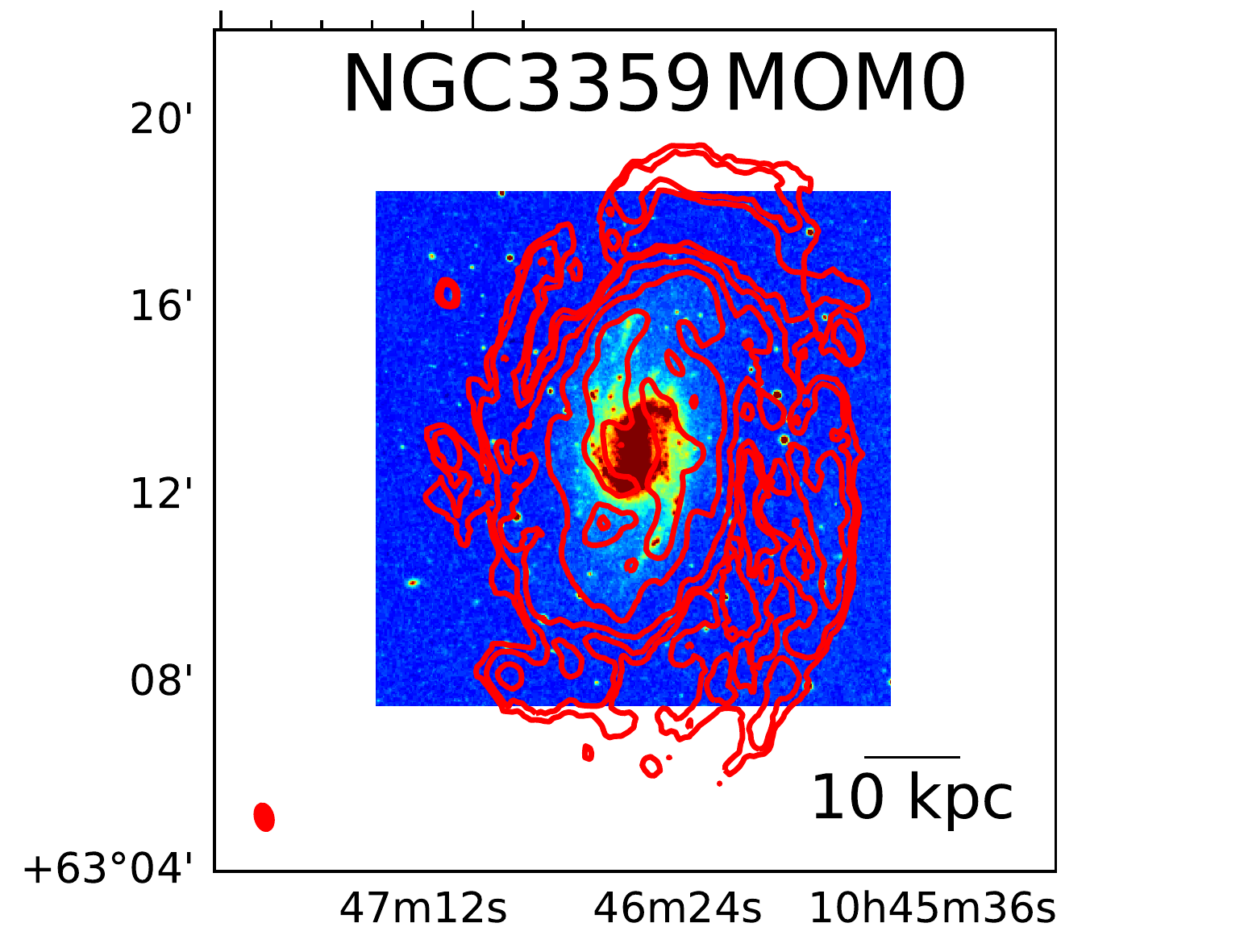} & 
       \includegraphics[height=3.8cm]{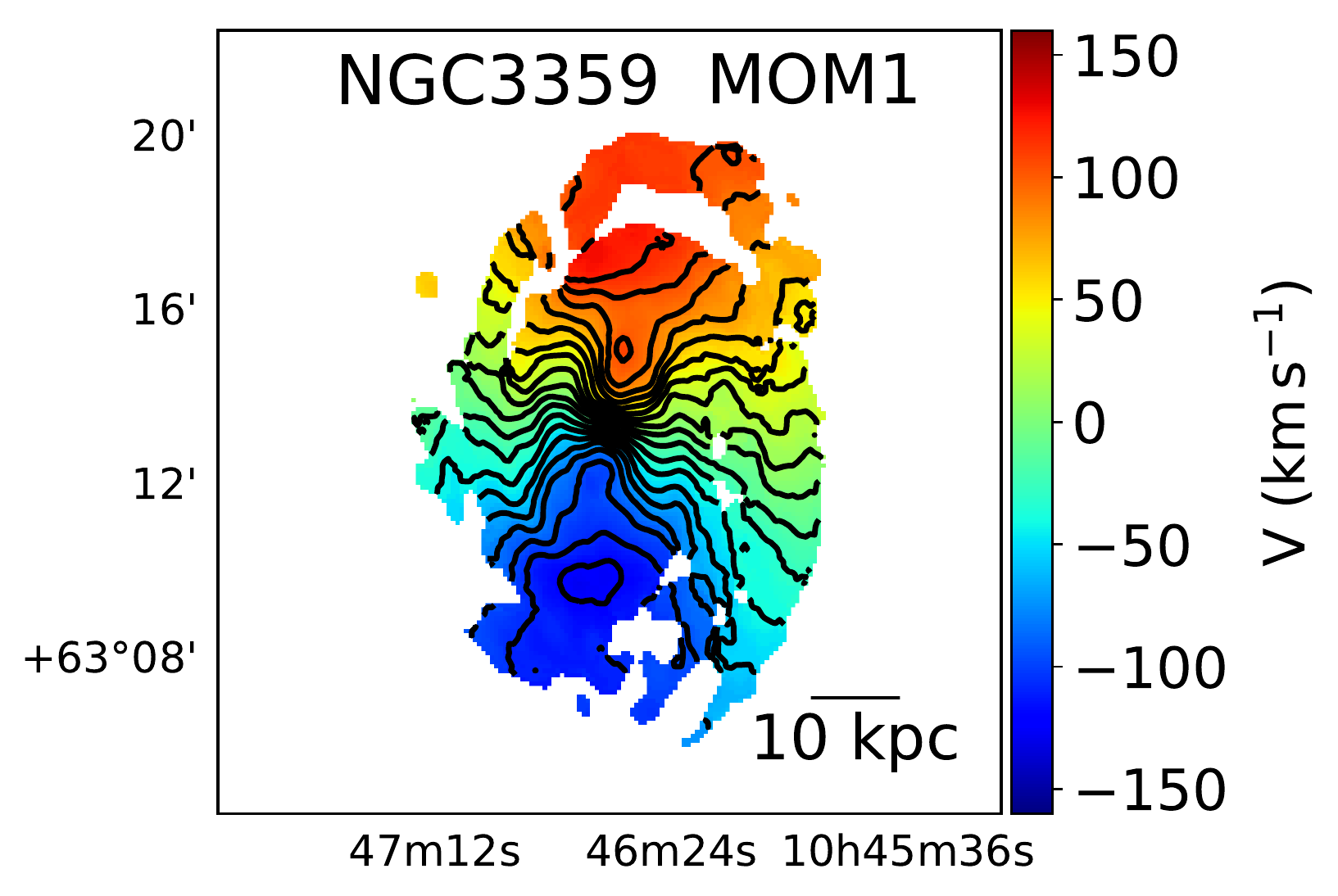} &
       \includegraphics[height=3.8cm]{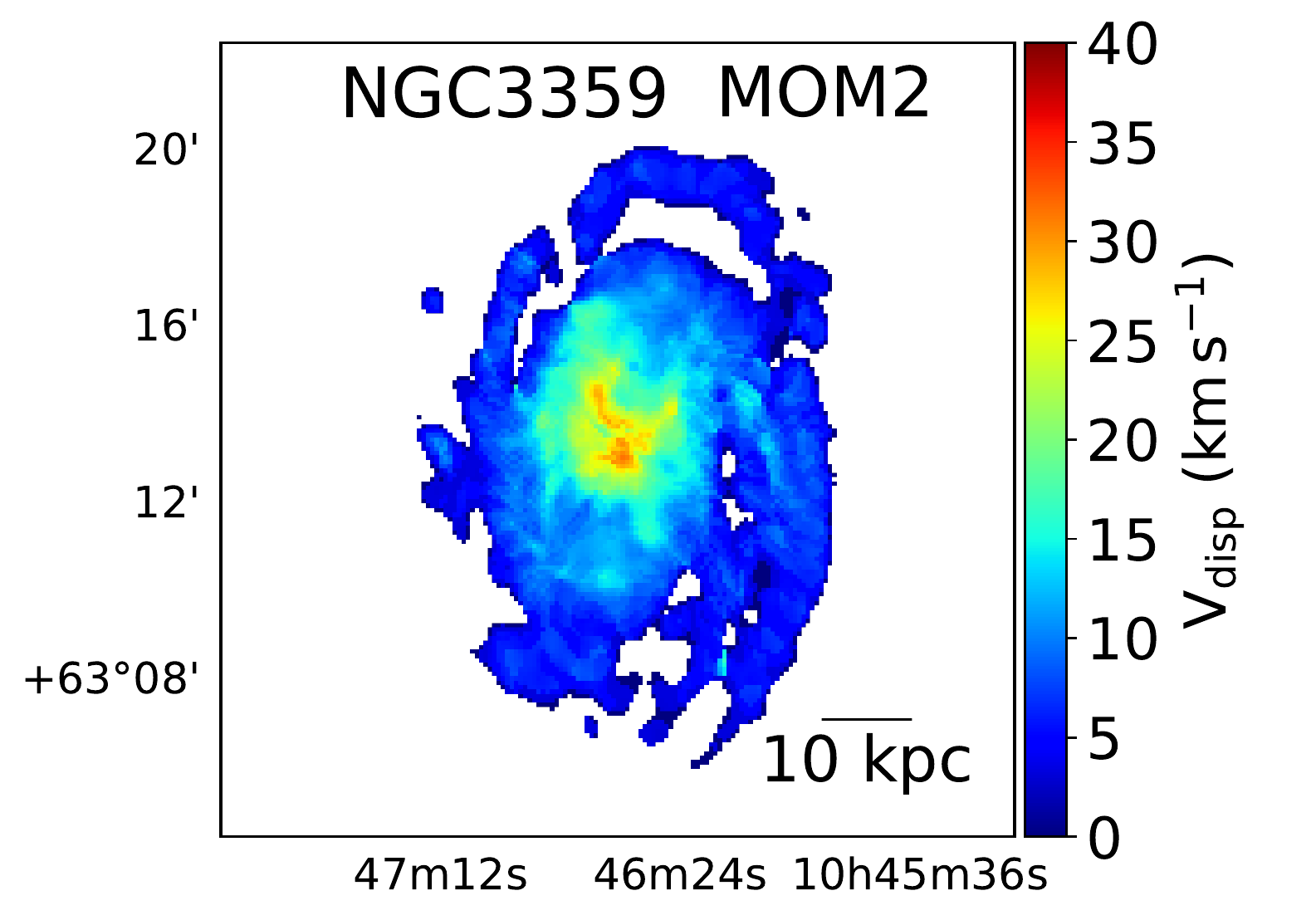} \\
       
       \includegraphics[height=4cm]{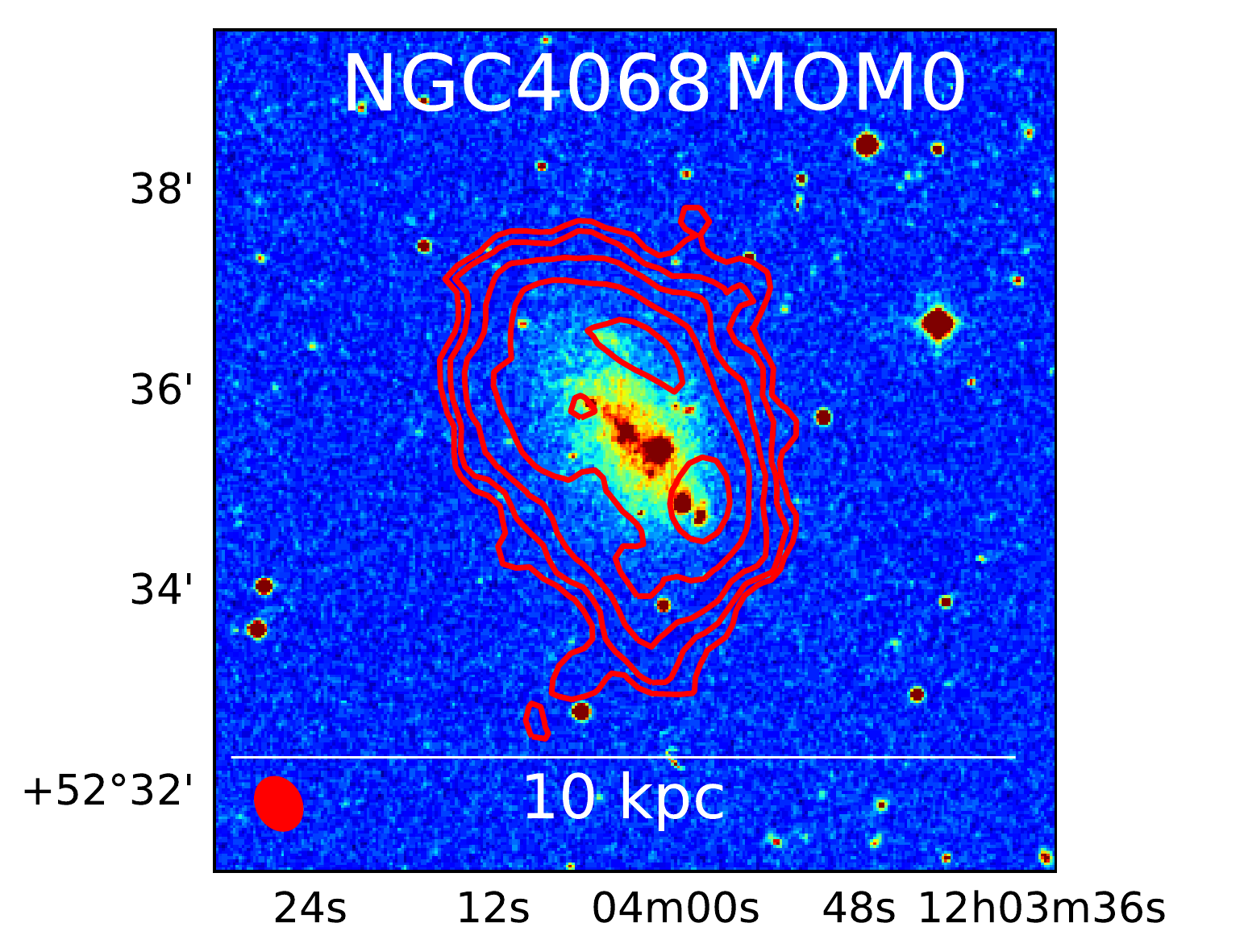} & 
       \includegraphics[height=3.8cm]{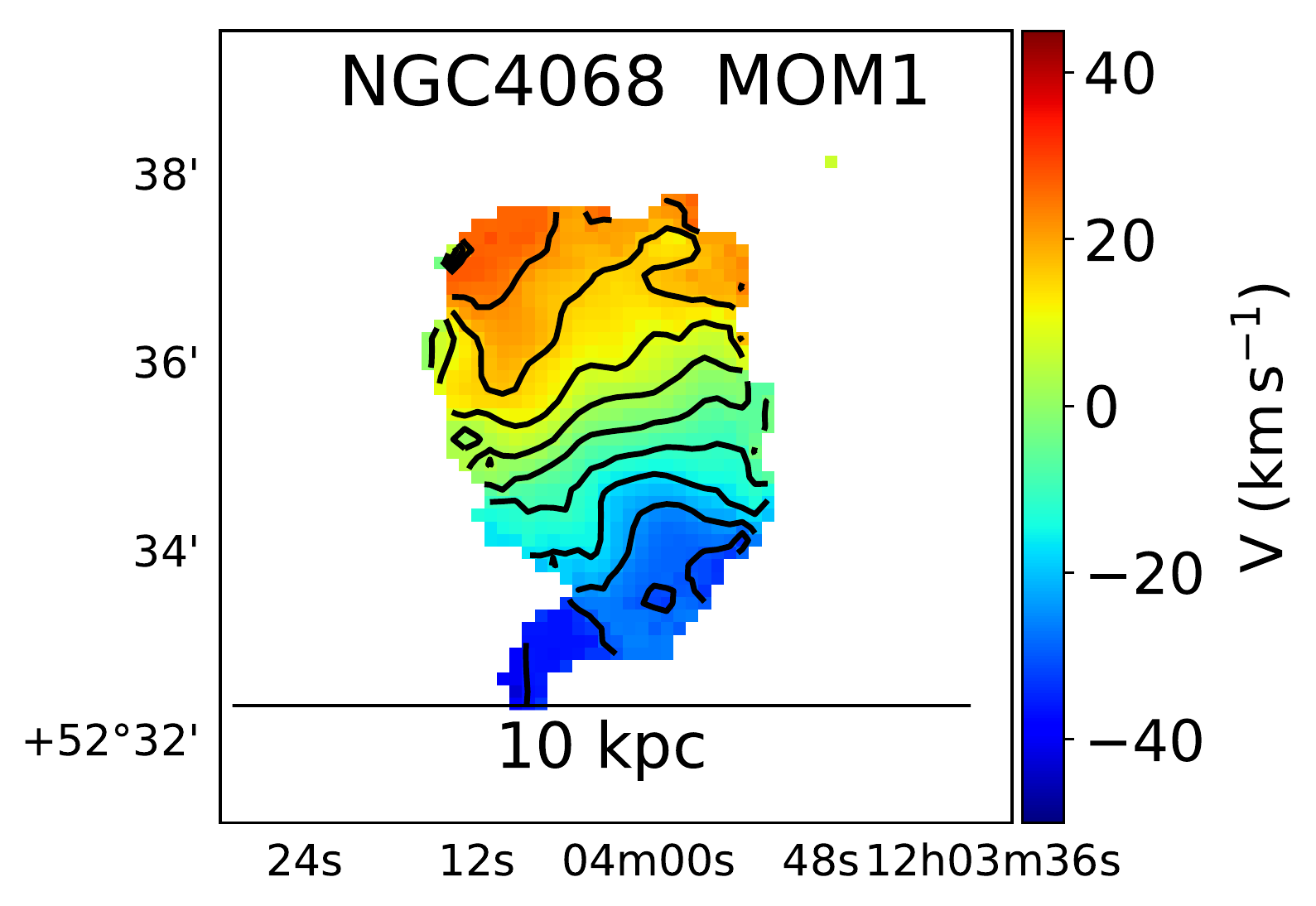} &
       \includegraphics[height=3.8cm]{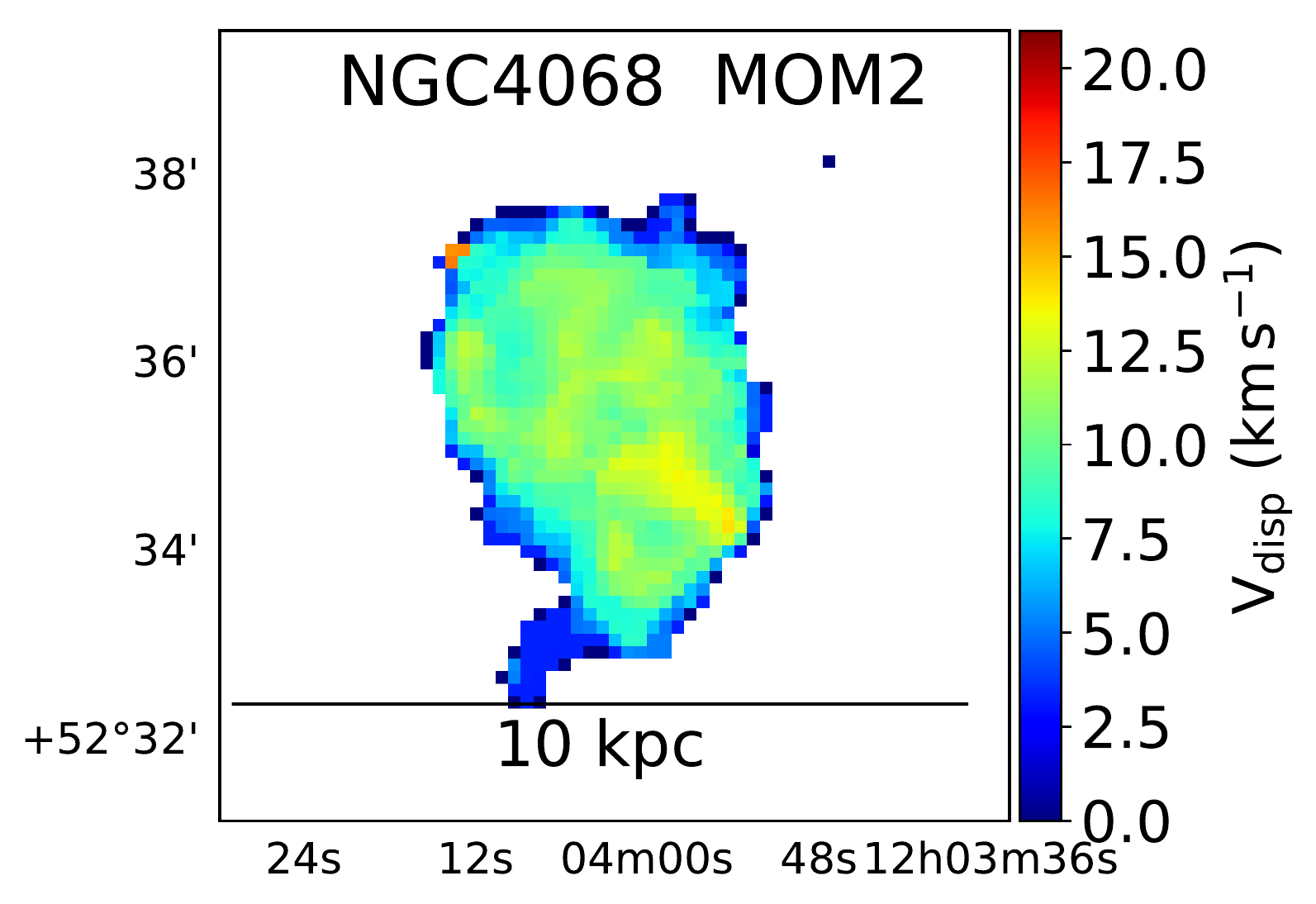} \\

     \end{tabular}
    \caption{Moment maps of the galaxies. The first, second and the third columns of the images show the Moment zero, Moment one and Moment two maps, respectively. Moment zero maps:  The contours of the images obtained from our analysis are plotted over the optical images of the respective galaxies. The contours represent the column density $[1, 2, 4, 8, 16]\times10^{20}$ cm$^{-2}$. The red patch at the bottom left corner represents the synthesised beam size and shape. Optical images of the galaxies are taken from The Digitized Sky Survey(DSS). \emph{(cont.)}}
    \label{fig:mom_maps}
\end{figure*}

 \begin{figure*}
 \ContinuedFloat
    \centering
    \begin{tabular}{cccc}
       \includegraphics[height=4cm]{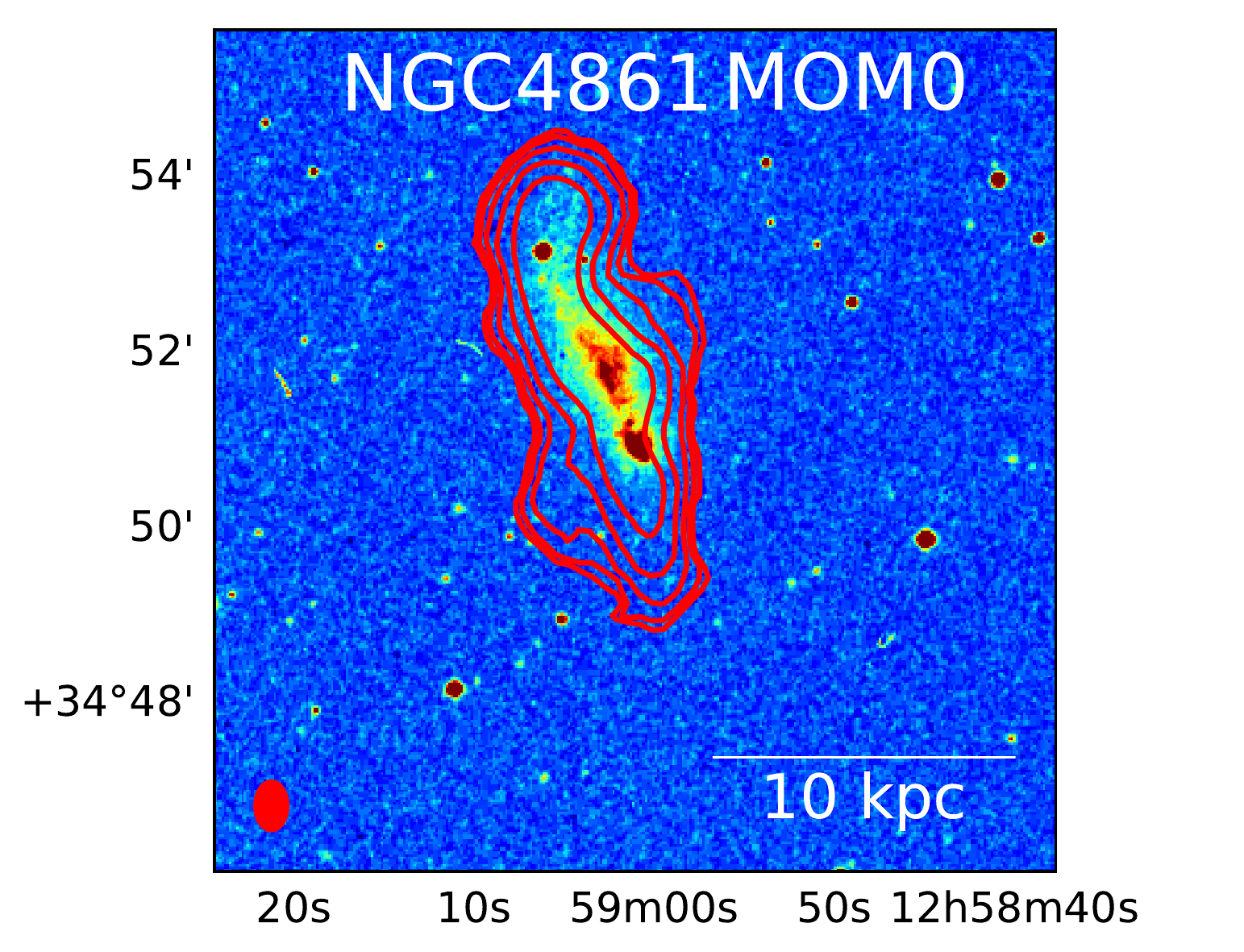} & 
       \includegraphics[height=3.8cm]{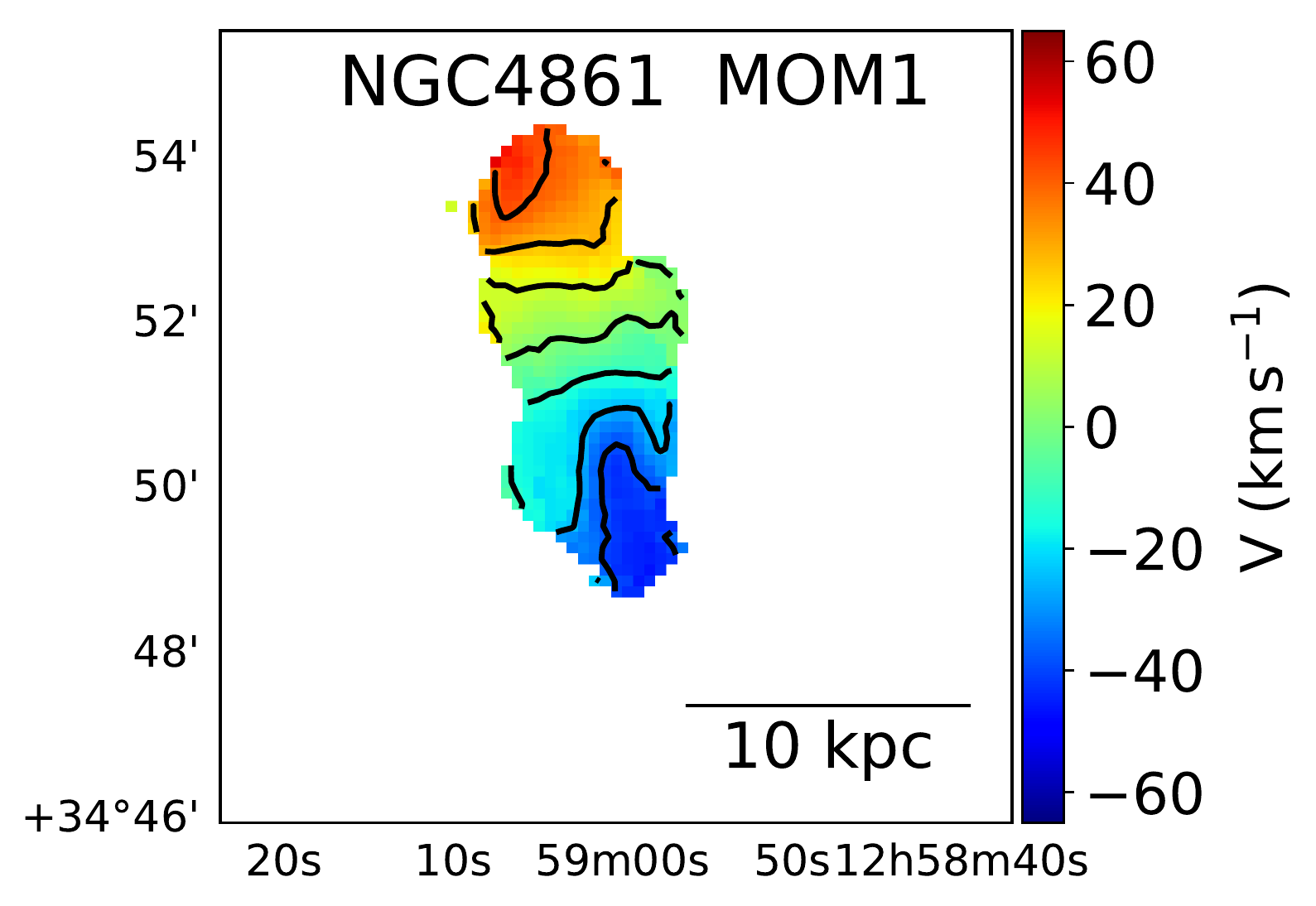} &
       \includegraphics[height=3.8cm]{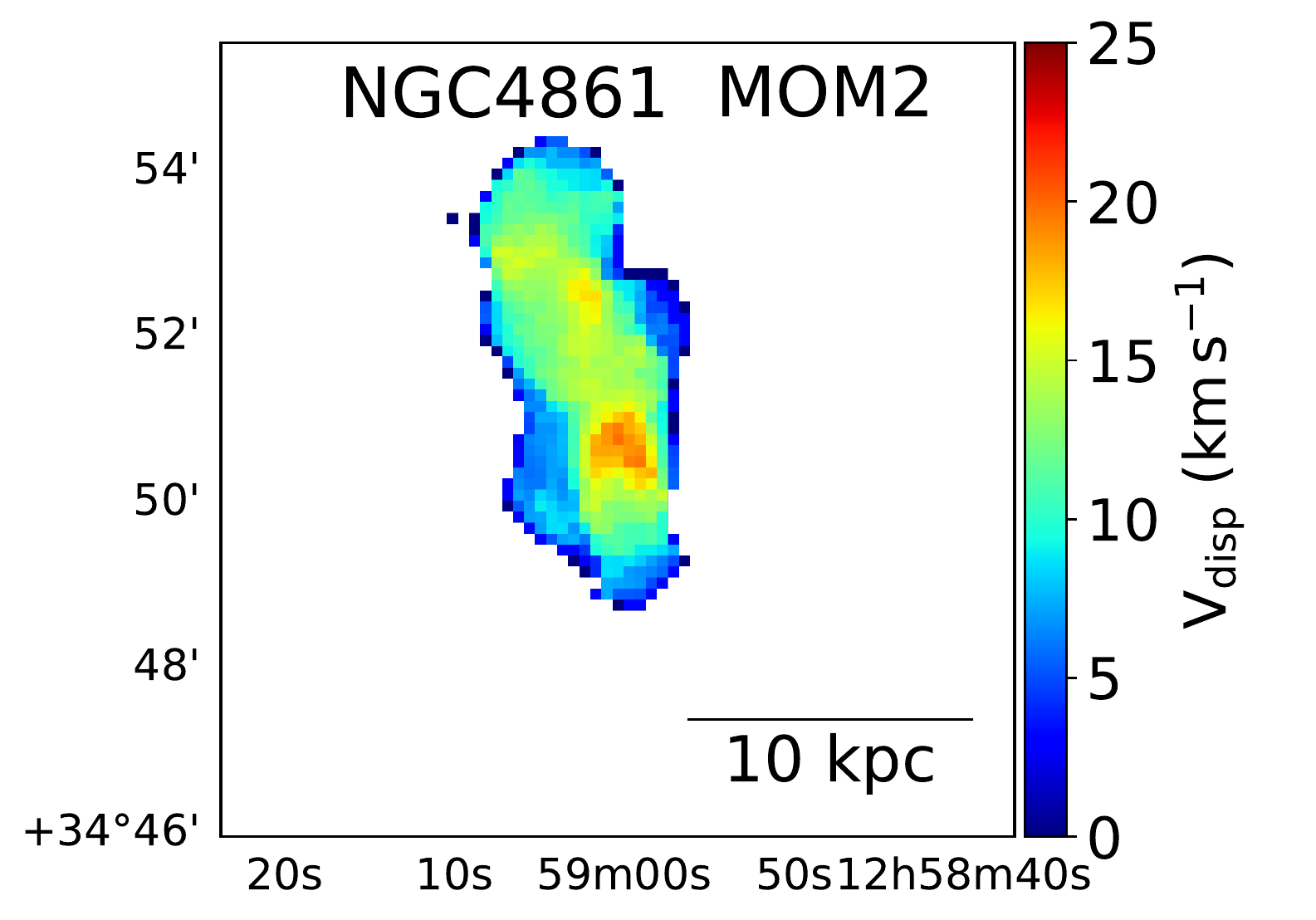} \\
       
       \includegraphics[height=4cm]{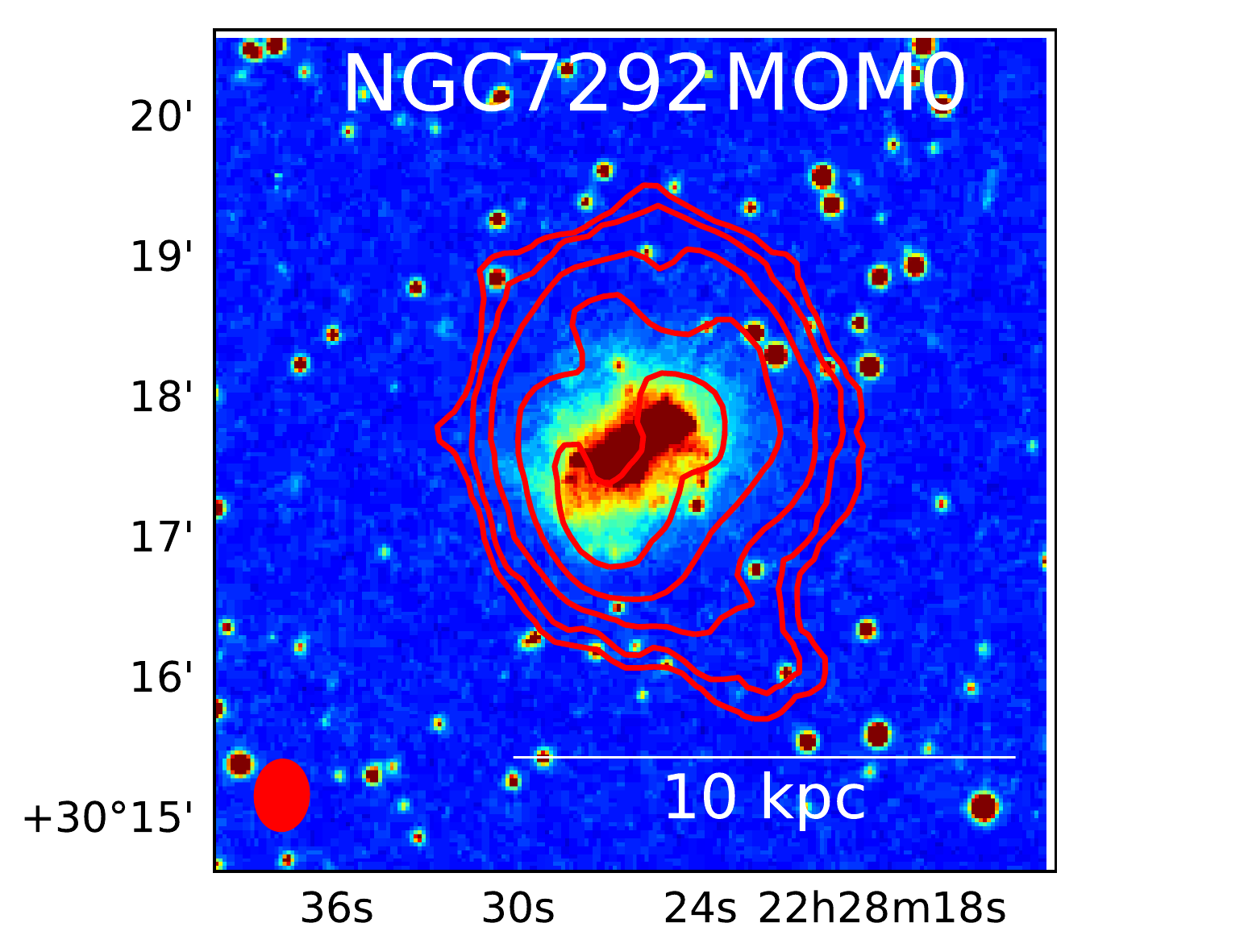} & 
       \includegraphics[height=3.8cm]{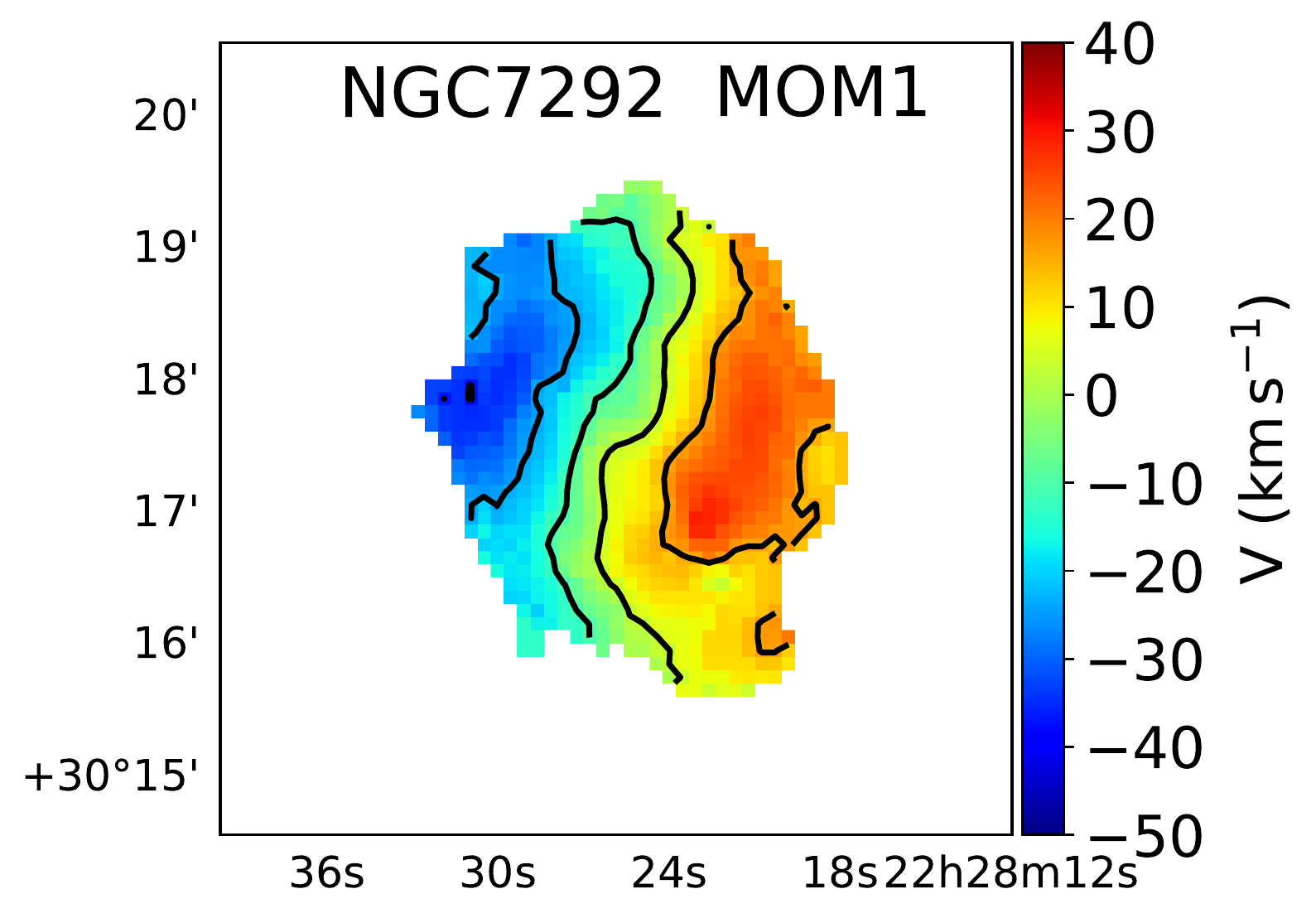} &
       \includegraphics[height=3.8cm]{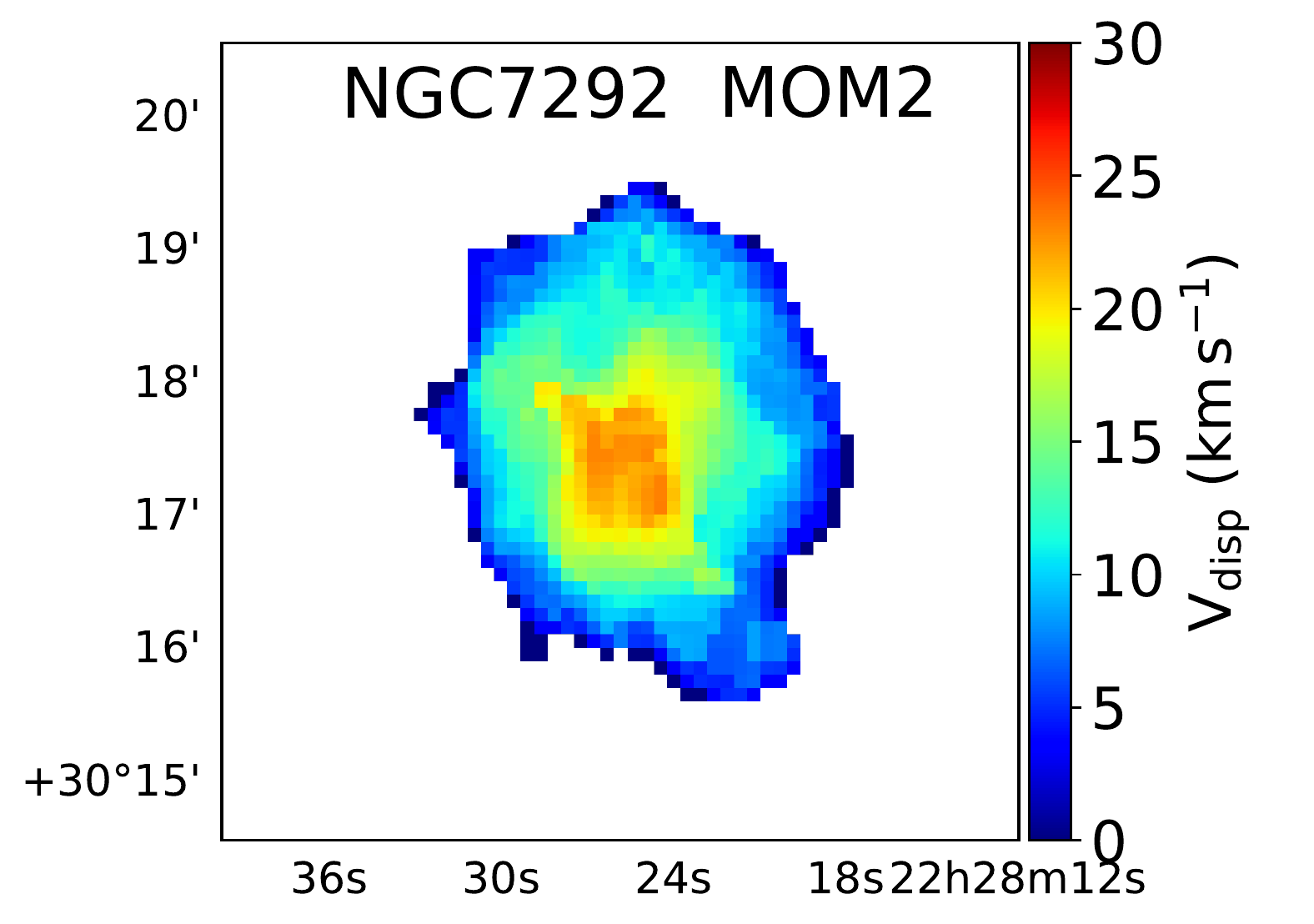} \\
       
       \includegraphics[height=4cm]{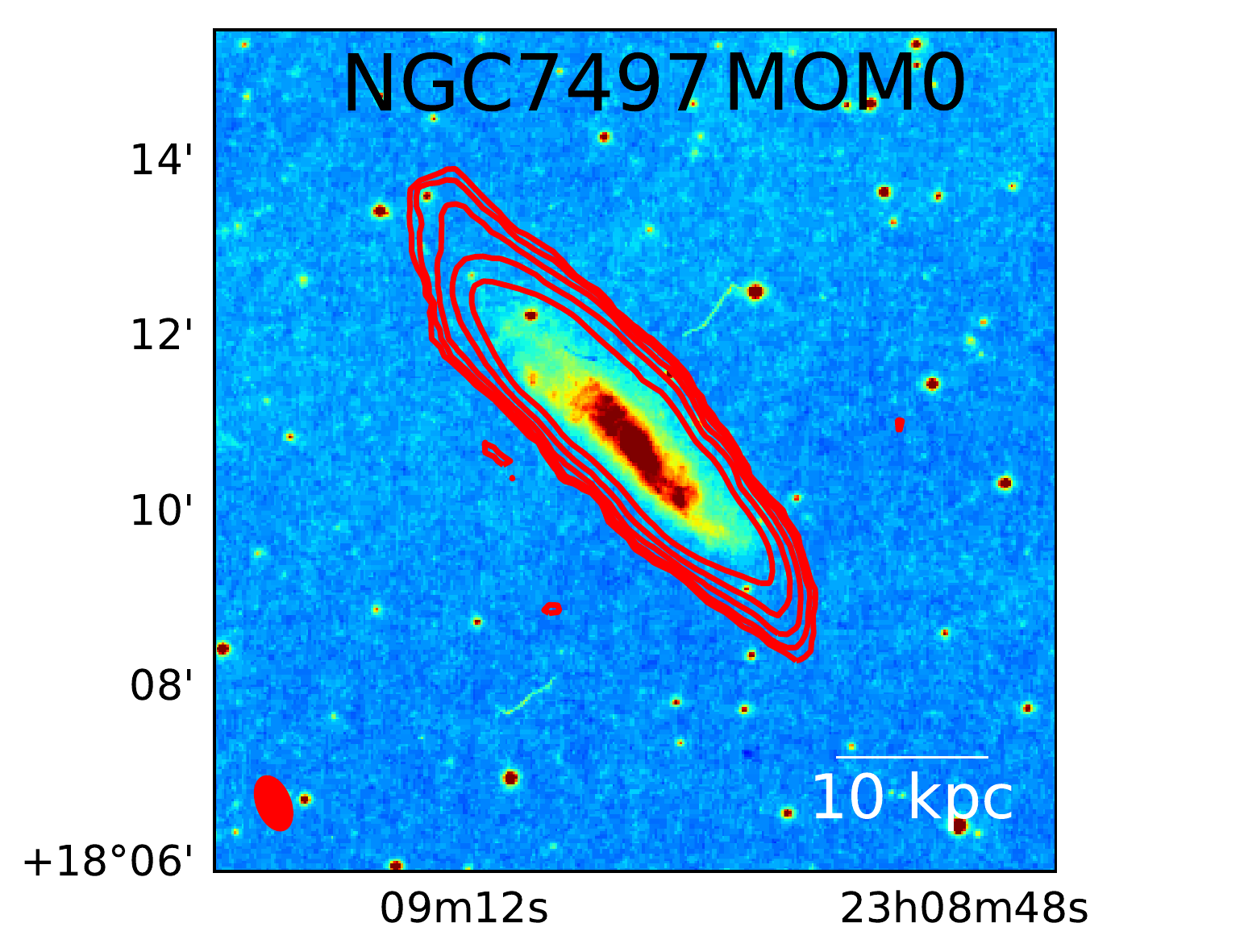} & 
       \includegraphics[height=3.8cm]{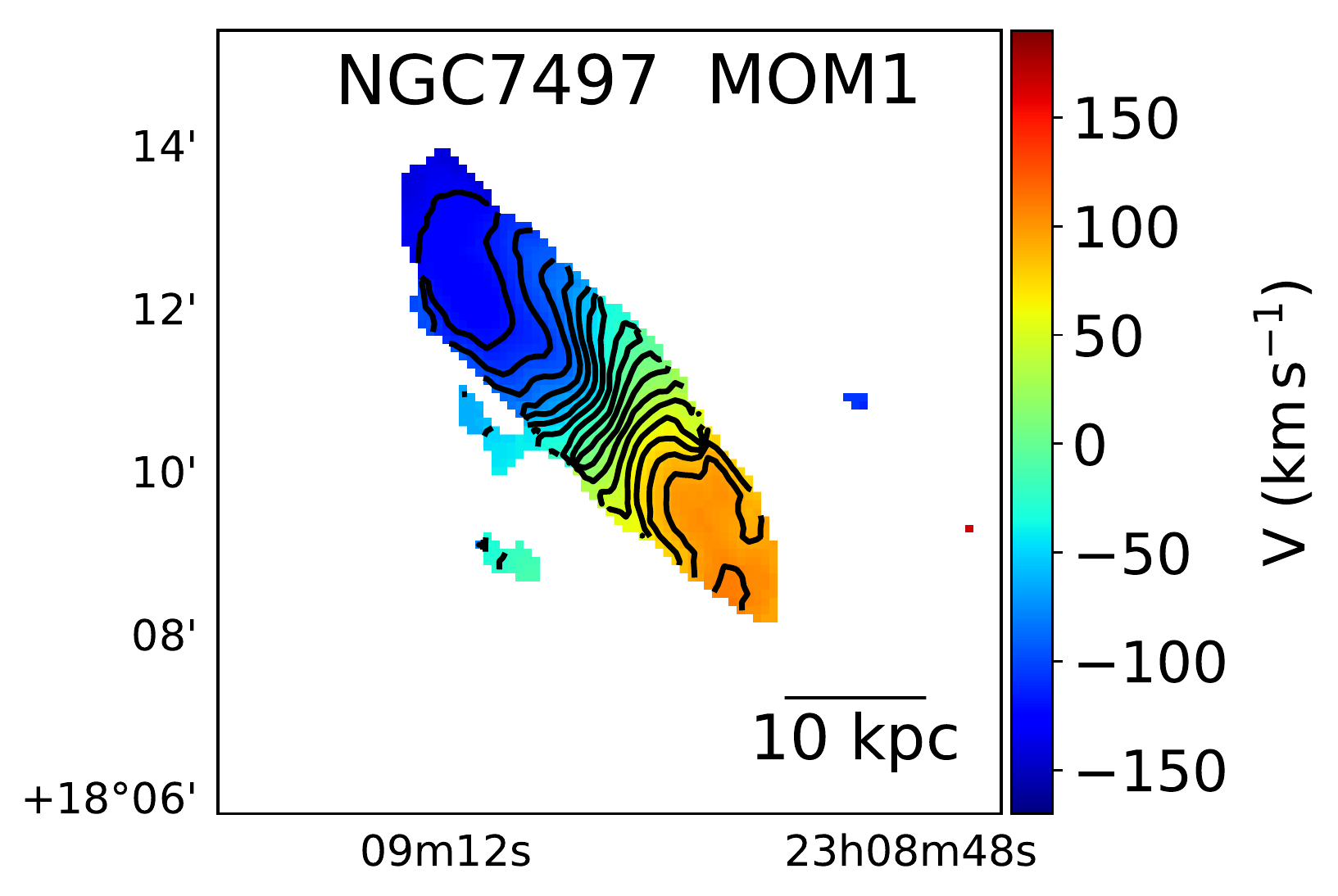} &
       \includegraphics[height=3.8cm]{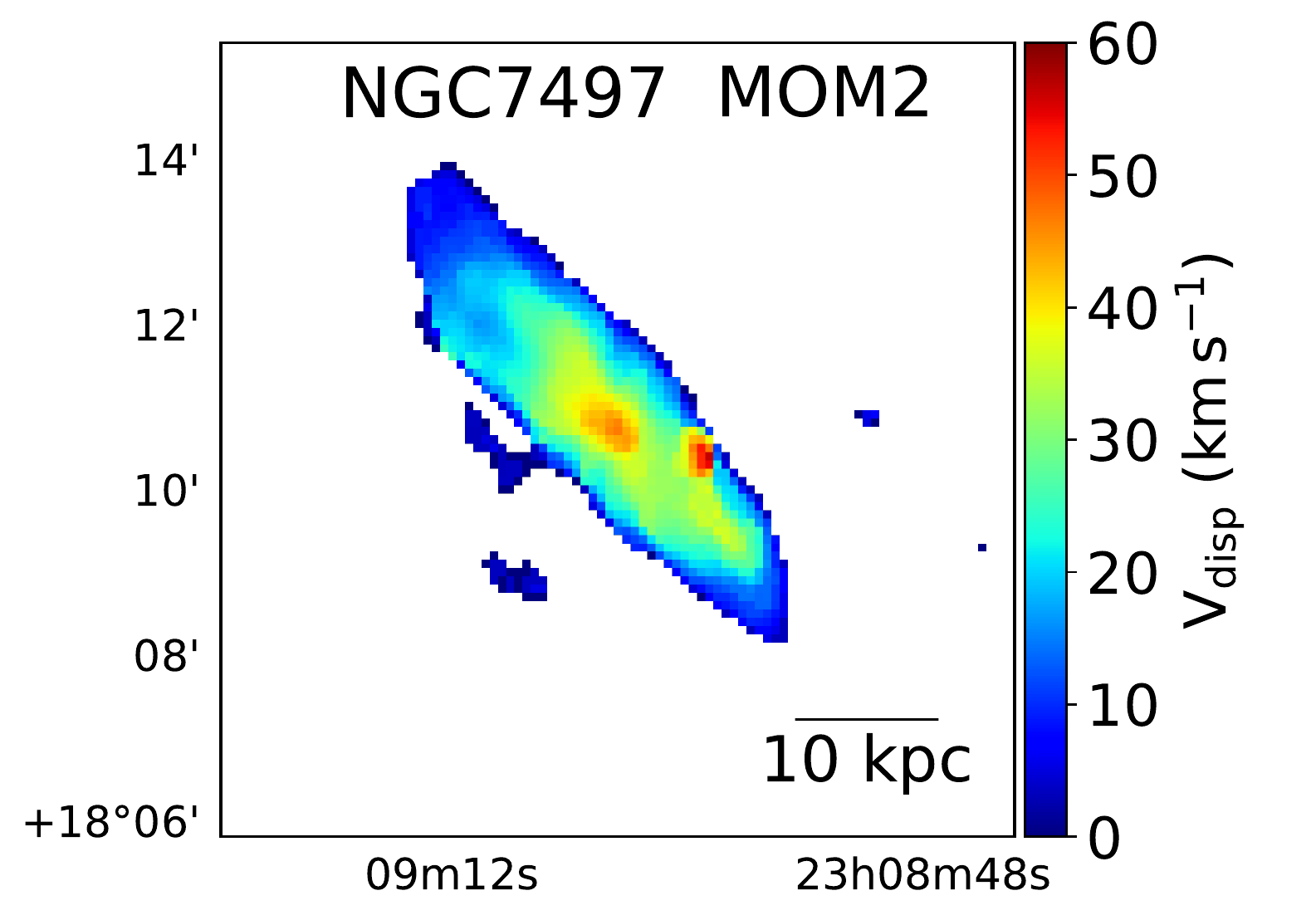} \\
  
       \includegraphics[height=4cm]{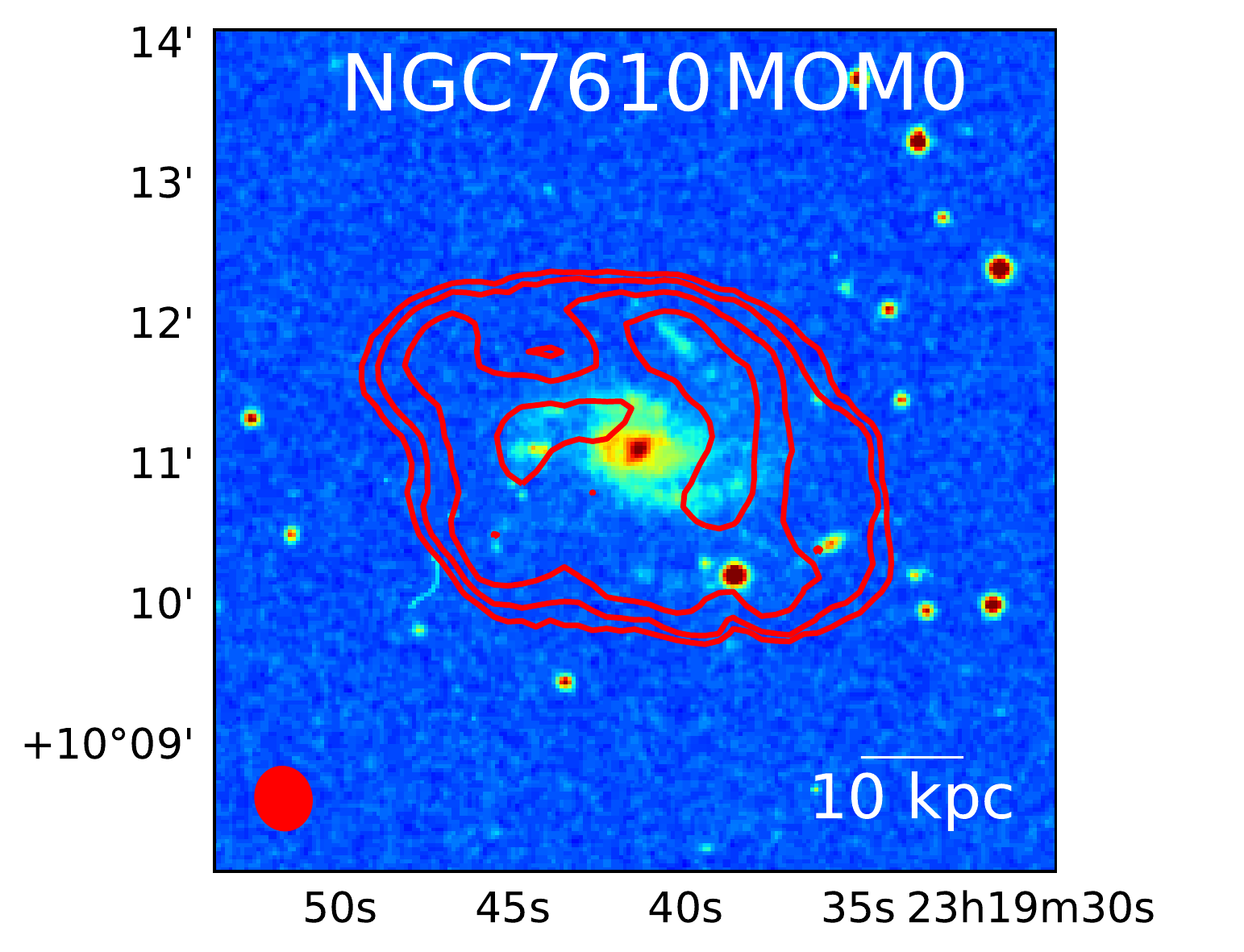} & 
       \includegraphics[height=3.8cm]{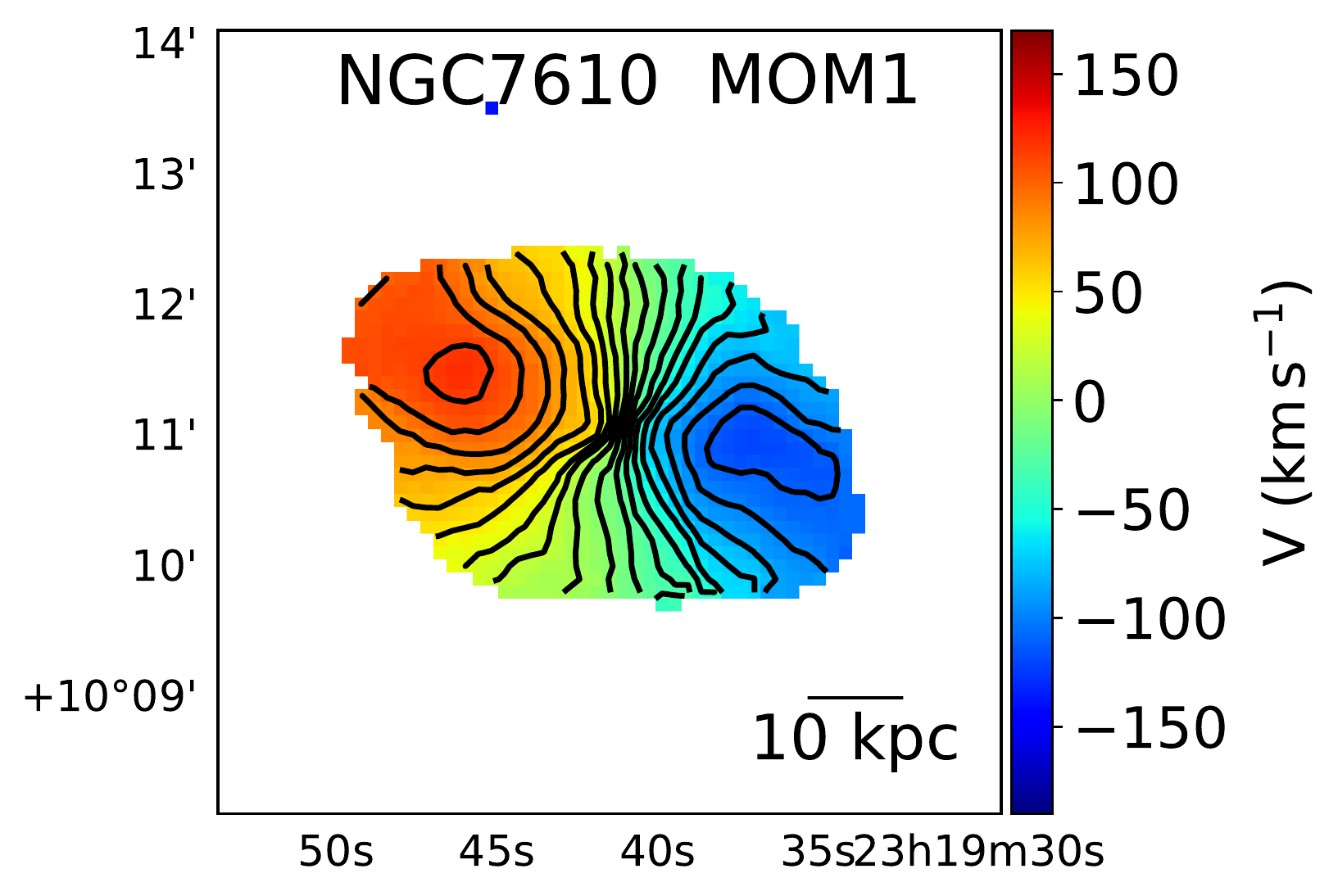} &
       \includegraphics[height=3.8cm]{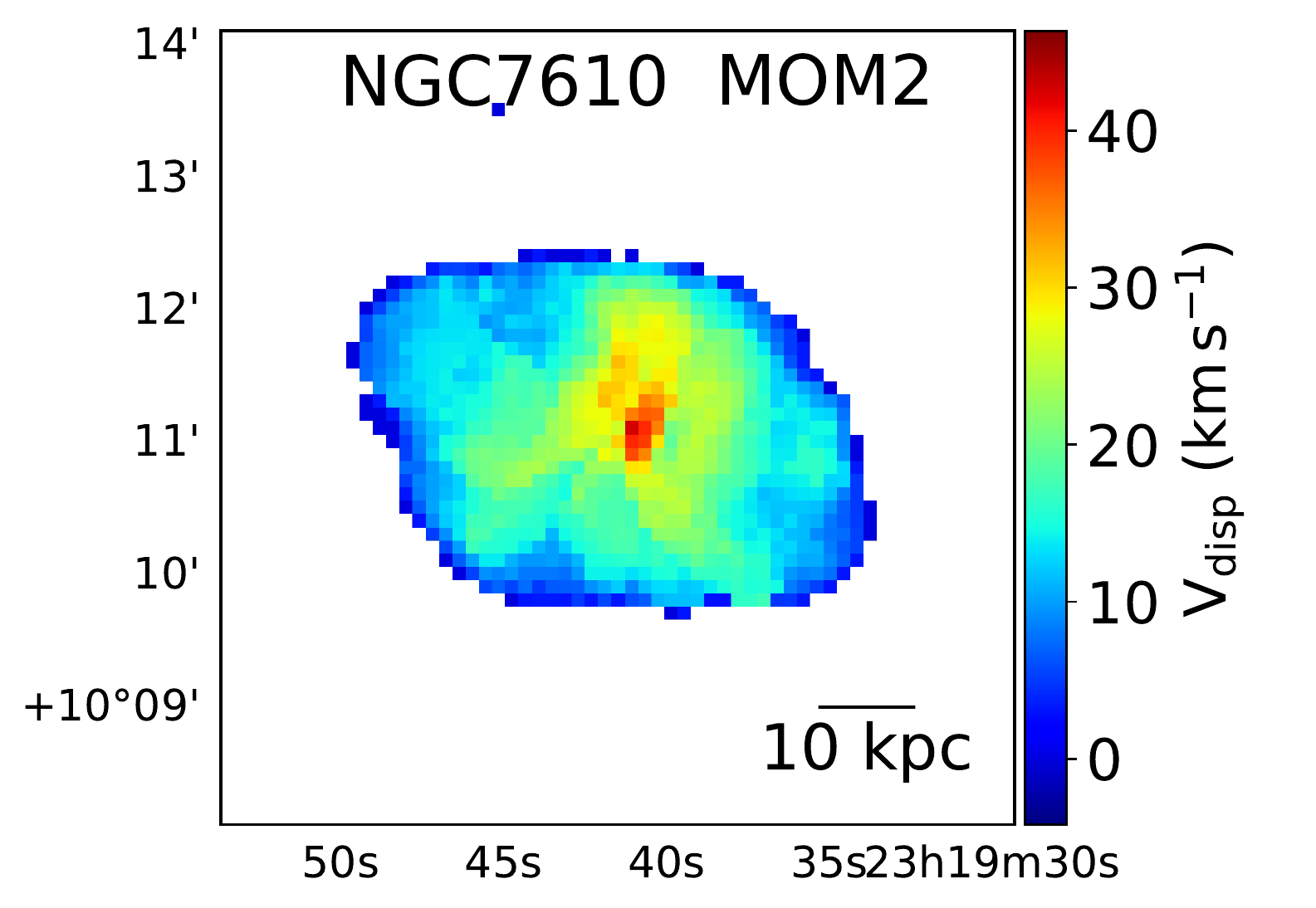} \\
       
       \includegraphics[height=4cm]{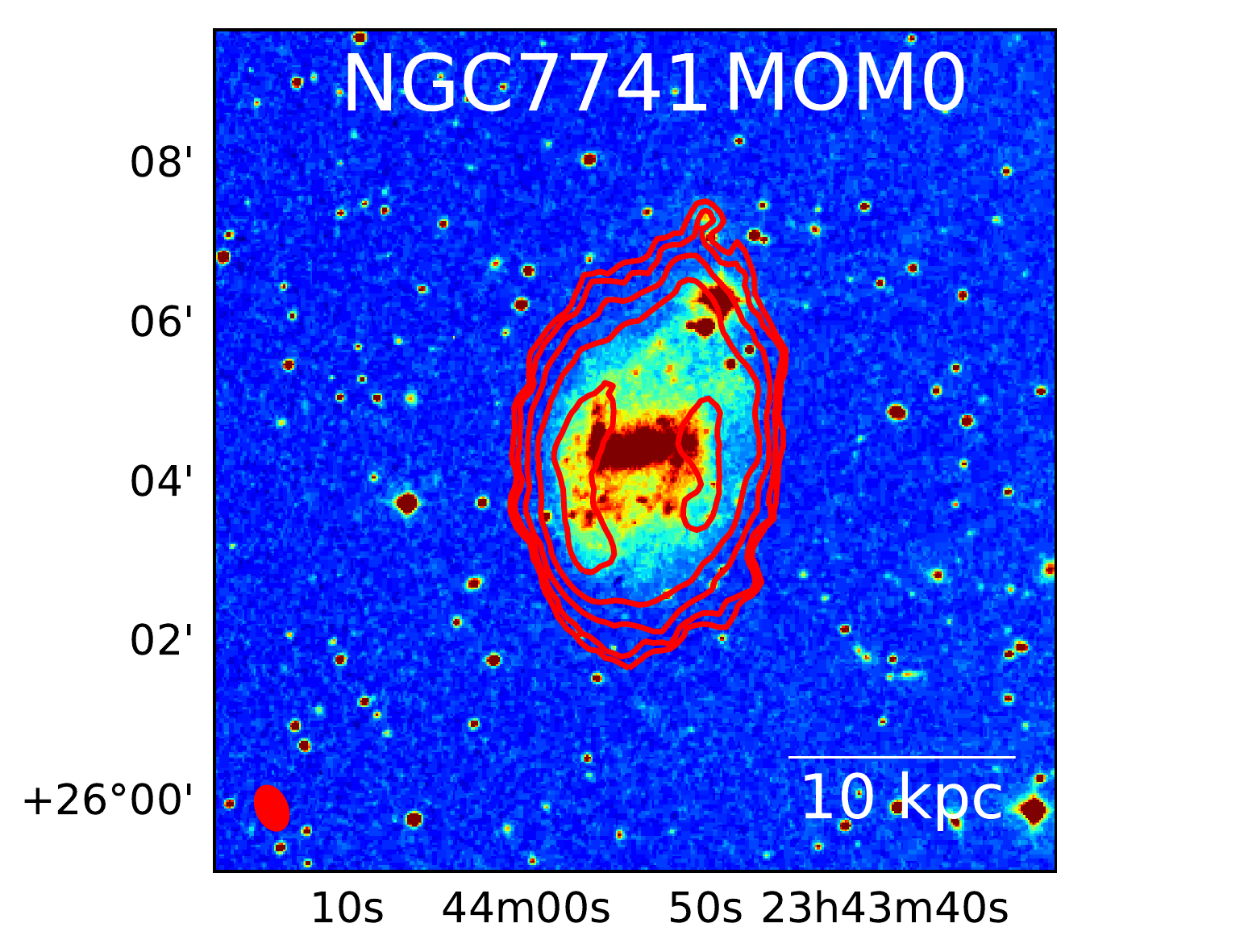} & 
       \includegraphics[height=3.8cm]{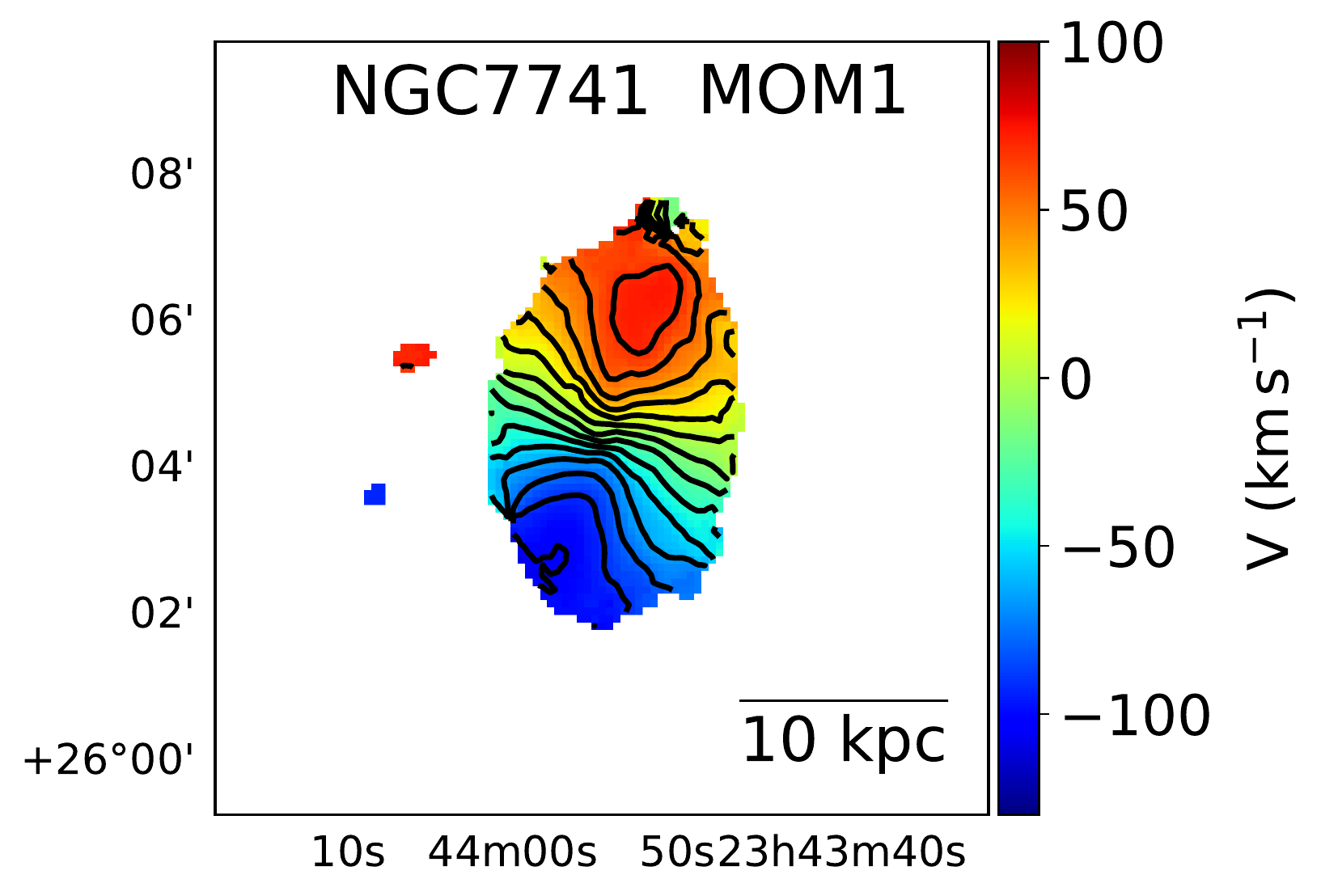} &
       \includegraphics[height=3.8cm]{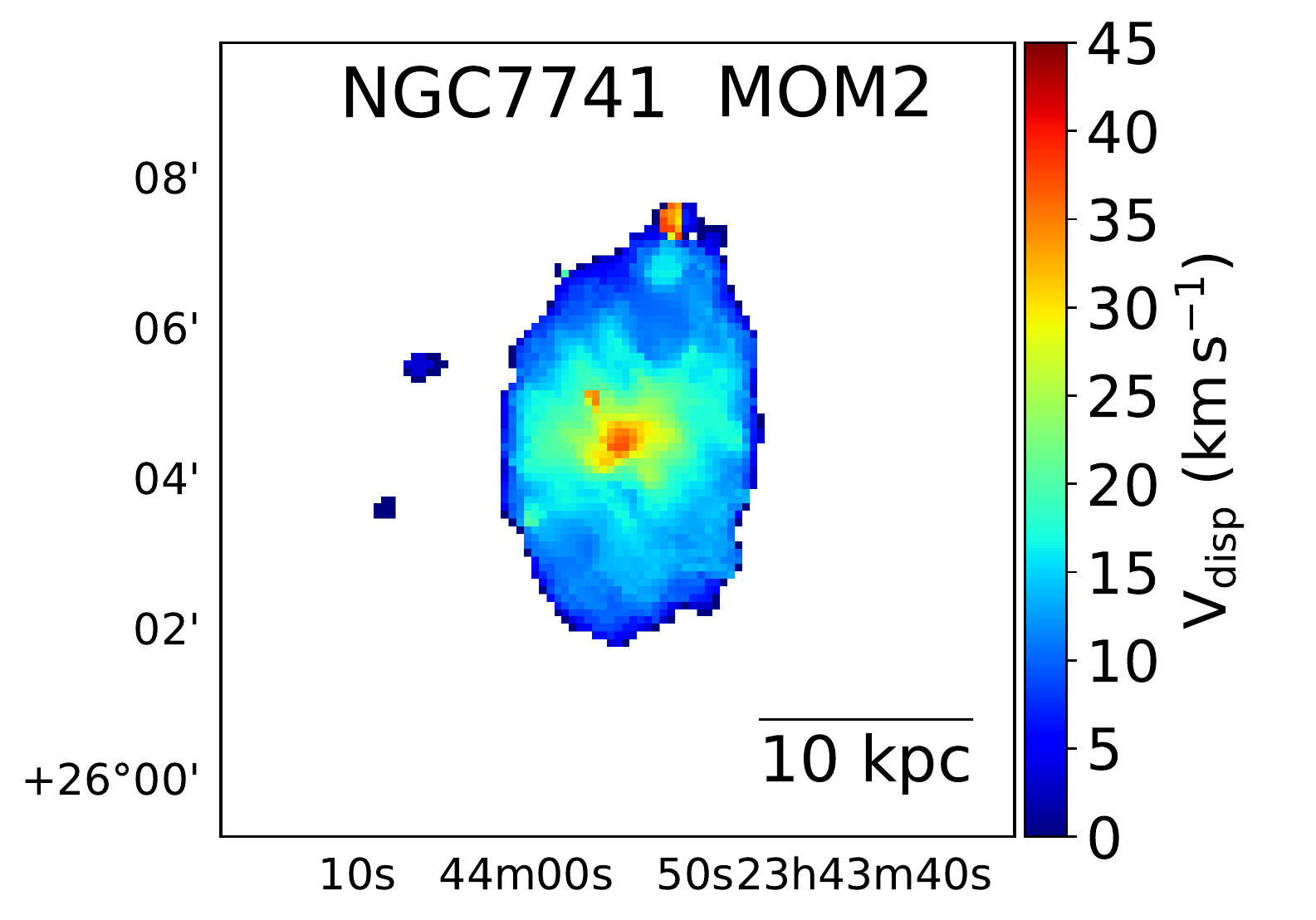} \\
        \end{tabular}
    \caption{Moment maps of the galaxies. The first, second and the third columns of the images show the Moment zero, Moment one and Moment two maps, respectively. \emph{(cont.)}}
%    \label{fig:mom_maps}
\end{figure*}
 
\begin{figure*}
\ContinuedFloat
    \centering
    \begin{tabular}{cccc}
       \includegraphics[height=4cm]{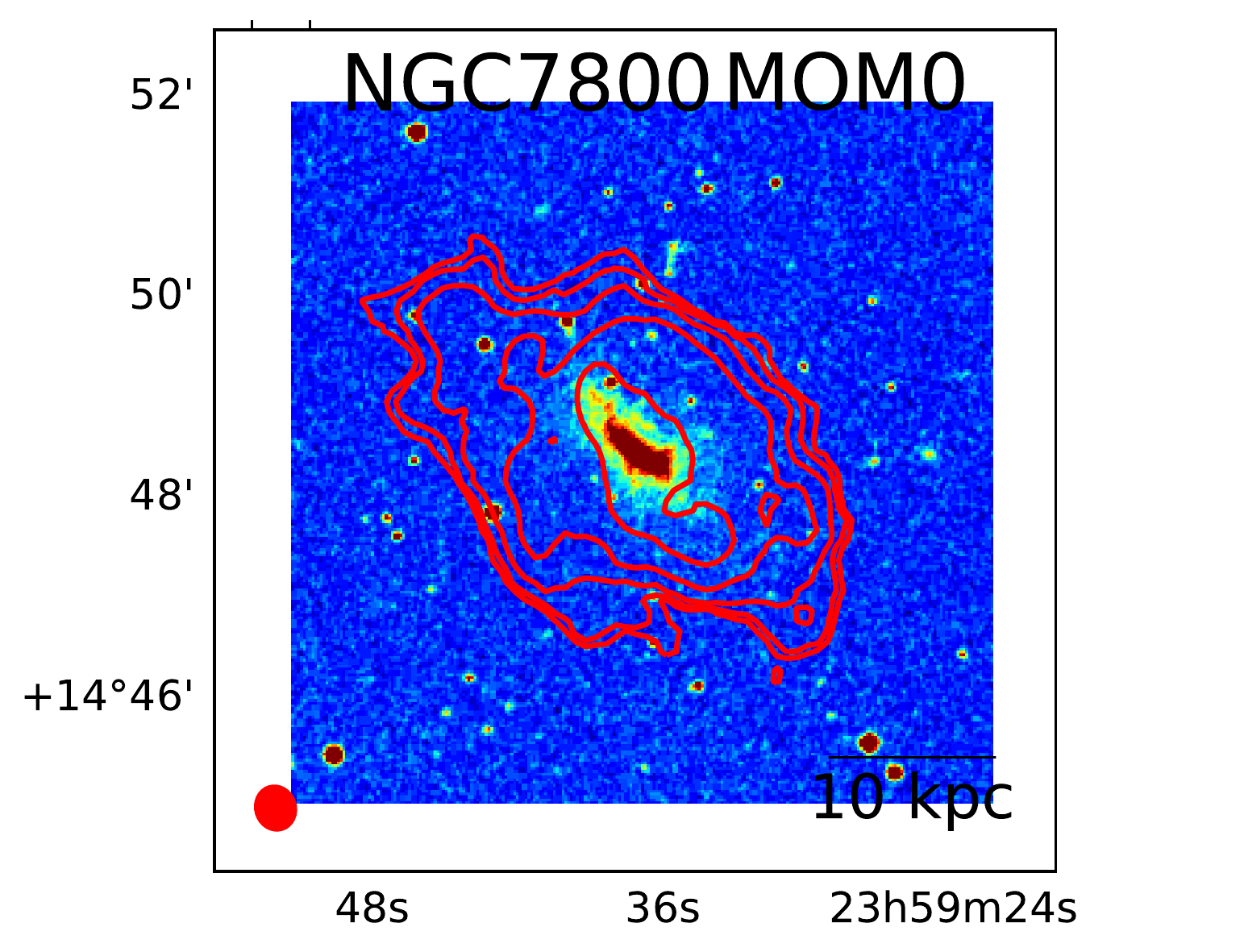} & 
       \includegraphics[height=3.8cm]{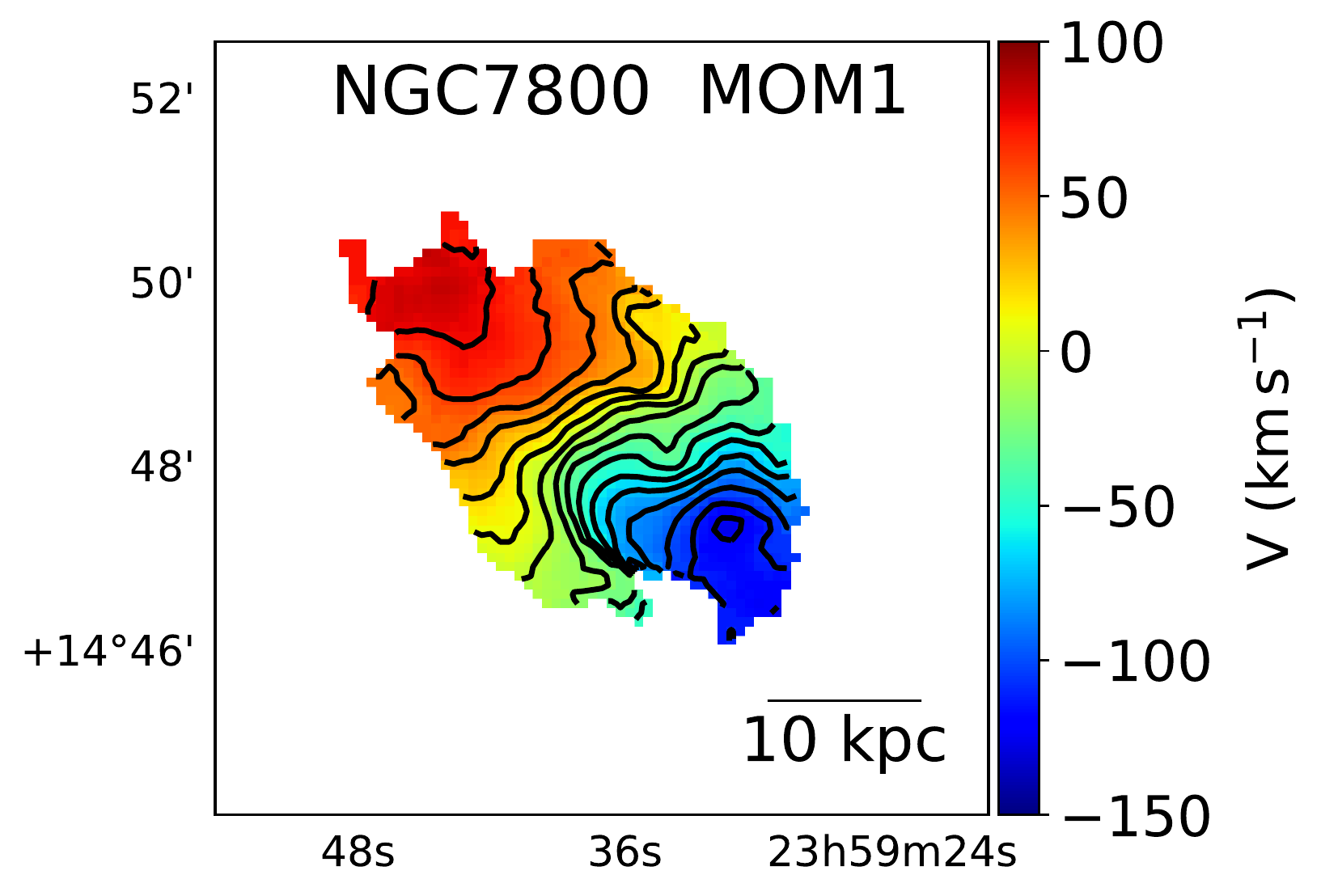} &
       \includegraphics[height=3.8cm]{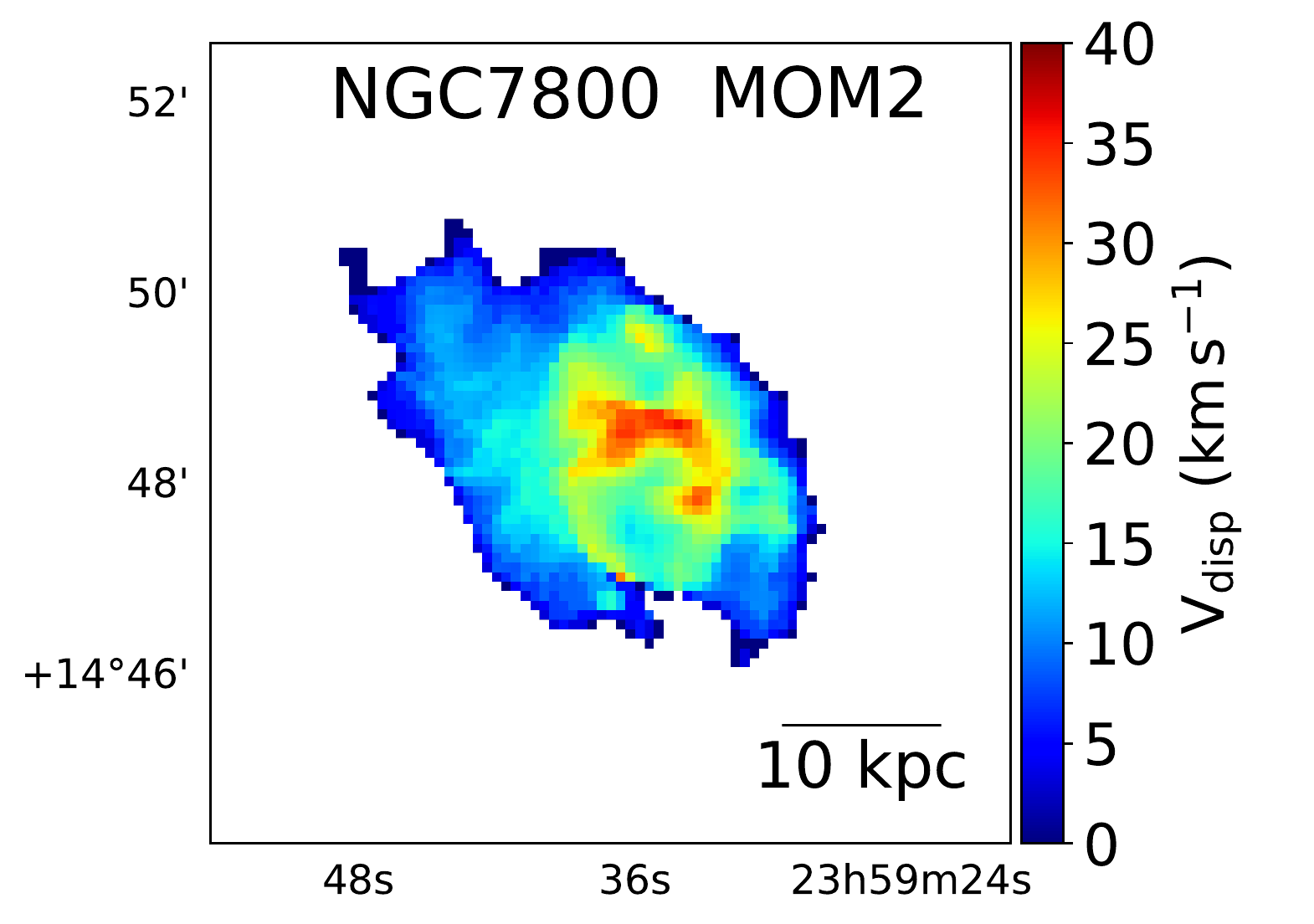} \\
     \end{tabular}
    \caption{Moment maps of the galaxy. The first, second and the third columns of the images show the Moment zero, Moment one and Moment two maps, respectively. }
%    \label{fig:mom_maps}
\end{figure*}

\subsection{The HI mass-size relation}
The total mass of the atomic gas in galaxies correlates tightly with the diameter of the HI disk. \citet{mass_size_first} first pointed out this relation and later \citet{mass_size} revisited this relation in great detail with a sample of more than 500 nearby galaxies over more than ten orders in $B$-band magnitudes and five orders of magnitudes in atomic gas mass. For both the studies, the atomic gas diameter ($a_{HI}$) is defined at a surface density ($\Sigma_{HI}$) of $1$ M$_{\sun}$\thinspace pc$^{-2}$. The relation obtained by \citet{mass_size} is given by the following equation:

\begin{equation}
    \log a_{HI} = (0.506 \pm 0.003) \log M_{HI}  -(3.293 \pm 0.009) 
\label{eqn:mass_size_ori}    
\end{equation}

\begin{figure}
    \centering
    \includegraphics[height=5cm]{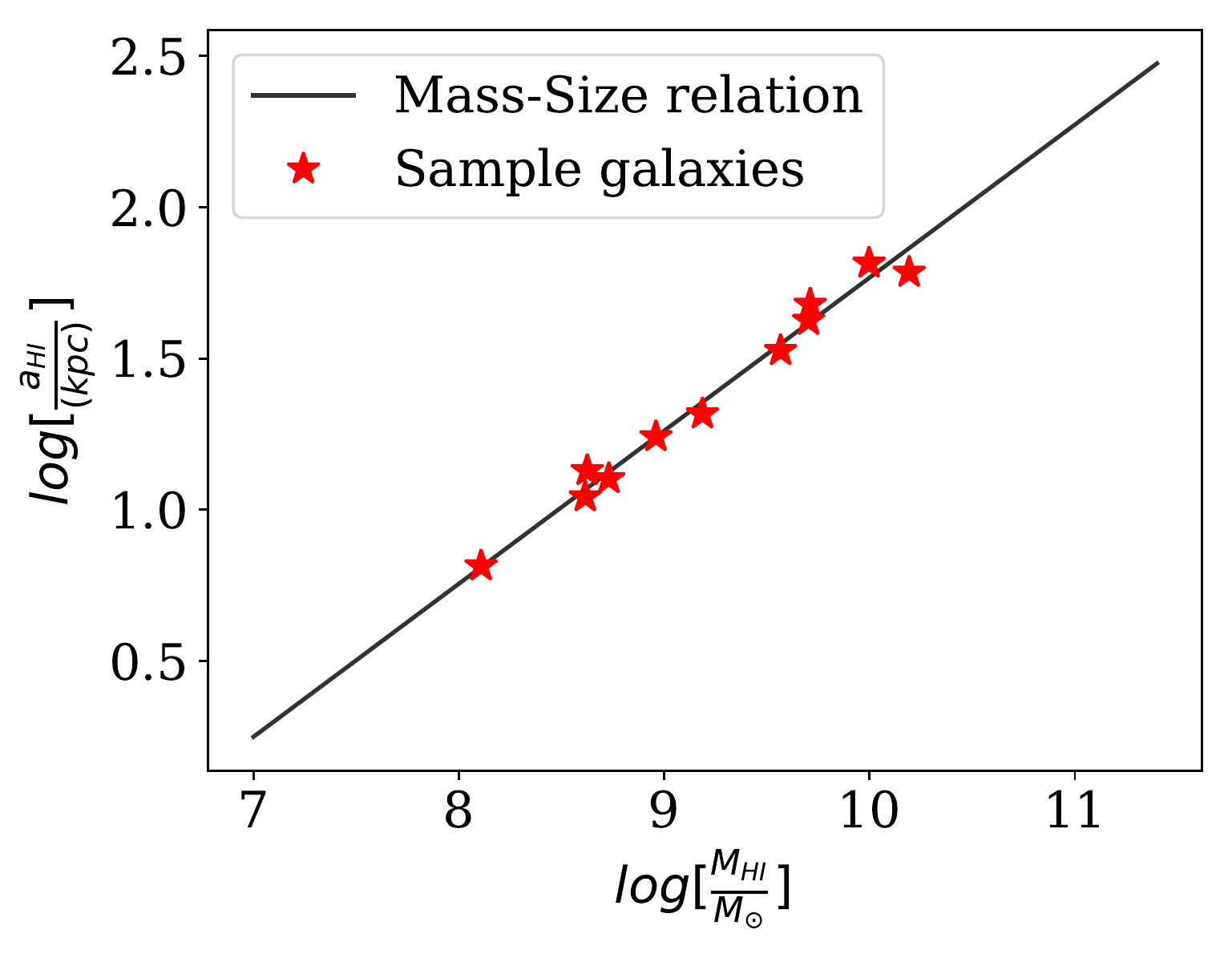}
    \caption{HI mass-size relation. The red stars denote the eleven sample galaxies, and the black solid line denote the mass-size correlation given by equation \ref{eqn:mass_size_ori}.}
    \label{fig:mass_size}
\end{figure}

 Now,  $1$ M$_{\sun}$\thinspace pc$^{-2}$ surface density is equivalent to the column density of $1.25\times10^{20}$ cm$^{-2}$. Thus, we fit an ellipse at $1.25\times10^{20}$ cm$^{-2}$ column density contour level of moment zero map for each of the eleven galaxies; and by measuring the length of the major axis of the ellipse, we estimate $a_{HI}$.  We then checked the compatibility of the measured HI masses ($M_{HI}$) and sizes ($a_{HI}$) of the eleven galaxies from our sample with this existing relation (equation \ref{eqn:mass_size_ori}) \citep[see][]{mass_size}. Figure \ref{fig:mass_size} shows that the HI masses and sizes of our sample of eleven galaxies are in good agreement with the existing relation. The HI mass used in these plots is derived from the line flux obtained in the global HI spectra from interferometric data of our analysis through equation \ref{hi_mass_ori}.

\section{The different phases of ISM}
\label{phase_ism_intro}
As mentioned in the introduction part, various data products produced through our analysis are helpful in studying several interesting science cases. To explore one of the science cases, in this section, we have further extended our study to separate the cold and the warm component of the atomic HI inside these galaxies.  Further in our upcoming paper (Biswas et al., in prep.), we are exploring the kinematic modeling and mass modeling of the present GARCIA sub-sample.

Depending upon the local environment, i.e., heating, cooling, and ionization, the gas in the ISM exist in different thermal phases. Molecular clouds represent the coldest phase of the ISM with a temperature of $\sim 10-20$ K and densities $\rm \gtrsim 10^3 \thinspace cm^{-3}$. These clouds are mostly gravitationally bound stable systems. The Cold Neutral Medium (CNM) consists of HI with temperature $\rm \sim 20 - 100 $ K and densities $\rm \sim 50 \thinspace cm^{-3}$. The Warm Neutral Medium (WNM) exists in much higher temperature ($\sim 5000-8000$ K) with much lower densities ($\rm \sim 0.5 \thinspace cm^{-3}$). Other phases of the ISM include the Warm Ionized Medium (WIM) and the Hot Ionized Medium (HIM). Theoretical calculation shows that the CNM and the WNM are the two stable phases of the neutral medium. Any gas in between these two phases quickly moves to one of them by thermal runaway processes \citep{Field2_1965, Field1969, wolfire1995, Wolfire2003}. The CNM and the WNM exist in two different temperatures; thus, any HI spectrum would contain their signature. In such a case, the HI spectral width would have narrow and wide components corresponding to the CNM and the WNM, respectively. Thus, decomposing an HI spectrum into multiple Gaussian components would provide the details of the neutral ISM phases along that line of sight.   

\subsection{Gaussian decomposition method:}

  The HI spectral cubes are used to identify the different phases of the neutral ISM in our sample galaxies. We extract the HI spectra from each pixel on the galaxy and perform a multi-Gaussian decomposition to identify narrow and wide components. We use the method developed by \citet{naren2016} for decomposing the spectra. Starting with a single Gaussian component, we add successively more components till the residual of the fit does not improve further. To test the quality of the fit (and the convergence on the number of components), we employ a statistical F-test to determine if the residual of a fit with (N+1) component is statistically better than the residual of a fit with just N component (see, \citet{naren2016} for more details). If not, then we conclude that N Gaussian components best fit the spectrum. However, we find that all our spectra were best fitted by either a single or a double Gaussian, which is consistent with what was found by other previous studies as well \citep[see][]{cnm_young96, Warren_2012, naren2016}.

\begin{figure}
    \centering
    \begin{tabular}{c}
        \includegraphics[height=6cm]{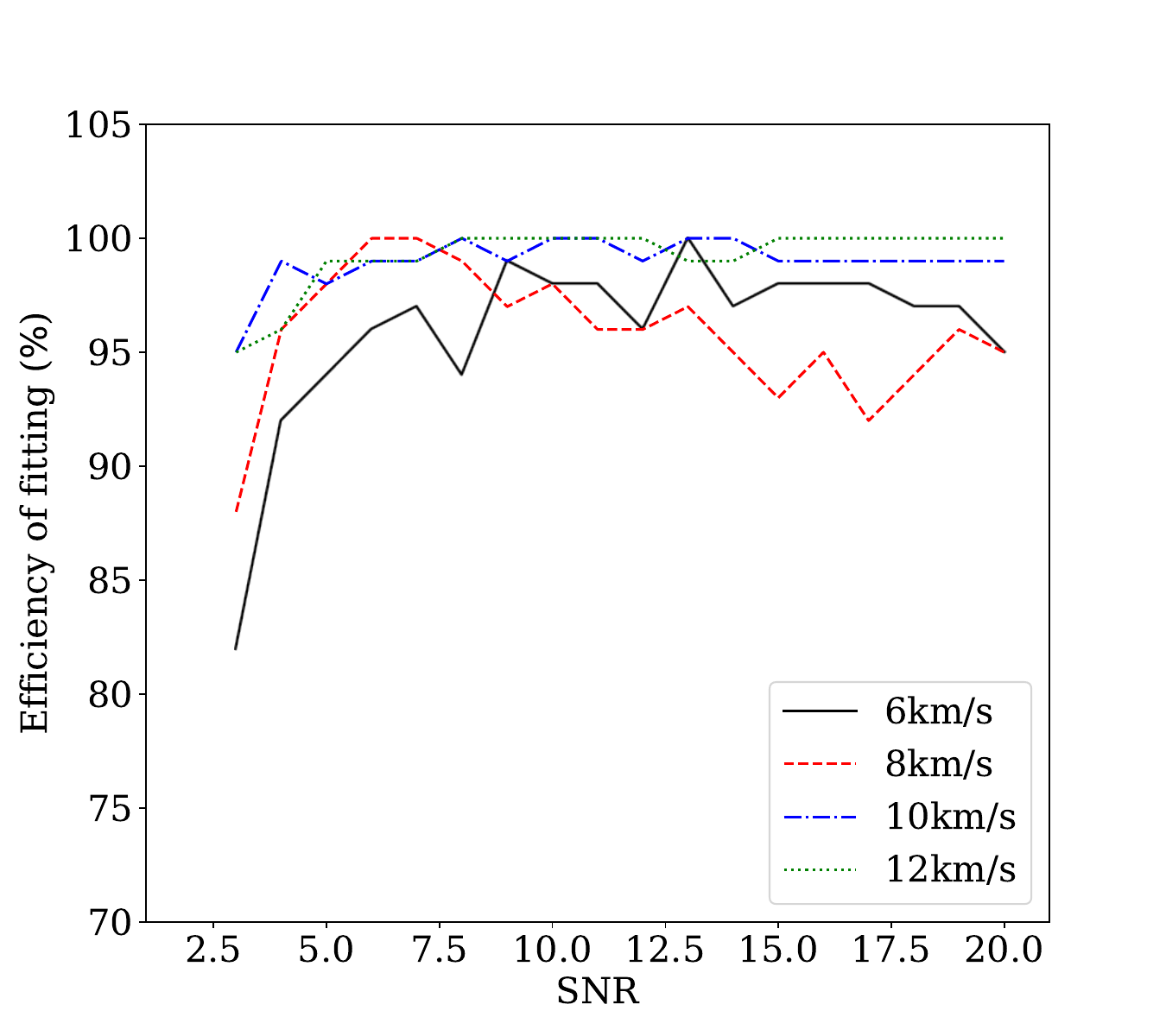} \\ 
        \includegraphics[height=6cm]{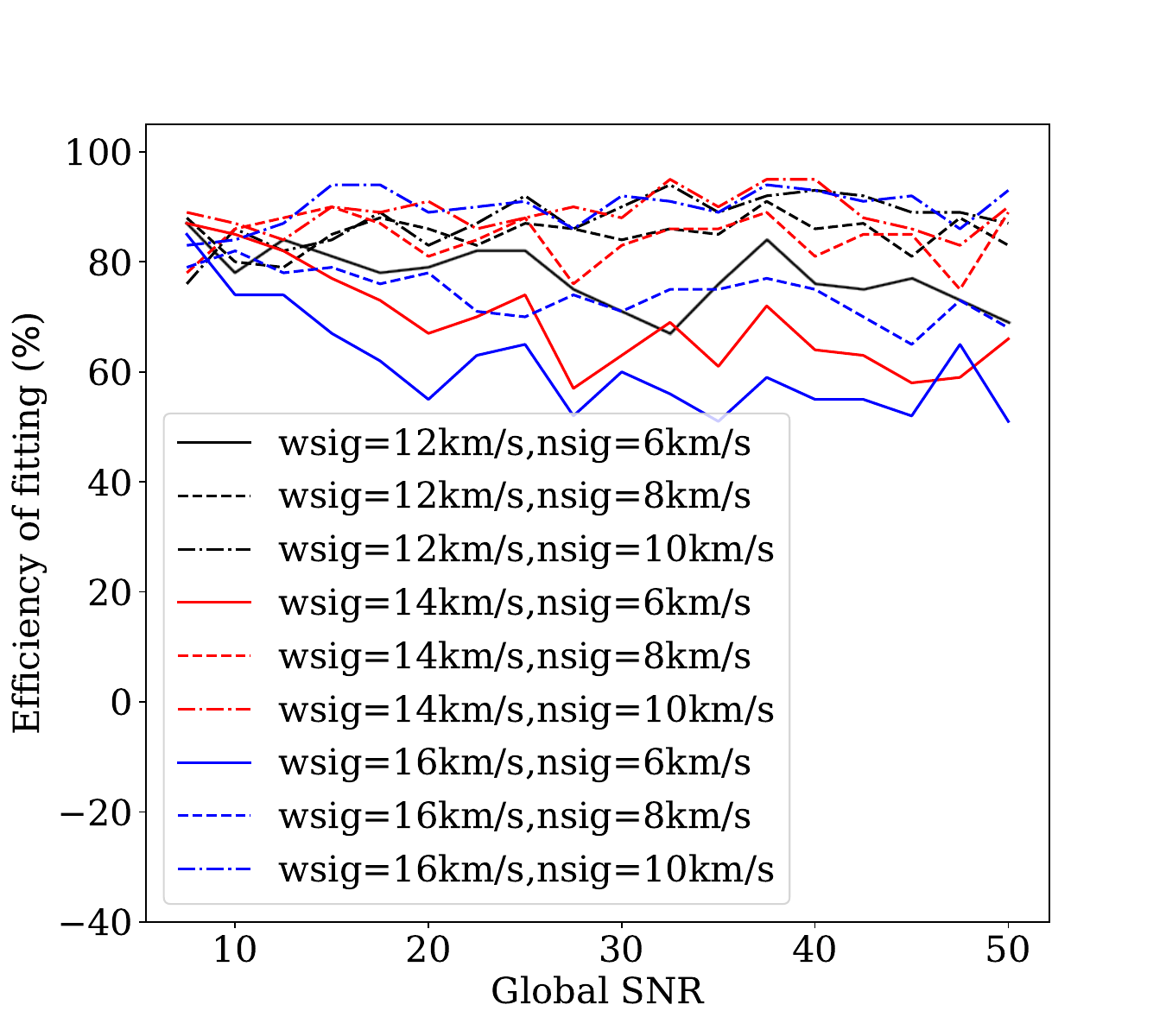}
    \end{tabular}
    \caption{Efficiency of recovering input sigma values. Figure on the top has single component input sigma and figure on the bottom has double component input sigma.}
    \label{fig:simspec}
\end{figure}

\begin{figure*}
    \centering
    \begin{tabular}{c c c c}
        \includegraphics[height=3.4cm]{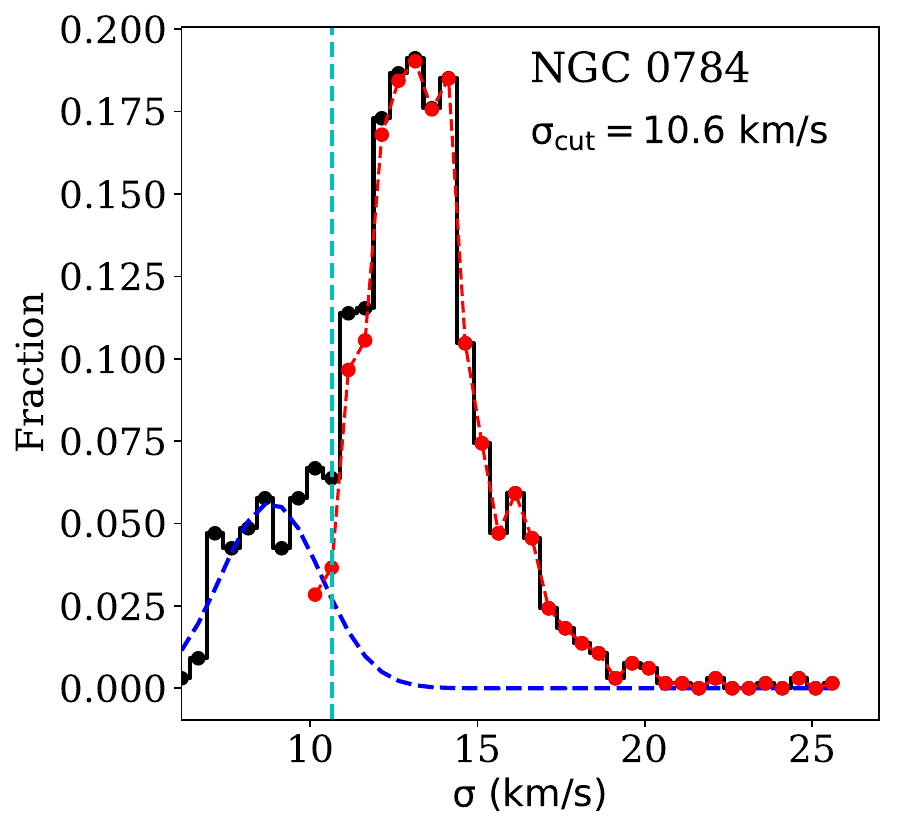} &
        \includegraphics[height=3.4cm]{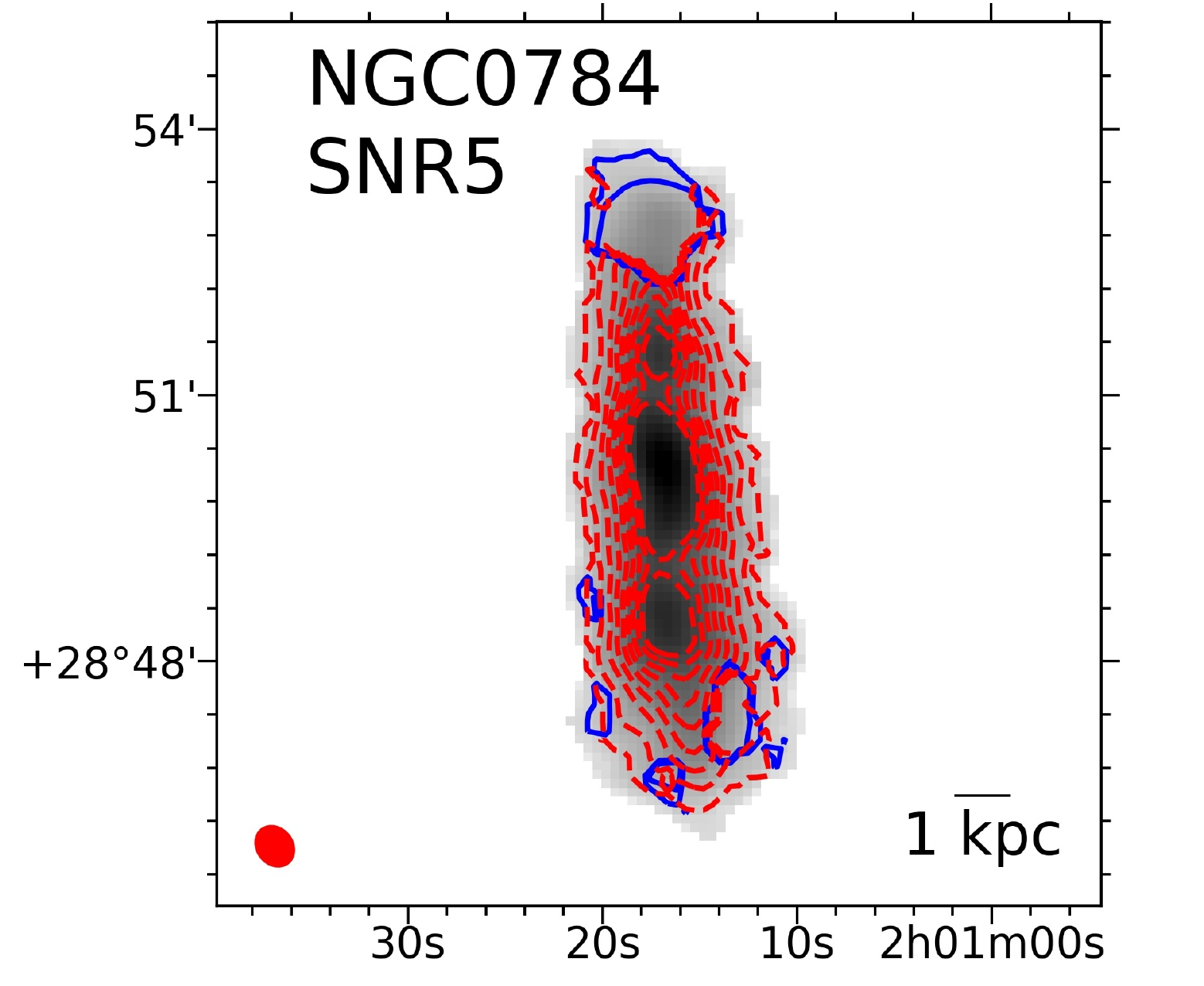} &
        \includegraphics[height=3.4cm]{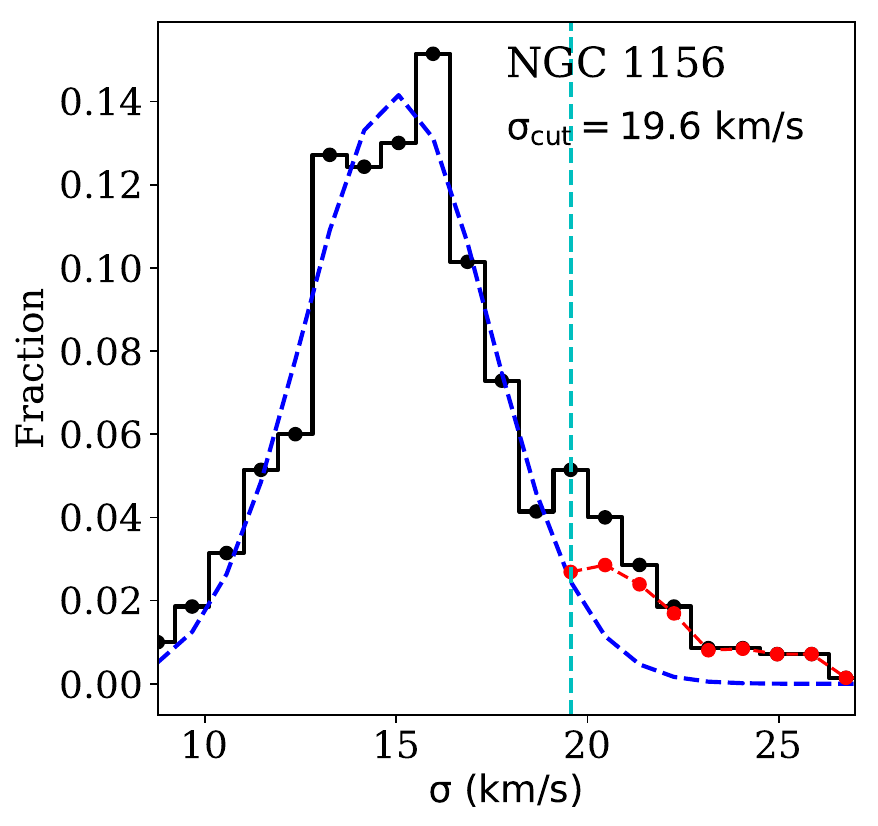} &
        \includegraphics[height=3.4cm]{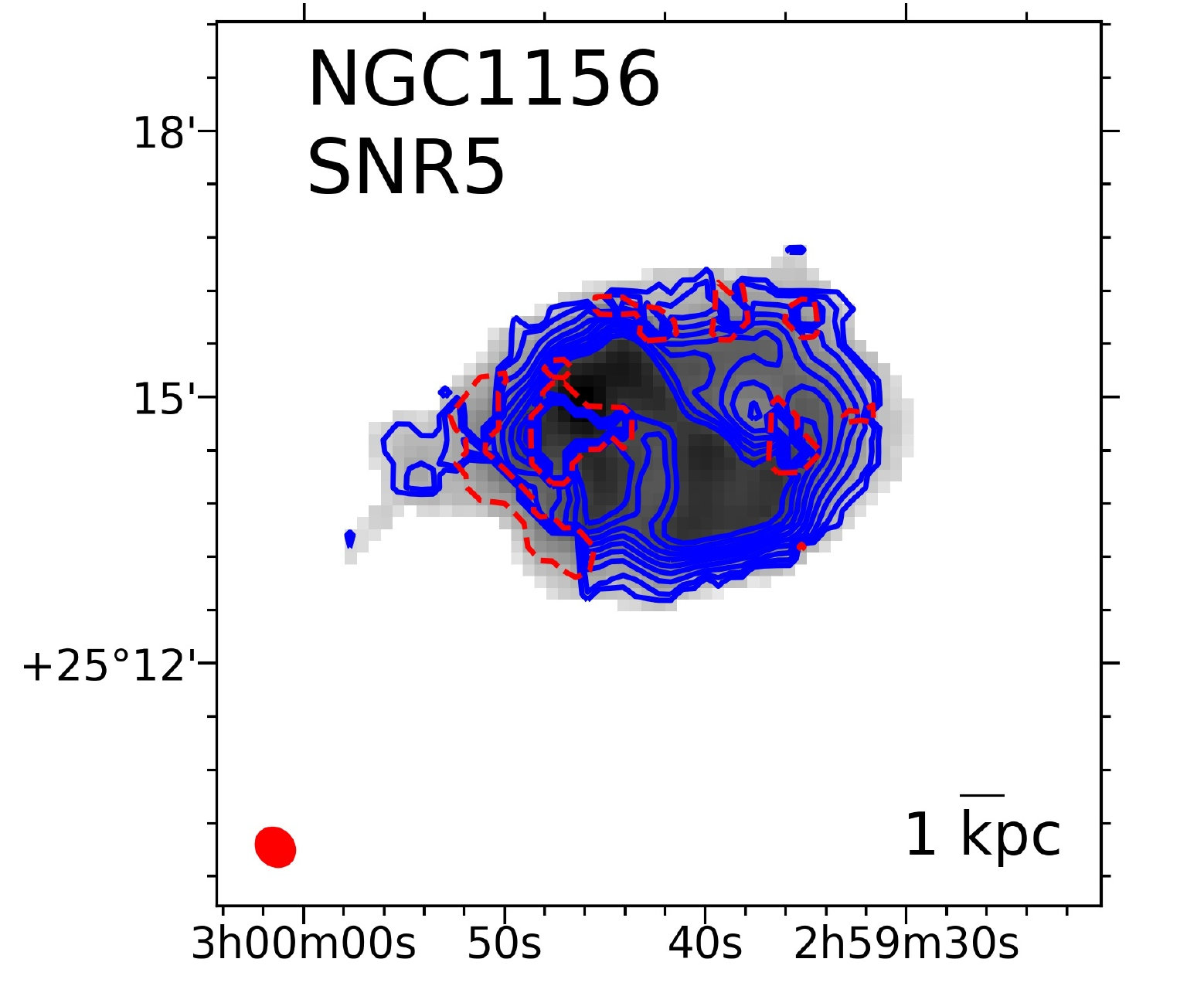} \\
        
        \includegraphics[height=3.4cm]{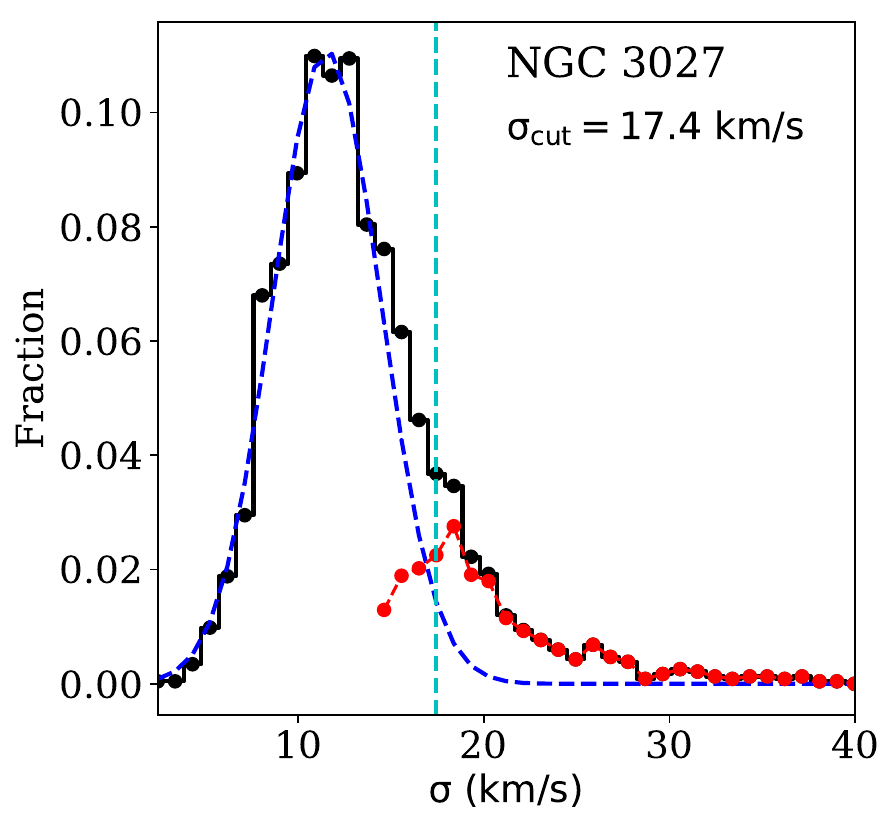} &
        \includegraphics[height=3.4cm]{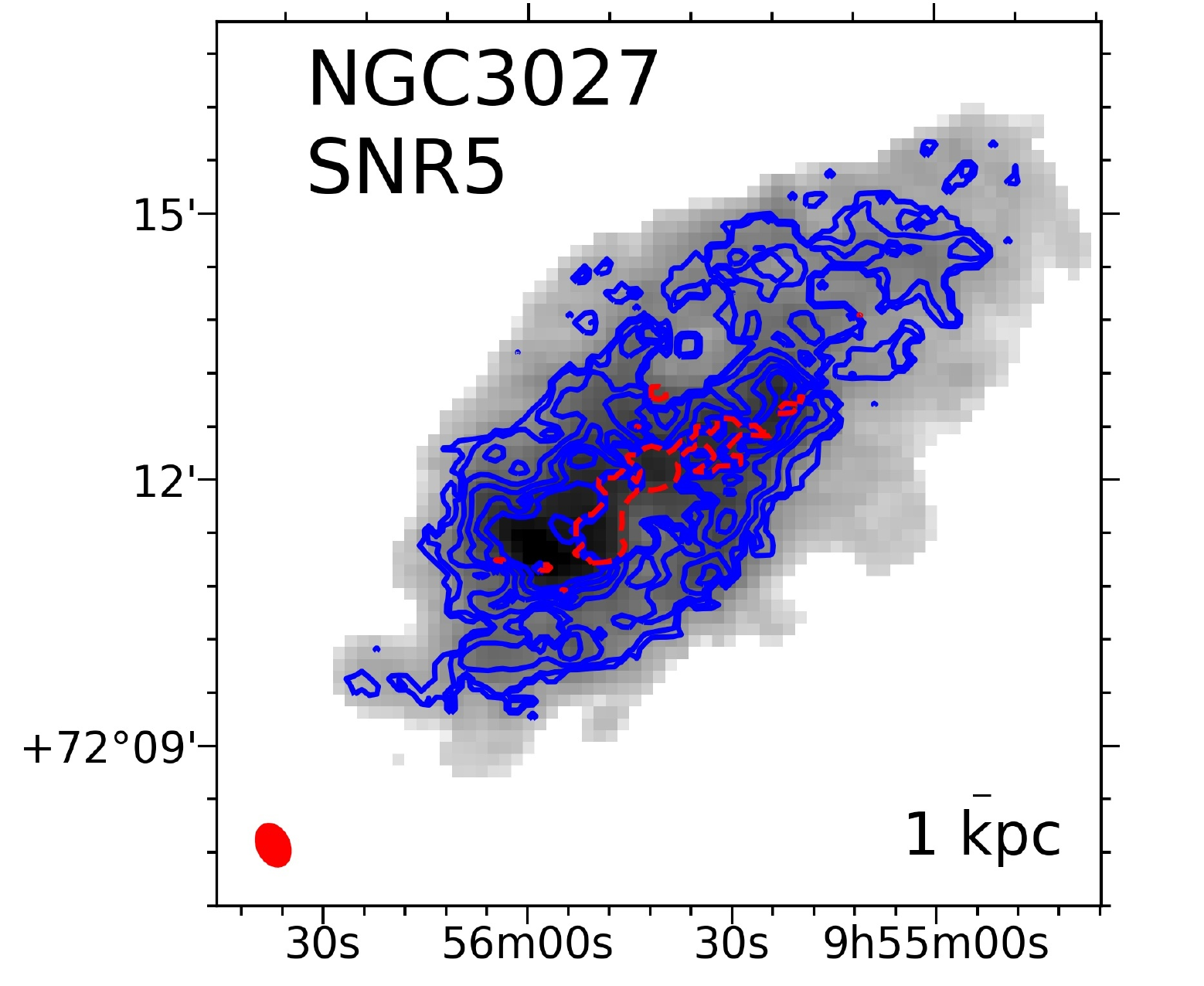} &
        \includegraphics[height=3.4cm]{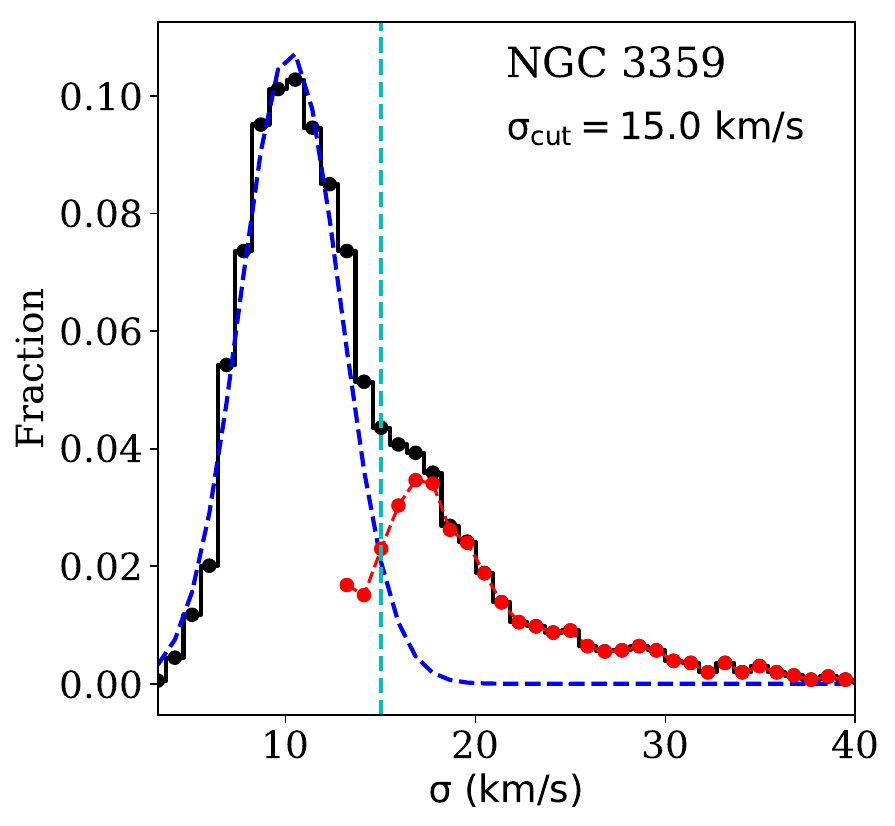} &
        \includegraphics[height=3.4cm]{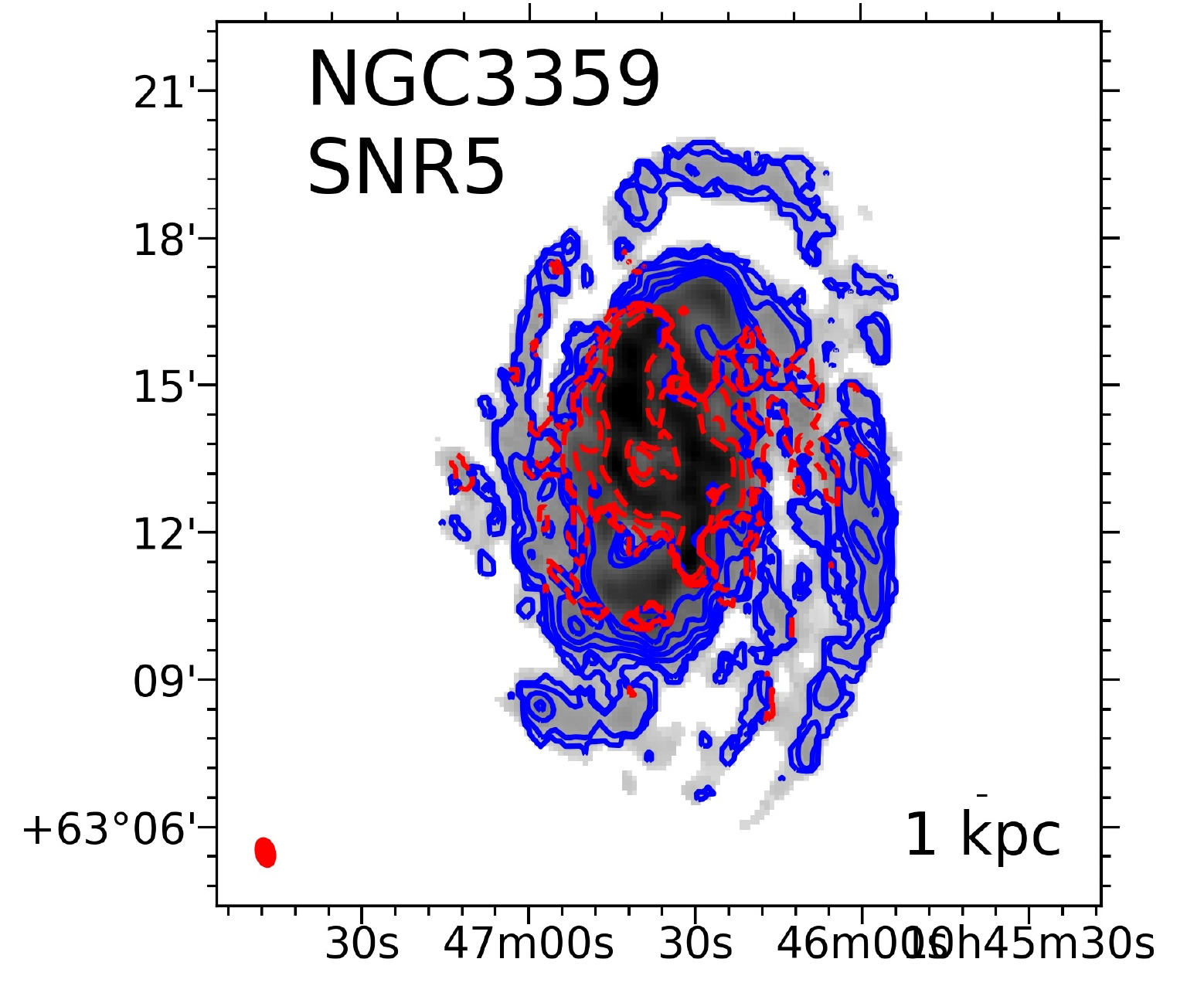} \\
        
        \includegraphics[height=3.4cm]{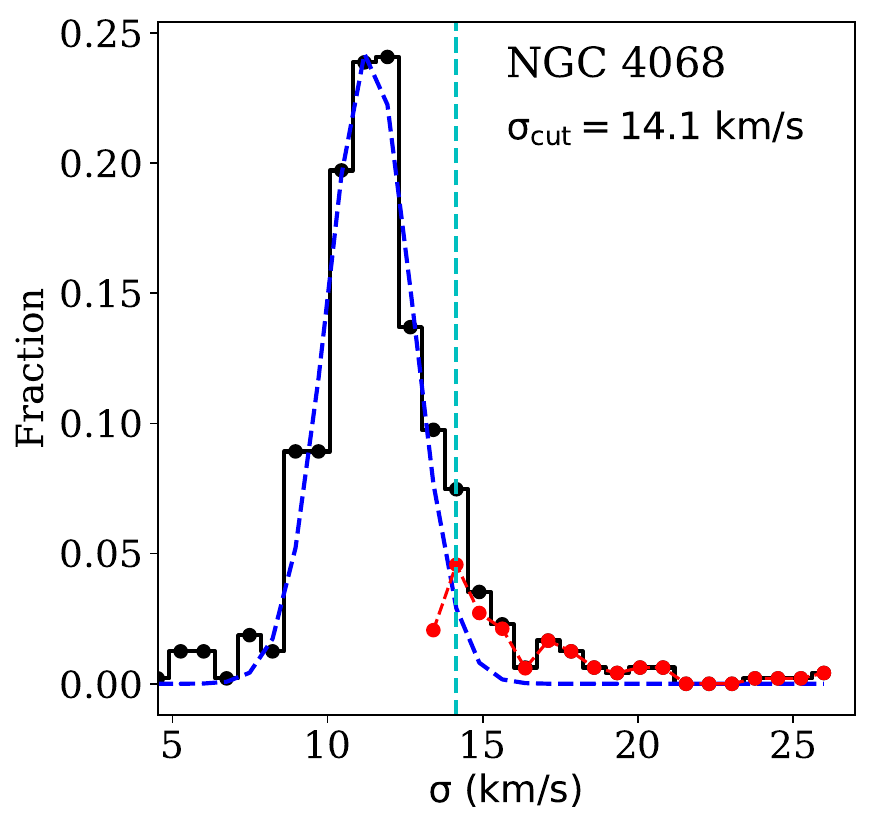} &
        \includegraphics[height=3.4cm]{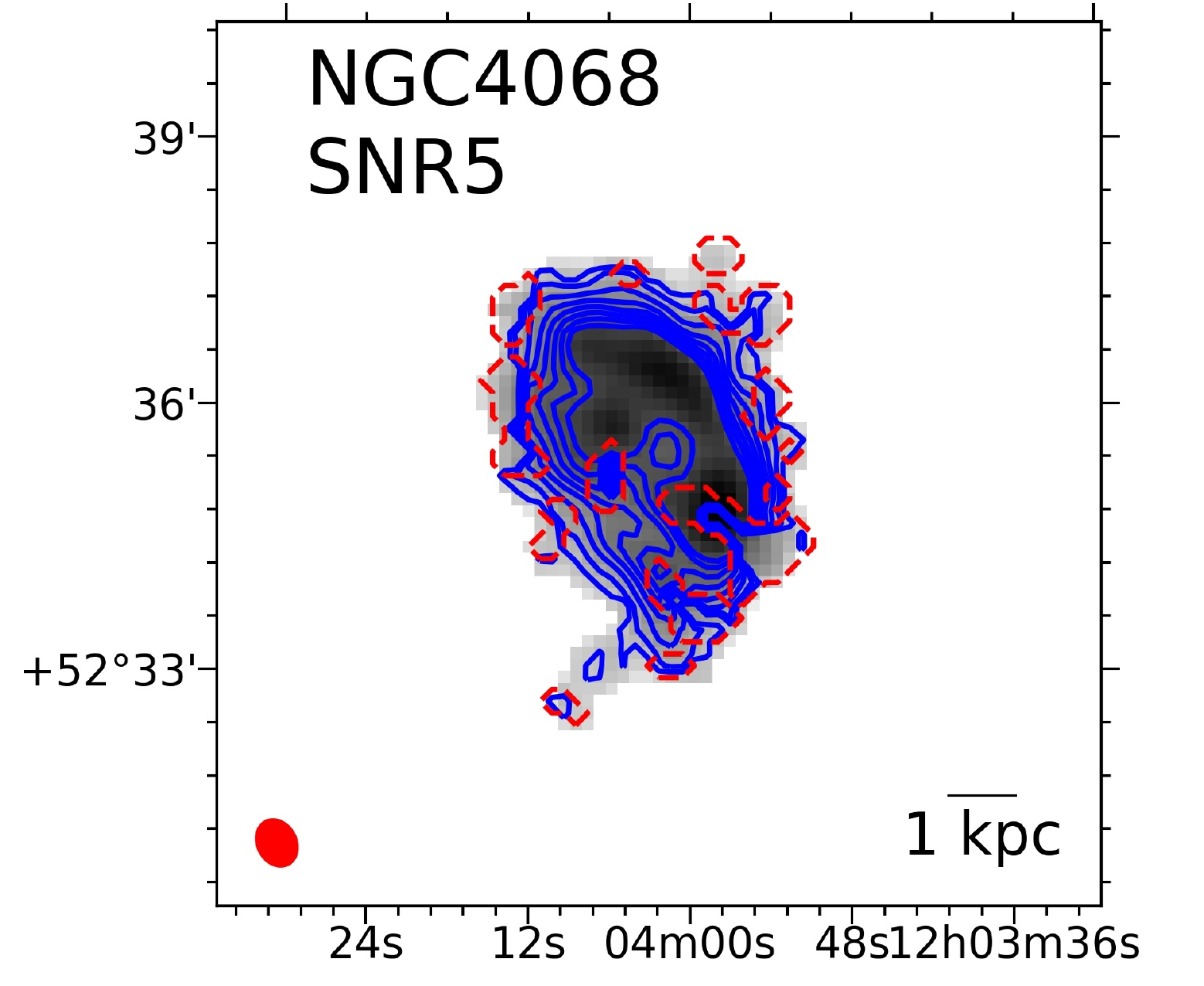} &
        \includegraphics[height=3.4cm]{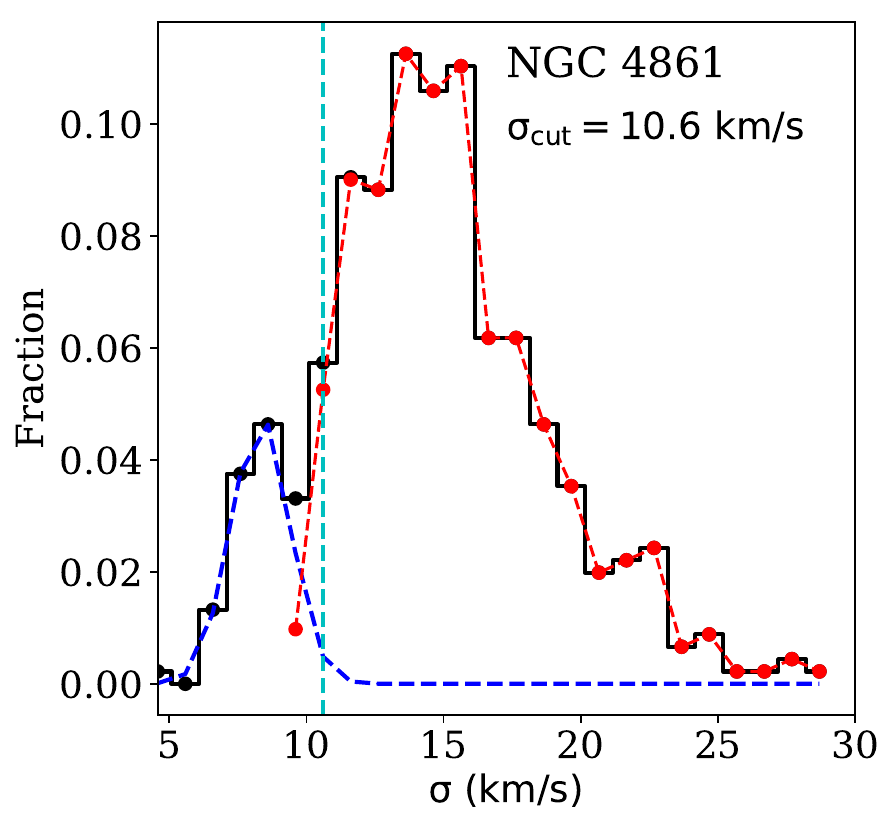} &
        \includegraphics[height=3.4cm]{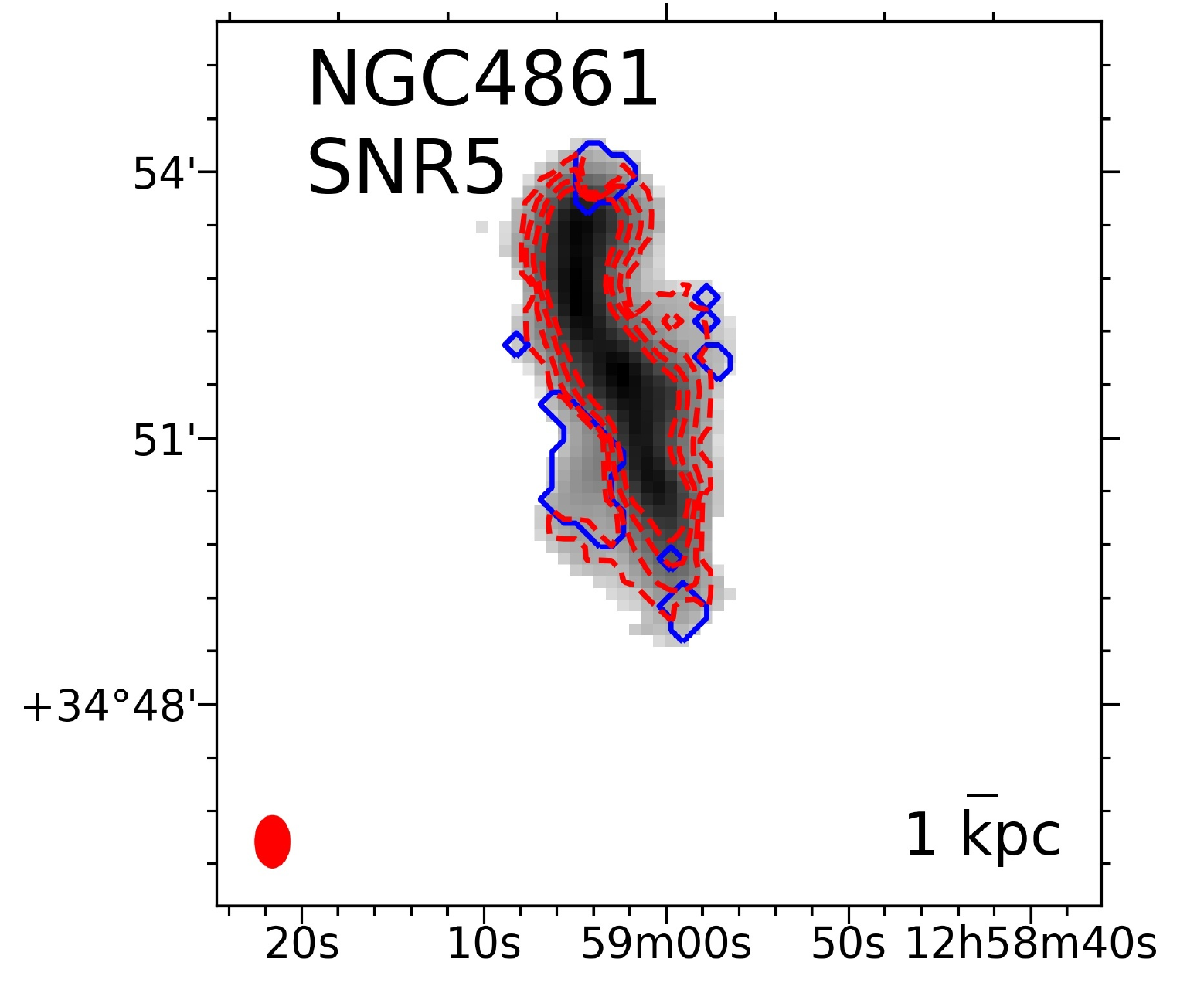} \\
        
        \includegraphics[height=3.4cm]{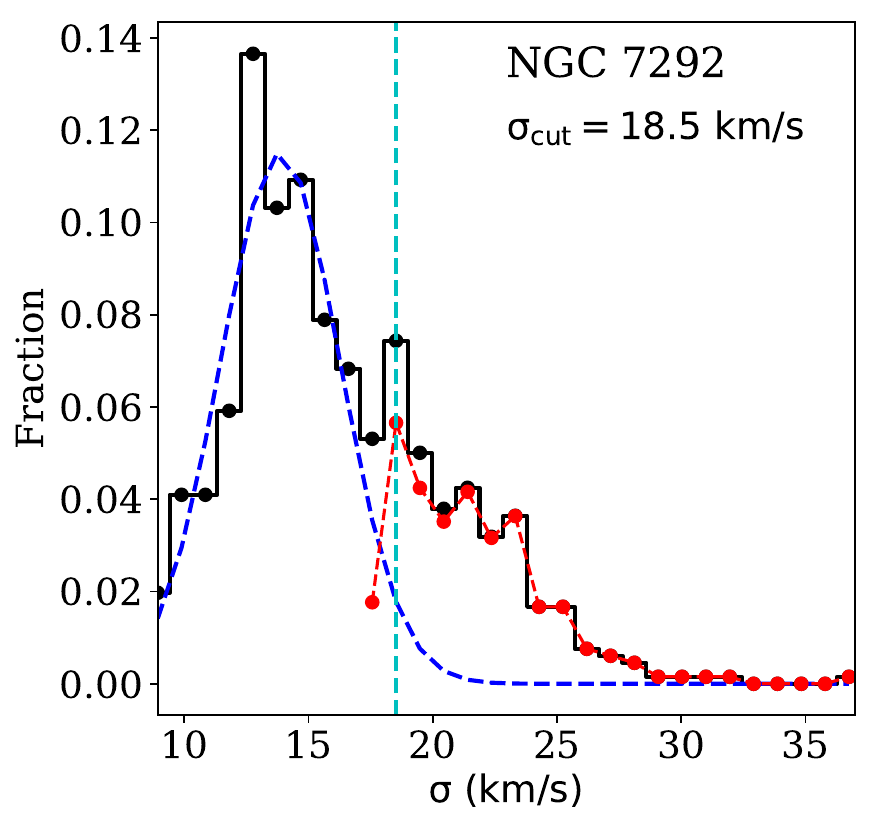} &
        \includegraphics[height=3.4cm]{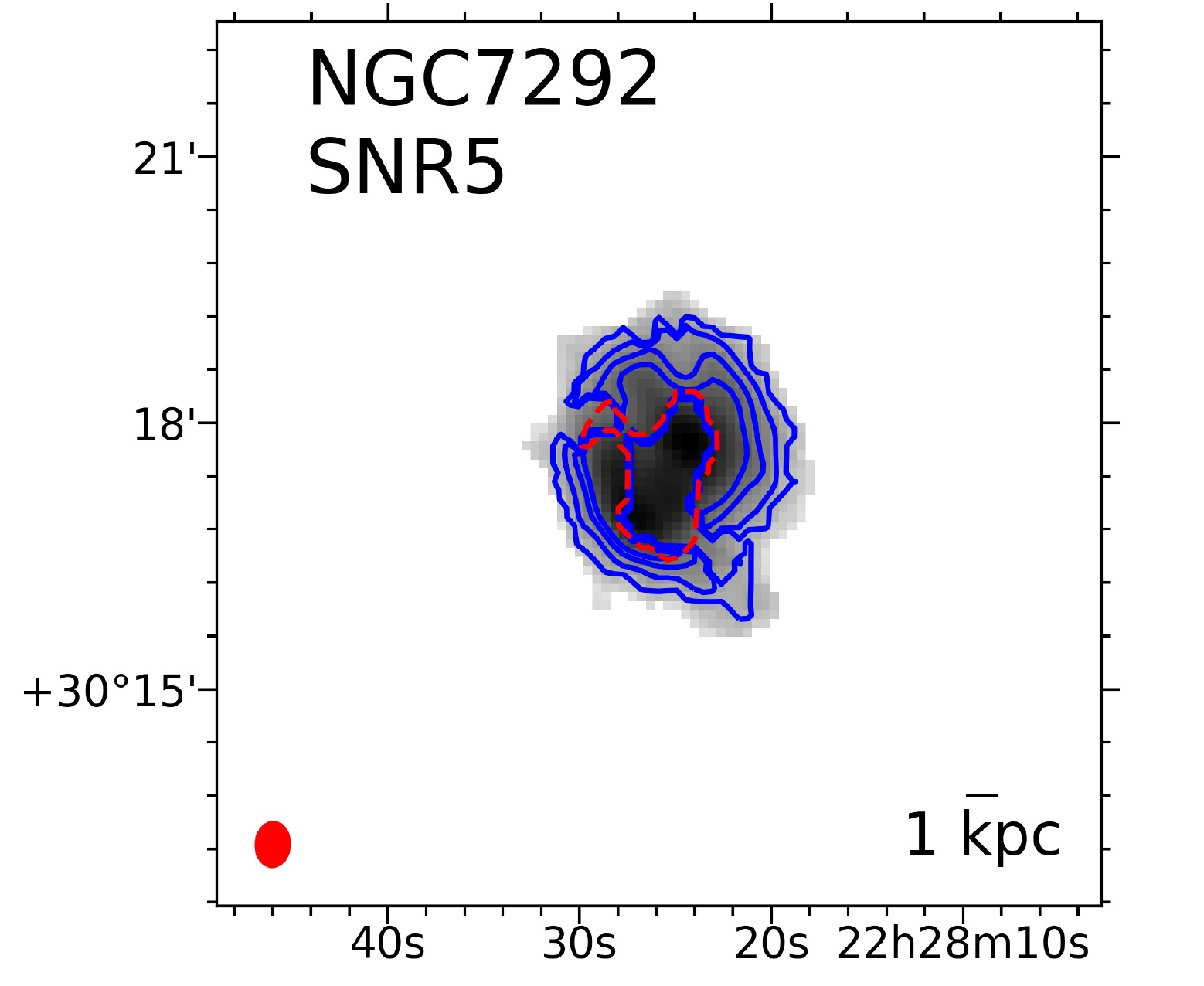} &
        \includegraphics[height=3.4cm]{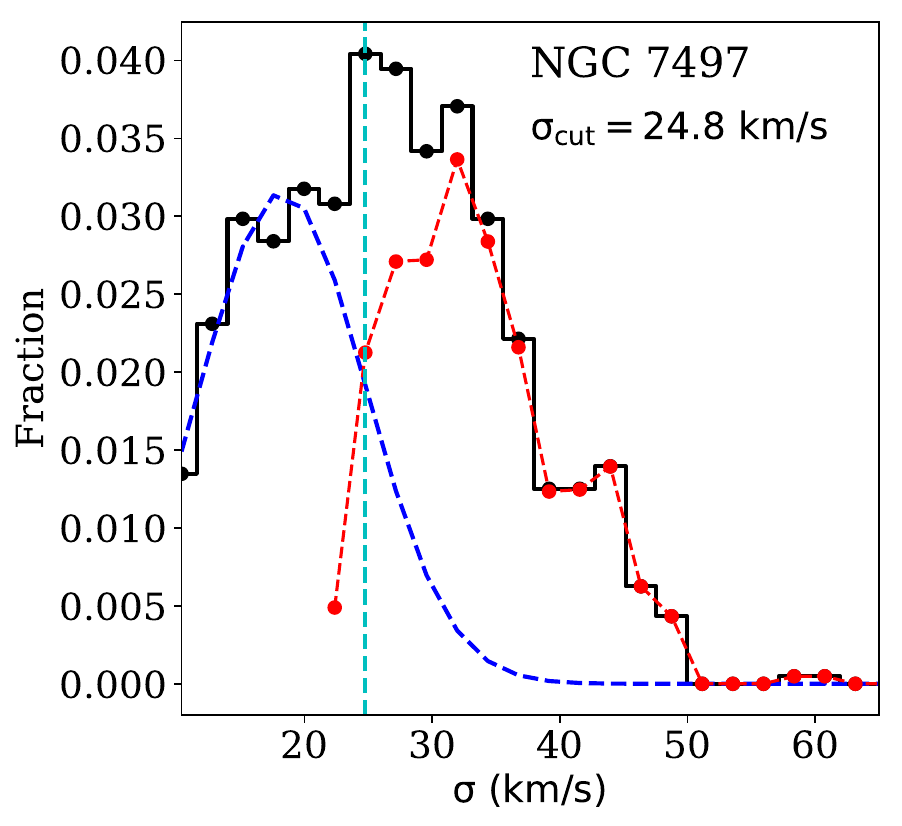} &
        \includegraphics[height=3.4cm]{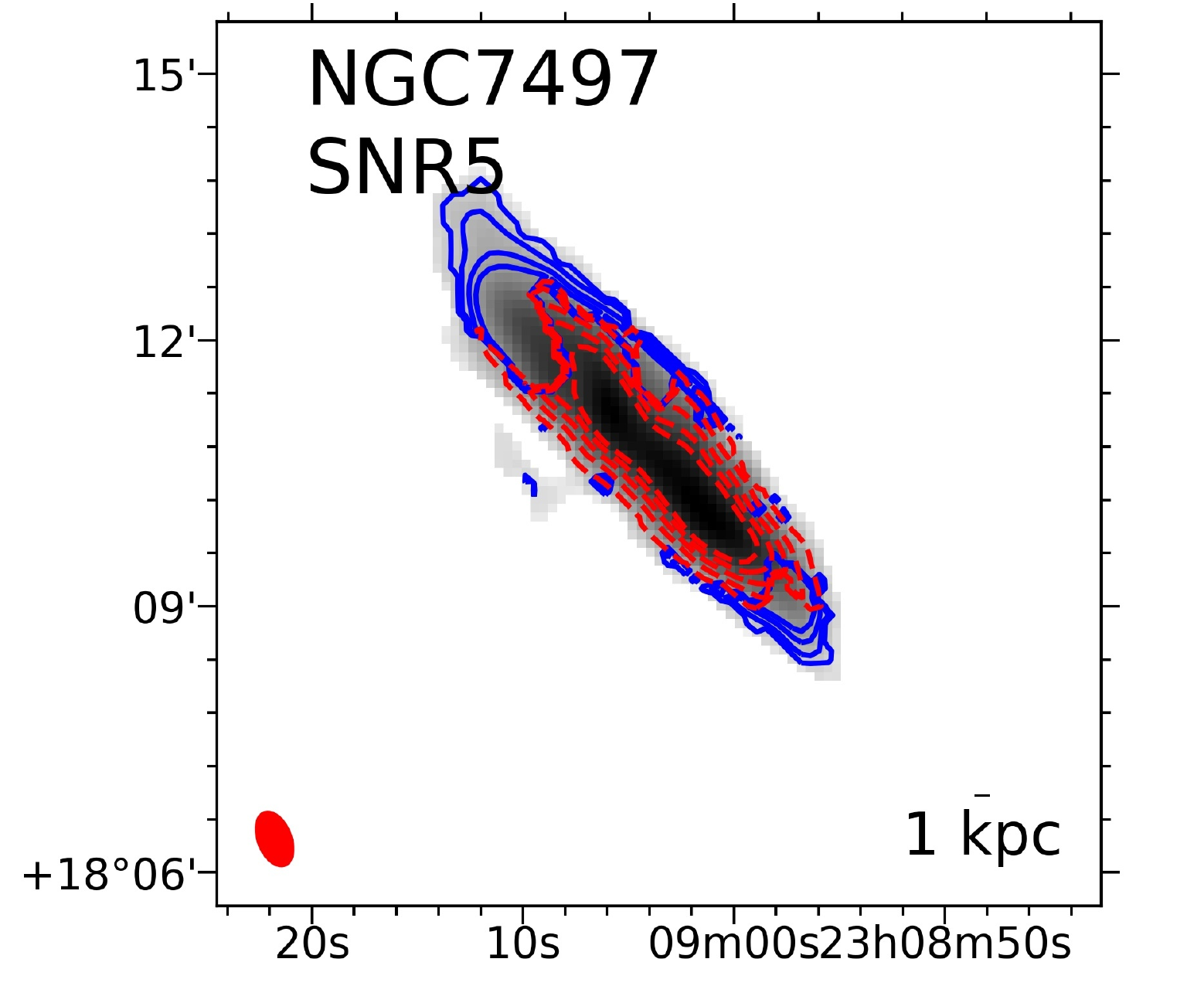} \\
        
        \includegraphics[height=3.4cm]{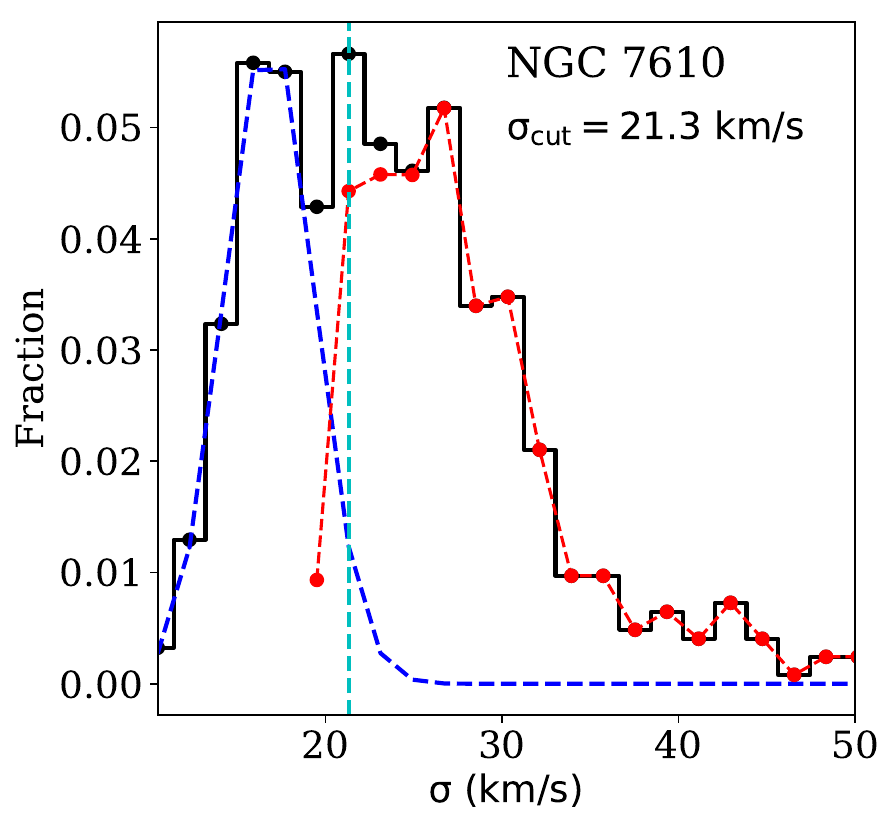} &
        \includegraphics[height=3.4cm]{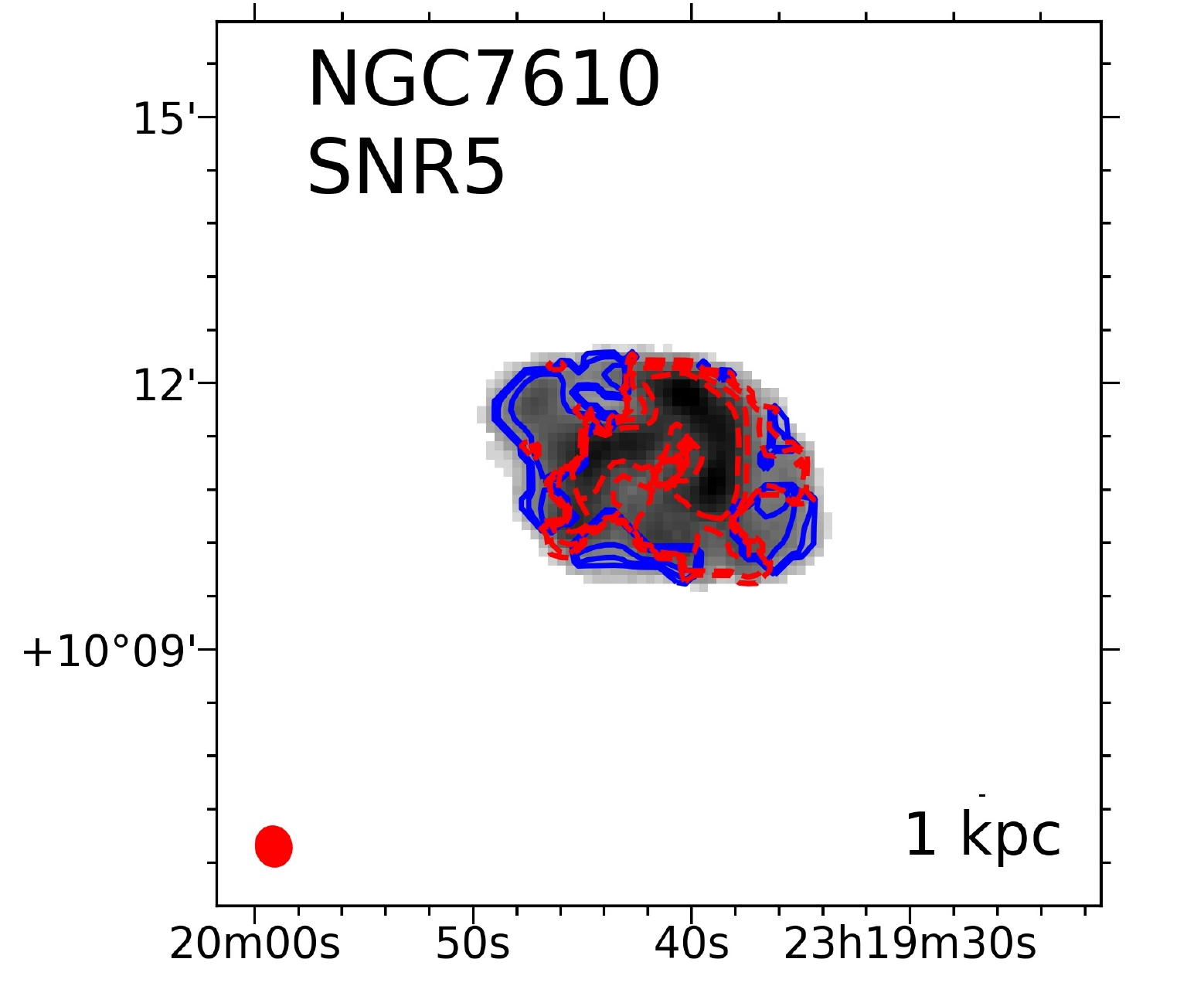} &
        \includegraphics[height=3.4cm]{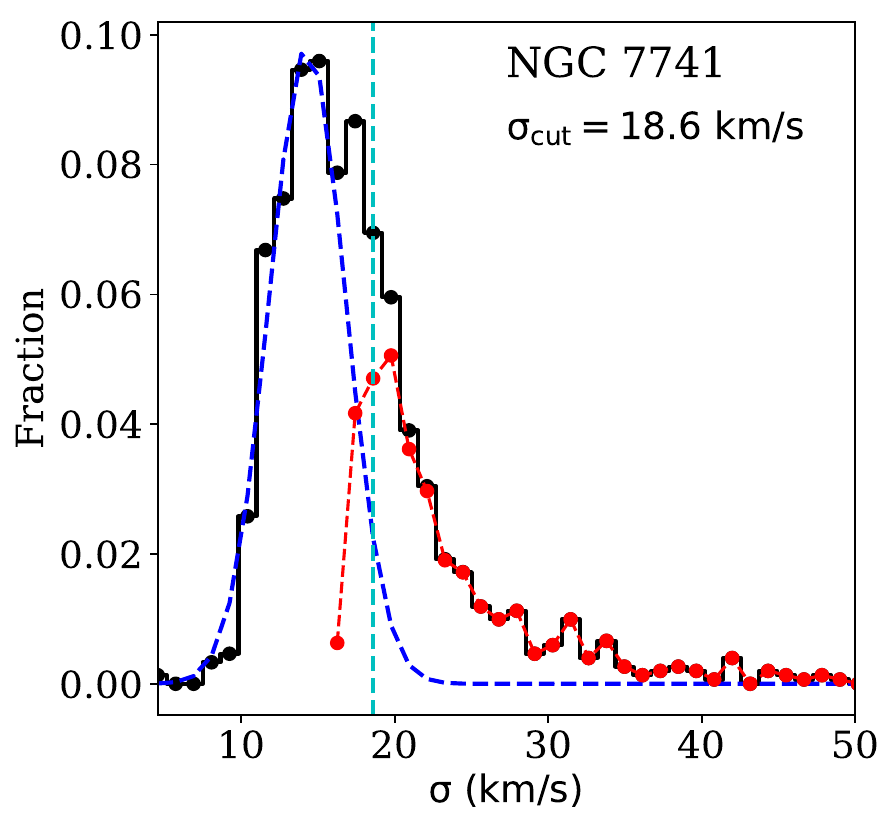} &
        \includegraphics[height=3.4cm]{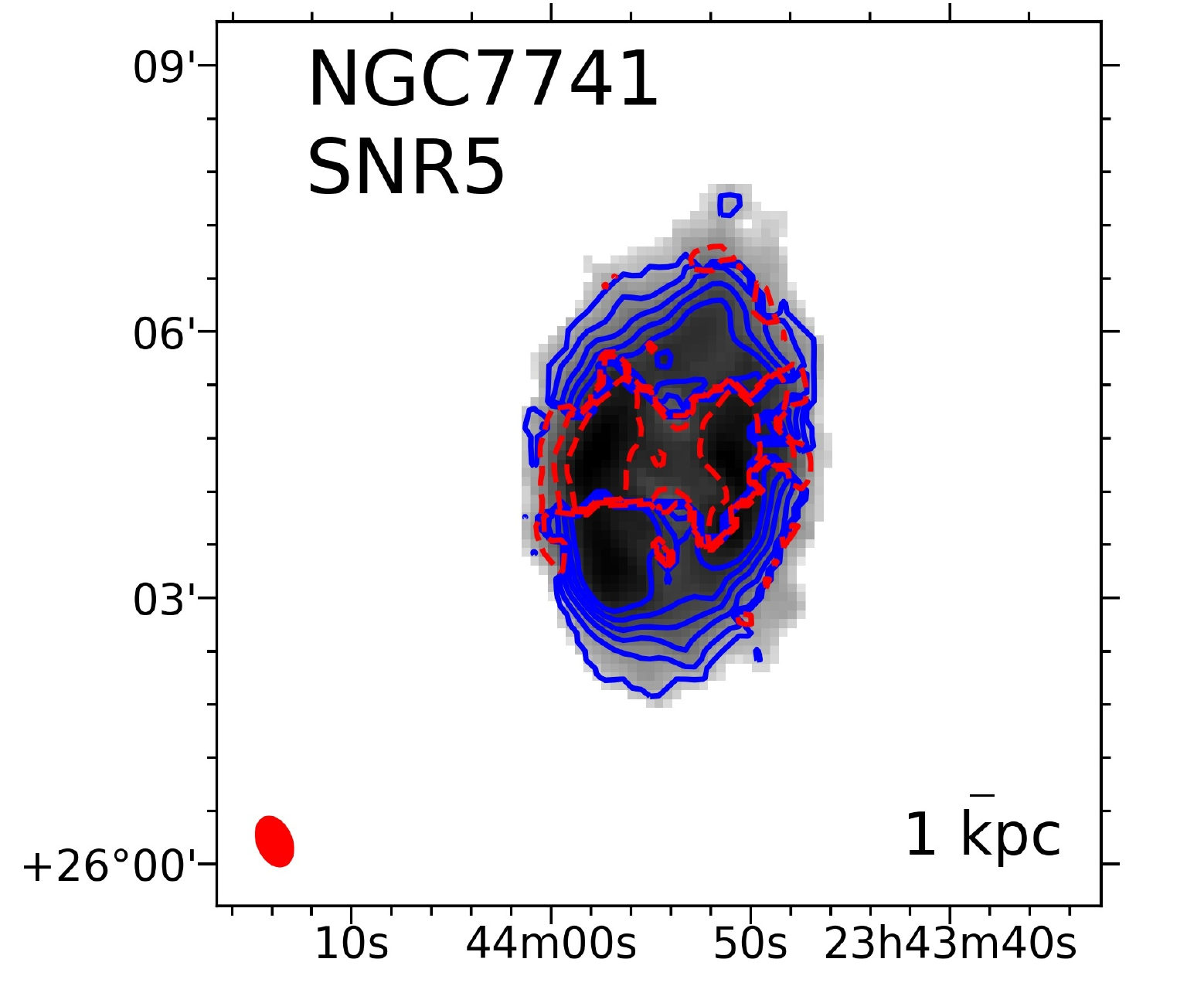} \\
     \end{tabular}
     \begin{tabular}{cc}
          \includegraphics[height=3.5cm]{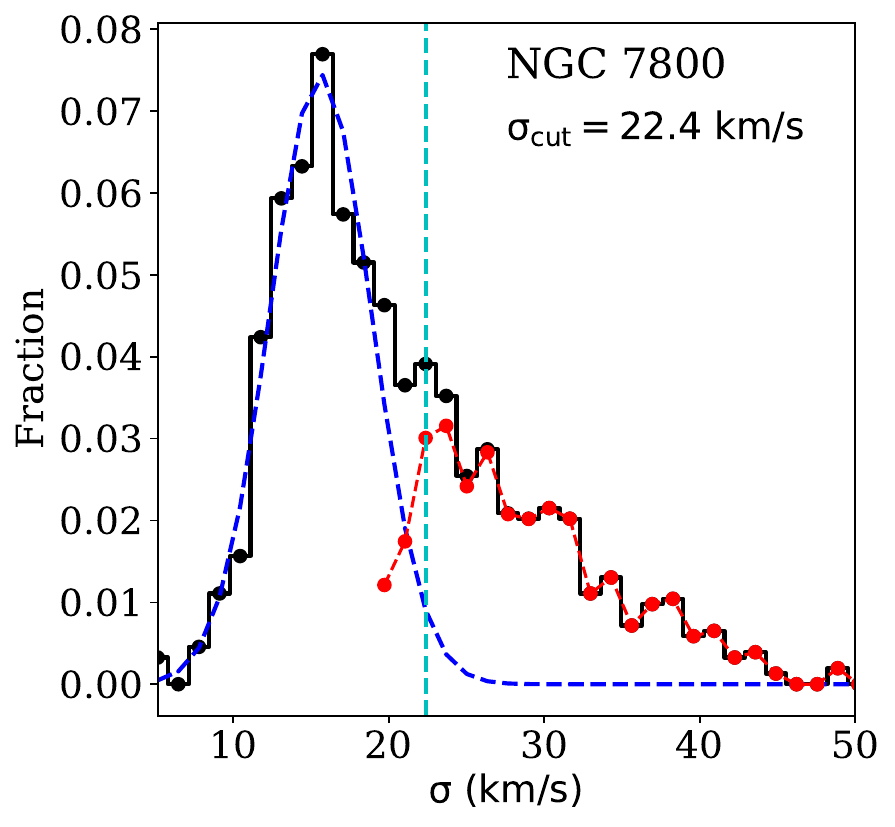} &
          \includegraphics[height=3.5cm]{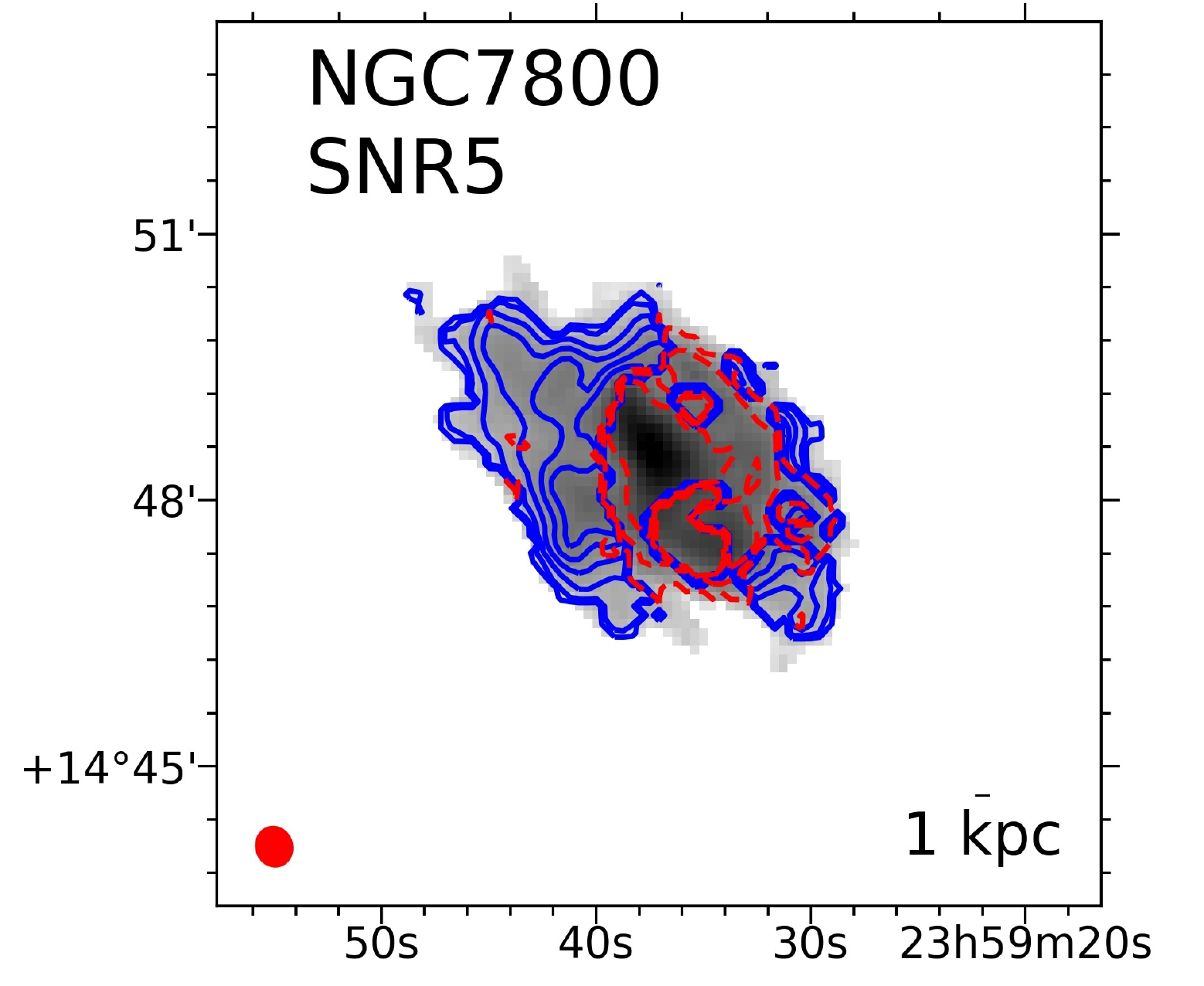} \\
     \end{tabular}
    
    \caption{The histograms represent the distribution of all the recovered velocity widths ($\sigma_{HI}$) from both single and double component Gaussian fitted to the spectra with SNR cutoff five.  The bimodal distribution of the velocity width is evident in these histograms. The blue dashed line in the histograms represents the fitted Gaussian and the red dashed line represents the data subtracted from the fitted model Gaussian; the cyan dashed vertical line shows the velocity cutoff line for distinguishing the cold and the warm gas.  The maps next to each histogram represent the CNM and WNM distribution over the Moment zero map of individual galaxies. The solid blue contours represent the cold neutral medium (WNM), and the dashed red contours represent the warm neutral medium (CNM) detected at SNR 5. }
    \label{fig:snr5}
\end{figure*}

\subsection{The Efficiency of the fitting}
\label{subsec:effciency_of_fitting}
 The spectral resolutions used in most of the previous studies to distinguish between the cold and the warm phases of the ISM were better than $\rm 4 \ km \thinspace s^{-1}$. For our sample galaxies, the spectral resolution is $\sim 6.6 \ km \thinspace s^{-1}$. Theoretically,  in the absence of turbulence in the ISM, the HI spectral width is expected to be $\sim$ few km/s solely due to kinetic temperature. However, the turbulence significantly increases this width, such as the signature of a two-phase medium can be observed at much higher spectral widths \citep{gautam2020}. \citet{gautam2020} used the HI spectral cubes of two galaxies, NGC 3184 and NGC 7793, and decomposed them into multiple Gaussian components (using the same technique we use here). They found a clear bimodal distribution in the widths of the Gaussian components (see Fig. 3 in their paper). The bimodal distribution they found was divided into narrow and wide regimes by a spectral width of $10$ km\thinspace s$^{-1}$. They identified any gas with a spectral width of less than $10$ km\thinspace s$^{-1}$ as CNM and more than $10$ km\thinspace s$^{-1}$ with WNM. In such cases, a velocity resolution of $6.6$ km\thinspace s$^{-1}$ could be sufficient to identify the CNM and WNM in our sample galaxies. 

\citet{naren2016} used a spectral resolution of $\sim 1.7$ km\thinspace s$^{-1}$ to characterize the efficiency of the Gaussian decomposition routine. As the spectral resolution of our sample galaxies is $\sim$ $6.6$ km\thinspace s$^{-1}$, we attempt to characterize the efficiency of the routine at this velocity resolution. To do that, we first produce a set of synthetic spectra with different SNR and spectral widths. For the first set, we use only single-component Gaussian spectra. Then we pass these spectra through our routine and compare the recovered widths with the assumed ones. We define efficiency as what percentage of the recovered widths matched the input ones within 3-sigma error. We repeat this exercise with spectra with different SNRs by adding appropriate noises. At each SNR, we generate 100 synthetic spectra for our characterization. In the top panel of Fig. \ref{fig:simspec} we plot the efficiency of the routine as a function of SNR for different assumed input spectral widths. As can be seen, at SNR $\sim$ 5, the routine achieves $\sim 90\%$ efficiency even with the lowest spectral width of 6 km/s. Secondly, we perform the same exercise with double-component Gaussian profiles. We produce synthetic HI spectra at different SNRs with narrow (width between $6$-$12$ km\thinspace s$^{-1}$) and broad (width between $12$-$16$ km\thinspace s$^{-1}$) Gaussian components. Adopting a similar criterion, we estimate the efficiency of the routine by matching both the narrow and the broad components with the assumed one. A spectrum is correctly recovered when both the widths of the narrow and the broad components match the input within 3-sigma error. As can be seen from the bottom panel of Fig. \ref{fig:simspec}, the efficiency increases with the increasing width of the input spectra (both narrow and broad). We also find that at the lowest input SNR of this exercise, i.e., 7.5, the efficiency of the routine is $\sim$ 80\%. This efficiency is consistent for any choice of combination for the input spectral widths (of the narrow and broad components). It should be noted that, for a double Gaussian spectrum, each component is made to have a significance of at least 3, leading to a total global SNR of 6 or more. Our exercise suggests that the efficiency of our routine is reasonable and can indeed be used for a velocity resolution of 6.6 km\thinspace s$^{-1}$.

\subsection{Results of decomposition}
We use our routine to decompose the line of sight HI spectra in our sample galaxies to identify the narrow (CNM) and the broad components (WNM). We tried image cubes with different spatial resolutions for this study and found that cubes with spatial resolution $\sim 15^{\prime\prime} \times 15^{\prime\prime}$ to $\sim 25^{\prime\prime} \times 35^{\prime\prime}$ show the most distinguishable bimodality in the distribution of recovered velocity widths ($\sigma_{HI}$). It should be mentioned here that while decomposing, we use the MOM0 maps as masks to avoid passing only noise (in all channels) to our decomposition routine.  Next, we extract the spectra at each pixel within the galaxy and pass them through our routine for Gaussian decomposition. We find that at high SNR of 10, all the spectra were best fitted by a single Gaussian. However, if we keep decreasing the SNR cutoff, then a tiny fraction of spectra are fitted with the double Gaussian; none of the spectra is fitted with more than two Gaussian even at the lowest SNR cutoff value of 3 . We then plot the histogram of the widths ($\sigma_{HI}$) of all the recovered components and notice bimodality in the histograms  (a peaked distribution at narrow line width and a tail extended to broad line width) , indicating the presence of a two-phase medium in almost all the galaxies from our sample.  As the recovery efficiency of the routine is high at the SNR cutoff of 10, we first use this cutoff in SNR 10 so that the demarcation between the narrow and the broad component is obtained with high confidence. However, an SNR cutoff of 10 excludes a significant number of the spectra in the cubes. Consequently, a significant amount of HI gas stays undistinguished inside the galaxy at this high signal-to-noise ratio.  As tested earlier, our routine works at a reasonable efficiency even at lower SNRs;  we run the routine on our galaxies at lower SNRs: at SNR 7, 5 and 3. As we keep lowering the SNR cutoff, more cold and warm gases are detected. However, with the decreasing of SNR cutoff, the efficiency of our routine decreases also. Comparing the distribution of $\sigma_{HI}$ and the maps of the cold and warm gas detected at different SNR, we choose to use the optimum SNR at SNR 5 (where the efficiency of the routine $\sim 90\%$). The comparison of the detected cold and warm gas at different SNR cutoff values is presented in Appendix \ref{diff_snr}.

 We use a Gaussian fit to the peaked distribution of $\sigma_{HI}$ at SNR 5 to approximately identify the bimodal cutoff velocity or the width separating narrow and wide components where the contribution of the peaked and long-tailed distribution is the same. The gas found with $\sigma_{HI}$ smaller than this separating velocity width is distinguished as cold gas and those with $\sigma_{HI}$ greater than this velocity width as warm gas.
The histograms in figure \ref{fig:snr5} represent the distribution of the recovered widths ($\sigma_{HI}$) of the velocity components using an SNR cutoff 5 in the case of these galaxies. The dashed vertical lines indicate the widths we choose to demarcate the narrow and the broad components. The maps next to each histogram show the cold gas and the warm gas distribution over the Moment zero maps of the individual galaxies. Further, at SNR 5, we calculate the mass of the cold HI gas and the cold gas mass fraction for all the galaxies. The cold gas mass, cold gas mass fraction and the bimodal cutoff velocities of each of the galaxies are listed in Table \ref{tab:cnm_wnm}. 

In Fig.~\ref{fig:snr5} maps (panels next to histograms), we can see the distribution of cold and warm gas in each galaxy. We see that some galaxies are cold gas dominated and some are warm gas dominated. Traditionally, the cutoff used in these kinds of studies was high (10 or more, see, \citep{Warren_2012,naren2016}). As a result, the amount of cold gas recovered was low. However, in our method, we detect far more cold gas. From the maps, we can see that the outskirts are dominated by cold gas for most of our galaxies, whereas warm gas dominates the central region. This could happen because in the central region, star formation activity is higher, and hence, there exists an increased amount of turbulence. This, in turn, blends the HI spectra, significantly increasing the width. On the other hand, low star formation and turbulence introduce minimal blending at the outskirt. This trend, however, has been seen in previous studies as well \citep[see, e.g.,][]{Warren_2012}. As we recover a significant amount of cold and warm gas in these galaxies, these cold and warm maps can further be used to investigate the connection between star formation and different phases of the ISM.

 \begin{table}
 \addtolength{\tabcolsep}{-2pt}
    \centering
    \begin{tabular}{|c|c|c|c|}
        \hline
         Galaxy    & Cold gas        &  Cold Gas   & Bimodal cutoff\\
         name      & mass ($M_{\sun}$) & Fraction  &  velocity($kms^{-1}$) \\
         \hline
NGC0784 & $(4.3\pm0.5)\times10^7$ & $0.101\pm0.013$ & 10.6 \\
NGC1156 & $(4.7\pm0.5)\times10^8$ & $0.85\pm0.09$ & 19.6 \\
NGC3027 & $(3.7\pm2.0)\times10^9$ & $0.7\pm0.4$ & 17.4 \\
NGC3359 & $(4.8\pm0.8)\times10^9$ & $0.47\pm0.08$ & 15.0 \\
NGC4068 & $(1.23\pm0.14)\times10^8$ & $0.92\pm0.10$ & 14.1 \\
NGC4861 & $(6.5\pm1.1)\times10^7$ & $0.069\pm0.012$ & 10.6 \\
NGC7292 & $(2.5\pm0.4)\times10^8$ & $0.58\pm0.09$ & 18.5 \\
NGC7497 & $(1.13\pm0.14)\times10^9$ & $0.217\pm0.027$ & 24.8 \\
NGC7610 & $(4.2\pm0.7)\times10^9$ & $0.26\pm0.04$ & 21.3 \\
NGC7741 & $(9.8\pm1.0)\times10^8$ & $0.61\pm0.06$ & 18.6 \\
NGC7800 & $(1.87\pm0.29)\times10^9$ & $0.48\pm0.08$ & 22.4 \\

         \hline
    \end{tabular}
    \caption{ Mass of the cold atomic gas of the galaxies  and the bimodal cut-off velocity determined at SNR 5.}
    \label{tab:cnm_wnm}
\end{table}

\section{Conclusions}
\label{conclusion}
In conclusion, we construct a sample of all galaxies to date observed in HI with the GMRT. This resulted in a sample of 515 galaxies, for which the data is available in the GMRT archive. This is the largest sample of galaxies, having HI interferometric observations. We show several scaling relations for this sample, span several orders of magnitudes, indicating a comprehensive range of different physical properties of galaxies we want to explore. 

As a pilot survey, we select 11 galaxies from the full sample and analyze their data. We present the data analysis and several data products which could be used for different science problems. We find that the data quality is reasonably good, and we could achieve a typical RMS of $\sim$ $1$ mJy/bm per $6.6$ km\thinspace s$^{-1}$ channel. We show that this sensitivity is sufficiently adequate to detect HI in these galaxies far beyond the optical radius. 

We use the resolved HI maps of our pilot sample galaxies to estimate several physical properties of the HI disks, e.g., HI diameter, inclination, position angle, etc. We use these estimated global parameters to investigate how our sample galaxies fit into several scaling relations. We find that our pilot sample spans across the scaling relations indicating it to be a representative sample. However, we plan to use the full sample to investigate the scaling relations in unprecedented detail. 

Further, we use the resolved HI maps to identify the warm and the cold ISM phases in our pilot sample galaxies. We use a multi-Gaussian decomposition method to decompose line-of-sight HI spectra into multiple components to identify the narrow and the broad component. For all our galaxies, we observe a bimodality in the spectral widths of the decomposed Gaussian components. This is an indicator of a two-phase medium. We identify the narrow and broad components and produce the cold and warm HI maps of our sample galaxies. We further find that the amount of cold gas in these galaxies ranges from $\rm \sim 4 \times 10^7 - 5 \times 10^9 \ M_{\odot}$. The cold gas fraction ranges from $\sim 6\% - 90\%$ . We note that this cold gas fraction could be one of the key factors in instigating star formation in these galaxies, which we intend to study in detail in the future. Our pilot survey demonstrates that the quality of the archival data is reasonably good to address several compelling science cases using the full sample.

\section*{Acknowledgements}
We thank Prof. Mousumi Das and Prof. Yogesh Wadadekar for the interesting and helpful discussion regarding this project during the 38th Annual Meeting of Astronomical Society of India, IISER Tirupati, in 2020. We also thank the anonymous reviewers for their suggestions and comments that have helped to improve this paper. This research has made use of the GMRT archival data and the HyperLeda database. GMRT online archive includes more than 120 terabytes of radio interferometric data. It is maintained by National Centre for Radio Astrophysics (NCRA)(Pune, India), while the HyperLeda database is maintained by the collaboration between Observatoire de Lyon (France) and the Special Astrophysical Observatory (Russia).  The research has also utilized the NASA/IPAC Extragalactic Database (NED), operated by the Jet Propulsion Laboratory, California Institute of Technology, under contract with the National Aeronautics and Space Administration.

%%%%%%%%%%%%%%%%%%%%%%%%%%%%%%%%%%%%%%%%%%%%%%%%%%
\section*{Data Availability}

All data used in this study are available publicly from the GMRT online archive. All reduced data (low and high-resolution images, rotation curve, spectra) will be made available for the entire sample in a phased manner. All reduced data from the current sub-sample will be shared at reasonable request to the corresponding author.



\bibliographystyle{mnras}
\bibliography{example} % if your bibtex file is called example.bib

%\begin{thebibliography}{99}
%\end{thebibliography}

%%%%%%%%%%%%%%%%%%%%%%%%%%%%%%%%%%%%%%%%%%%%%%%%%%

%%%%%%%%%%%%%%%%% APPENDICES %%%%%%%%%%%%%%%%%%%%%

\appendix

\section{Results of decomposition of spectra at different SNR cutoff}
\label{diff_snr}
The efficiency of the routine is the best for the spectra with a signal-to-noise ratio greater than equals to ten. Nevertheless, we find that a significant portion of gas inside the galaxy stays undistinguished at this higher SNR cutoff. Now, as described in subsection \ref{subsec:effciency_of_fitting}, the routine works reasonably well at lower SNRs also;  we study this process of differentiating the CNM and WNM at different SNR cutoff values: SNR 7, 5 and 3. The comparison of the detected cold and warm gas for each of the galaxies is presented in figure \ref{fig:snr_comp}.

\begin{figure*}
    \centering
    \begin{tabular}{c c c c}
    \includegraphics[height=3.3cm]{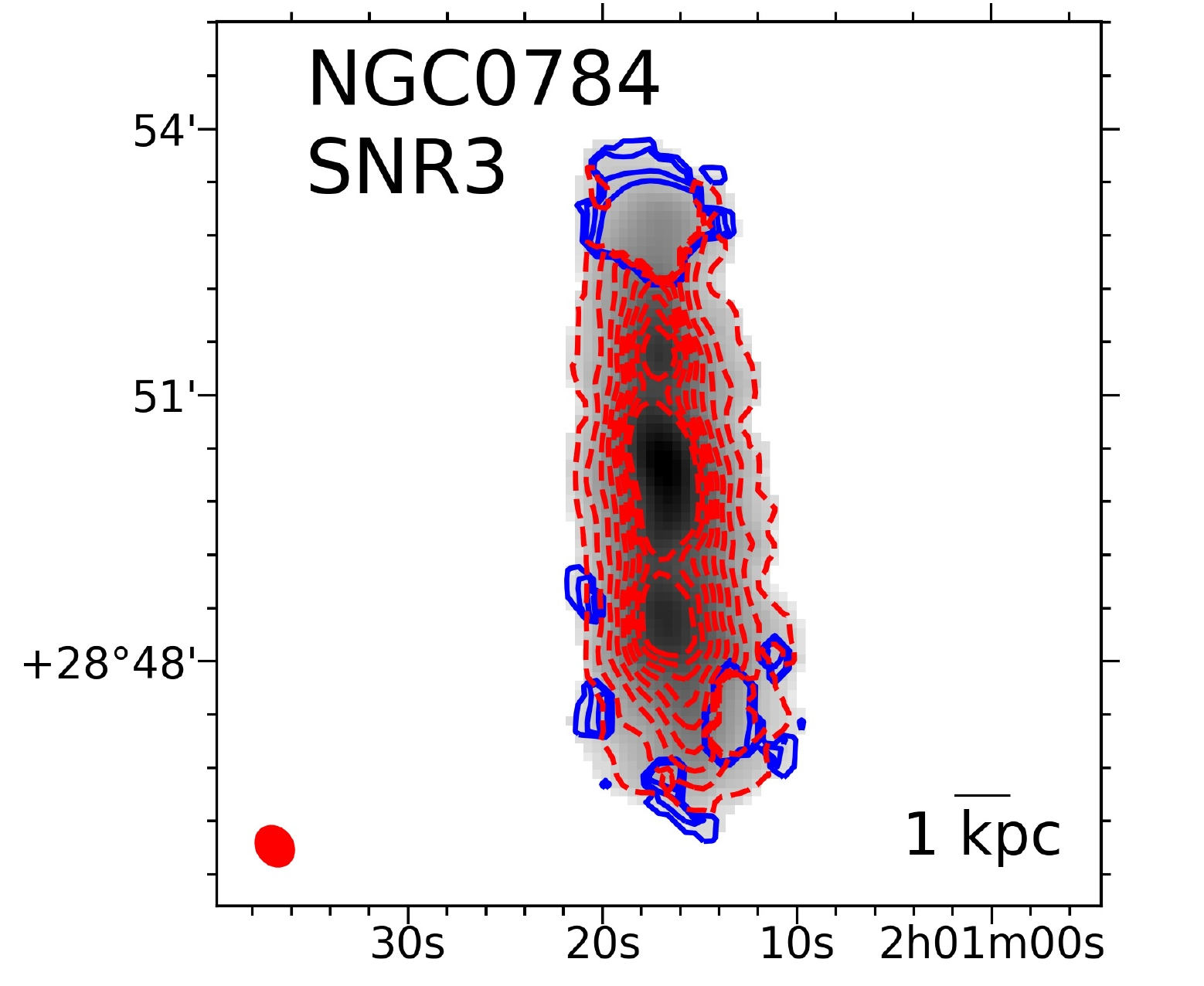} &
    \includegraphics[height=3.3cm]{figs/cnm_wnm_figs/NGC0784SNR5_cnm_wnm.pdf} &
    \includegraphics[height=3.3cm]{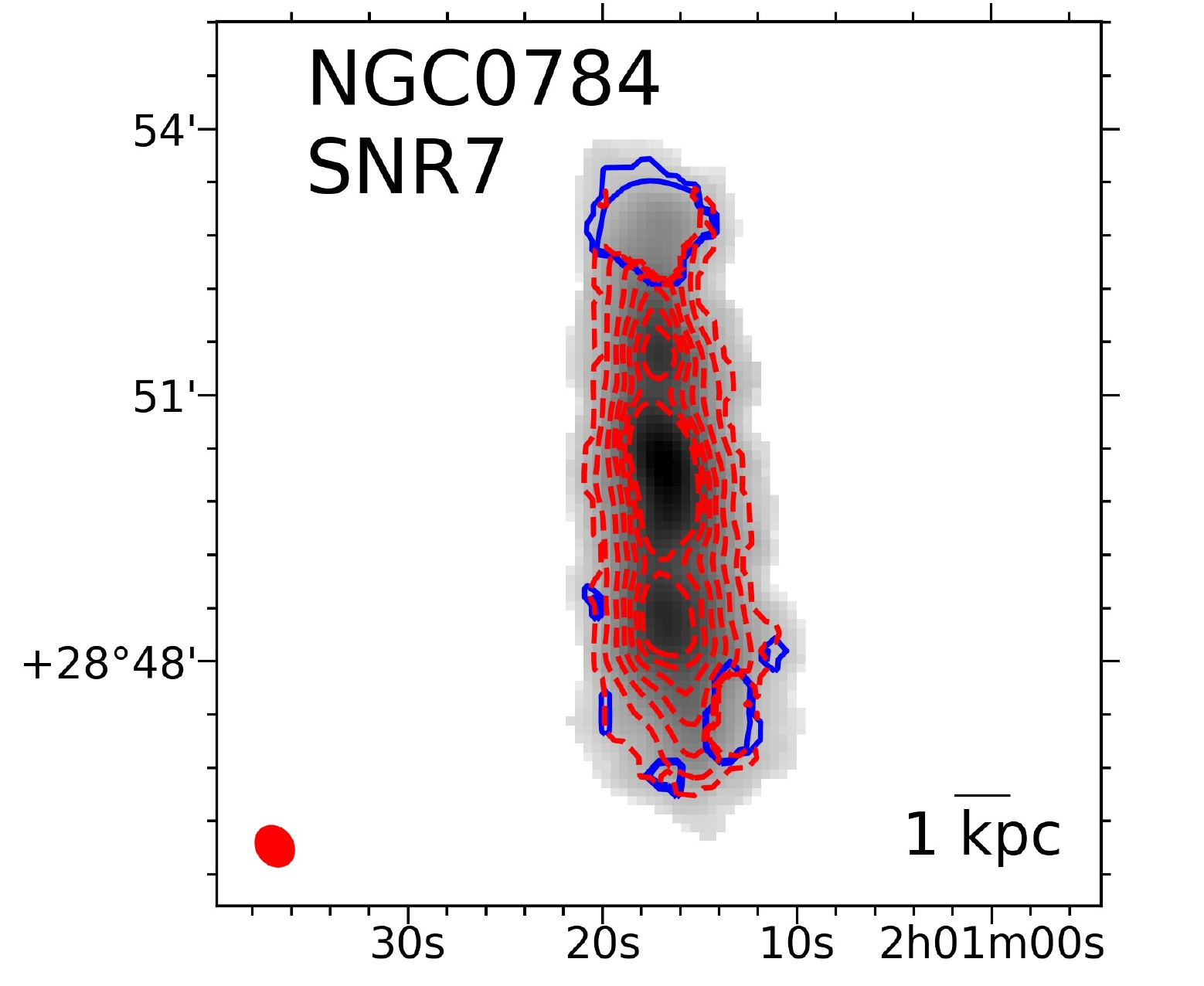} &
    \includegraphics[height=3.3cm]{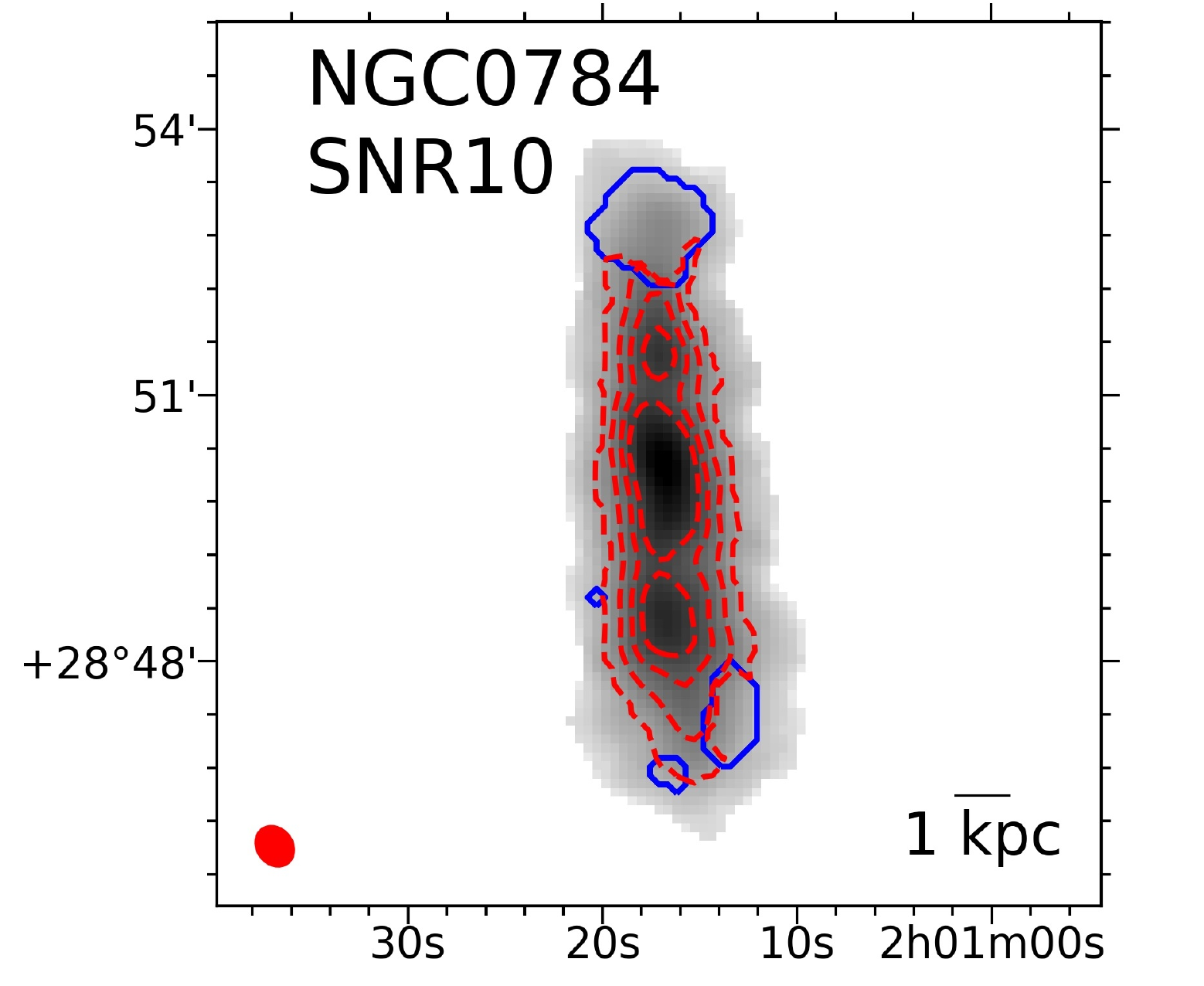} \\
    
    \includegraphics[height=3.3cm]{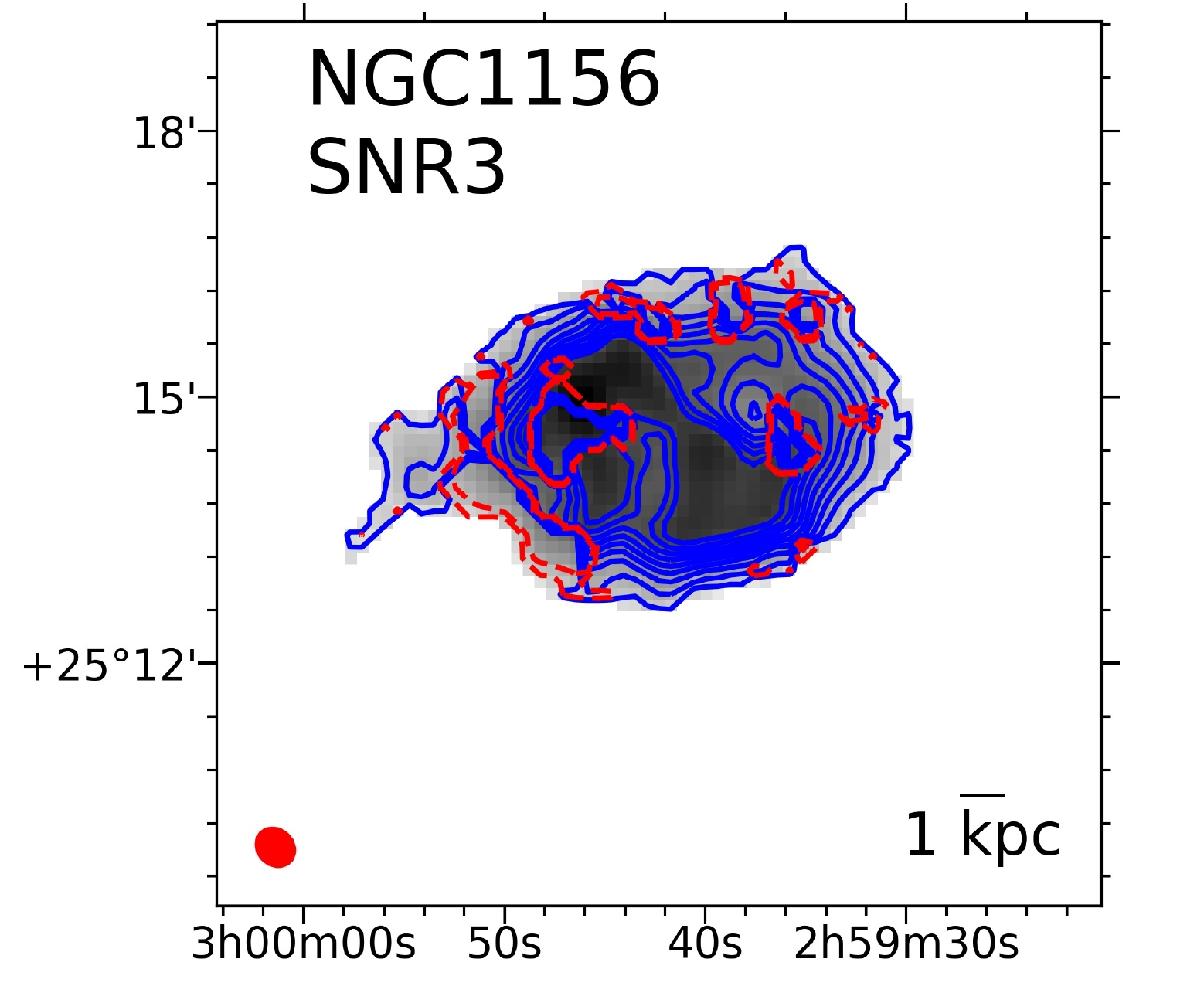} &
    \includegraphics[height=3.3cm]{figs/cnm_wnm_figs/NGC1156SNR5_cnm_wnm.pdf} &
    \includegraphics[height=3.3cm]{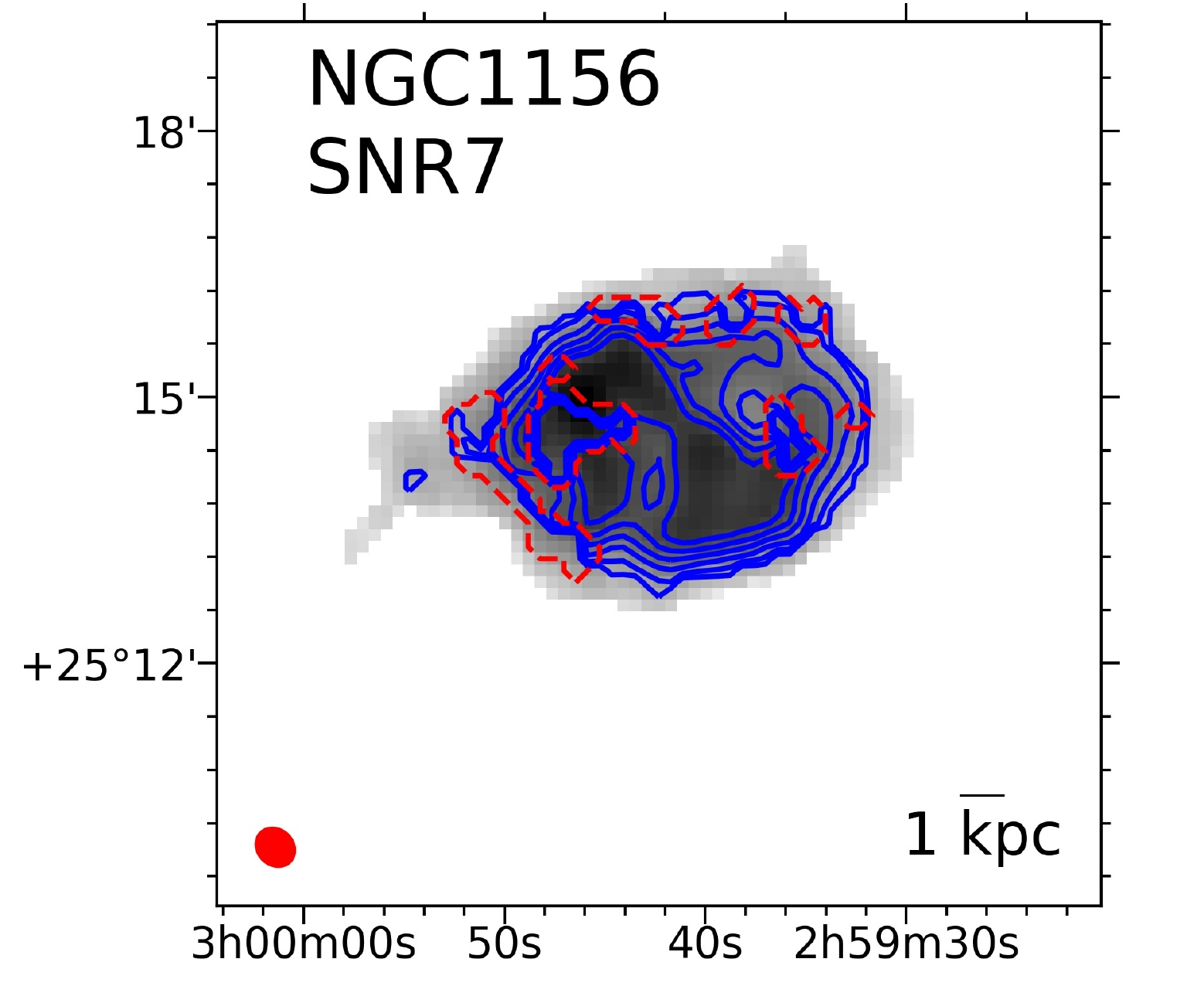} &
    \includegraphics[height=3.3cm]{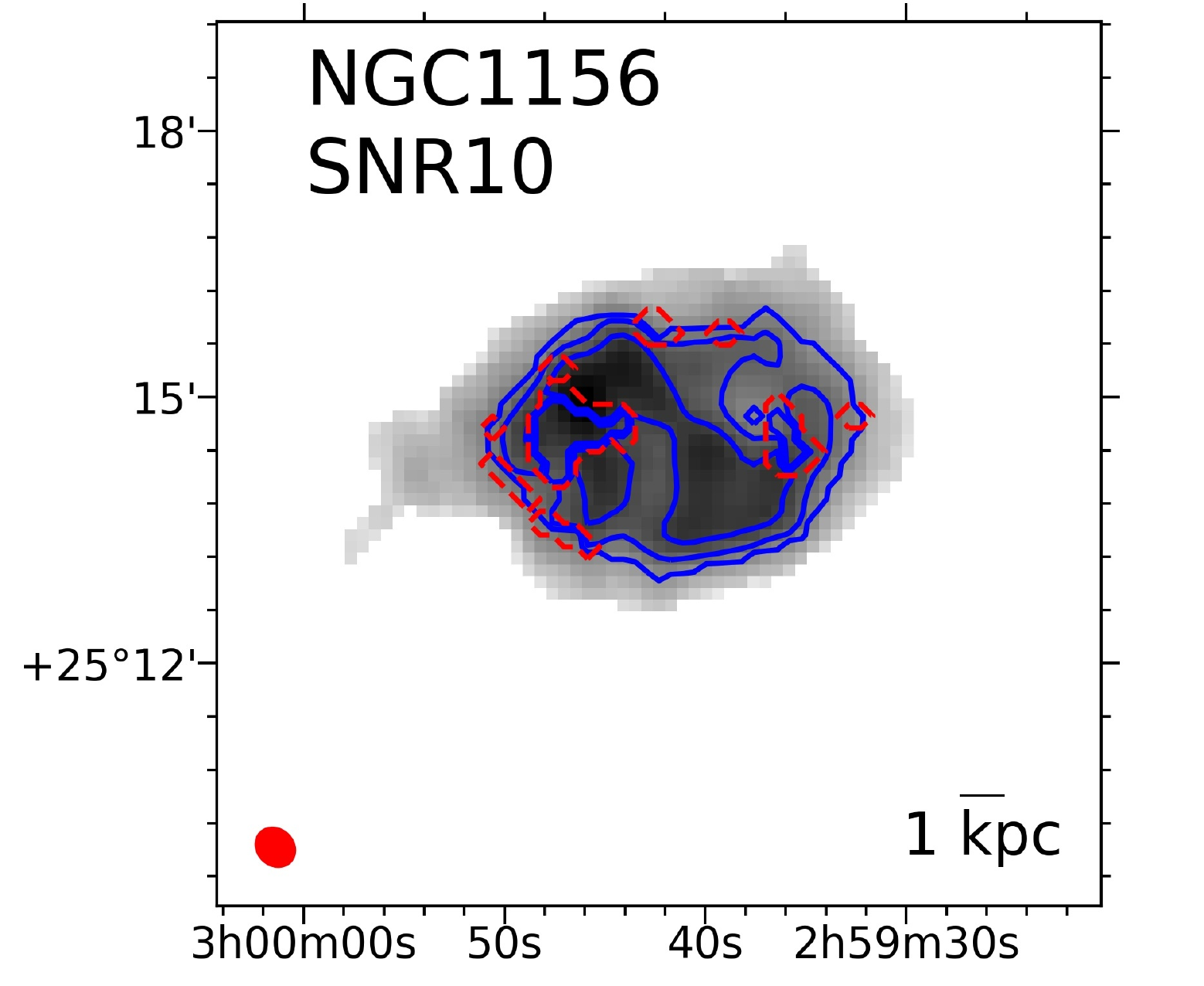} \\
    
    \includegraphics[height=3.3cm]{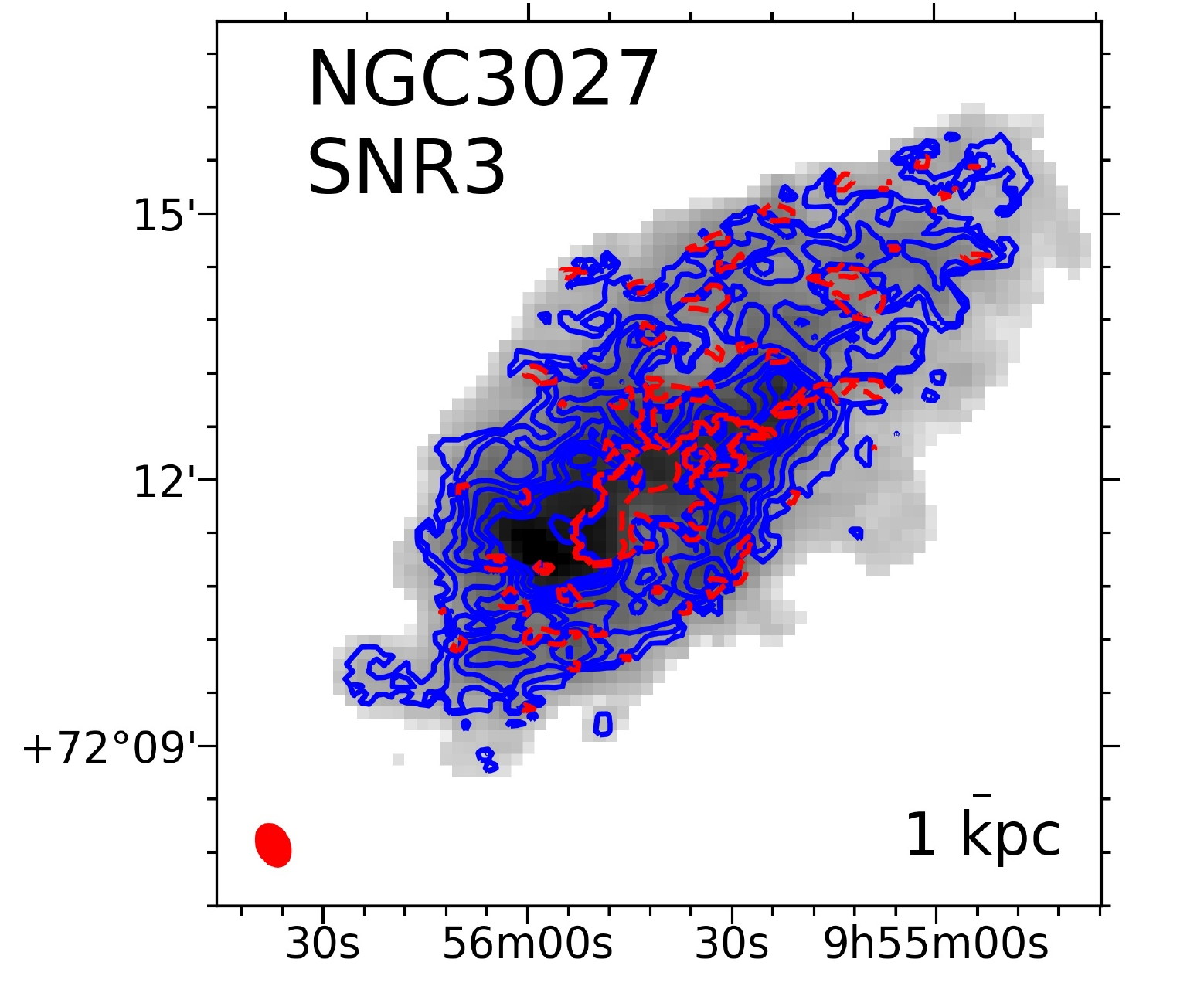} &
    \includegraphics[height=3.3cm]{figs/cnm_wnm_figs/NGC3027SNR5_cnm_wnm.pdf} &
    \includegraphics[height=3.3cm]{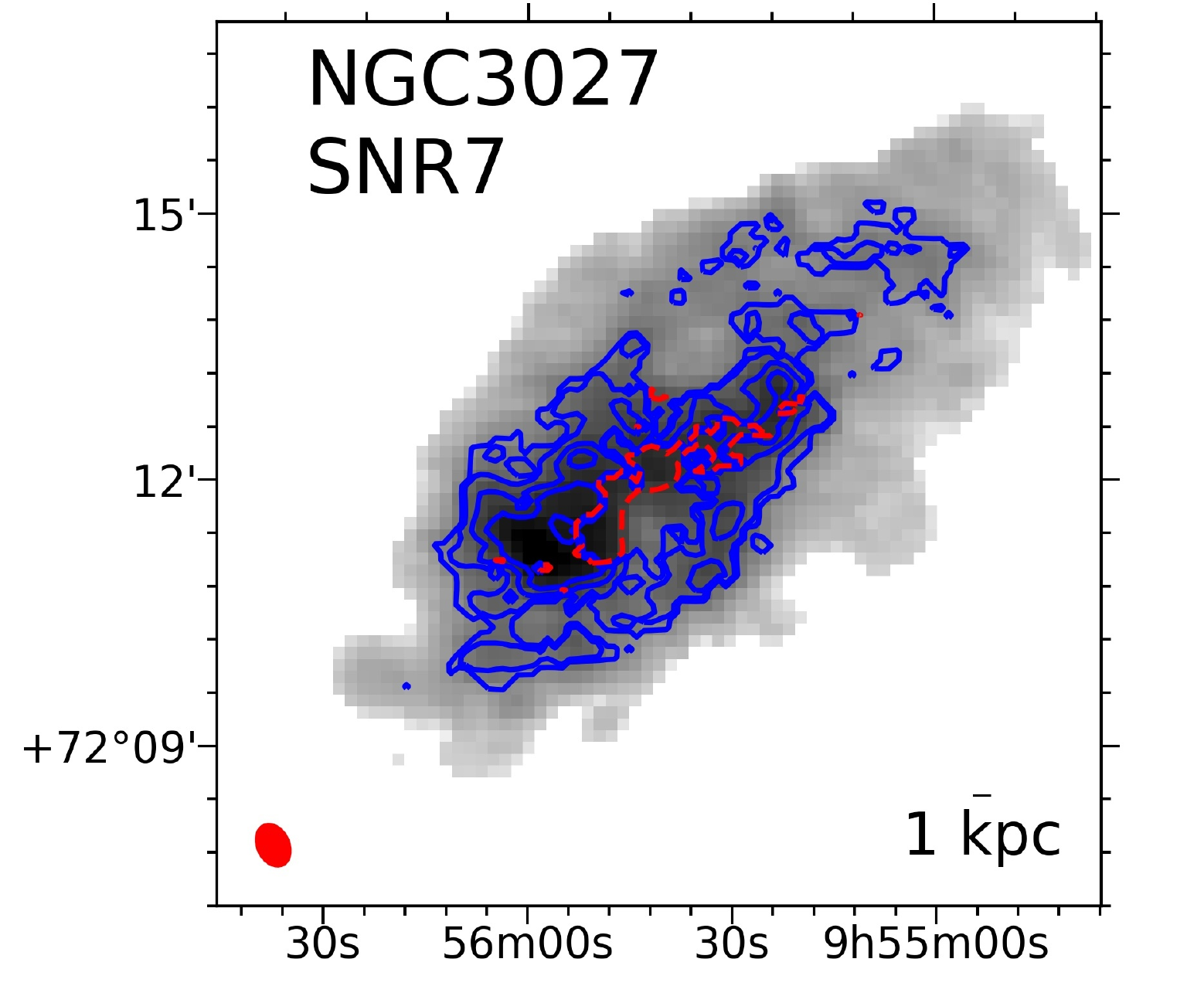} &
    \includegraphics[height=3.3cm]{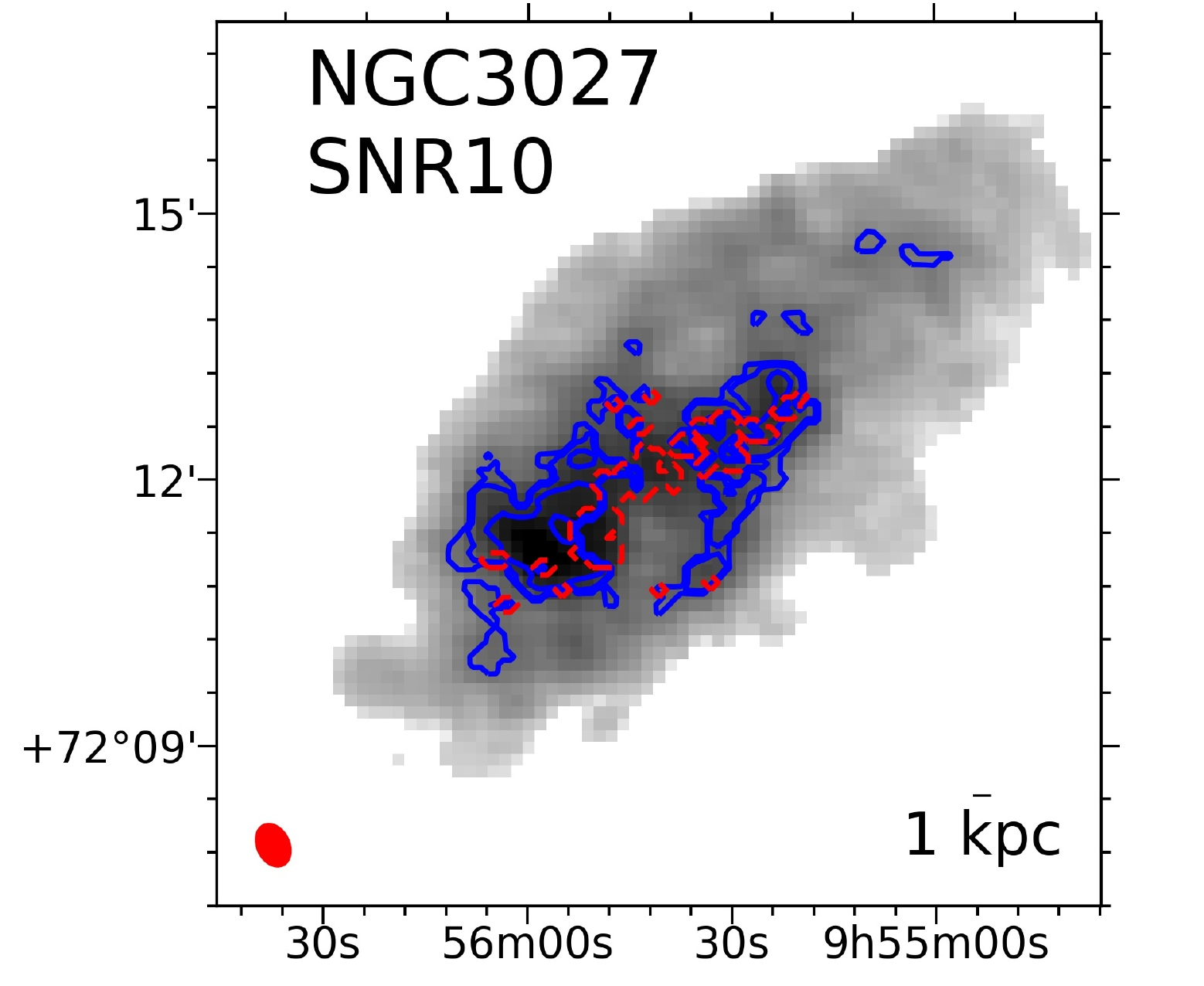} \\
    
    \includegraphics[height=3.3cm]{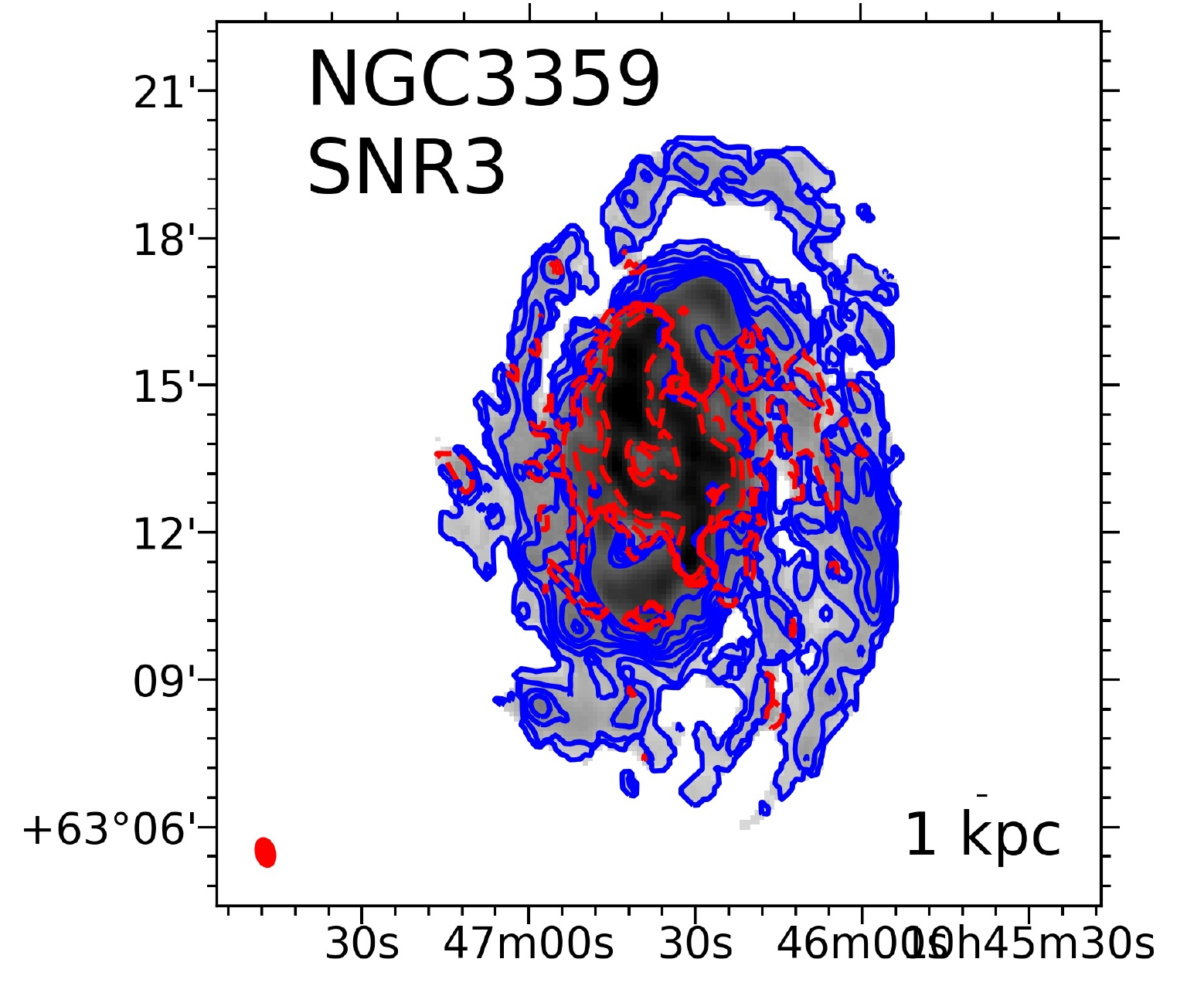} &
    \includegraphics[height=3.3cm]{figs/cnm_wnm_figs/NGC3359SNR5_cnm_wnm.pdf} &
    \includegraphics[height=3.3cm]{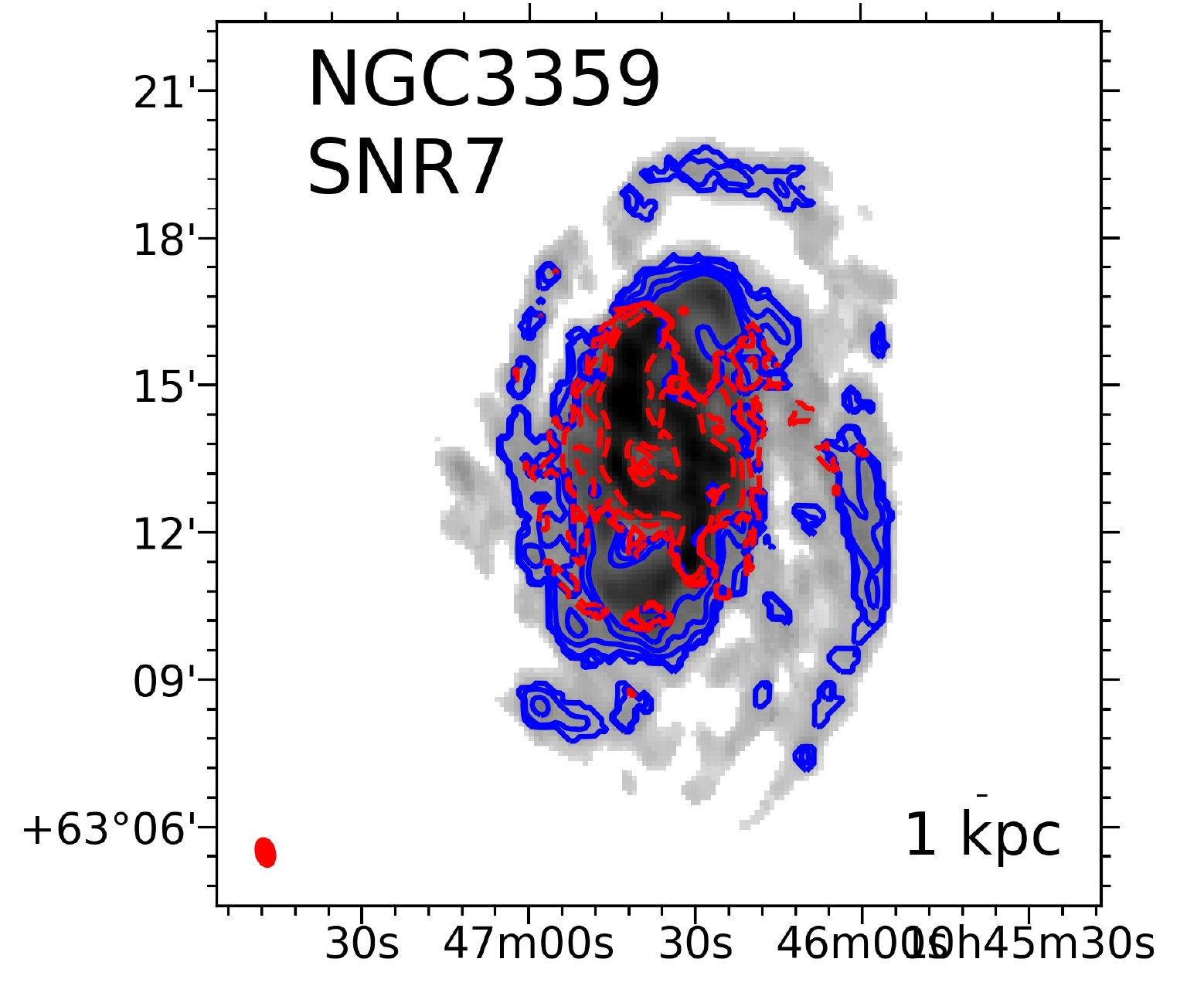} &
    \includegraphics[height=3.3cm]{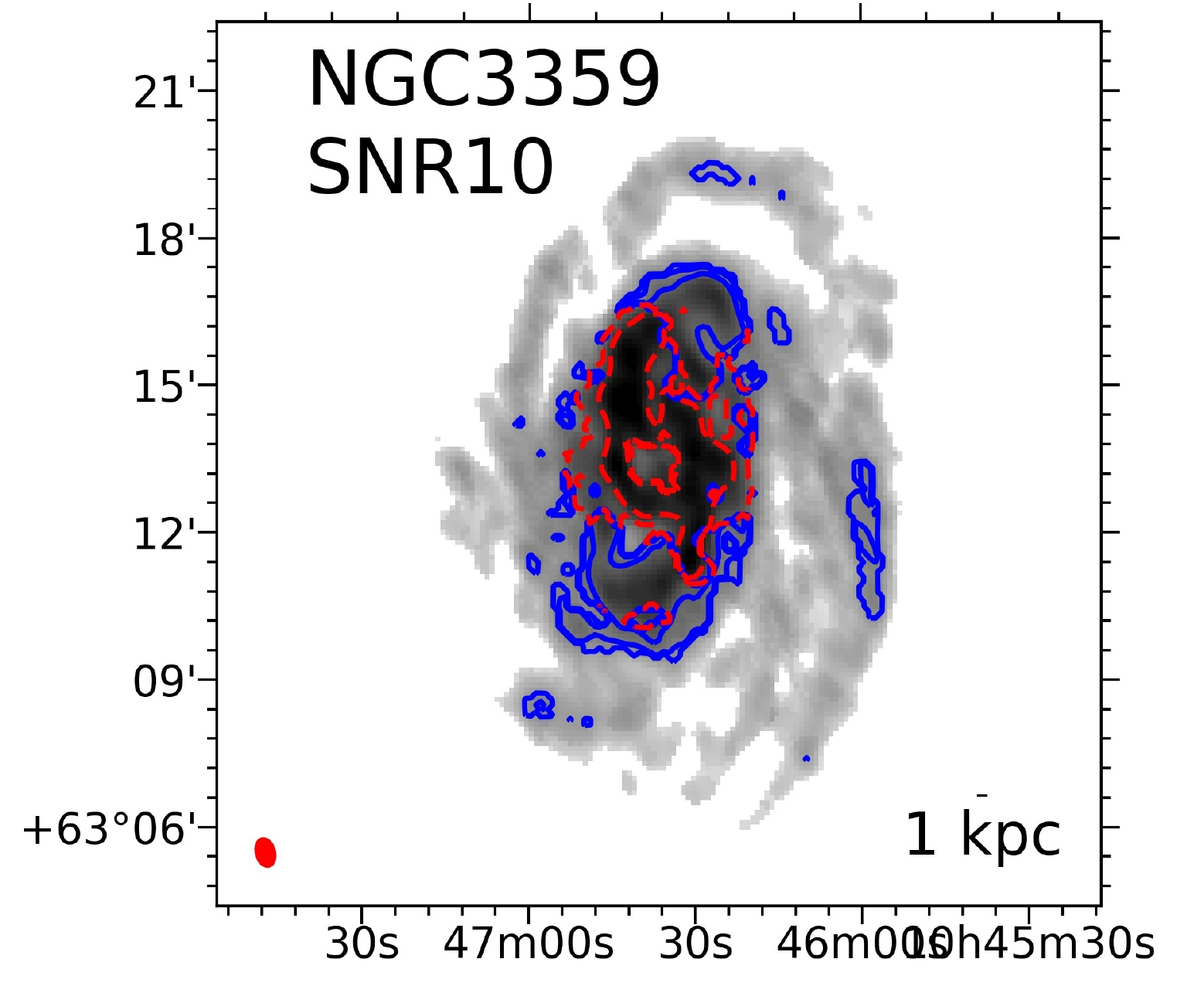} \\
    
    \includegraphics[height=3.3cm]{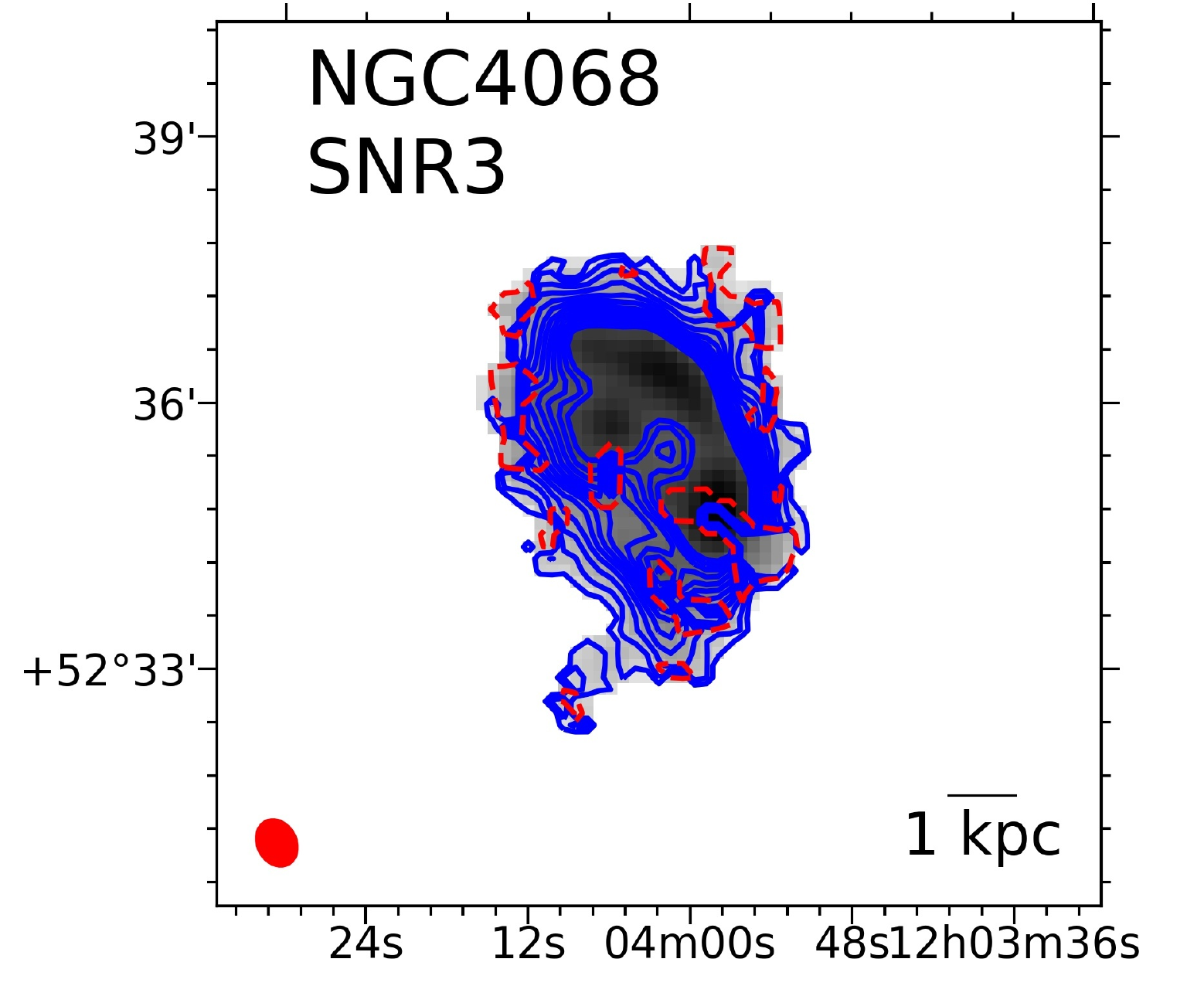} &
    \includegraphics[height=3.3cm]{figs/cnm_wnm_figs/NGC4068SNR5_cnm_wnm.pdf} &
    \includegraphics[height=3.3cm]{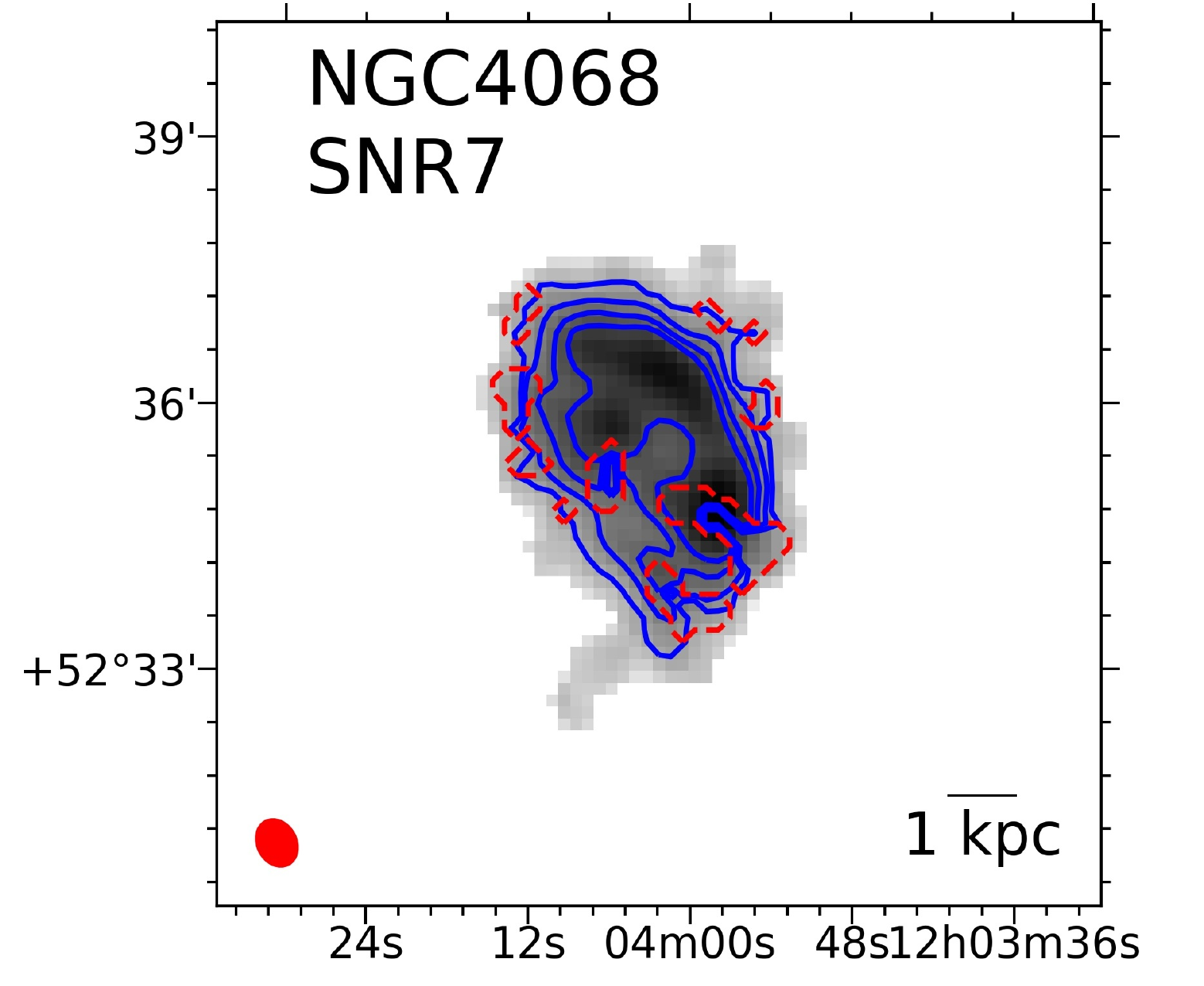} &
    \includegraphics[height=3.3cm]{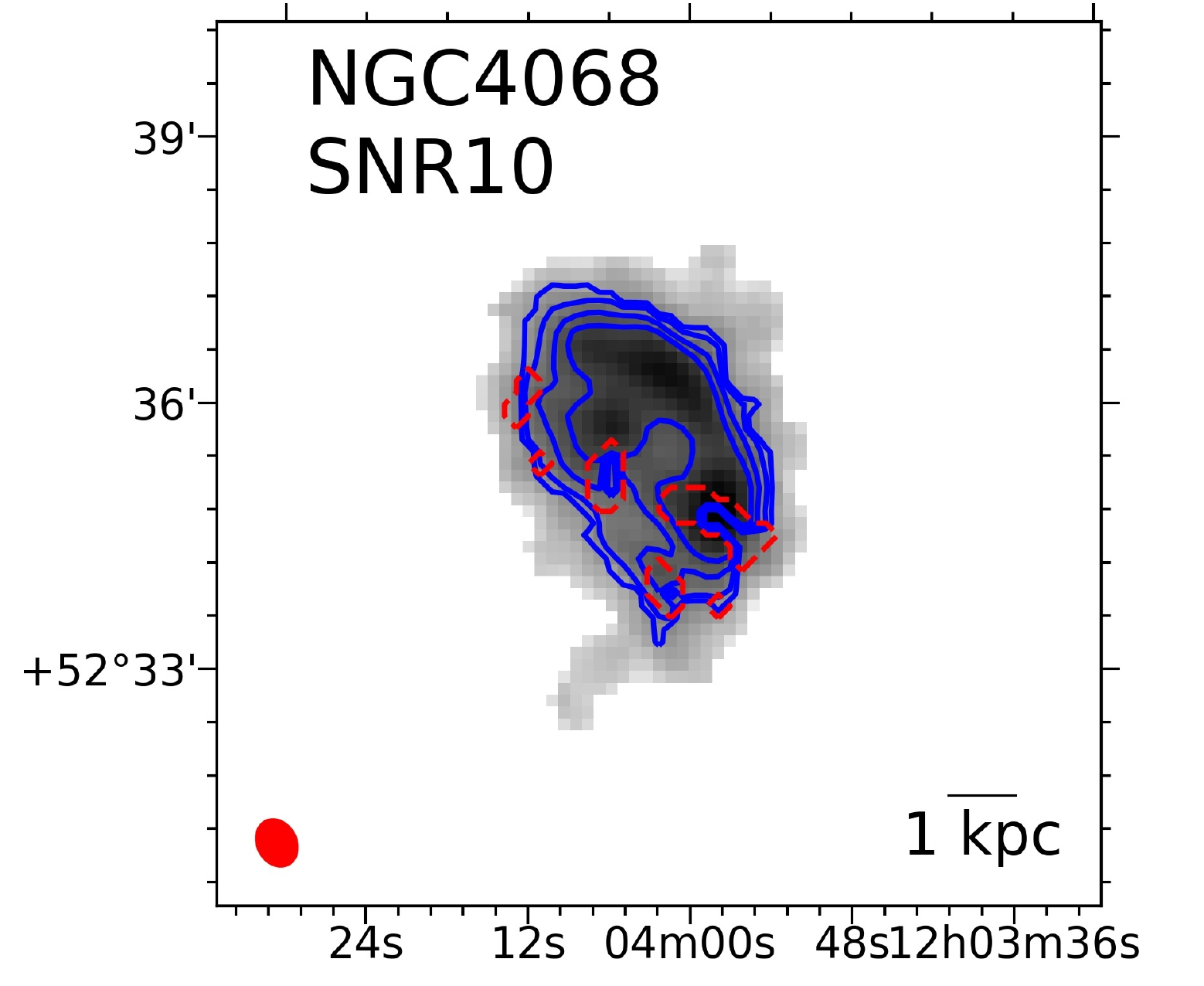} \\
    
    \includegraphics[height=3.3cm]{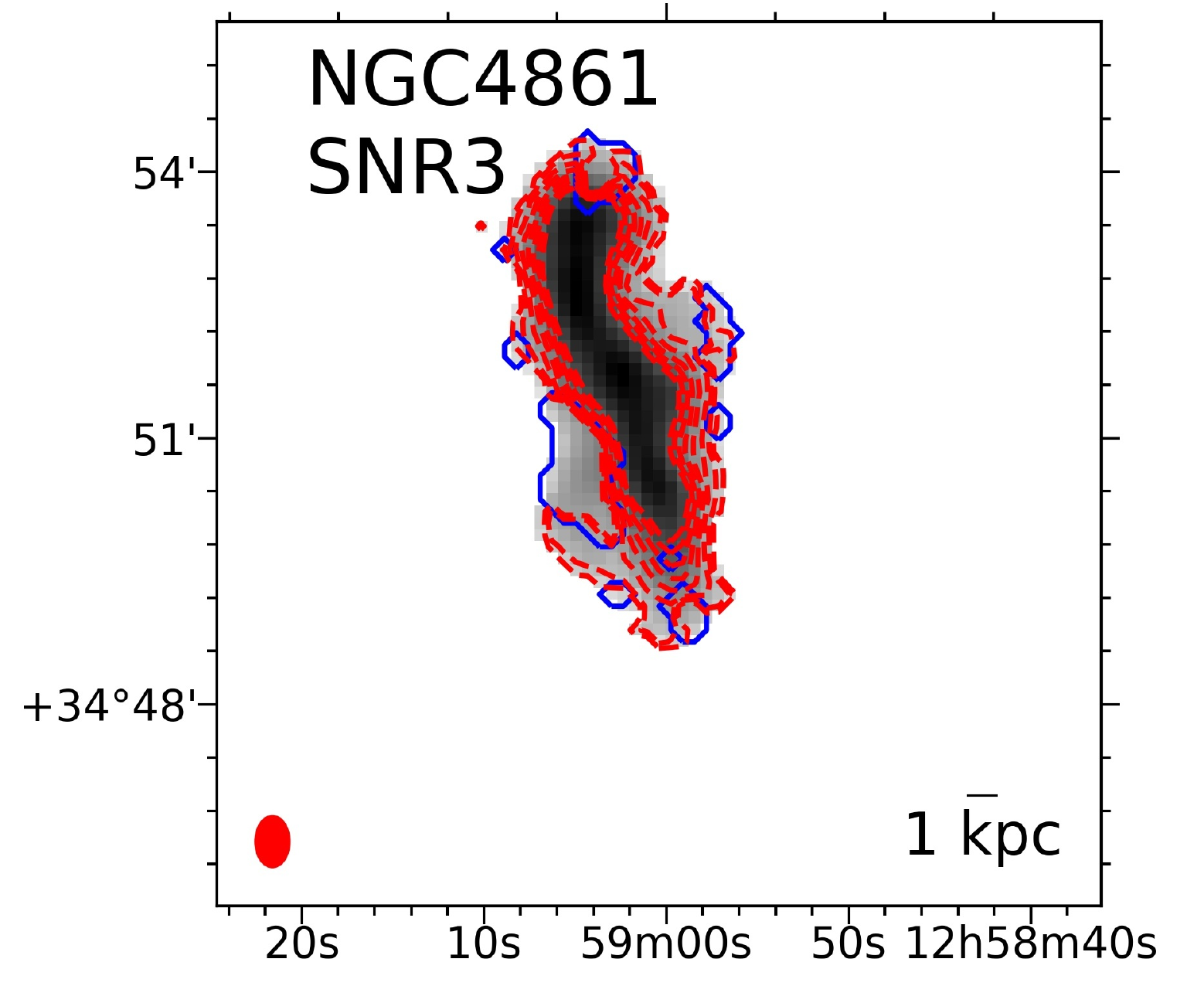} &
    \includegraphics[height=3.3cm]{figs/cnm_wnm_figs/NGC4861SNR5_cnm_wnm.pdf} &
    \includegraphics[height=3.3cm]{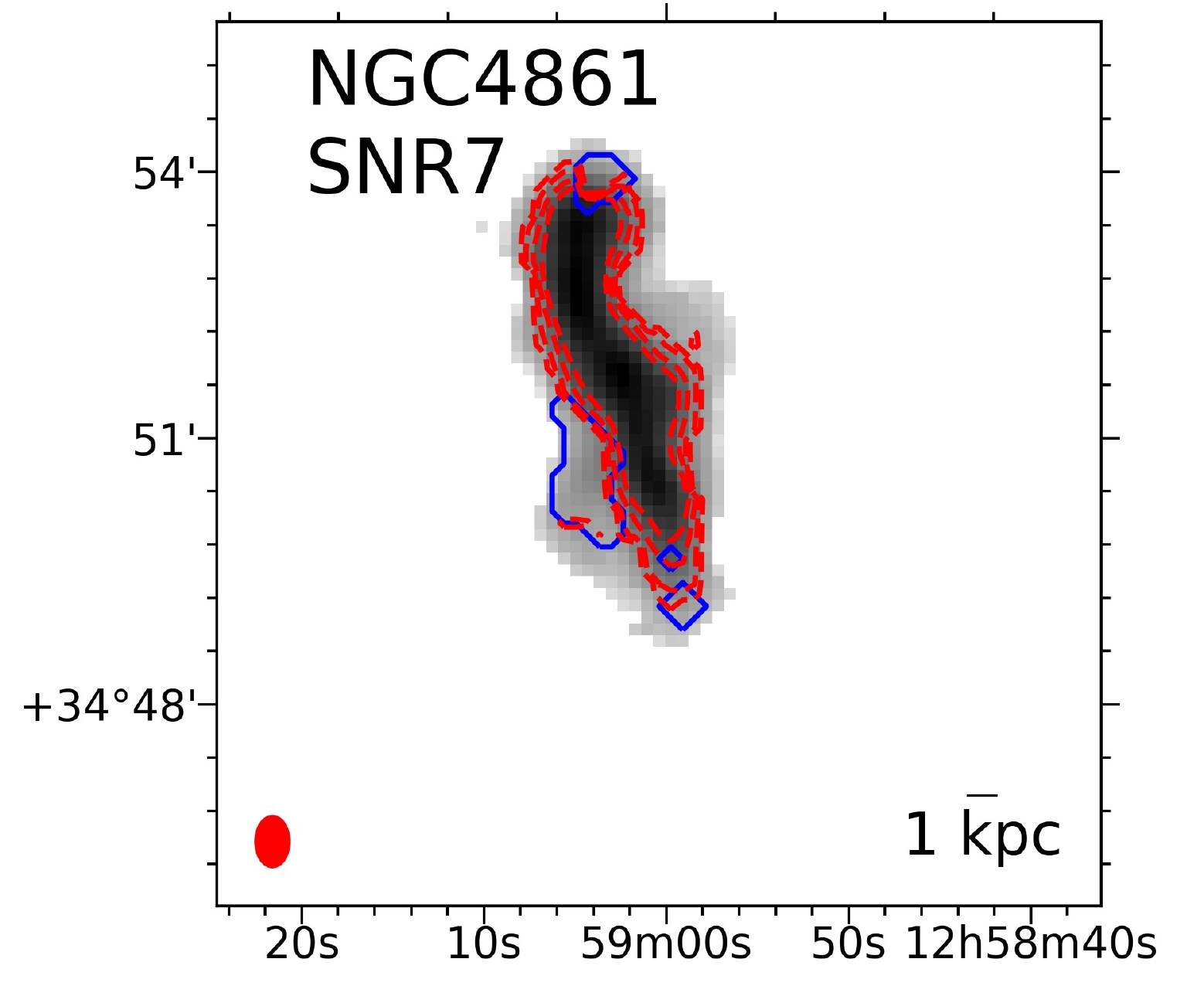} &
    \includegraphics[height=3.3cm]{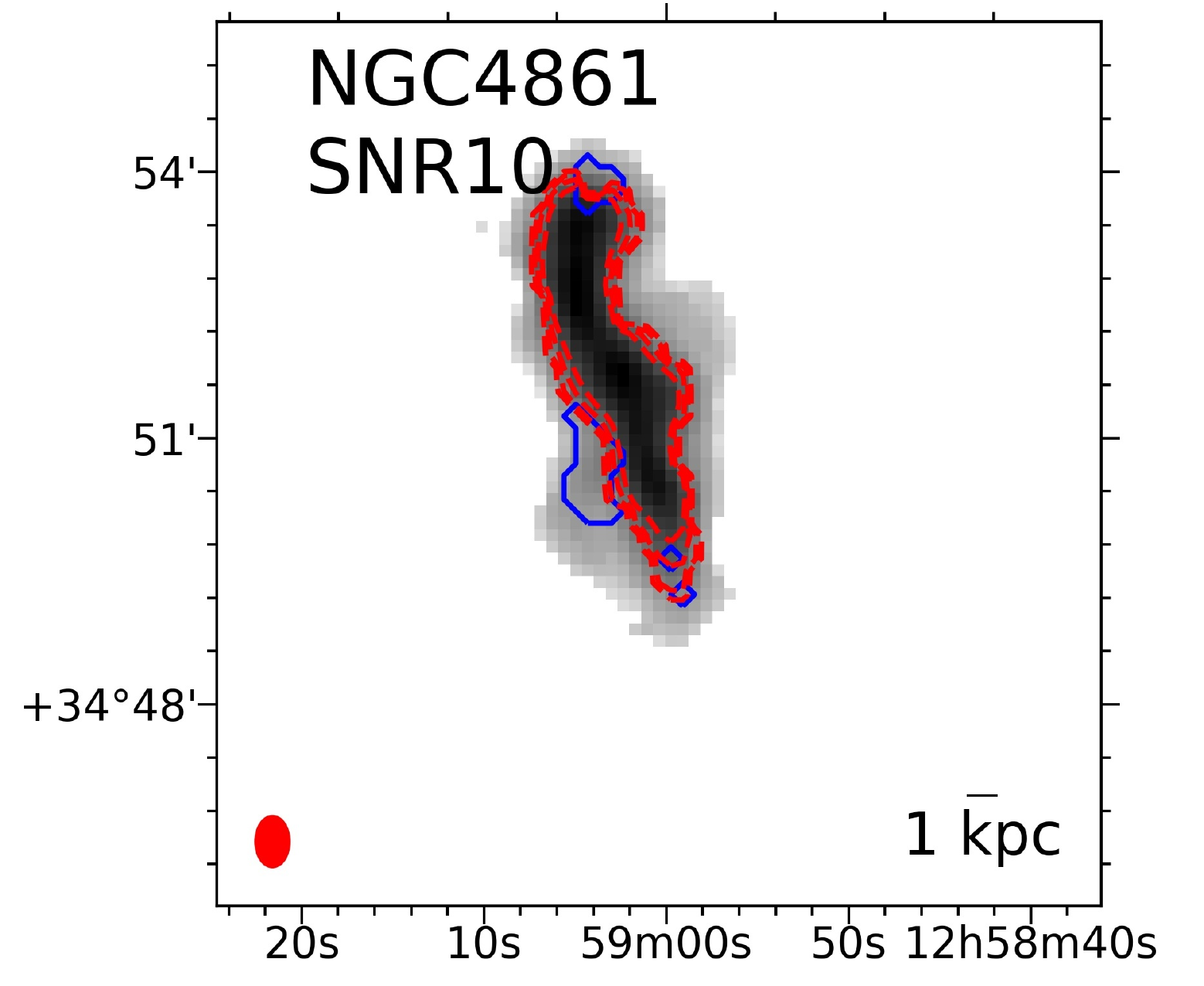} \\
    
    \end{tabular}
    \caption{The maps represent the CNM and WNM over the Moment zero map of individual galaxies at the different SNR cutoff values. The solid blue contours represent the cold neutral medium (WNM), and the dashed red contours represent the warm neutral medium (CNM).\emph{(cont.)} }
    \label{fig:snr_comp}
\end{figure*}

\begin{figure*}
\ContinuedFloat
    \centering
    \begin{tabular}{c c c c}
    \includegraphics[height=3.3cm]{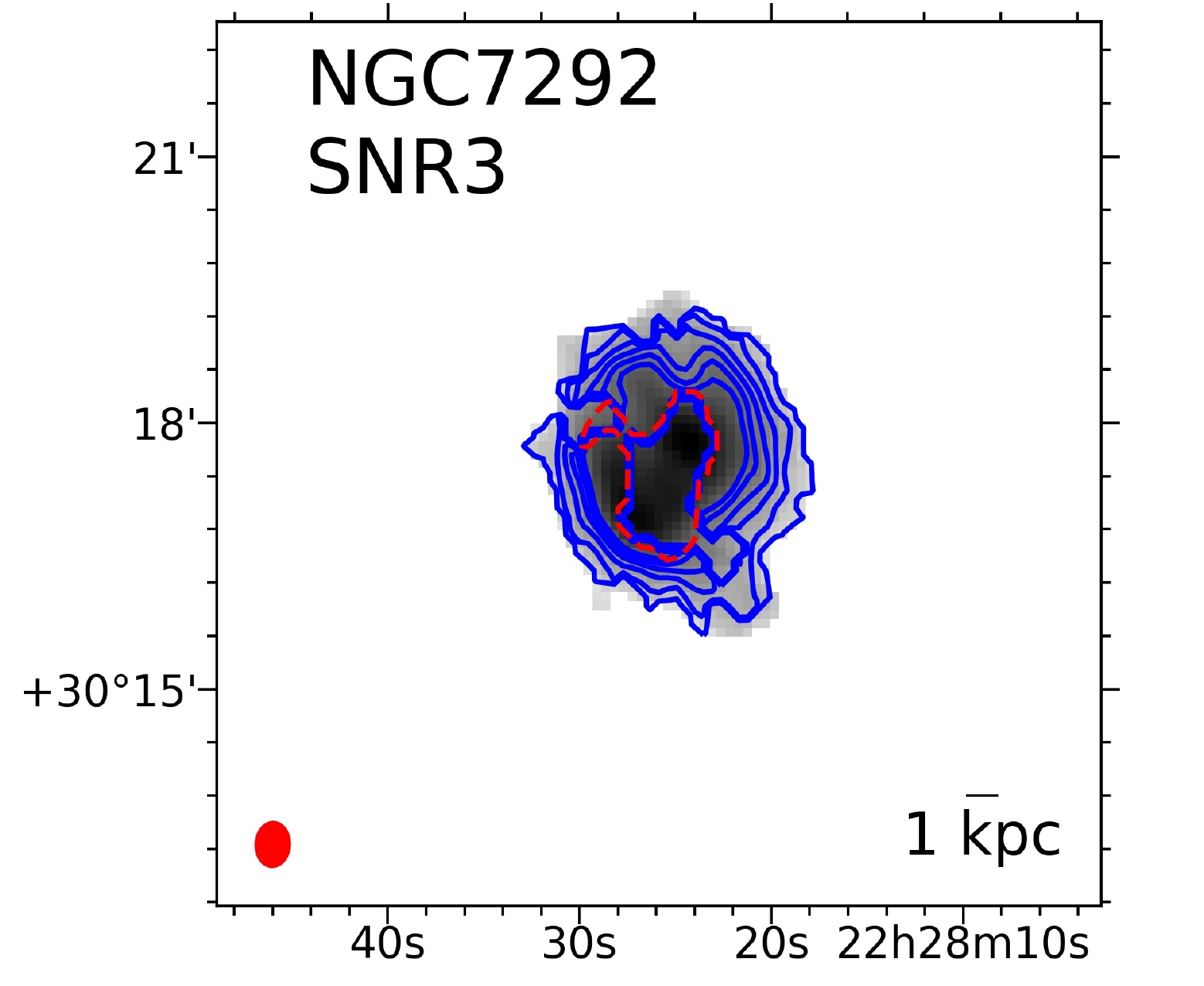} &
    \includegraphics[height=3.3cm]{figs/cnm_wnm_figs/NGC7292SNR5_cnm_wnm.pdf} &
    \includegraphics[height=3.3cm]{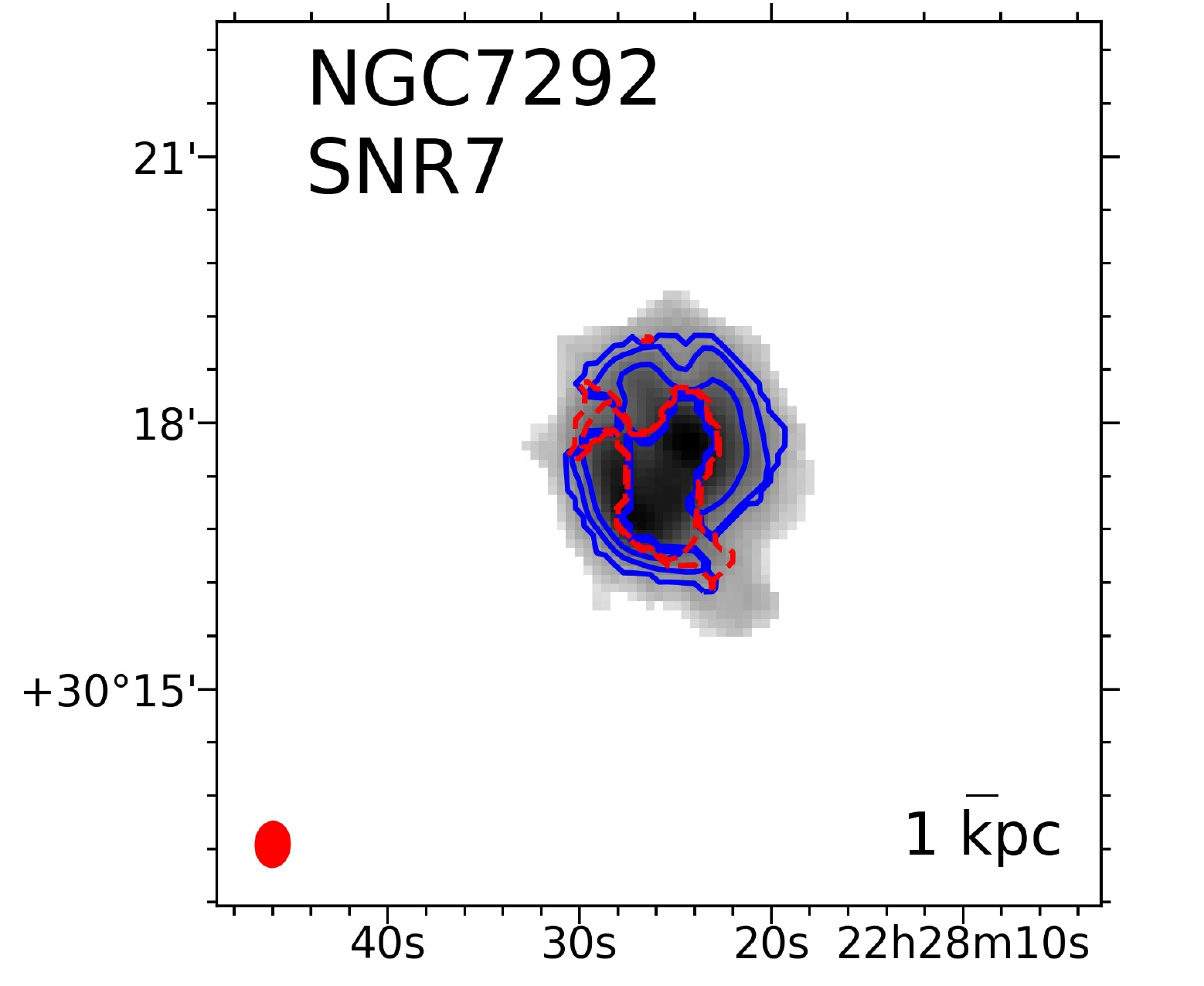} &
    \includegraphics[height=3.3cm]{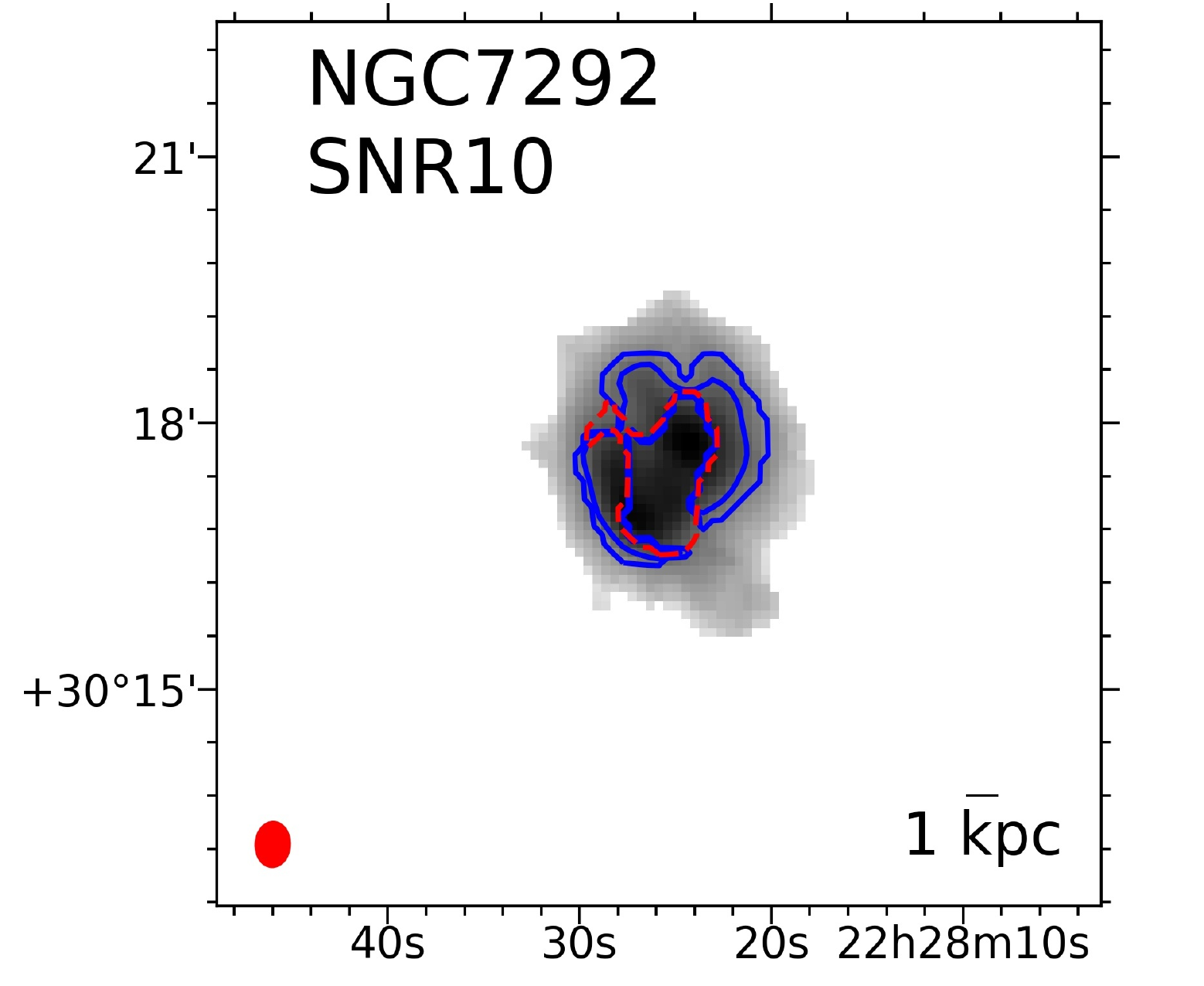} \\
   
   \includegraphics[height=3.3cm]{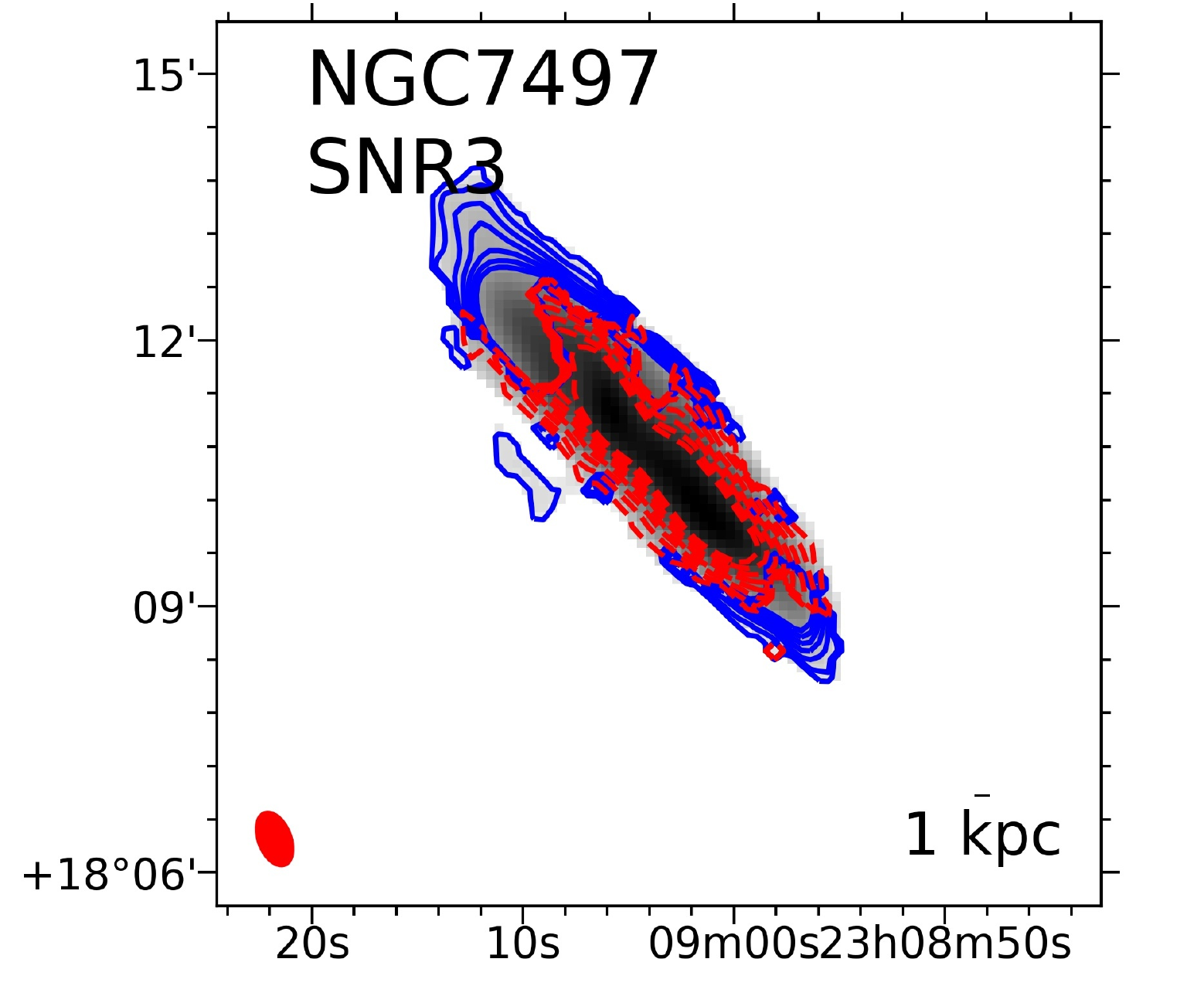} &
   \includegraphics[height=3.3cm]{figs/cnm_wnm_figs/NGC7497SNR5_cnm_wnm.pdf} &
   \includegraphics[height=3.3cm]{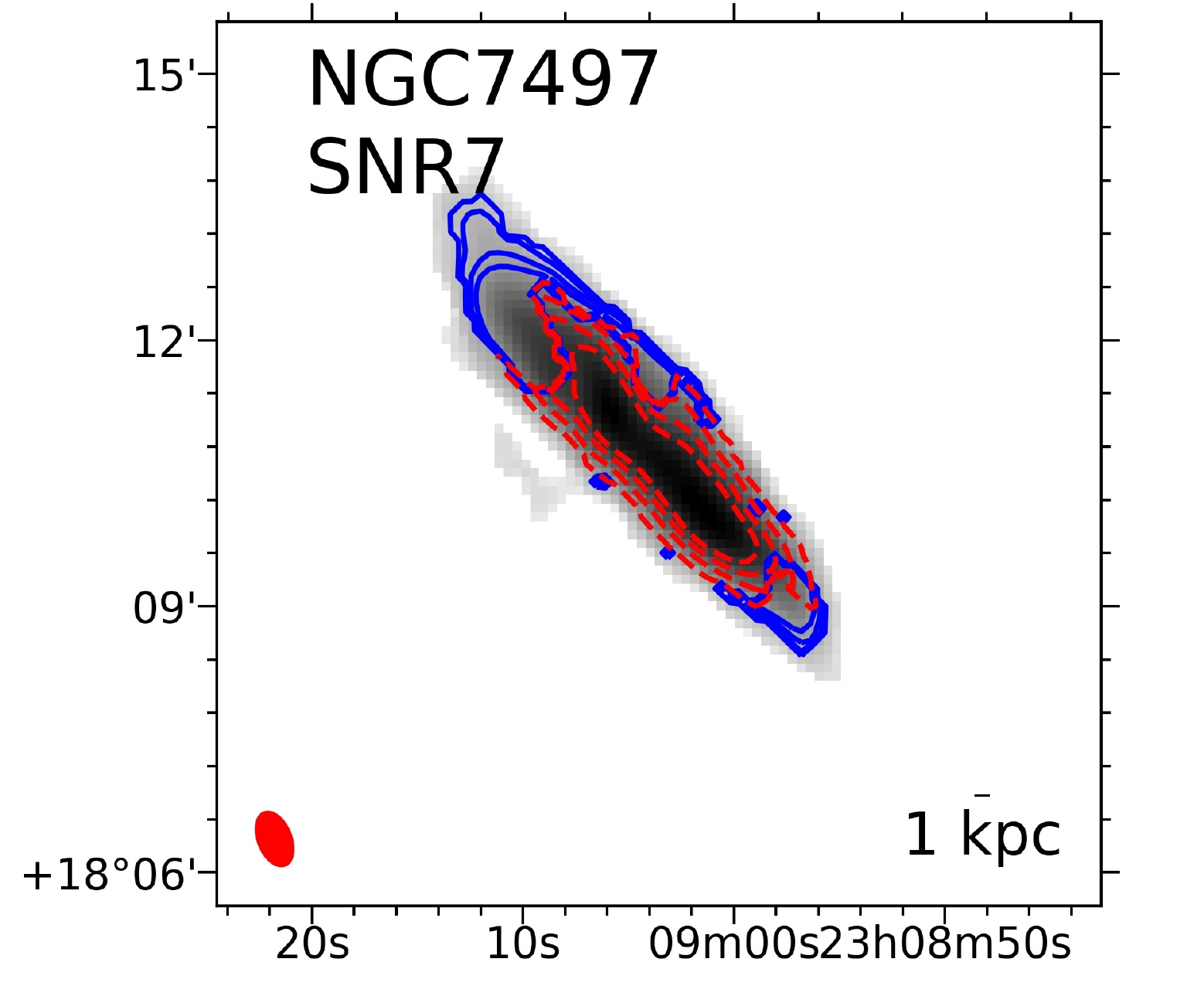} &
   \includegraphics[height=3.3cm]{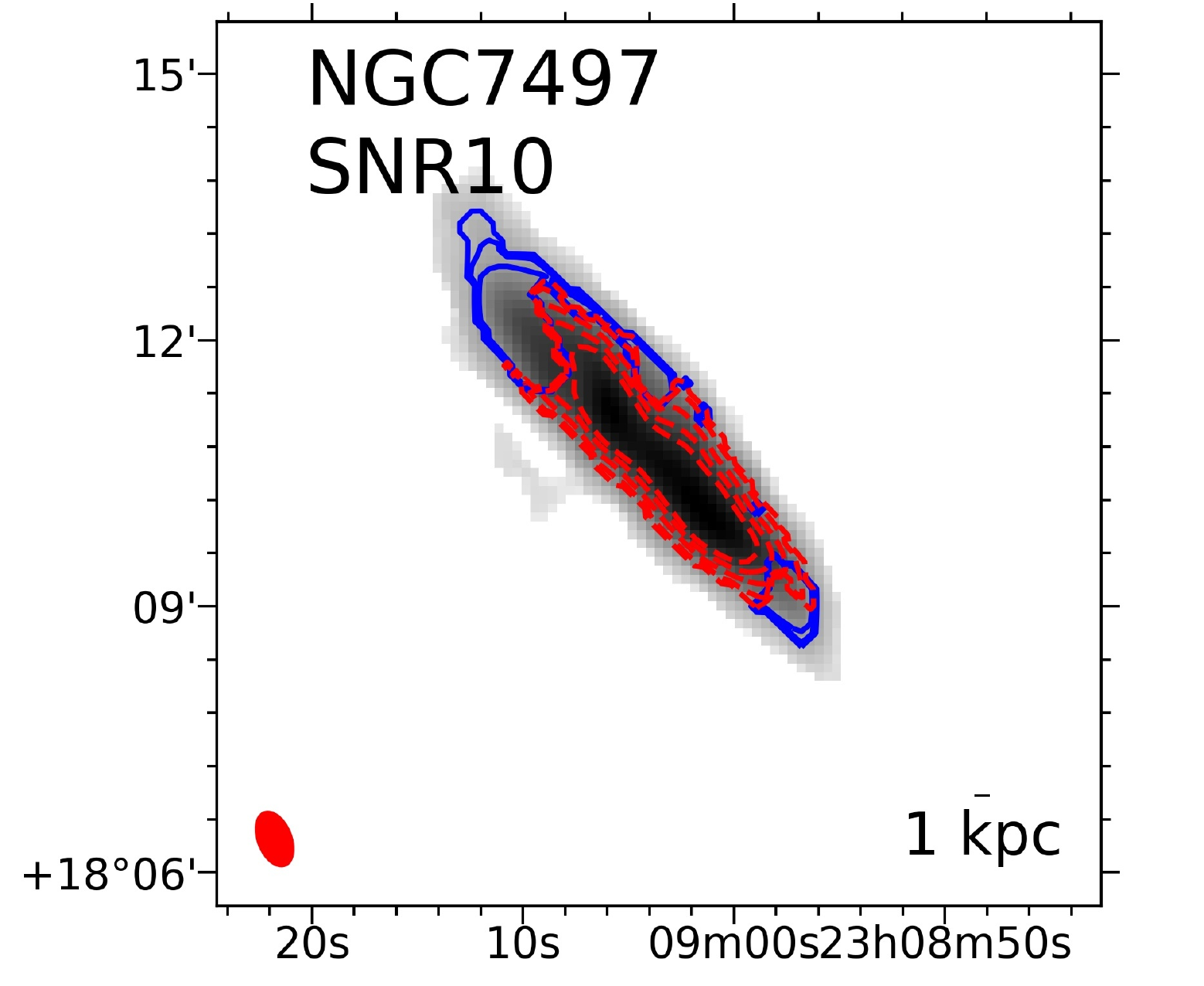} \\
   
   \includegraphics[height=3.3cm]{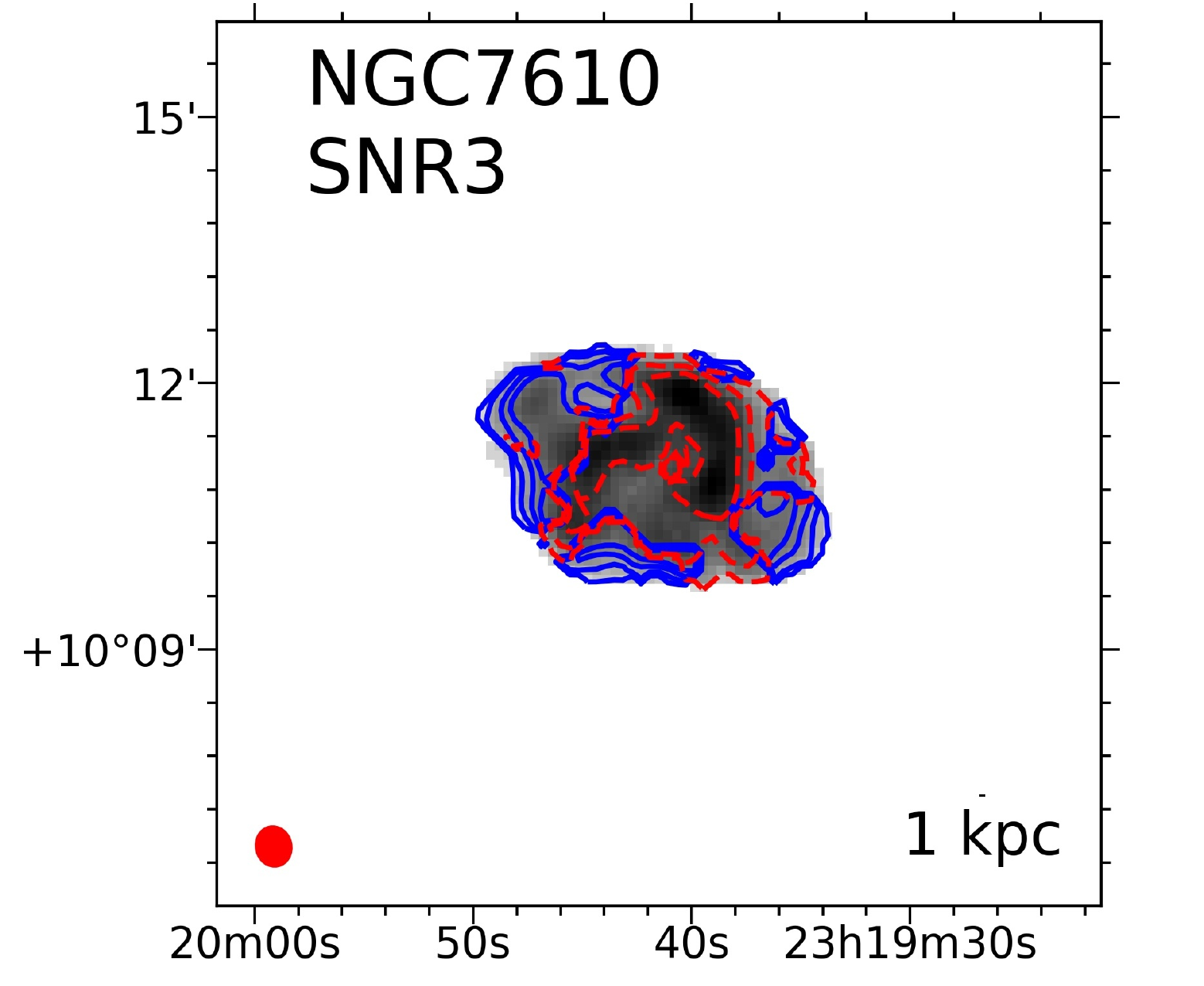} &
   \includegraphics[height=3.3cm]{figs/cnm_wnm_figs/NGC7610SNR5_cnm_wnm.pdf} &
   \includegraphics[height=3.3cm]{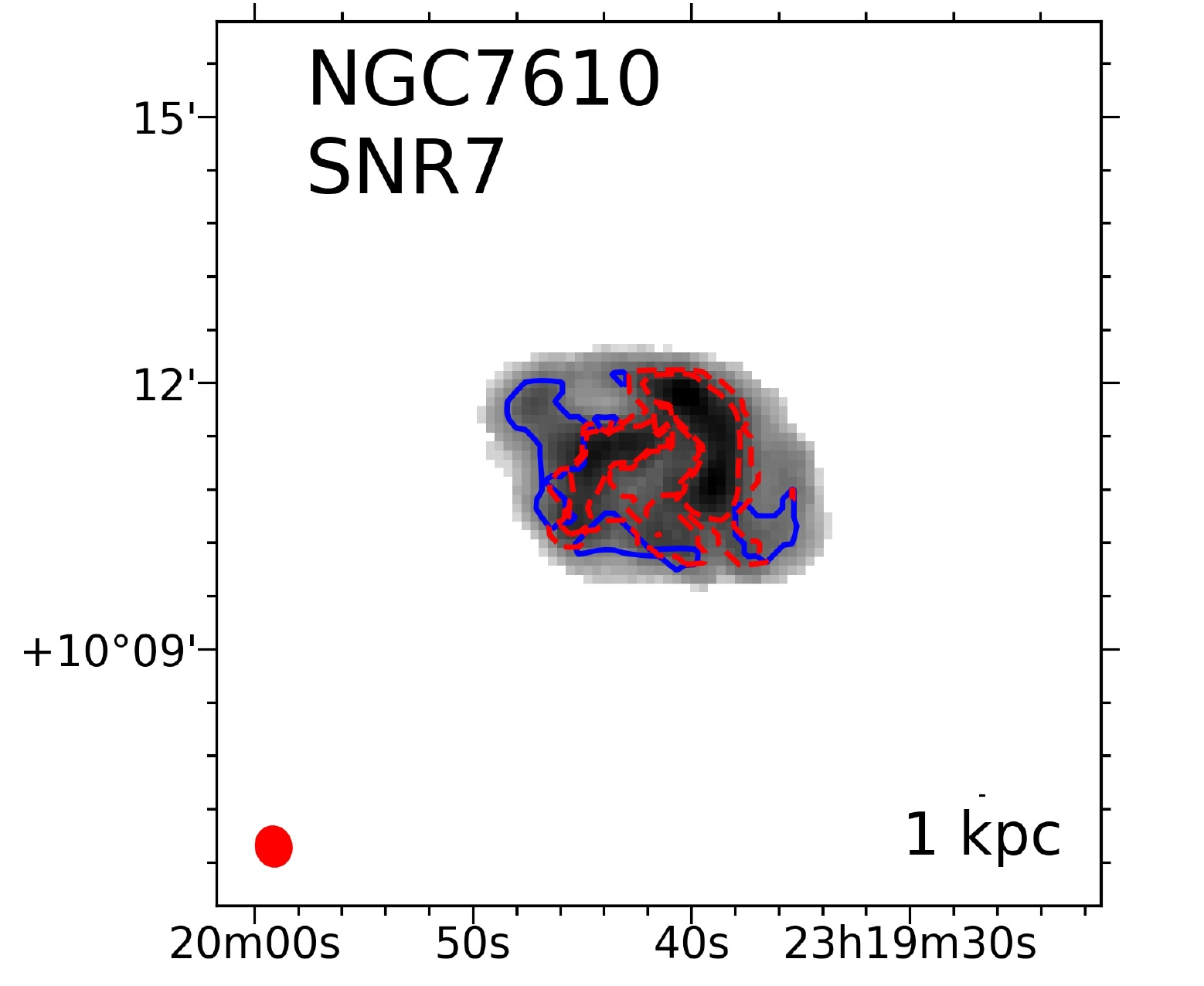} &
   \includegraphics[height=3.3cm]{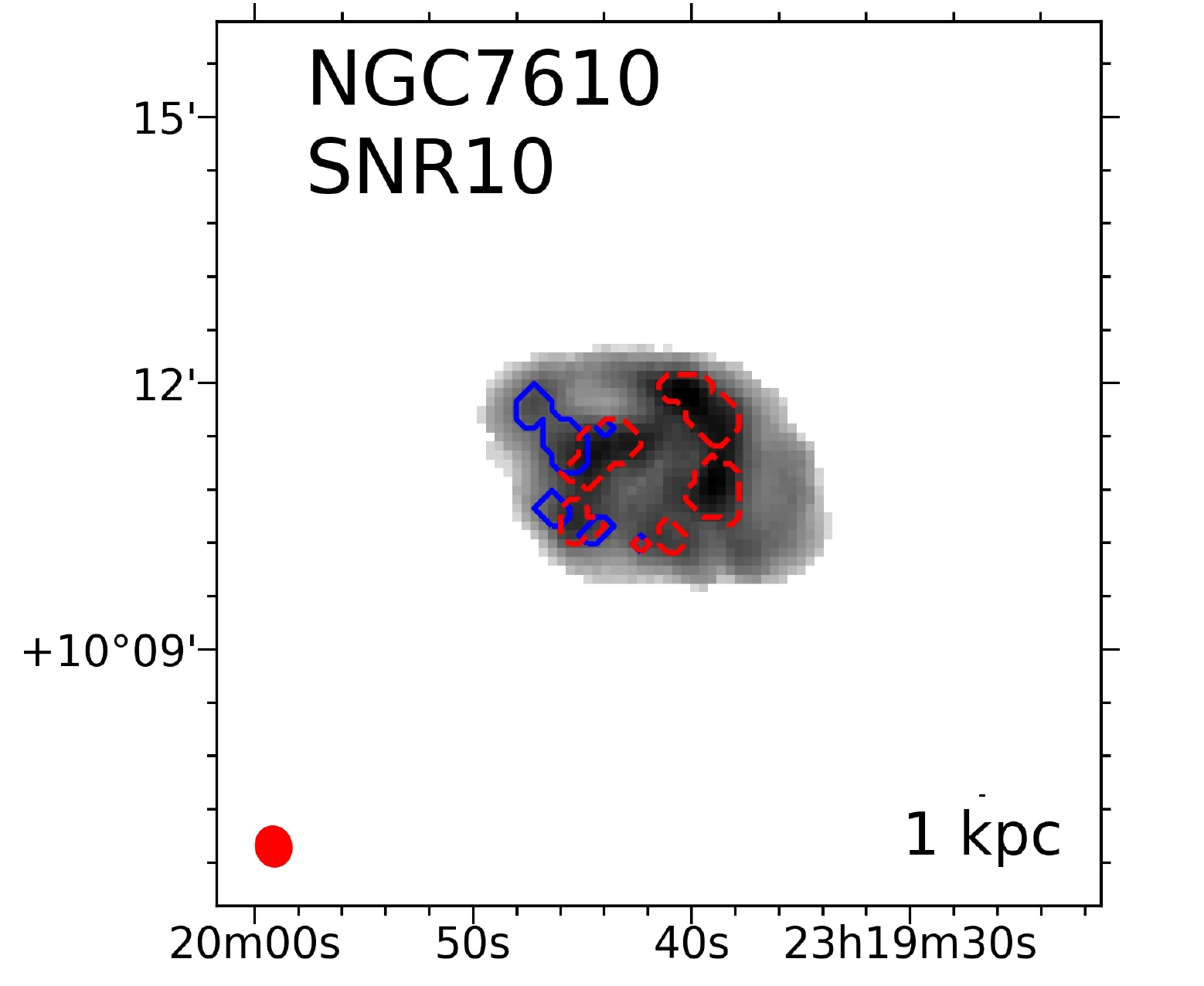} \\
   
   \includegraphics[height=3.3cm]{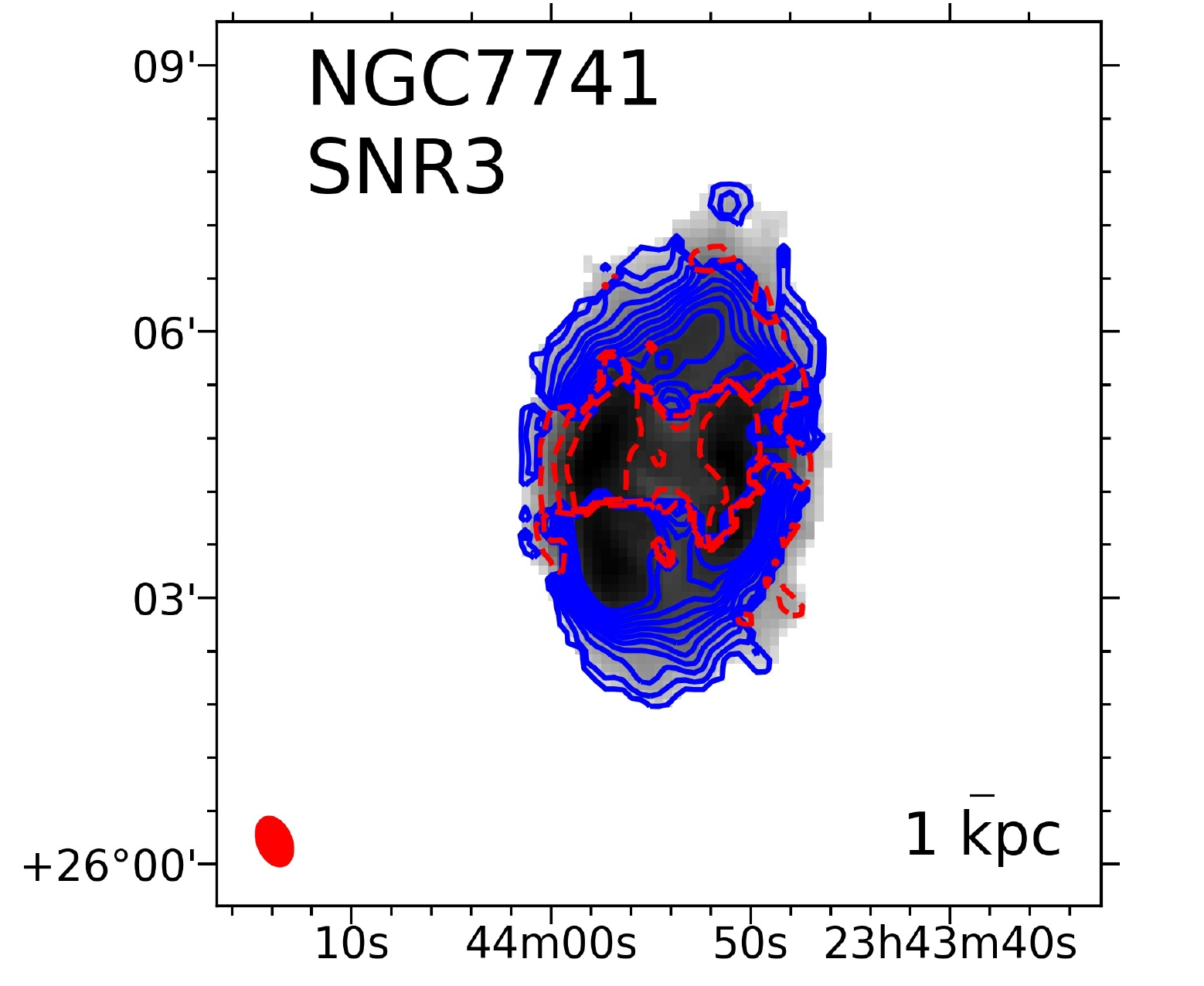} &
   \includegraphics[height=3.3cm]{figs/cnm_wnm_figs/NGC7741SNR5_cnm_wnm.pdf} &
   \includegraphics[height=3.3cm]{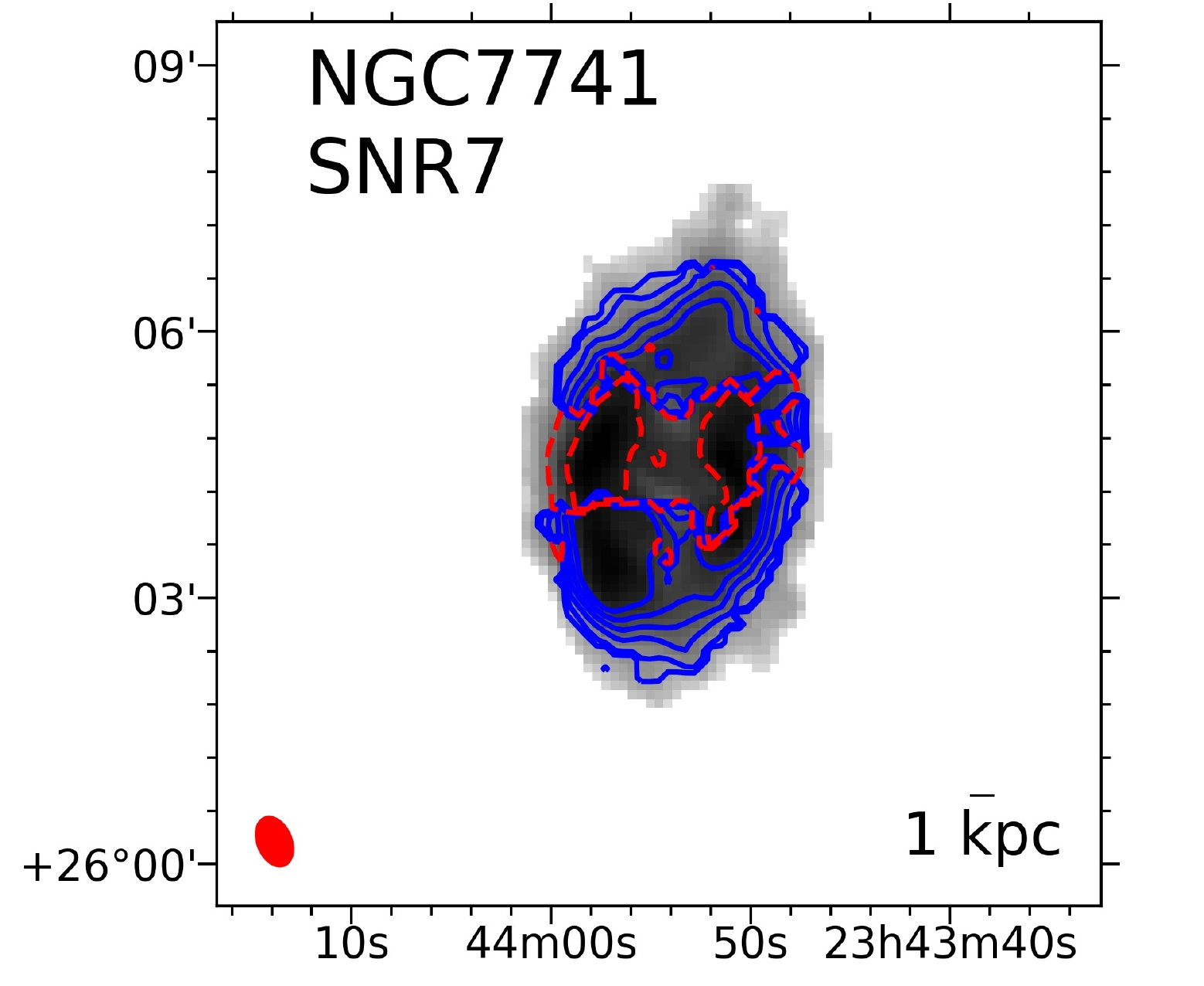} &
   \includegraphics[height=3.3cm]{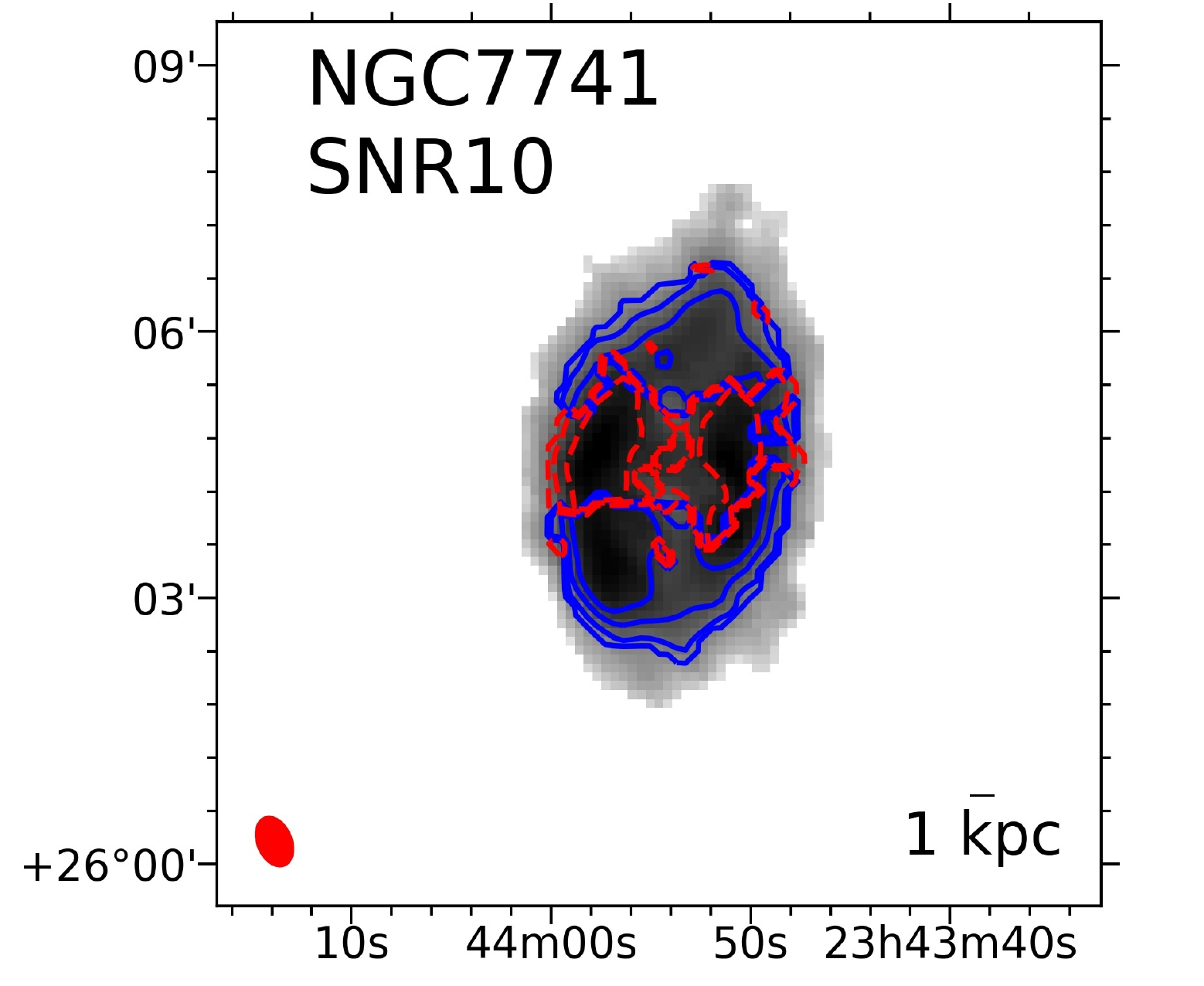} \\
   
   \includegraphics[height=3.3cm]{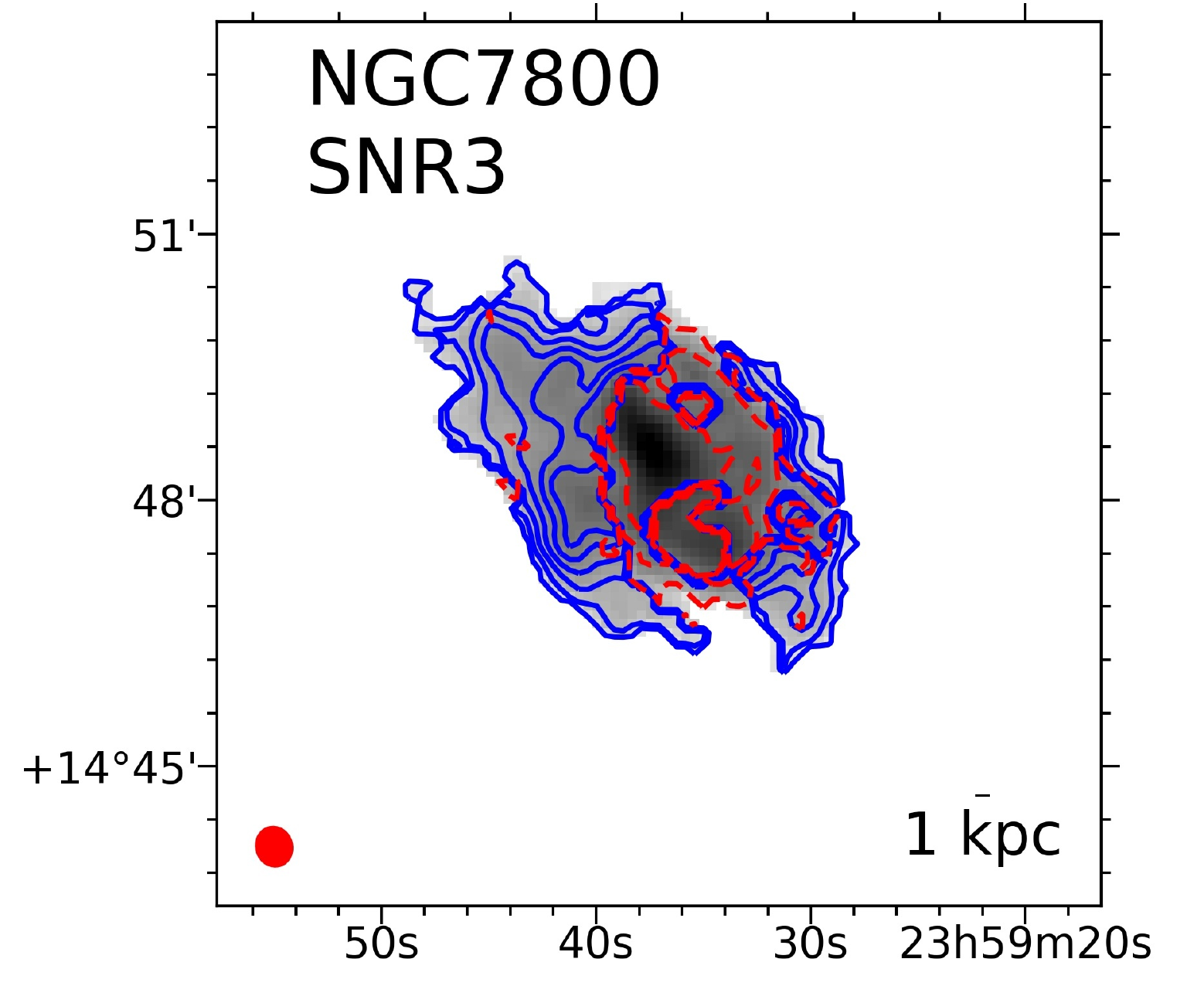} &
   \includegraphics[height=3.3cm]{figs/cnm_wnm_figs/NGC7800SNR5_cnm_wnm.pdf} &
   \includegraphics[height=3.3cm]{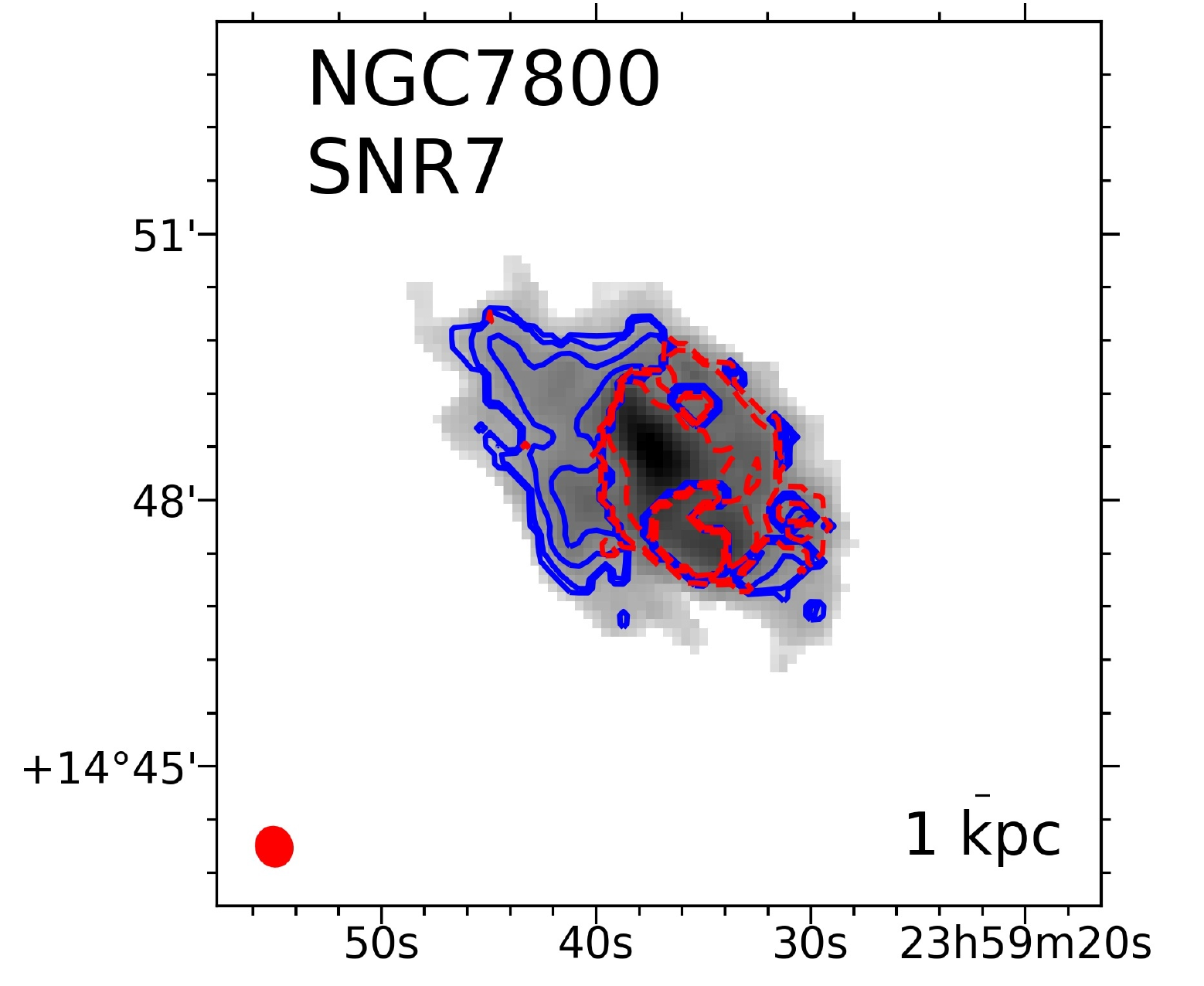} &
   \includegraphics[height=3.3cm]{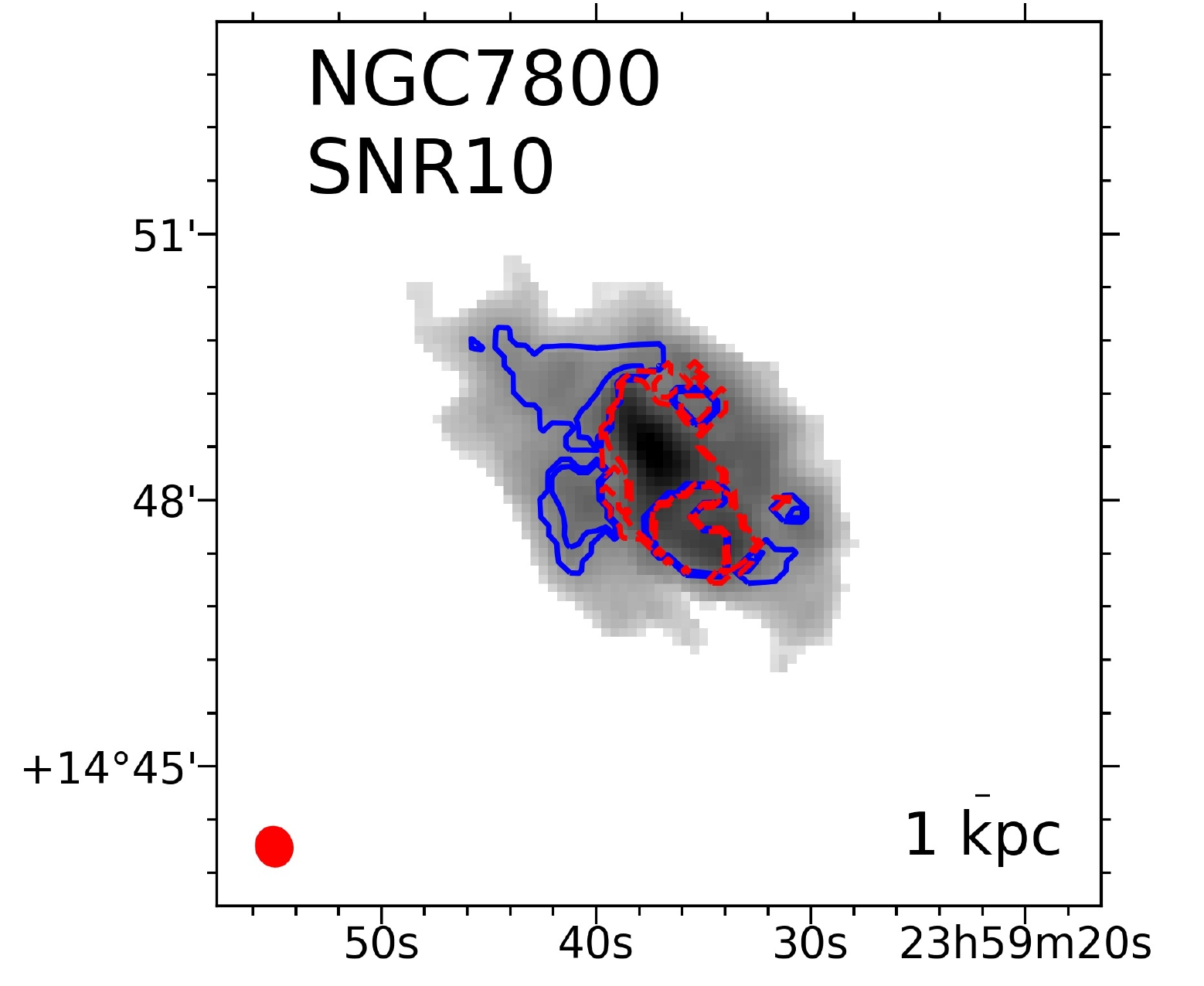} \\
   
    \end{tabular}
    
    \caption{The maps represent the CNM and WNM over the Moment zero map of individual galaxies at the different SNR cutoff values. The solid blue contours represent the cold neutral medium (WNM), and the dashed red contours represent the warm neutral medium (CNM).}
%    \label{fig:snr_comp}
\end{figure*}

% Don't change these lines
\bsp	% typesetting comment
\label{lastpage}

\end{document}